\documentclass[english,12pt]{article}
\usepackage{graphicx}
\usepackage{amsmath}
\usepackage{amssymb}
\usepackage{bm}
\usepackage{slashed}
\usepackage{dcolumn}
\usepackage{epsfig}
\usepackage{times}
\usepackage{cite}
\usepackage{afterpage}
\usepackage{dsfont}
\usepackage{color}
\usepackage[T1]{fontenc}
\usepackage[latin9]{inputenc}
\usepackage{babel}
\usepackage{lipsum}
\usepackage{ulem} % for the command \sout{}
\usepackage{url}
\usepackage[breaklinks]{hyperref}
\usepackage{hyperref}
\hypersetup{
    colorlinks,
    citecolor=black,
    filecolor=black,
    linkcolor=black,
    urlcolor=black
}

\allowdisplaybreaks[1]

\newcommand{\bea}{\begin{eqnarray}}
\newcommand{\eea}{\end{eqnarray}}
\newcommand{\beq}{\begin{equation}}
\newcommand{\eeq}{\end{equation}}

\newcommand{\eq}{\begin{eqnarray}}
\newcommand{\en}{\end{eqnarray}}
\newcommand*{\wave}[6]{\ensuremath{{#1}^{{#2}}\allowbreak{#3}^{#4}\allowbreak{#5}\allowbreak{\pi}{#6}}}
\newcommand{\dvec}{\bm{d}}
\newcommand{\Pvec}{\bm{P}}
\newcommand{\svec}{\bm{s}}

\newcommand{\Ecm}{E_{\rm cm}}
\newcommand{\nvec}{\bm{n}}
\newcommand{\zvec}{\bm{z}}
\newcommand{\qcma}{\bm{q}_{{\rm cm},a}}
\newcommand{\sveca}{\bm{s}_a}
\newcommand{\cm}{{\mathrm {cm}}}

\newcommand{\beqs}{\begin{eqnarray}}
\newcommand{\eeqs}{\end{eqnarray}}
\newcommand{\nn}{\nonumber}
\newcommand{\wvec}{\mathbf{w}}
\newcommand{\xvec}{\mathbf{x}}
\newcommand{\ket}[1]{\vert #1\rangle}

\newcommand{\me}[3]{\langle #1\vert\ #2\ \vert #3\rangle}
\newcommand{\mpi}{m_{\pi}}
\newcommand{\mK}{m_K}
\newcommand{\Fpi}{f_\pi}
\newcommand{\Order}{\mathcal{O}}
\newcommand{\barr}[1]{\bar{\bar{#1}}}
\newcommand{\mpid}{{m_\pi^2}}
\newcommand{\mkd}{m_K^2}
\renewcommand{\Re}{\text{Re}\,}

\def\lambdabar{\lambda\kern-1ex\raise0.65ex\hbox{-}}

\def\ket#1{\left| #1\right\rangle}

\def\Kp    {\ensuremath{K^+}\xspace}
\def\Km    {\ensuremath{K^-}\xspace}

\def\KpKm  {\ensuremath{\Kp \kern -0.16em \Km}\xspace}

\DeclareMathOperator{\im}{Im}

\newcommand{\diff}{\ensuremath{\mathrm{d}}}
\usepackage[margin=1in]{geometry}
\begin{document}
\hfill{\bf Date:} {\today}
\begin{center}
{\huge\bf Workshop on Pion-Kaon Interactions \\
	(PKI2018) \\
	Mini-Proceedings} 
\end{center}
\hspace{0.1in}
\begin{center}
{\large 14th - 15th February, 2018}
{\large Thomas Jefferson National Accelerator Facility, Newport 
	News, VA, U.S.A.}
\end{center}
\hspace{0.1in}
\begin{center}
M.~Amaryan,
M.~Baalouch,
G.~Colangelo,
J.~R.~de~Elvira, 
D.~Epifanov, 
A.~Filippi, 
B.~Grube, 
V.~Ivanov, 
B.~Kubis,
P.~M.~Lo,
M.~Mai,
V.~Mathieu,
S.~Maurizio,
C.~Morningstar,
B.~Moussallam,
F.~Niecknig,
B.~Pal,
A.~Palano,
J.~R. Pelaez,
A.~Pilloni,
A.~Rodas, 
A.~Rusetsky, 
A.~Szczepaniak, and
J.~Stevens
\end{center}
\hspace{0.1in}
\begin{center}
\textbf{Editors}: 
M.~Amaryan, 
Ulf-G.~Mei{\ss}ner, 
C.~Meyer, 
J.~Ritman, and 
I.~Strakovsky
\end{center}

\noindent

%%%%%%%%%%%%%%%%%%%%%%%%%%%%%%%%%%%%%%%%%%%%%%%%%%%%%%%%%%%%%%%%%%%%%%%%%
\begin{center}{\large\bf Abstract}\end{center}

This volume is a short summary of talks given at the PKI2018 Workshop organized to 
discuss current status and future prospects of $\pi-K$ interactions. The precise 
data on $\pi_K$ interaction will have a strong impact on strange meson spectroscopy 
and form factors that are important ingredients in the Dalitz plot analysis of a 
decays of heavy mesons as well as precision measurement of $V_{us}$ matrix element and 
therefore on a test of unitarity in the first raw of the CKM matrix. The workshop has 
combined the efforts of experimentalists, Lattice QCD, and phenomenology communities. 
Experimental data relevant to the topic of the workshop were presented from the broad 
range of different collaborations like CLAS, GlueX, COMPASS, BaBar, BELLE, BESIII, 
VEPP-2000, and LHC$_b$. One of the main goals of this workshop was to outline a need 
for a new high intensity and high precision secondary $K_L$ beam facility at JLab 
produced with the 12~GeV electron beam of CEBAF accelerator. 

This workshop is a successor of the workshops
\textit{Physics with Neutral Kaon Beam at JLab}~\cite{Albrow:2016ibsa} held at JLab, February, 2016;
\textit{Excited Hyperons in QCD Thermodynamics at Freeze-Out}~\cite{Alba:2017cbra} held at JLab, November, 2016;
\textit{New Opportunities with High-Intensity Photon Sources}~\cite{Horn:2017vkza} held at CUA, February, 2017.
Further details about the PKI2018 Workshop can be found on the web page of 
the conference: http://www.jlab.org/conferences/pki2018/ .

%%%%%%%%%%%%%%%%%%%%%%%%%%%%%%%%%%%%%%%%%%%%%%%%%%%%%%%%%%%%%%%%%%%%%%%%%
\newpage
\tableofcontents

%%%%%%%%%%%%%%%%%%%%%%%%%%%%%%%%%%%%%%%%%%%%%%%%%%%%%%%%%%%%%%%%%%%%%%%%%
\newpage
\pagenumbering{arabic}
\setcounter{page}{1}
\section{Preface}
\halign{#\hfil&\quad#\hfil\cr
}

%%%-----------------------------------
\begin{enumerate}
\item 
%\textbf{Preface}
From February 14-15, 2018, the Thomas Jefferson Laboratory in Newport News, 
Virginia hosted the PKI2018, an international workshop to explore the 
physics potential to investigate $\pi$-K interactions. This was the fourth 
of a series of workshops held to establish a neutral kaon beam facility at 
JLab Hall~D with a neutral kaon flux which will be three orders of 
magnitude higher than was available at SLAC. This facility will enable 
scattering experiments of $K_L$ off both proton and neutron (for the first 
time) targets in order to measure differential cross section distributions 
with the GlueX detector.

The combination of data from this facility with the self-analyzing power of 
strange hyperons will enable precise partial-wave analyses (PWA) in order 
to determine dozens of predicted $\Lambda^\ast$, $\Sigma^\ast$, $\Xi^\ast$, 
and $\Omega^\ast$ resonances up to 2.5~GeV. Furthermore, the KLF will enable 
strange meson spectroscopy by studies of the $\pi$-K interaction to locate 
pole positions in the I = 1/2 and 3/2 channels. Detailed study of $\pi$-K 
system with PWA will allow to observe and measure quantum numbers of missing 
kaon states, which in turn will also impact Dalitz plot analyses of heavy 
meson decays, as well as tau-lepton decay the with $\pi$-K in the final 
state.

%The organizing committee was chaired by Moskov Amaryan (ODU) along with 
%Ulf-G. Mei{\ss}ner (U. Bonn/FZ J\"ulich), Curtis Meyer (CMU), James 
%Ritman (Ruhr-Univ.-Bochum/FZ J\"ulich) and Igor Strakovsky (GWU). The 
%workshop attracted 48 registered participants. Most of the young 
%scientists were provided with financial support to defray costs for 
%lodging and the conference registration fee. The program consisted of 
%26 invited plenary talks over 8 sessions.

The program of the workshop had special emphasis on topics connected to 
the proposed KLF experiments. A detailed description of the workshop, 
including the scientific program, can be found on the workshop web page, 
https://www.jlab.org/conferences/pki2018/. 

The talks presented at this workshop were grouped into the following 
categories: 
\begin{enumerate}
\item The KL Facility at JLab, 
\item Lattice QCD approaches to $\pi$-K interactions,
\item Results from Chiral Effective Theories
\item Results from Dispersion Relations
\item $\pi$-K formfactor and heavy meson and tau decay
\item Hadron Spectroscopy at GlueX, CLAS, CLAS12, BaBar, and COMPASS
\end{enumerate}

%%%-----------------------------------
\item \textbf{Acknowledgments}

The workshop would not have been possible without dedicated work of many
people. First, we would like to thank the service group and the staff
of JLab for all their efforts. We would like to thank JLab management, 
especially Robert~McKeown for their help and encouragement to organize 
this workshop. Financial support was provided by the JLab, J\"ulich 
Forschungszentrum, The George Washington and Old Dominion universities.

\vskip 1cm
\leftline {Newport News, March 2018.}

%\rightline {M.~Amaryan}
%\rightline {Ulf-G.~Mei{\ss}ner}
%\rightline {C.~Meyer}
%\rightline {J.~Ritman}
%\rightline {I.~Strakovsky}
\end{enumerate}

%%%-----------------------------------

%%%%%%%%%%%%%%%%%%%%%%%%%%%%%%%%%%%%%%%%%%%%%%%%%%%%%%%%%%%%%%%%%%%%%%%%%
\newpage
\section{Program}
%%%%%%%%%%%%%%%%%%%%%%%%%%%%%%%%%%%%%%%%%%%%%%%%%%%%%%%%%%%%%%%%%%%%%%%%%

\textbf{Wednesday, February 14, 2018}

 8:15am - 8:45am:	 Registration and coffee

\underline{Session 1}: Chair: Rolf Ent / Secretary: Stuart Fegan

 8:45am - 9:00am:	\textit{Welcome and Introductory Remarks} -- Jianwei Qiu (JLab)

 9:00am - 9:25am:	\textit{KL Facility at JLab}	          -- Moskov Amaryan (ODU)

 9:25am - 9:50am:	\textit{Kaon-pion scattering from lattice QCD}	-- Colin Morningstar (CMU)

 9:50am -10:15am:	\textit{Study of $k-\pi$ interaction with KLF}	-- Marouen Baalouch (ODU)

10:15am -10:45am:	Coffee break

\underline{Session 2}: Chair: Eugene Chudakov  / Secretary: Chan Kim

10:45am -11:15am:	\textit{Dalitz plot analysis of three-body charmonium decays at BaBar} -- Antimo Palano  (INFN/Bari U.)

11:15am -11:45am:	\textit{Kaon and light-meson resonances at COMPASS} -- Boris Grube  (TUM)

11:45am -12:15pm:	\textit{Recent Belle results related to pion-kaon interactions} -- Bilas Pal (Cincinnati U.)

12:15pm - 2:00pm:	Conference Photo \& Lunch break - on your own

\underline{Session 3}: Chair: Curtis Meyer / Secretary: Torry Roak

 2:00pm - 2:25pm:	\textit{Study of $\tau\to K\pi\nu$ decay at the B factories} -- Denis Epifanov (BINP, NSU) 

 2:25pm - 2:50pm:	\textit{From $\pi-K$ amplitudes to $\pi-K$ form factors and back} -- Bachir Moussallam (Paris-Sud U.)

 2:50pm - 3:15pm:	\textit{Three-body interactions in isobar formalism} Maxim Mai (GW)

 3:15pm - 3:40pm:	\textit{Study of the processes $e^+e^-\to K\bar K n\pi$ with the CMD-3 detector at VEPP-2000 collider} -- Vyacheslav Ivanov (BINP)

 3:40pm - 4:10pm:	Coffee break

\underline{Session 4}: Chair: Charles Hyde / Secretary: Tyler Viducic

 4:10pm - 4:35pm:	\textit{The GlueX Meson Program}	-- Justin Stevens  (W\&M)

 4:35pm - 5:00pm:	\textit{Strange meson spectroscopy at CLAS and CLAS12}	-- Alessandra Filippi (INFN Torino) 

 5:00pm - 5:25pm:	\textit{Non-leptonic charmless three body decays at LHCb} -- Rafael Silva Coutinho (Zuerich U.)

 5:25pm- 5:50pm:	\textit{Dispersive determination of the $\pi-K$ scattering lengths} -- Jacobo Ruiz de Elvira  (Bern U.)

 5:50pm:	Adjourn	 
 	 	 
 6:10pm:	Networking Reception - CEBAF Center Lobby

\textbf{Thursday, February 15, 2018}

 8:15am - 8:45am:	Coffee

\underline{Session 5}: Chair: David Richards / Secretary: Will Phelps

 8:45am - 9:15am:	\textit{Meson-meson scattering from lattice QCD} -- Jo Dudek (W\&M)

 9:15am - 9:45am:	\textit{Dispersive analysis of pion-kaon scattering} -- Jose R. Pelaez  (U. Complutense de Madrid)

 9:45am -10:15am: \textit{Analyticity Constraints for Exotic Mesons} -- Vincent Mathieu (JLab) 

10:15am -10:55am:	Coffee break

\underline{Session 6}: Chair: Jacobo Ruiz / Secretary: Wenliang Li	 

10:55am -11:25am: \textit{Pion-kaon scattering in the final-state interactions of heavy-meson decays} -- Bastian Kubis (Bonn U.)

11:25am -11:55am:	\textit{Using $\pi K$ to understand heavy meson decays}	-- Alessandro Pilloni (JLab)

11:55am -12:25pm:	\textit{Three particle dynamics on the lattice}	-- Akaki Rusetsky (Bonn U.)   

12:25pm - 2:00pm:	Lunch break - on your own

\underline{Session 7}:  Chair: James Ritman / Secretary: Amy Schertz	 

 2:00pm - 2:30pm:	\textit{S-matrix approach to the thermodynamics of hadrons} -- Pok Man Lo (Wroclaw U.)  

 2:30pm - 3:00pm:	\textit{Measurement of hadronic cross sections with the BaBar detector}	-- Alessandra Filippi (INFN Torino)

 3:00pm - 3:30pm:	\textit{A determination of the pion-kion scattering length from 2+1 flavor lattice QCD} -- Daniel Mohler (Helmholtz-Inst. Mainz)

 3:30pm - 4:10pm:	Coffee break

\underline{Session 8}: Chair: Bachir Moussallam / Secretary: Nilanga Wickramaarachch

 4:10pm - 4:35pm:	\textit{Strangeness-changing scalar form factor from scattering data and CHPT} -- Michael D\"oring (GW/JLab)

 4:35pm - 4:50pm:	\textit{Closing Remarks}	-- Bachir Moussallam (Paris-Sud U.)

 4:50pm:	Closing

%%%%%%%%%%%%%%%%%%%%%%%%%%%%%%%%%%%%%%%%%%%%%%%%%%%%%%%%%%%%%%%%%%%%%%%%%
\newpage
\section{Summaries of Talks}
%%%%%%%%%%%%%%%%%%%%%%%%%%%%%%%%%%%%%%%%%%%%%%%%%%%%%%%%%%%%%%%%%%%%%%%%%

%%%%%%%%%%%%%%%%%%%%%%%%%%%%%%%%%%%%%%%%%%%%%%%%%%%%%%%%%%%%%%%%%%%%%%%%%
\subsection{Secondary $K_L^0$  Beam Facility at JLab for Strange Hadron Spectroscopy}
\addtocontents{toc}{\hspace{2cm}{\sl Moskov~Amaryan}\par}
\setcounter{figure}{0}
\setcounter{table}{0}
\setcounter{equation}{0}
\halign{#\hfil&\quad#\hfil\cr
\large{Moskov Amaryan}\cr
\textit{Department of Physics}\cr
\textit{Old Dominion University}\cr
\textit{Norfolk, VA 23529, U.S.A.}\cr}

%%%-----------------------------------
\begin{abstract}
In this talk, I discuss the photoproduction of a secondary $K_L^0$ beam  
at JLab to be used with the GlueX detector in Hall-D for a strange hadron 
spectroscopy.
\end{abstract}
\vskip 10mm

\rightline{ \it It is comforting to reflect that the disproportion}
\rightline  {\it of things in the world seems to be only arithmetical.}
\vskip 3mm
\rightline{Franz Kafka}

%%%-----------------------------------
\begin{enumerate}
\item \textbf{Introduction}

Current status of our knowledge about the strange hyperons and mesons is 
far from being satisfactory. One of the main reasons for this is that 
first of all dozens of strange hadron states predicted by Constituent 
Quark Model (CQM) and more recently by Lattice QCD calculations are still 
not observed. The detailed discussions about the missing hyperons {\it 
per se} and in particular their connection to thermodynamics of the Early 
Universe at freeze-out were performed respectively in a three preceding 
workshops~\cite{Albrow:2016ibso,Alba:2017cbro,Horn:2017vkzo}. Topics
discussed in these workshops were significant part of the  proposal 
submitted to the JLab PAC45~\cite{Amaryan:2017ldwo}.

This is a fourth workshop in this series devoted to the physics program 
related to the strange meson states and $\pi-K$ interactions. As it has 
been summarized in all four workshops it is not only the disproportion
between the number of currently observed and CQM and LQCD predicted 
states that makes experimental studies of the strange quark sector to 
be of high priority. These experiments are crucially important to 
understand QCD at perturbative domain and the dynamics of strange hadron 
production using hadronic beam with the strange quark in the projectile.

Many aspects of $\pi-K$ interactions and their impact on different 
important problems in particle physics have been discussed in this 
workshop.

Below we describe conceptually the main  steps needed to produce 
intensive $K_L$ beam. We discuss momentum resolution of the beam using 
time-of-flight technique, as well as the ratio of $K_L$ over neutrons as 
a function of their momenta simulated based on well known production 
processes.  In some examples the quality of expected experimental data 
obtained by using GlueX setup in Hall-D will be demonstrated using 
results of Monte Carlo studies.

%%%-----------------------------------
\item \textbf{The $K_L^0$ Beam in Hall~D}

In this chapter we describe photo-production of secondary $K_L^0$ beam in 
Hall~D.  There are few points that need to be decided. To produce intensive 
photon beam one needs to increase radiation length of the radiator up to 
10$\%$ radiation length.  In a first scenario, $E_e=12$~GeV electrons 
produced at CEBAF will scatter in a radiator in the tagger vault,  
generating  intensive beam of bremsstrahlung photons. This may will then 
require removal of all tagger counters and electronics and very careful 
design of radiation shielding, which is very hard to optimize and design.

In a second scenario. one may use Compact Photon Source design (for more 
details see a talk by Degtiarenko in Ref.~\cite{Albrow:2016ibso}) 
installed after the tagger magnet, which will produce bremsstrahlung 
photons and dump electron beam inside the source shielding the radiation 
inside.  At the second stage, bremsstrahlung photons interact with Be 
target placed on a distance 16~m upstream of liquid hydrogen ($LH_2$) 
target  of GlueX experiment in Hall D producing $K_L^0$ beam. To stop 
photons a 30 radiation length lead absorber will be installed in the 
beamline followed by a sweeping magnet to deflect the flow of charged 
particles. The flux of $K_L$ on ($LH_2$) target of GlueX experiment in
Hall~D will be measured with pair spectrometer upstream the target. For  
details of this part of the beamline see a talk by Larin in 
Ref.~\cite{Albrow:2016ibso}. Momenta of $K_L$ particles will be measured 
using the time-of-flight between RF signal of CEBAF and start counters 
surrounding $LH_2$ target. Schematic view of beamline is presented in 
Fig.~\ref{fig:setup}. The bremsstrahlung photons, created by electrons
at a distance about 75~m upstream, hit the Be target and produce $K_L^0$ 
mesons along with neutrons and charged particles. The lead absorber of 
$\sim$30 radiation length is installed to absorb photons  exiting Be 
target. The sweeping magnet deflects any remaining charged particles 
(leptons or hadrons) remaining after the absorber. The pair spectrometer 
will monitor the flux of $K_L^0$ through the decay rate of kaons at 
given distance about 10~m from Be target. The beam flux could also be 
monitored by installing nuclear foil in front of pair spectrometer
to measure a rate of $K^0_S$ due to regeneration process $K_L +p \to 
K_S +p$ as it was done at NINA  (for a details see a talk my Albrow at 
this workshop).
%%%-----------------------------------
\begin{figure}[htb!]
\centering
{
	\includegraphics[width=4.5in]{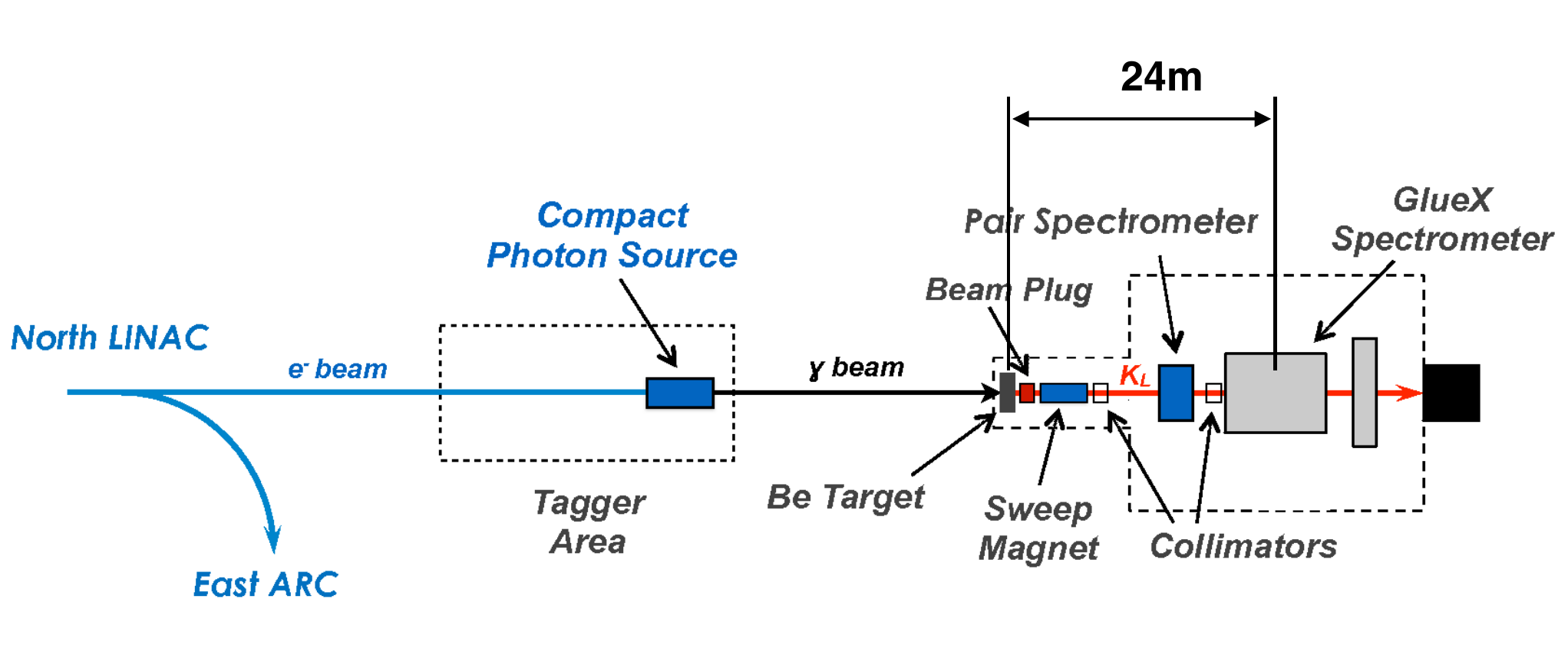}
}

	\caption{ Schematic view of Hall~D beamline. See a text for 
	explanation. \label{fig:setup}}
\end{figure}
%%%-----------------------------------

Here we outline experimental conditions and simulated flux of $K_L^0$ 
based on GEANT4 and known cross sections of underlying 
subprocesses~\cite{Seraydaryan:2013ijao,Titov:2003bko,Mcclellan:1971tko}.

The expected flux of $K_L^0$ mesons integrated in the range of momenta 
$P=0.3-10 GeV/c$ will be on the order of  $\sim 10^4$~$K_L^0/s$ on the 
physics target of the GlueX setup under the following conditions:
\begin{itemize}
	\item A thickness of the radiator 10$\%$.
	\item The distance between Be and $LH_2$ targets in the range of 24~m.
	\item The Be target with a length $L=40$~cm.
\end{itemize}

In addition, the lower repetition rate of electron beam with 64~ns spacing 
between bunches will be required to have enough time to measure 
time-of-flight of the beam momenta and to avoid an overlap of events produced 
from alternating pulses. Low repetition rate was already successfully used 
by G0 experiment in Hall C at JLab~\cite{Androic:2011rhao}.

The final flux of $K_L^0$ is presented with 10$\%$ radiator, corresponding 
to maximal rate .

In the production of a beam of neutral kaons, an important factor is the 
rate of neutrons as a background. As it is well known, the ratio 
$R=N_n/N_{K_L^0}$ is on the order $10^3$ from primary proton 
beams~\cite{Cleland:1975exo}, the same ratio with primary electromagnetic
interactions is much lower. This is illustrated in Fig.~\ref{fig:ratio}, 
which  presents the rate of kaons and neutrons as a function of the momentum, 
which resembles similar behavior as it was measured at 
SLAC~\cite{Brandenburg:1972pmo}.
%%%-----------------------------------
\begin{figure}[htb!]
\centering
{
	\includegraphics[width=2.50in]{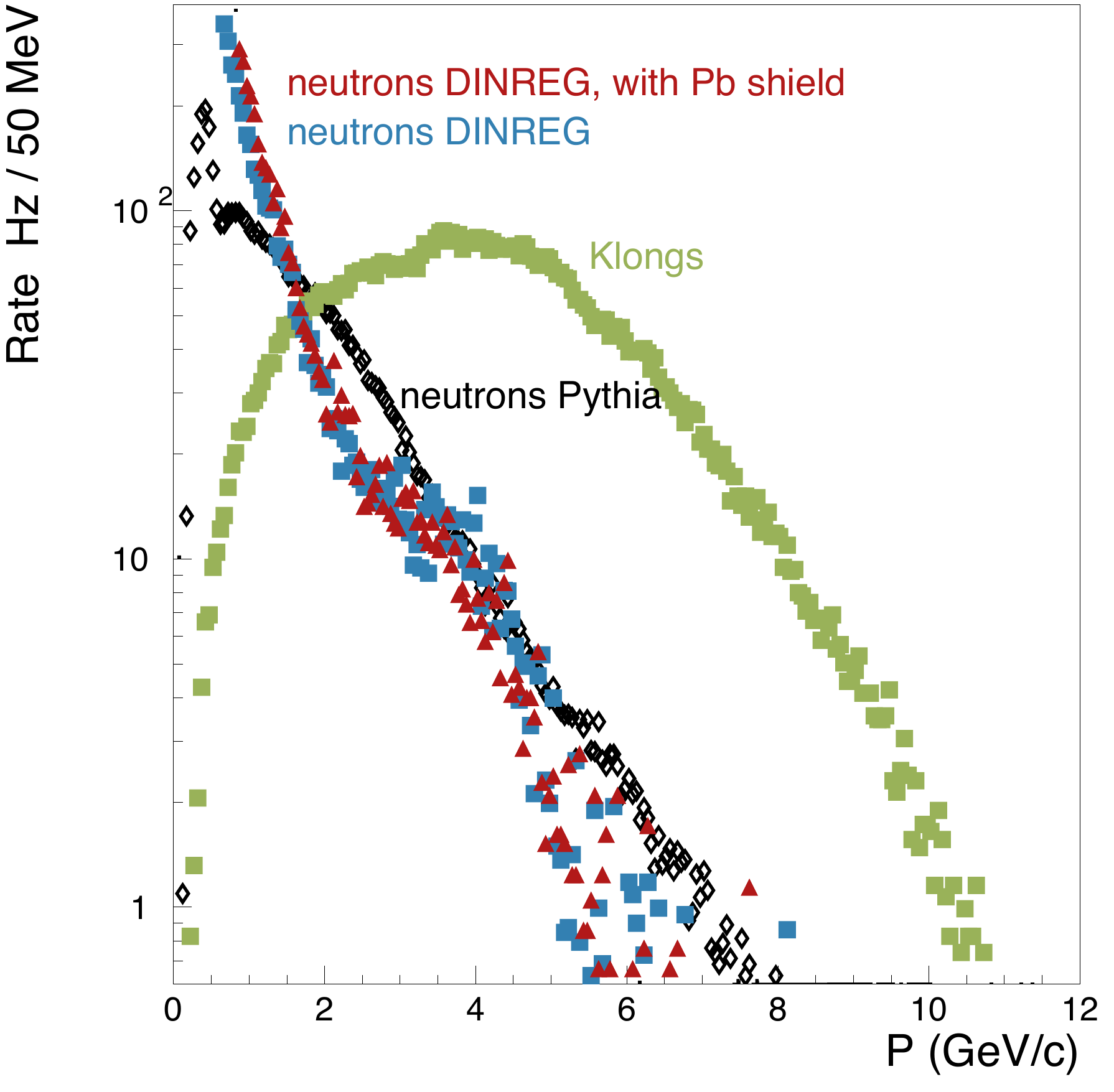}
}

	\caption{\baselineskip 13pt The rate of neutrons (open symbols)
	and $K_L^0$ (full squares) on $LH_2$ target of Hall~D as 
	a function of their momenta simulated with different MC 
	generators with $10^4K^0_L$/sec. \label{fig:ratio}}
\end{figure}
%%%-----------------------------------

Shielding of the low energy neutrons in the collimator cave and flux of 
neutrons has been estimated to be affordable, however detailed simulations 
are under way to show the level of radiation along the beamline.

Th response of GlueX setup, reconstruction efficiency and resolution are 
presented in a talk by Taylor in Ref.~\cite{Albrow:2016ibso}.

%%%-----------------------------------
\item \textbf{Expected Rates}

In this section, we discuss expected rates of events for some selected 
reactions. The production of $\Xi$ hyperons has been measured only with 
charged kaons with very low statistical precision and never with primary 
$K_L^0$ beam. In Fig.~\ref{fig:xi_prod}, panel (a) shows existing data 
for the octet ground state $\Xi$'s with theoretical model predictions 
for $W$ (the reaction center of mass energy) distribution, panel (b) 
shows the same model prediction~\cite{Jackson:2015dvao} presented with 
expected experimental points and statistical error for 10 days of 
running with our proposed setup with a beam intensity $2\times 
10^3~K_L$/sec using missing mass of $K^+$ in the reaction $K_L^0+p\to
K^+\Xi^0$ without detection of any of decay products of $\Xi^0$ (for 
more details on this topic see a talk by Nakayama in 
Ref.~\cite{Albrow:2016ibso}).
%%%-----------------------------------
\begin{figure}[htb!]
\centering
{
	\includegraphics[width=2.50in]{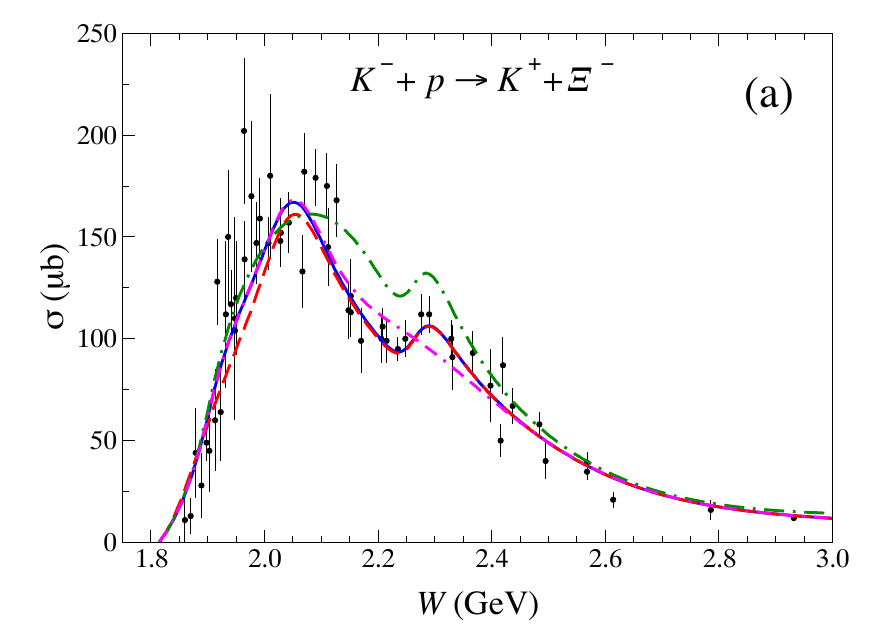}
	\includegraphics[width=2.5in,height=2.1in]{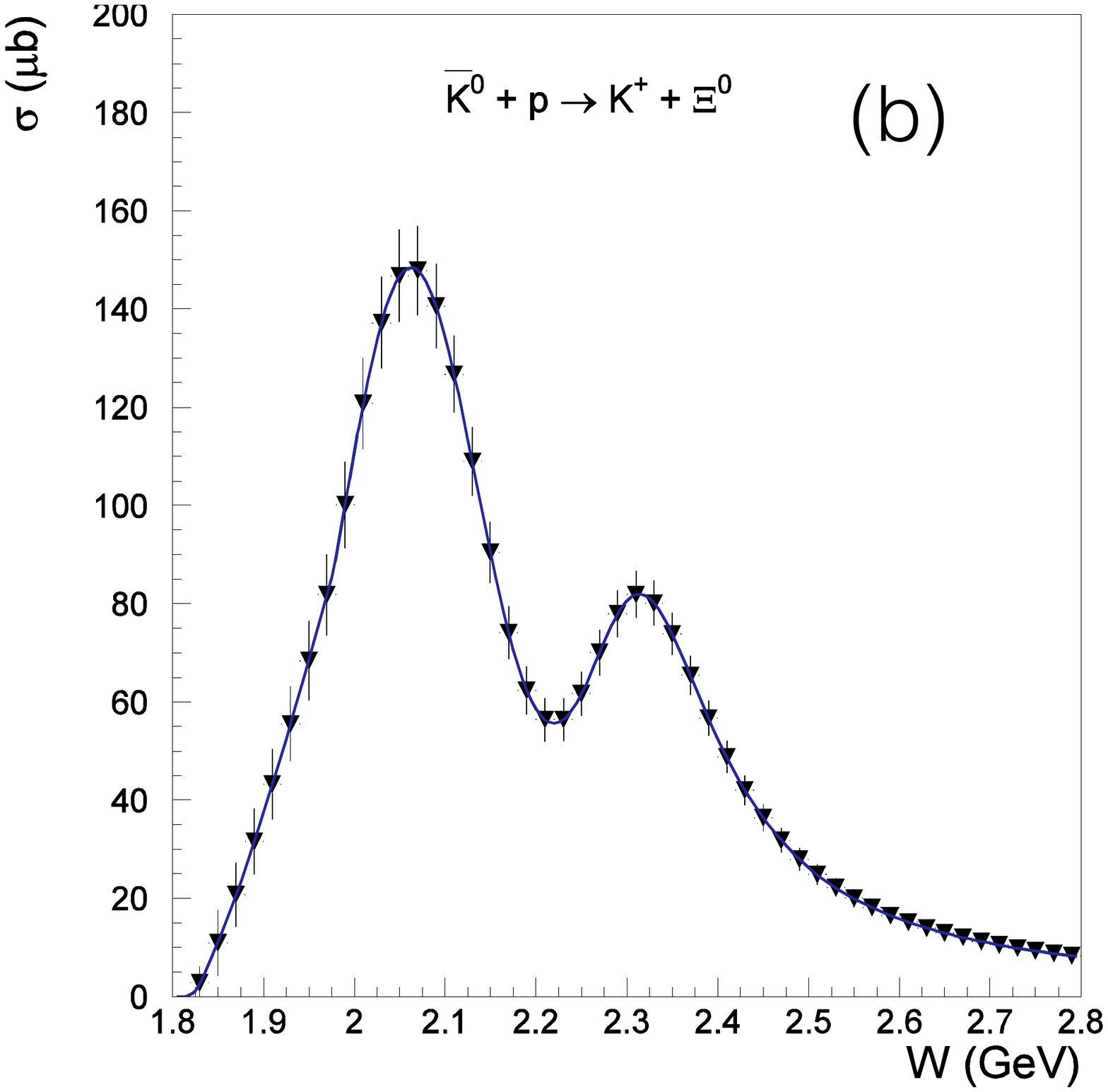}
}

	\caption{\baselineskip 13pt (a) Cross section for existing
	world data on $K^-+p\to K^+\Xi^-$ reaction with model 
	predictions from Ref.~\protect\cite{Jackson:2015dvao};
	(b) expected statistical precision for the reaction 
	$K_L^0+p\to K^+\Xi^0$ in 10 days of running with a beam 
	intensity $2\times 10^3 K_L$/sec overlaid on theoretical
        prediction~\protect\cite{Jackson:2015dvao}.
        \label{fig:xi_prod}}
\end{figure}
%%%-----------------------------------

The physics of excited hyperons is not well explored, remaining essentially 
at the pioneering stages of '70s-'80s. This is especially true for 
$\Xi^\ast(S=-2)$ and $\Omega^\ast(S=-3)$ hyperons.  For example, the $SU(3)$ 
flavor symmetry allows as many $S=-2$ baryon resonances, as there are $N$ 
and $\Delta$ resonances combined ($\approx 27$); however, until now only 
three [ground state $\Xi(1382)1/2^+$, $\Xi(1538)3/2^+$, and $\Xi(1820)3/2^-$] 
have their quantum numbers assigned and few more states have been 
observed~\cite{Patrignani:2016xqpo}. The status of $\Xi$ baryons is 
summarized in a table presented in Fig.~\ref{fig:table1} together with the 
quark model predicted states~\cite{Chao:1980emo}.
%%%-----------------------------------
\begin{figure}[htb!]
\centering
{
	\includegraphics[width=4.0in,height=3.0in]{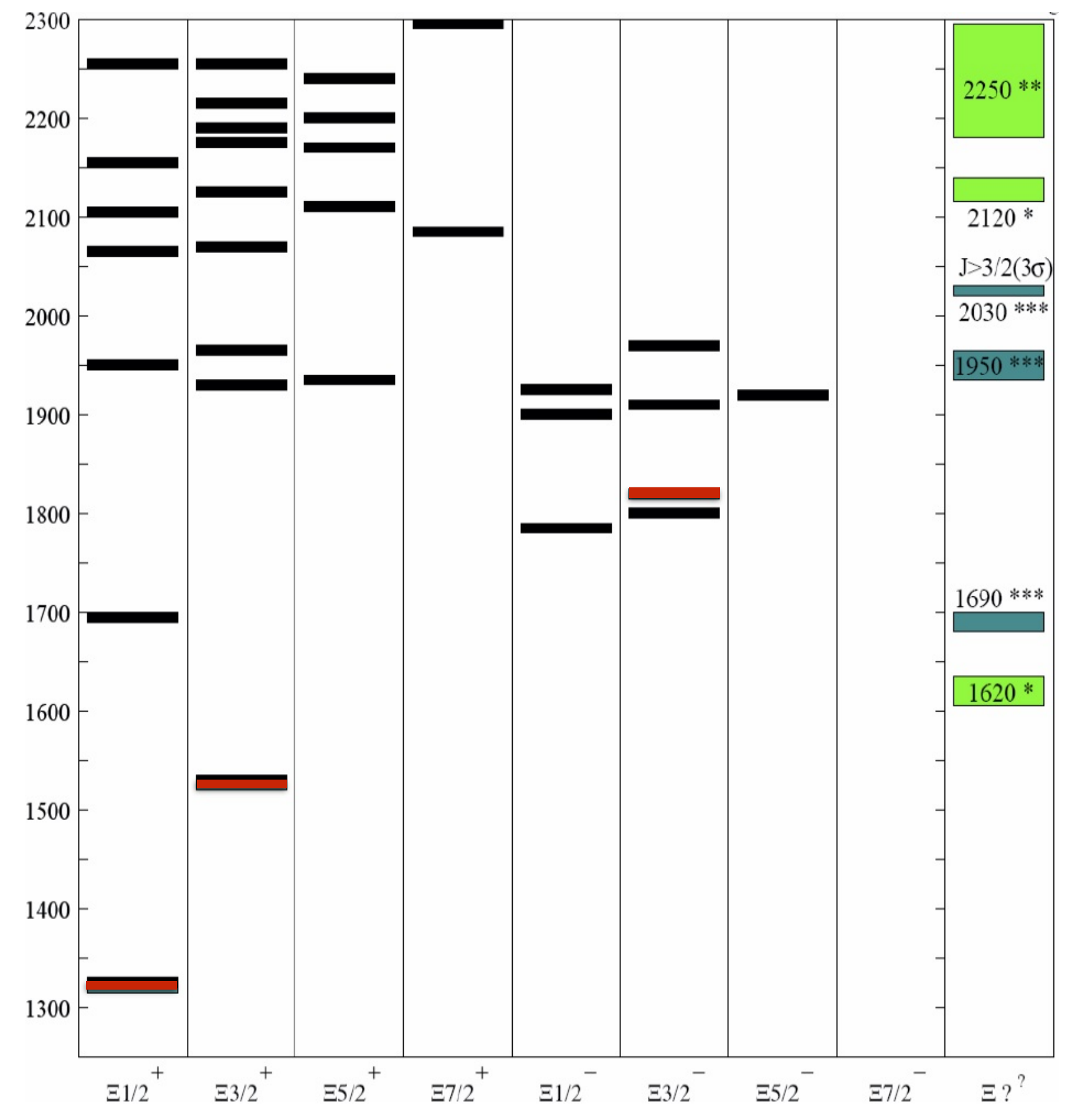}
}

	\caption{\baselineskip 13pt Black bars: Predicted $\Xi$ spectrum based
	on the quark model calculation~\protect\cite{Chao:1980emo}. 
	Colored bars: Observed states.  The two ground octet and decuplet 
	states together with $\Xi(1820)$ in the column $J^P=3/2^-$ are 
	shown in red color. Other observed states with unidentified 
	spin-parity are plotted in the rightest column. \label{fig:table1}}
\end{figure}
%%%-----------------------------------

Historically the $\Xi^\ast$ states were intensively searched for mainly in 
bubble chamber experiments using the $K^-p$ reaction in '60s-'70s. The 
cross section was estimated to be on the order of 1-10~$\mu b$ at the beam 
momenta up to ~10~GeV/c. In '80s-'90s, the mass or width of ground and
some of excited states were measured with a spectrometer in the CERN 
hyperon beam experiment. Few experiments have studied cascade baryons with 
the missing mass technique. In 1983, the production of $\Xi^\ast$ 
resonances up to 2.5~GeV were reported from $p(K^-,K^+)$ reaction from the 
measurement of the missing mass of $K^+$~\cite{Jenkins:1983pmo}. The 
experimental situation with $\Omega^{-\ast}$'s is even worse than the 
$\Xi^\ast$ case, there are very few data for excited states. The main 
reason for such a scarce dataset in multi strange hyperon domain is mainly 
due to very low cross section in indirect production with pion or in 
particular- photon beams.  Currently only ground state $\Omega^-$ quantum 
numbers are identified. Recently significant progress is made in lattice 
QCD calculations of excited baryon states~\cite{Edwards:2012fxo,Engel:2013igo} 
which poses a challenge to experiments to map out all predicted states  
(for more details see a talk by Richards at this workshop). The advantage 
of baryons containing one or more strange quarks for lattice calculations 
is that then number of open decay channels is in general smaller than for 
baryons comprising only the light u and d quarks. Moreover, lattice 
calculations show that there are many states with strong gluonic content 
in positive parity sector for all baryons.  The reason why hybrid baryons
have not attracted the same attention as hybrid mesons is mainly due to 
the fact that they lack manifest ``exotic" character. Although it is 
difficult to distinguish hybrid baryon states, there is significant 
theoretical insight to be gained from studying spectra of excited baryons,
particularly in a framework that can simultaneously calculate properties 
of hybrid mesons. Therefore this program will be very much complementary 
to the GlueX physics program of hybrid mesons.

The proposed experiment  with a beam intensity $10^4 K_L$/sec will result 
in about $2\times 10^5$~$\Xi^\ast$'s and $4\times 10^3$~$\Omega^\ast$'s 
per month.

A similar program for KN scattering is under development at J-PARC with 
charged kaon beams. The current maximum momentum of 
secondary beamline of 2~GeV/c is available at the K1.8 beamline. The 
beam momentum of 2~GeV/c corresponds to $\sqrt{s}$=2.2~GeV in the $K^-p$ 
reaction which is not enough to generate even the first excited 
$\Xi^\ast$ state predicted in the quark model. However, there are plans 
to create high energy beamline in the momentum range 5-15~GeV/c to be 
used with the spectrometer commonly used with the J-PARC P50 experiment 
which will lead to expected yield of $(3-4)\times10^5$~$\Xi^\ast$'s and 
$10^3$~$\Omega^\ast$'s per month.

Statistical power of proposed experiment with $K_L$ beam at JLab will be 
of the same order  as that in J-PARC with charged kaon beam.

An experimental program with kaon beams will be much richer and allows 
to perform a complete experiment using polarized target and measuring 
recoil polarization of hyperons. This studies are under way to find an 
optimal solution for the GlueX setup.

The strange meson spectroscopy is another important subject for $K_L$ 
Facility at JLab. The high intensity $K_L$ beam will allow to study 
final state $K-\pi$ system. In particluar to perform phase shift 
analysis for different partial-waves, which may have significant imact 
to all systems having $K-\pi$ in the final state. This includes heavy 
$D$- and $B$-meson decays as well as $\tau\to K\pi \nu_{\tau}$ decay.

%%%-----------------------------------
\item \textbf{Summary}

In summary we intend to propose production of high intensity $K_L$ beam 
using photoproduction processes from a secondary Be target. A flux as high 
as $\sim 10^4 K_L$/sec could be achieved. Momenta of $K_L$ beam particles 
will be measured with time-of-flight. The flux of kaon beam will be 
measured through partial detection of $\pi^+\pi^-$ decay products from 
their decay to $\pi^+\pi^-\pi^0$ by exploiting similar procedure used by 
LASS experiment at SLAC~\cite{Brandenburg:1972pmo}. Besides using 
unpolarized liquid hydrogen target currently installed in GlueX experiment 
the unpolarized deuteron target may be installed. Additional studies are 
needed to find an optimal choice of polarized targets. This proposal will 
allow to measure $KN$ scattering with different final states including 
production of strange and multi strange baryons with unprecedented 
statistical precision to test QCD in non perturbative domain. It has a 
potential to distinguish between different quark models and test lattice 
QCD predictions for excited baryon states with strong hybrid content. It 
will also be used to study $\pi-K$ interactions, which is the topic of the 
current workshop.

%%%-----------------------------------
\item \textbf{Acknowledgments}

My research is supported by DOE Grant-100388-150.
\end{enumerate}

%%%-----------------------------------

%%%%%%%%%%%%%%%%%%%%%%%%%%%%%%%%%%%%%%%%%%%%%%%%%%%%%%%%%%%%%%%%%%%%%%%%%
\newpage
\subsection{Kaon-pion Scattering from Lattice QCD}
\addtocontents{toc}{\hspace{2cm}{\sl Colin Morningstar}\par}
\setcounter{figure}{0}
\setcounter{table}{0}
\setcounter{equation}{0}
\setcounter{footnote}{0}
\halign{#\hfil&\quad#\hfil\cr
\large{Colin Morningstar}\cr
\textit{Carnegie Mellon University}\cr
\textit{Pittsburgh, PA 15213, U.S.A.}\cr}

%%%-----------------------------------
\begin{abstract}
Recent progress in determining scattering phase shifts, and hence, 
resonance properties from lattice QCD in large volumes with nearly 
realistic pion masses is presented. A crucial ingredient in carrying 
out such calculations in large volumes is estimating quark propagation
using the stochastic LapH method. The elastic $I = 1/2$, $S$- and 
$P$-wave kaon-pion scattering amplitudes are calculated using an ensemble 
of anisotropic lattice QCD gauge field configurations with $N_f = 2 + 1$ 
flavors of dynamical Wilson-clover fermions at $m_\pi = 240$~MeV. The 
$P$-wave amplitude is well described by a Breit-Wigner shape with
parameters which are insensitive to the inclusion of $D$-wave mixing and 
variation of the $S$-wave parametrization.
\end{abstract}

%%%-----------------------------------
\begin{enumerate}
\item A key goal in lattice QCD is the determination of the spectrum of 
hadronic resonances from first principles. One of the best methods of 
computing the masses and other properties of hadrons from QCD involves 
estimating the QCD path integrals using Markov-chain Monte Carlo methods,
which requires formulating the theory on a space-time lattice. Such 
calculations are necessarily carried out in finite volume. However, most 
of the excited hadrons we seek to study are unstable resonances. 
Fortunately, it is possible to deduce the masses and widths of resonances
from the spectrum determined in finite volume. The method we use is 
described in Ref.~\cite{morn:kmatrixg} and the references contained 
therein.

To study low-lying resonances in lattice QCD, we first use lattice QCD 
methods to calculate the interacting two-particle lab-frame energies $E$ 
in a spatial $L^3$ volume with periodic boundary conditions.  Our 
stationary states have total momentum $\Pvec=(2\pi/L)\dvec$, where $\dvec$ 
is a vector of integers.  We boost the lab-frame energies to the
center-of-momentum frame using
\beq
   \Ecm = \sqrt{E^2-\Pvec^2},\qquad
   \gamma = \frac{E}{\Ecm}.
\eeq
Assuming there are $N_d$ channels with particle masses $m_{1a}, m_{2a}$ 
and spins $s_{1a}, s_{2a}$ of particle 1 and 2 for each decay channel, we 
can define
\beqs
   \qcma^{2} &=& \frac{1}{4} \Ecm^{2}
   - \frac{1}{2}(m_{1a}^2+m_{2a}^2) + 
   \frac{(m_{1a}^2-m_{2a}^2)^2}{4\Ecm^{2}},\\
   u_a^2&=& \frac{L^2\qcma^2}{(2\pi)^2},\qquad
   \sveca= \left(1+\frac{(m_{1a}^2-m_{2a}^2)}{\Ecm^{2}}\right).
\eeqs
The relationship between the finite-volume energies $E$  and the 
scattering matrix$S$ is given by
\beq
   \det[1+F^{(\Pvec)}(S-1)]=0,
\label{morneq:quant1}
\eeq
where the $F$-matrix in the $JLSa$ basis states is given by
\beqs
  &&\langle  J'm_{J'}L'S'a'\vert F^{(\Pvec)}\vert Jm_JLSa\rangle
  =\delta_{a'a}\delta_{S'S}\ \frac{1}{2}
  \Bigl\{\delta_{J'J}\delta_{m_{J'}m_J}\delta_{L'L}\nn\\
  &&\qquad + \langle J'm_{J'}\vert L'm_{L'} Sm_{S}\rangle
  \langle Lm_L Sm_S\vert Jm_J\rangle
  W_{L'm_{L'};\ Lm_L}^{(\Pvec a)}\Bigr\},
\eeqs
and where $J,J'$ indicate total angular momenta, $L,L'$ are orbital angular 
momenta, $S,S'$ are intrinsic spins, and $a,a'$ denote all other identifying 
information, such as decay channel.  The $W$-matrix is given by
\beqs
  &&-iW^{(\Pvec a)}_{L'm_{L'};\ Lm_L}
  = \sum_{l=\vert L'-L\vert}^{L'+L}\sum_{m=-l}^l
  \frac{ {\cal Z}_{lm}(\svec_a,\gamma,u_a^2) }{\pi^{3/2}\gamma u_a^{l+1}}
  \sqrt\frac{(2{L'}+1)(2l+1)}{(2L+1)}\nonumber\\
  &&\qquad\qquad\times \langle {L'} 0,l 0\vert L 0\rangle
  \langle {L'} m_{L'},  l m\vert  L m_L\rangle,
\eeqs
and the Rummukainen-Gottlieb-L\"uscher (RGL) shifted zeta functions ${\cal 
Z}_{lm}$ are evaluated using
\beqs
  &&{\cal Z}_{lm}(\svec,\gamma,u^2)=\sum_{\nvec\in \mathbb{Z}^3}
  \frac{{\cal Y}_{lm}(\zvec)}{(\zvec^2-u^2)}e^{-\Lambda(\zvec^2-u^2)}
  +\delta_{l0}\frac{\gamma\pi}{\sqrt{\Lambda}} F_0(\Lambda u^2)\nn\\
  &&+\frac{i^l\gamma}{\Lambda^{l+1/2}} \int_0^1\!\!dt
  \left(\frac{\pi}{t}\right)^{l+3/2}\! e^{\Lambda t u^2}
  \sum_{\nvec\in \mathbb{Z}^3\atop \nvec\neq 0}
  e^{\pi i \nvec\cdot\svec}{\cal Y}_{lm}(\wvec)
  \  e^{-\pi^2\wvec^2/(t\Lambda)} \label{morneq:rglzeta}
\eeqs
where
\beqs
  && \zvec= \nvec -\gamma^{-1} \bigl[\textstyle\frac{1}{2}
  +(\gamma-1)s^{-2}\nvec\cdot\svec \bigl]\svec,\\
  && \wvec=\nvec - (1  - \gamma) s^{-2}
  \svec\cdot\nvec\svec,\qquad
  {\cal Y}_{lm}(\xvec)=\vert \xvec\vert^l\ Y_{lm}(\widehat{\xvec})\\
  &&   F_0(x) =  -1+\frac{1}{2}
  \int_0^1\!\! dt\ \frac{e^{tx}-1 }{t^{ 3/2}}.
\eeqs
We choose $\Lambda\approx 1$ for fast convergence of the summation, and the 
integral in Eq.~(\ref{morneq:rglzeta}) is done using Gauss-Legendre 
quadrature.  $F_0(x)$ given in terms of the Dawson or erf function.

The quantization condition in Eq.~(\ref{morneq:quant1}) relates a single 
energy $E$ to the entire $S$-matrix.  This equation cannot be solved for 
$S$, except in the case of a single channel and single partial wave. To 
proceed, we approximate the $S$-matrix using functions depending on a 
handful of fit parameters, then obtain estimates of these parameters
using fits involving as many energies as possible.  It is easier to 
parametrize a Hermitian matrix than a unitary matrix, so we use the 
$K$-matrix defined as usual by
\beq
   S = (1+iK)(1-iK)^{-1} = (1-iK)^{-1}(1+iK).
\eeq
The Hermiticity of the $K$-matrix ensures the unitarity of the $S$-matrix.  
With time reversal invariance, the $K$-matrix must be real and symmetric. 
The multichannel effective range expansion suggests writing
\beq
  K^{-1}_{L'S'a';\ LSa}(E)=u_{a'}^{-L'-\frac{1}{2}}\ 
  {\widetilde{K}}^{-1}_{L'S'a';\ LSa}(\Ecm)
  \ u_a^{-L-\frac{1}{2}}
\eeq
since ${\widetilde{K}}^{-1}_{L'S'a';\ LSa}(\Ecm)$ should behave
smoothly with $\Ecm$, then the quantization condition can be written
\beq
  \det(1-B^{(\Pvec)}\widetilde{K})=\det(1-\widetilde{K}B^{(\Pvec)})=0,
  \label{morneq:quant2}
\eeq
where we define the important \textit{box matrix} by
\beqs
  && \me{J'm_{J'}L'S'a'}{B^{(\Pvec)}}{Jm_JLS a} =
  -i\delta_{a'a}\delta_{S'S} \ u_a^{L'+L+1}\ W_{L'm_{L'};\ Lm_L}^{(\Pvec a)}  
  \nn\\
  &&\qquad\qquad \times\langle J'm_{J'}\vert L'm_{L'},Sm_{S}\rangle
  \langle Lm_L,Sm_S\vert Jm_J\rangle.
\eeqs
The box matrix is Hermitian for $u_a^2$ real.  The quantization condition 
can also be expressed as
\beq
  \det(\widetilde{K}^{-1}-B^{(\Pvec)})=0,
  \label{morneq:quant3}
\eeq
and the determinants in Eqs.~(\ref{morneq:quant2}) and (\ref{morneq:quant3})
are real.

The quantization condition involves a determinant of an infinite matrix.  To 
make such determinants practical for use, we first transform to a 
block-diagonal basis, and then truncate in orbital angular momentum. For a 
symmetry operation $G$, define theunitary matrix
\beq
      \me{J'm_{J'}L'S'a'}{Q^{(G)}}{ Jm_JLS a}
      = \left\{\begin{array}{ll}
      \delta_{J'J}\delta_{L'L}\delta_{S'S}\delta_{a'a}
      D^{(J)}_{m_{J'}m_{J}}(R),  & (G=R),\\
      \delta_{J'J}\delta_{m_{J'}m_J}\delta_{L'L}\delta_{S'S}\delta_{a'a}
      (-1)^{L}, & (G=I_s), \end{array}\right.
\eeq
where $D^{(J)}_{m'm}(R)$ are the Wigner rotation matrices, $R$ is an ordinary 
rotation, and $I_s$ is spatial inversion.  One can show that the box matrix 
satisfies
\beq
   B^{(G\Pvec)} = Q^{(G)}\ B^{(\Pvec)}\ Q^{(G)\dagger}.
\eeq
If $G$ is in the little group of $\Pvec$, then $G\Pvec=\Pvec,\  
G\svec_a=\svec_a$ and
\beq
   [B^{(\Pvec)}, Q^{(G)}] = 0,
   \qquad\mbox{($G$ in little group of $\Pvec$).}
\eeq
This means we can use the eigenvectors of $Q^{(G)}$ to block diagonalize 
$B^{(\Pvec)}$. The block-diagonal basis can be expressed using
\beq
  \ket{\Lambda\lambda n JLS a}= \sum_{m_J} c^{J(-1)^L;\,\Lambda\lambda 
  n}_{m_J} \ket{Jm_JLS a}
\eeq
for little group irrep $\Lambda$, irrep row $\lambda$, and occurrence index 
$n$. The transformation coefficients depend on $J$ and $(-1)^L$, but not on 
$S,a$. Essentially, the transformation replaces $m_J$ by 
$(\Lambda,\lambda,n)$. Group theoretical projections with Gram-Schmidt are 
used to obtain the basis expansion coefficients.  In this block-diagonal 
basis, the box matrix has the form
\beq
  \me{\Lambda'\lambda' n'J'L'S' a'}{B^{(\Pvec)}}{\Lambda\lambda nJLS a}
  = \delta_{\Lambda'\Lambda}\delta_{\lambda'\lambda}\delta_{S'S}
  \delta_{a'a}\ B^{(\Pvec\Lambda_B Sa)}_{J'L'n';\ JLn}(\Ecm),
\eeq
and the $\widetilde{K}$-matrix for $(-1)^{L+L'}=1$ has the form
\beq
  \me{\Lambda'\lambda' n'J'L'S' a'}{\widetilde{K}}{\Lambda\lambda nJLS a}
  = \delta_{\Lambda'\Lambda}\delta_{\lambda'\lambda}\delta_{n'n} \delta_{J'J}
  \ {\cal K}^{(J)}_{L'S'a';\ LS a}(\Ecm).
\eeq
$\Lambda$ is the irrep for the $K$-matrix, but we need $\Lambda_B$ for the
box matrix.  When $\eta^P_{1a}\eta^P_{2a}=1$, then $\Lambda_B=\Lambda$,
but they differ slightly when $\eta^P_{1a}\eta^P_{2a}=-1$.
%%%-----------------------------------
\begin{figure}
\begin{center}
\includegraphics[width=6.0in]{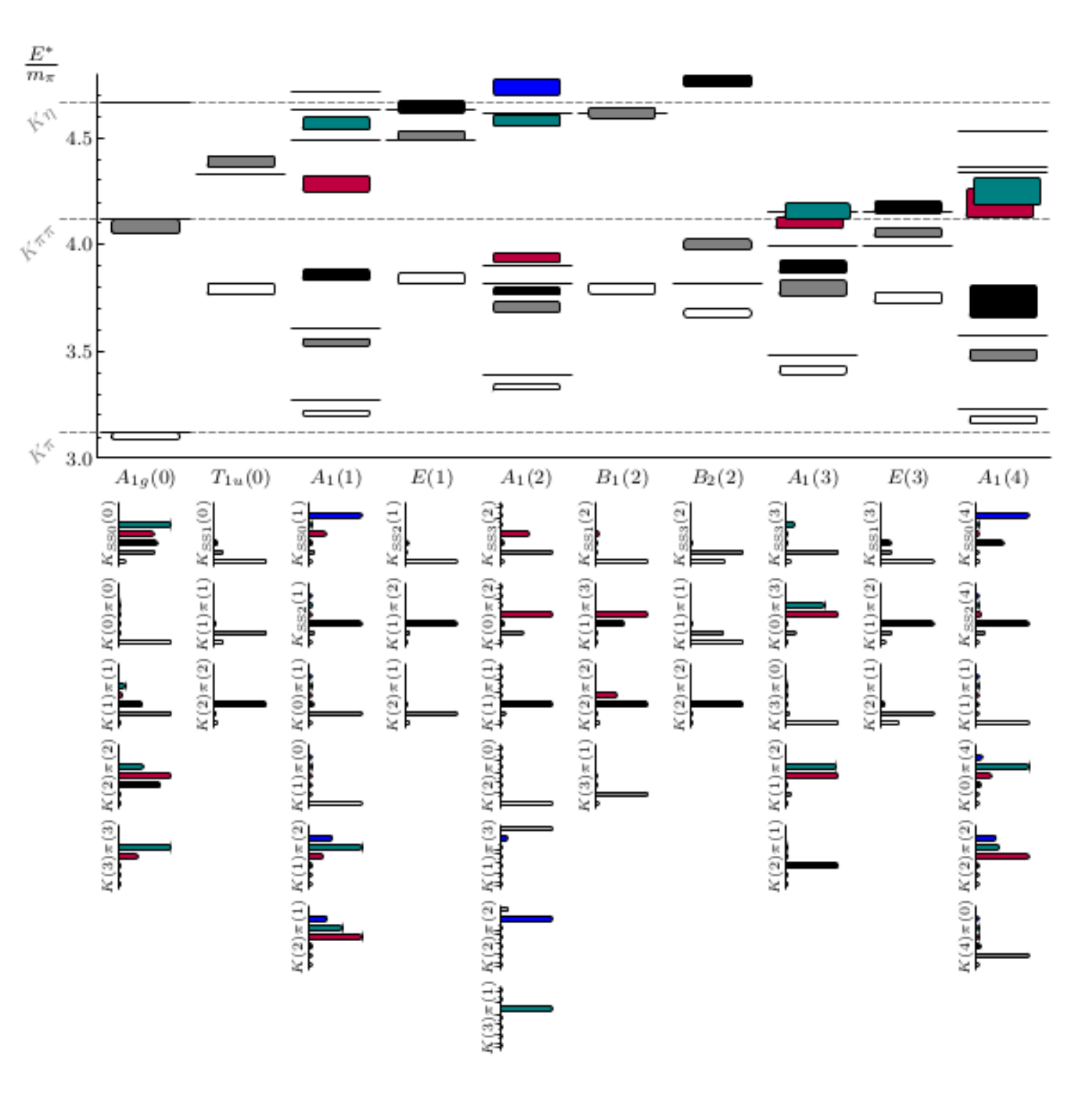}
\end{center}

\caption[boxplot]{
	Center-of-momentum energies over the pion mass in the isodoublet 
	strange mesonic sector for a $32^3\times 256$ anisotropic lattice 
	with $m_\pi\sim 240$~MeV. Each irrep is located in one column, 
	where the energy ratios are shown in the upper panel with a vertical 
	thickness showing the statistical error. The solid horizontal lines 
	should the two-hadron noninteracting energies, and the gray dashed 
	lines show relevant thresholds. The corresponding columns in the 
	lower panel indicate overlaps of each interpolating operator onto 
	the finite-volume Hamiltonian eigenstates. \label{mornfig:boxplot}}
\end{figure}
%%%-----------------------------------
%%%-----------------------------------
\begin{table}
\caption[results]{Best-fit results for various parameters related
	to the $K^\ast(892)$ calculations.
	\label{morntab:kstarfit}}
\begin{center}
\begin{tabular}{|cccccc|} \hline
Fit &   $s$-wave par. & $m_{K^\ast}/m_{\pi}$ & $g_{K^\ast K\pi}$ & $m_{\pi}a_0$& $\chi^2/\mathrm{d.o.f.}$ \\ \hline\hline
(1a,1b) &        \textsc{lin}   & $3.819(20)$ & $5.54(25)$ & $-0.333(31)$ & ($1.04,$\textendash) \\
2 &       \textsc{lin}  & $3.810(18)$ & $5.30(19)$ & $-0.349(25)$ & $1.49$ \\
3 &       \textsc{quad} & $3.810(18)$ & $5.31(19)$ & $-0.350(25)$ & $1.47$ \\
4 &       \textsc{ere}  & $3.809(17)$ & $5.31(20)$ & $-0.351(24)$ & $1.47$ \\
5 &       \textsc{bw}   & $3.808(18)$ & $5.33(20)$ & $-0.353(25)$ & $1.42$ \\
6 &       \textsc{bw}   & $3.810(17)$ & $5.33(20)$ & $-0.354(25)$ & $1.50$ \\ 
\hline
\end{tabular}
\end{center}
\end{table}
%%%-----------------------------------

Using a large set of single- and two-hadron operators, as described in
Ref.~\cite{morn:operatorsg}, for several different total momenta,  we have
evaluated a large number of energies in the isodoublet $I=\frac{1}{2}$
strange $S = 1$ mesonic sector~\cite{morn:kstarg}. We used an anisotropic
$32^3\times 256$ lattice with a pion mass $m_\pi\sim 240$~MeV. The
determinations of these energies are possible since we use the stochastic
LapH method~\cite{morn:laphg} to estimate all quark propagation. The
center-of-momentum energies over the pion mass are shown in
Fig.~\ref{mornfig:boxplot}.  Horizontal lines indicates the non-interacting
two-hadron energies, and the dashed lines show the $K\pi$, $K\pi\pi$,
and $K\eta$ thresholds. Operator overlaps are shown in the lower panel
of the figure.

To extract the resonance properties of the $K^\ast(892)$, we included
the $L = 0, 1, 2$ partial waves. The fit forms used for the $P$ and $D$
partial waves were
\beqs
  (\widetilde{K}^{-1})_{11} &=& \frac{6\pi\Ecm}{g_{K^\ast\pi\pi}^2m_{\pi}}
  \left(\frac{m_{K^\ast}^2}{m_{\pi}^2} - \frac{\Ecm^2}{m_{\pi}^2} \right), 
\qquad
(\widetilde{K}^{-1})_{22} = \frac{-1}{m_\pi^5 a_2},
\eeqs
and for the $S$-wave, several different forms were tried:
\beqs
  (\widetilde{K}^{-1})_{00}^{\rm lin} &=& a_{\rm l}+b_{\rm l}\Ecm,\\
  (\widetilde{K}^{-1})_{00}^{\rm quad} &=& a_{\rm q}+b_{\rm q}\Ecm^2,\\
        (\tilde{K}^{-1})_{00}^\textsc{ERE}&=& \frac{-1}{m_{\pi}a_0}  +
        \frac{m_\pi r_0}{2} \frac{q^2_{\cm}}{m_{\pi}^2},\\
        (\tilde{K}^{-1})_{00}^{\textsc{bw}} &=&
        \left(\frac{m^2_{K_0^\ast}}{m^2_{\pi}} - 
\frac{E^2_{\cm}}{m_{\pi}^2}\right)
        \frac{6\pi m_{\pi} E_{\cm}}{g_{K_0^{\ast}\pi\pi}^2m^2_{K_0^\ast}}.
\eeqs
A summary of our fit results is presented in Table~\ref{morntab:kstarfit}.
Our determinations of the partial-wave scattering phase shifts are shown in
Fig.~\ref{mornfig:phaseshifts}.  We found that the $q\overline{q}$ operators
in the $A_{1g}(0)$ channel overlap many of the eigenvectors in this channel.
Better energy resolution is needed for a determination of the $K_0^\ast(800)$
parameters, which will be done in the future, but an NLO effective range
parametrization finds $m_{R}/m_{\pi} = 4.66(13) -0.87(18)i$, consistent with
a Breit-Wigner fit.  In Ref.~\cite{morn:kstarg}, our results are compared
to the few other recent lattice results~\cite{kstar:prelovsekg,kstar:wilsong,
kstar:balig} that are available. See also Ref.~\cite{k0star:doringg}.

We have also recently determined the decay width of the $\rho$-meson,
including $L = 1, 3, 5$ partial waves~\cite{morn:rhog}, as well as the
$\Delta$ baryon~\cite{morn:deltag}.
%%%-----------------------------------
\begin{figure}
\begin{center}
\includegraphics[width=3.0in]{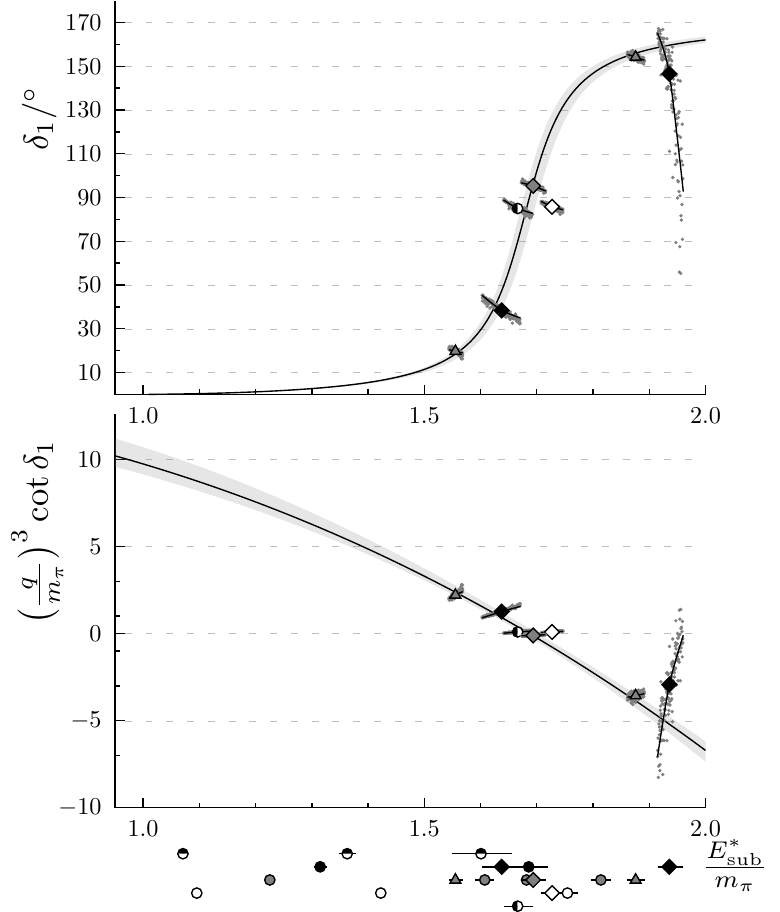}
\includegraphics[width=3.0in]{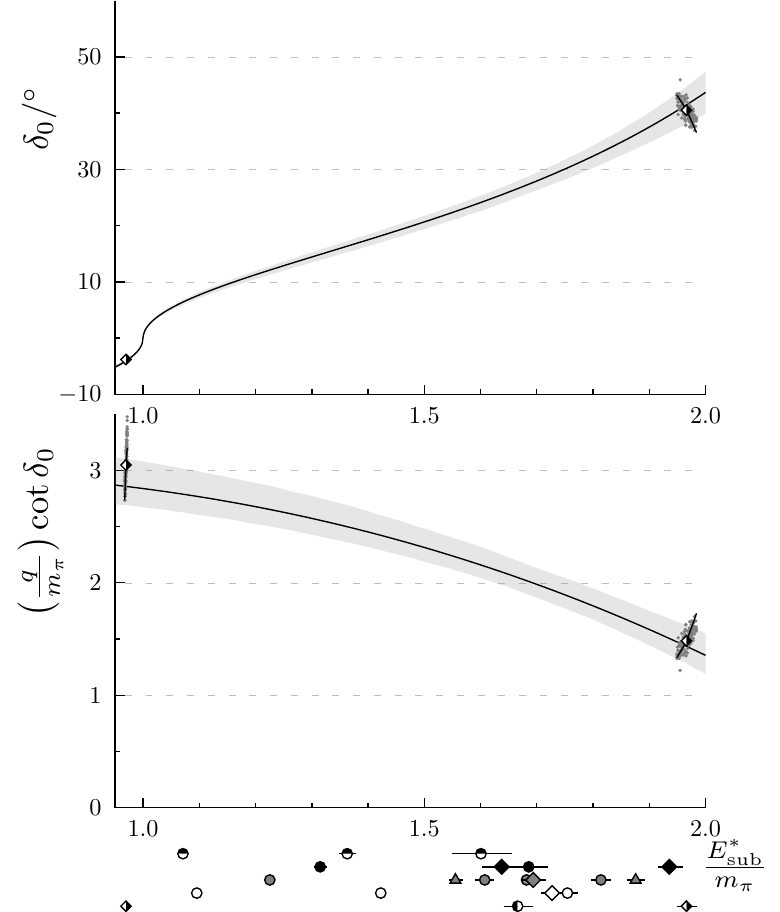}
\end{center}

\caption[shifts]{(Left) Our determination of the $P$-wave $I=\frac{1}{2}$
	$K\pi$ scattering phase shift $\delta_1$ and $q_{\cm}^3 \cot 
	\delta_1$ using a $32^3 \times 256$ anisotropic lattice with 
	$m_\pi\sim 240$~MeV. (Right) Our calculations of the $S$-wave 
	scattering phase shift $\delta_0$ (quadratic fit) and $q_{\cm} \cot 
	\delta_0$. Results are from Ref.~\protect\cite{morn:kstarg}.
	\label{mornfig:phaseshifts}}
\end{figure}
%%%-----------------------------------

%%%-----------------------------------
\newpage
\item \textbf{Acknowledgments}

This work was done in collaboration with John Bulava (U. Southern 
Denmark), Ruairi Brett (CMU), Daniel Darvish (CMU), Jake Fallica (U. 
Kentucky), Andrew Hanlon (U. Mainz), Ben Hoerz (U. Mainz), and Christian W 
Andersen (U. Southern Denmark). I thank the organizers, especially Igor 
Strakovsky, and JLab for the opportunity to participate in this Workshop. 
This work was supported by the U.S. National Science Foundation under 
award PHY--1613449. Computing resources were provided by the Extreme 
Science and Engineering Discovery Environment (XSEDE) under grant number 
TG-MCA07S017. XSEDE is supported by National Science Foundation grant 
number ACI-1548562.
\end{enumerate}

%%%-----------------------------------

%%%%%%%%%%%%%%%%%%%%%%%%%%%%%%%%%%%%%%%%%%%%%%%%%%%%%%%%%%%%%%%%%%%%%%%%%
\newpage
\subsection{K-$\pi$ Scattering with $K_L$ Beam Facility}
\addtocontents{toc}{\hspace{2cm}{\sl Marouen Baalouch}\par}
\setcounter{figure}{0}
\setcounter{table}{0}
\setcounter{equation}{0}
\halign{#\hfil&\quad#\hfil\cr
\large{Marouen Baalouch}\cr
\textit{Department of Physics}\cr
\textit{Old Dominion University}\cr
\textit{Norfolk, VA 23529, U.S.A.~\&}\cr
\textit{Thomas Jefferson National Accelerator Facility}\cr
\textit{Newport News, VA 23606, U.S.A.}\cr}

%%%-----------------------------------
\begin{abstract}
In this talk, I discuss the importance of the $K\pi$ scattering amplitude
analysis, its impact on other physics studies and the possible related
elements that need to be improved. Finally I discuss the feasibility of
performing a $K\pi$ amplitude scattering analysis within KLF.
\end{abstract}

%%%-----------------------------------
\begin{enumerate}
\item \textbf{Introduction}

The $K_L$ Beam Facility can offer a good opportunity to study the kaon-pion
interaction experimentally, by producing the final state $K\pi$ using the
scattering of a neutral kaon off proton or neutron as $K_L N \rightarrow
[K\pi]N^\prime$. The analysis of the kaon-pion interaction experimentally
has several implication on the imperfect phenomenological studies, as the
test of the Chiral Perturbation Theory, Strange Meson Spectroscopy and
Physics beyond Standard Model. These phenomenological studies require more
data to improve the precision on the observable of interest. So far, the
main experimental data used to study kaon-pion scattering at low energy
comes from kaon beam experiments at SLAC in the 1970s and 1980s.

%%%-----------------------------------
\item \textbf{Chiral Perturbation Theory}

In Quantum Chromodynamics (QCD), the strong interaction coupling constant
increases with decreasing energy. This means that the coupling becomes
large at low energies, and one can no longer rely on perturbation theory.
Few phenomenological approaches can be used at this energy level, as 
Lattice QCD or the Chiral Perturbation Theory (ChPT)~\cite{chpt}. The 
purpose of the ChPT is to use an effective Lagrangian where the mesons 
$\pi$, $\eta$, and $K$, called also Goldstone Bosons, are the fundamental 
degrees of freedom.  The ChPT studies on on the $\pi\pi$ scattering 
amplitude shows a good agreement with the experimental studies ($e.g.$, 
see reference~\cite{pipi}). However, this theory is less successful with 
the $K\pi$ scattering amplitude~\cite{kpichpt1,kpichpt2,kpichpt3, 
kpichpt4,kpichpt5,kpichpt6,kpichpt7}, and so far no accurate experimental 
data is available at low energy.

%%%-----------------------------------
\item \textbf{Strange Meson Spectroscopy}

Hadron Spectroscopy plays an important role to understand QCD in the non
perturbative domain by performing a quantitative understanding of quark 
and gluon confinement, and validate Lattice QCD prediction.  In the last 
years, an important number of resonances have been identified,
especially resonances with heavy flavored quark. However, the sector of
strange baryons and mesons was not significantly improved and several
estimated states by lattice QCD and quark model still not yet observed.
Moreover, the identification of the scalar strange light mesons, as
$\kappa$ and $K^{\ast}_{0}(1430)$, still a long-standing puzzle because 
of their large decay width that causes an overlap between resonances at 
low Lorentz-invariant mass. The indications on the presence of $\kappa$
resonance have been reported based on the data of the E791~\cite{e791}
and BES~\cite{bes} Collaborations and several phenomenological
studies~\cite{kappa1,kappa2,kappa3} have been made to measure the pole
of $\kappa$ resonance. However, the results from Roy-Steiner dispersive
representation~\cite{kappa1} not in good agreement with low energy
experimental data, and the confirmation of this pole in the amplitude
for elastic $K\pi$ scattering requires more data at low energy.
The $K^\ast_0(1430)$ is the second scalar strange resonance which is also
not well understood. And recently the $K\pi$ $S$-wave amplitude extracted
from $\eta_c \to KK\pi$~\cite{palano} found to be very different with
respect to the amplitude measured by LASS and E791. Fig~\ref{pal}, taken
from the reference~\cite{palano}, shows the comparison of the amplitude
extracted from $\eta_c \to KK\pi$, $Kp \to K\pi n$ and $D \to KK\pi$.

The light strange scalar mesons can be produced in $KN$ scattering, and
more data from these type of reactions will certainly improve the
understanding of the non well identified strange resonances.
%%%-----------------------------------
\begin{figure}[htbp]
\centering
	\includegraphics[height=7.4cm]{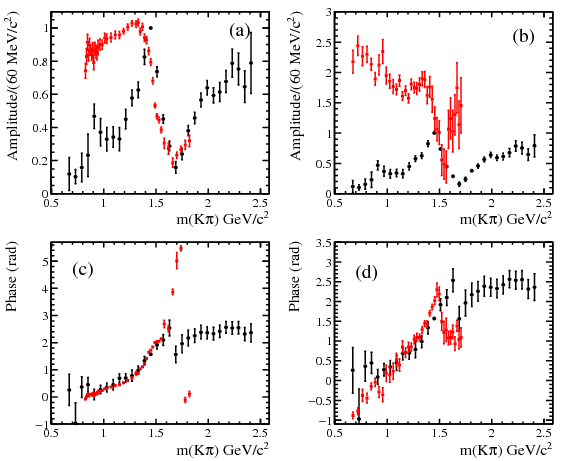}

	\caption{Figure taken from ref~~\cite{palano}:  The 
	I=1/2 $K\pi$ S-wave amplitude measurements from $\eta_c 
	\to KK\pi$ compared to the (a) LASS and (b) E791 
	results: the corresponding I=1/2 $K\pi$ S-wave phase 
	measurements compared to the (c) LASS and (d) E791
	measurements. Black dots indicate the results from the 
	present analysis; square (red) points indicate the LASS 
	or E791 results. The LASS data are plotted in the region 
	having only one solution.} \label{pal} 
\end{figure}
%%%-----------------------------------

%%%-----------------------------------
\item \textbf{$K\pi$ Scattering Amplitude and Physics Beyond 
	Standard Model}

The determination of the CKM matrix elements $V_{us}$ is mainly performed
using $\tau$ or kaon decays. As an example, the $V_{us}$ matrix element is
accessible from the $K_{l3}$ decays using the braching ratio function
\begin{equation}
	\Gamma(K \to \pi l \nu) \propto N |V_{us}|^{2}|f^{K\pi}_+(0)|^{2} I_K^{l},
\end{equation}
\noindent
where
\begin{equation}
	I_K = \int {\rm d}t \frac{1}{m_K^8}\lambda^{3/2}F(\widetilde{f}_+(t),
	\widetilde{f}_0(t)).
\end{equation}
\noindent

In this function $f^{K\pi}_0$ and $f^{K\pi}_+$ represent the form factors
of the strangeness changing scalar and vector, respectively. These form
factors in the low energy region, can be obtained from Lattice QCD or from
the study of the $K\pi$ scattering using dispersion relation
analysis~\cite{ver}. The precision on $V_{us}$ depends strongly on the
precision of these strangeness changing form factors. And by improving
the precision of $V_{us}$ one can probe the physics beyond the standard
model indirectly thanks to the unitarity of the CKM matrix
\begin{equation}
	|V_{ud}|^{2} + |V_{us}|^{2} + |V_{ub}|^{2} = 1.
\end{equation}

Therefore, any shift from unitarity is a sign of physics beyond Standard
Model. Fig~\ref{ckm} shows the fit to the different CKM elements involved
in the unitarity equation.
%%%-----------------------------------
\begin{figure}[htbp]
\centering
	\includegraphics[height=7cm]{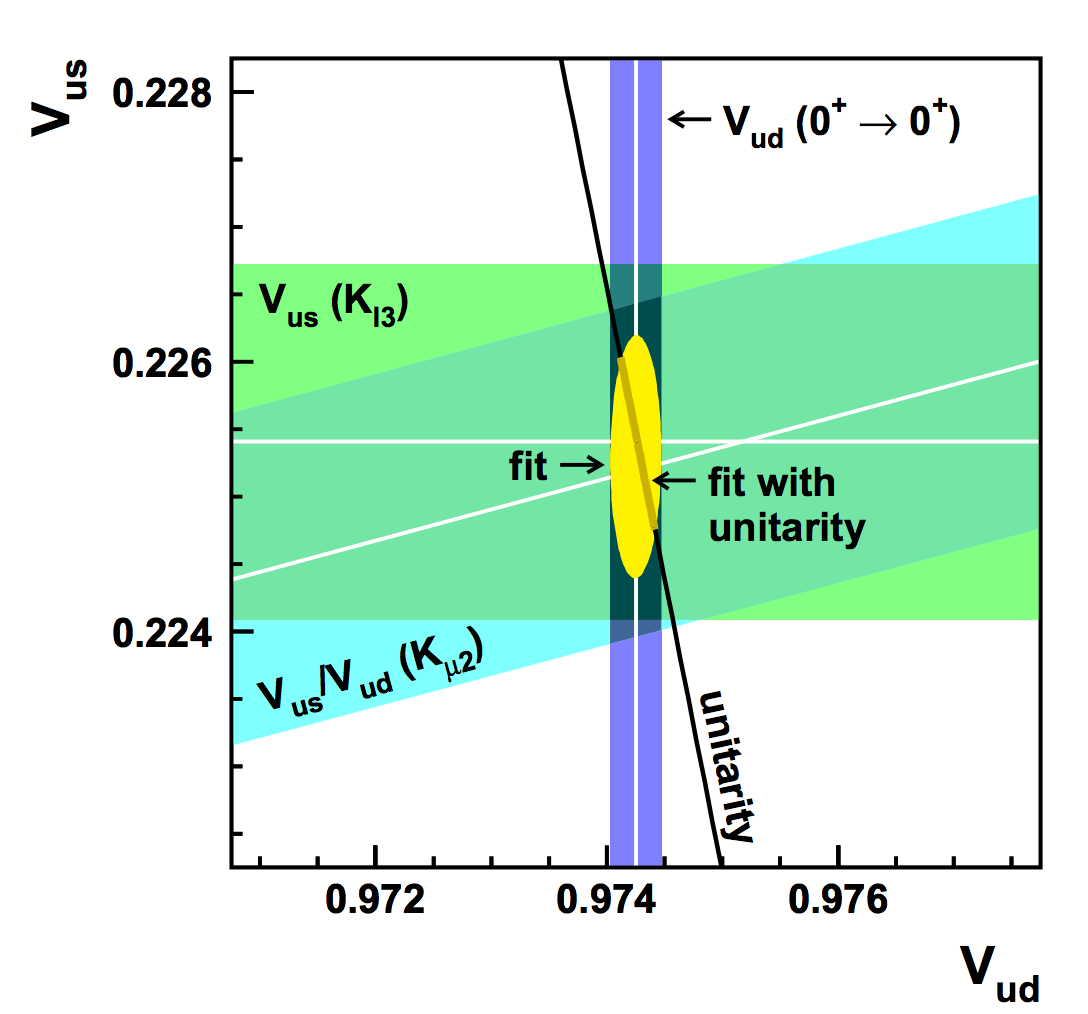}

	\caption{Figure taken from Ref.~\protect\cite{cckm}: Results of 
	fits to $|V_{us}|$, $|V_{us}|$, and $|V_{ud}/V_{us}|$.} 
	\label{ckm}
\end{figure}
%%%-----------------------------------

%%%-----------------------------------
\item \textbf{Kaon-Production and GlueX Detector}

The hadroproduction of the $K\pi$ system has been intensively studied with
charged Kaon beam~\cite{lass, kpip1, kpip2, kpip3, kpip4}. However, few
studies have been made using a neutral kaon beam. The production mechanism
of the $K\pi$ system with charged kaon beam is proportional to the
mechanism with neutral kaon, the main difference related to the
Clebsch-Gordan coefficients. In LASS analysis~\cite{lass}, the $K\pi$
production mechanism is parametrized using a model consisting of exchange
degenerate Regge poles together with non-evasive ``cut" contributions. 
These parameterization was extrapolated to the neutral kaon beam and used 
in the simulation of $K_{\rm L} p \to [K\pi]^0 p$ in KLF where the 
reconstruction of the events is made by GlueX spectrometer. The GlueX 
spectrometer is built in Hall~D at JLab and using photon beam scattering 
off proton to provide critical data needed to address one of the 
outstanding and fundamental challenges in physics the quantitative 
understanding of the confinement of quarks and gluons in QCD. The GlueX 
detector is azimuthally symmetric and nearly hermetic for both charged 
particles and photons, which make it a relevant detector for studying 
$K\pi$ scattering amplitude. Fig~\ref{kini} shows the kinematics region 
that can be covered by the detector using the simulation of $K_{\rm L} p 
\to [K\pi]^0 p$ and LASS parameterization. More details about the detector 
performance can be found in reference~\cite{gluex}.
%%%-----------------------------------
\begin{figure}[htbp]
\centering
	\includegraphics[height=4cm]{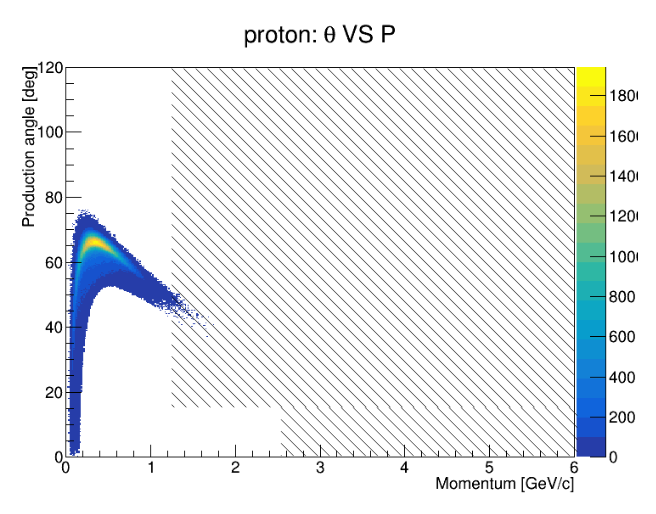}
	\includegraphics[height=4cm]{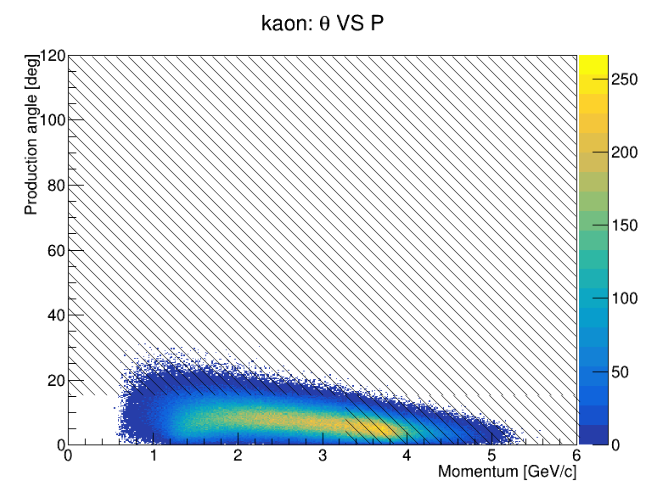}
	\includegraphics[height=4cm]{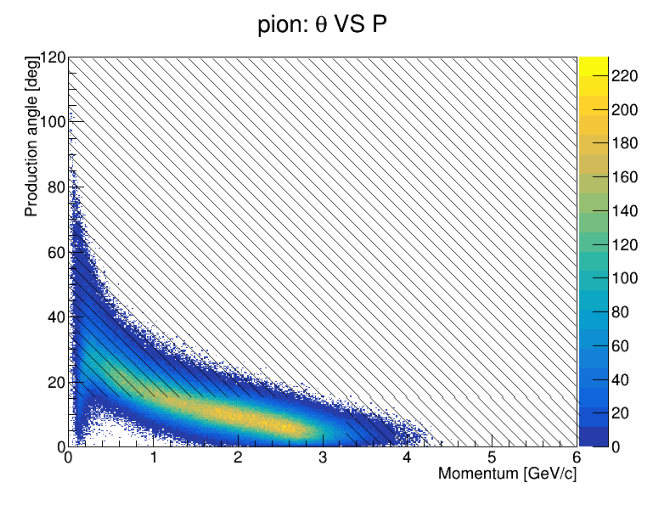}

	\caption{The simulated events  of the reaction 
	$K_{\rm L} p \to [K\pi]^0 p$ projected on the 
	plane of the production angle (in the Lab frame) 
	$\theta$ versus the magnitude of the momentum. 
	The dashed region represents the region where 
	the GlueX detector performance of Particle 
	Identification are low: on the top left for the 
	proton, on the top right for the kaon and on the 
	bottom for the pion.} \label{kini}
\end{figure}
%%%-----------------------------------

%%%-----------------------------------
\item \textbf{Acknowledgments}

My research is supported by Old Dominion University, Old Dominion
University Research Foundation and Jefferson Laboratory.
\end{enumerate}

%%%-----------------------------------

%%%%%%%%%%%%%%%%%%%%%%%%%%%%%%%%%%%%%%%%%%%%%%%%%%%%%%%%%%%%%%%%%%%%
\newpage
\subsection{Dalitz Plot Analysis of Three-body Charmonium Decays at BaBar}
\addtocontents{toc}{\hspace{2cm}{\sl Antimo Palano}\par}
\setcounter{figure}{0}
\setcounter{table}{0}
\setcounter{equation}{0}
\setcounter{footnote}{0}
\halign{#\hfil&\quad#\hfil\cr
\large{Antimo Palano (on behalf of the BaBar Collaboration)}\cr
\textit{I.N.F.N. and University of Bari}\cr
\textit{Bari 70125, Italy}\cr}

%%%-----------------------------------
\begin{abstract}
We present results on a Dalitz plot analysis of $\eta_c$ and $J/\psi$
decays to three-body. In particular, we report the first observation of 
the decay $K^\ast_0(1430)\to K\eta$ in the $\eta_c$ decay to $K^+K^-\eta$. 
We also report on a measurement of the I=1/2 $K\pi$ $\mathcal{S}$-wave 
through a model independent partial wave analysis of $\eta_c$ decays to 
$K_S^0K^\pm\pi^\mp$ and $K^+K^-\pi^0$. The $\eta_c$ resonance is produced 
in two-photon interactions. We perform the first Dalitz plot analysis of 
the $J/\psi$ decay to $K_S^0K^\pm\pi^\mp$ produced in the initial state 
radiation process.
\end{abstract}

%%%-----------------------------------
\begin{enumerate}
\item \textbf{Introduction}

Charmonium decays can be used to obtain new information on light meson 
spectroscopy.  In $e^+e^-$ interactions, samples of charmonium resonances 
can be obtained using different processes.
\begin{itemize}
\item{}
  In two-photon interactions we select events in which the $e^+$ and $e^-$  
  beam particles are scattered at small angles and remain undetected.
  Only resonances with $J^{PC}=0^{\pm+}, 2^{\pm+}, 3^{++}, 4^{\pm+}$....
 can be produced.
\item{}
  In the Initial State Radiation (ISR) process, we reconstruct events 
  having a (mostly undetected) fast forward $\gamma_{ISR}$ and, in this 
  case, only $J^{PC}=1^{--}$ states can be produced.
\end{itemize}

%%%-----------------------------------
\item \textbf{Selection of Two-Photon Production of $\eta_c\to K^+K^- 
\eta$, $\eta_c\to K^+K^-\pi^0$, and $\eta_c\to K_S^0K^\pm\pi^\mp$}

We study the reactions
%%%-----------------------------------
\footnote {Charge conjugation is implied through all this work.}
%%%-----------------------------------
\begin{center}
  $\gamma\gamma\to K^+K^-\eta$, ~~~~~$\gamma\gamma\to K^+K^-\pi^0$,~~~~~ 
  $\gamma\gamma\to K_S^0K^\pm\pi^\mp$ ,
\end{center}
where $\eta\to\gamma\gamma$ and 
$\eta\to\pi^+\pi^-\pi^0$~\cite{Lees:2014iuaf,Lees:2015zzrf}.
%----------------------------------
\begin{figure}[tbp]
    \centering
    \includegraphics[width=1\textwidth]{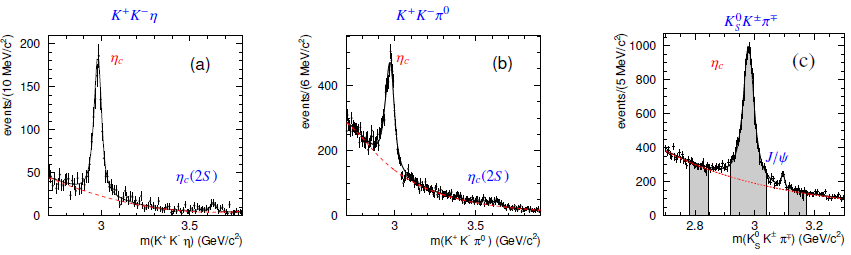}

    \caption{(a) $K^+K^+\eta$, (b) $K^+K^-\pi^0$, and 
	(c) $K_S^0K^\pm\pi^\mp$ mass spectra. The superimposed 
	curves are from the fit results. In
        (c) the shaded area evidences definition of the signal 
	and sidebands regions.} \label{fig:fig1}
\end{figure}
%-----------------------------------

%-------------------------------------------------
Two-photon candidates are reconstructed from the sample of events having
the exact number of charged tracks for that $\eta_c$ decay mode.
Since two-photon events balance the transverse momentum, we require 
$p_T$, the transverse momentum of the  system with respect to the beam 
axis, to be $p_T<0.05~GeV/c$ for $\eta_c\to K^+K^-\eta/\pi^0$ and 
$p_T<0.08~GeV/c$ for $\eta_c\to K_S^0K^\pm\pi^\mp$. We also define 
$M_{rec}^2\equiv(p_{e+e-}-p_{rec})^2$, where $p_{e+e-}$ is the 
four-momentum of the initial state and $p_{\mathrm{rec}}$ is the 
four-momentum of the three-hadrons system and remove ISR events requiring  
$M_{rec}^2>10~GeV^2/c^4$. Figure~\ref{fig:fig1} shows the $K^+K^-\eta$, 
$K^+ K^-\pi^0$, and $K_S^0K^\pm\pi^\mp$ mass spectra where signals of 
$\eta_c$ can be observed.

Selecting events in the $\eta_c$ mass region, the Dalitz plots for the 
three $\eta_c$ decay modes are shown in Fig.~\ref{fig:fig2}. The $\eta_c 
\to K\bar K\pi$ Dalitz plots are dominated by the presence of
horizontal and vertical uniform bands at the position of the 
$K_0^\ast(1430)$ resonance.
%----------------------------------
\begin{figure}[tbp]
    \centering
    \includegraphics[width=1\textwidth]{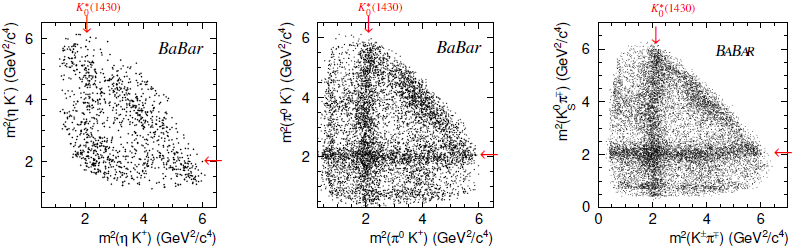}

    \caption{(Left) $K^+K^+\eta$, (Center) $K^+K^-\pi^0$, and (Right) 
	$K_S^0K^\pm\pi^\mp$ Dalitz plots. The arrows indicate the 
	positions of the $K^\ast_0(1430)$ resonance.} \label{fig:fig2}
\end{figure}
%-----------------------------------

The $\eta_c$ signal regions contain 1161 events with (76.1$\pm$1.3)\% 
purity for $\eta_c\to K^+K^-\eta$, 6494 events with (55.2$\pm$0.6)\% 
purity for $\eta_c\to K^+K^-\pi^0$, and 12849 events with (64.3 $\pm$ 
0.4)\% purity for  $\eta_c\to K_S^0 K^\pm\pi^\mp$. The backgrounds below 
the $\eta_c$ signals are estimated from the sidebands. We observe 
asymmetric $K^\ast$'s in the background to the $\eta_c\to 
K_S^0K^\pm\pi^\mp$ final state due to interference between I=1 and I=0 
contributions.

%------------------------------------------------------------
\item \textbf{Dalitz Plot Analysis of $\eta_c\to K^+K^-\eta$ and 
$\eta_c\to K^+K^-\pi^0$}

We first perform unbinned maximum likelihood fits using the Isobar
model~\cite{Eidelman:2004wyf}. Figure~\ref{fig:fig3} shows the $\eta_c
\to K^+K^-\eta$ Dalitz plot projections.
%----------------------------------
\begin{figure}[tbp]
    \centering
    \includegraphics[width=0.8\textwidth]{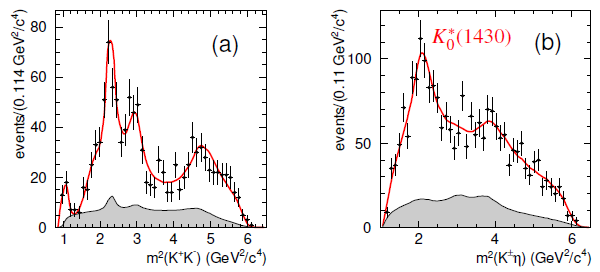}

    \caption{$\eta_c\to K^+K^-\eta$ Dalitz plot analysis. 
	(a) $K^+K^-$, and (b) $K^\pm\eta$ squared mass projections. 
	Shaded is the background contribution.} \label{fig:fig3}
\end{figure}
%-----------------------------------

The analysis of the $\eta_c\to K^+K^-\eta$ decay requires significant
contributions from $f_0(1500)\eta$ $(23.7\pm7.0\pm1.8)$\% and
$K^\ast_0(1430)^+K^-$ $(16.4\pm4.2\pm1.0)$\%, where $K^\ast_0(1430)^+ 
\to K^+\eta$: a first observation of this decay mode. It is found that 
the $\eta_c$ three-body hadronic decays proceed almost entirely through 
the intermediate production of scalar meson resonances. A similar 
analysis performed on the $\eta_c\to K^+K^-\pi^0$ allows to obtain
the corresponding contribution from $K^\ast_0(1430)^+K^-$ to be
$(33.8\pm1.9\pm0.4)$\%, where $K^\ast_0(1430)^+\to K^+\pi^0$.
Combining the above information with the measurement of the $\eta_c$ 
relative branching fraction
\begin{equation}
	\frac{BR(\eta_c\to K^+K^-\eta)}{BR(\eta_c\to K^+K^-\pi^0)} 
	= 0.571\pm 0.025\pm 0.051,
\end{equation}
we obtain
\begin{equation}
	\frac{BR(K^\ast_0(1430) \to \eta K)}{BR(K^\ast_0(1430) \to\pi 
	K)} = 0.092\pm 0.025^{+0.010}_{-0.025}.
\end{equation}
We perform a Likelihood scan and obtain a measurement of the 
$K^\ast_0(1430)$ parameters
\begin{equation}
	m(K^\ast_0(1430))=1438\pm 8\pm 4~MeV/c^2, ~~~~~~~ 
	\Gamma(K^\ast_0(1430)) = 210\pm 20\pm 12~MeV.
\end{equation}

%%%-----------------------------------
\newpage
\item \textbf{Model Independent Partial Wave Analysis of $\eta_c \to K^+
K^-\pi^0$ and $\eta_c\to K_S^0 K^\pm\pi^\mp$}

We perform a Model Independent Partial Wave Analysis 
(MIPWA)~\cite{Aitala:2005yhf} of $\eta_c\to K^+K^-\pi^0$ and $\eta_c\to 
K_S^0 K^\pm\pi^\mp$. In the MIPWA the $K\pi$ mass spectrum is divided 
into 30 equally spaced mass intervals 60~MeV/c$^2$ wide and for each bin 
we add to the fit two new free parameters, the amplitude and the phase of 
the $K\pi$ $\mathcal{S}$-wave (constant inside the bin).

We also fix the amplitude to 1.0 and its phase to $\pi/2$ in an arbitrary
interval of the mass spectrum (bin 11 which corresponds to a mass of 
1.45~GeV/c$^2$). The number of additional free parameters is therefore 58.
Due to isospin conservation in the decays, amplitudes are symmetrized with
respect to the two $K \pi$ decay modes.  The $K^\ast_2(1420)$, $a_0(980)$, 
$a_0(1400)$, $a_2(1310)$, ... contributions are modeled as relativistic 
Breit-Wigner functions multiplied by the corresponding angular functions.
Backgrounds are fitted separately and interpolated into the $\eta_c$
signal regions. The fits improves when an additional high mass $a_0(1950) 
\to K \bar K$, I=1 resonance, is included with free parameters in both 
$\eta_c$ decay modes. The weighted average of the two measurement is:
$m(a_0(1950))=1931\pm 14\pm 22$~MeV/c$^2$, $\Gamma(a_0(1950))=271\pm
22\pm 29$~MeV. The statistical significances for the $a_0(1950)$
effect (including systematics) are $2.5\sigma$ for $\eta_c\to K_S^0
K^\pm\pi^\mp$ and $4.0\sigma$ for $\eta_c\to K^+K^-\pi^0$.

The Dalitz plot projections with fit results for $\eta_c\to K_S^0 K^\pm
\pi^\mp$ and $\eta_c\to K^+K^-\pi^0$ are shown in Fig.~\ref{fig:fig4}.
We observe a good description of the data.
%----------------------------------
\begin{figure}[tbp]
    \centering
    \includegraphics[width=1\textwidth]{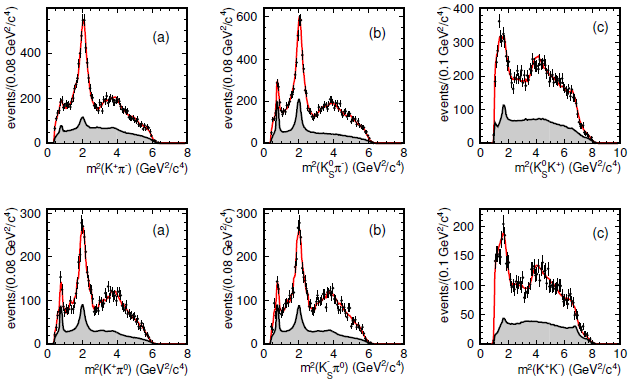}

    \caption{(Top) $\eta_c \to K_S^0K^\pm\pi^\mp$ and (Bottom) $\eta_c 
	\to K^+K^-\pi^0$ Dalitz plots projections. The superimposed 
	curves are from the fit results. Shaded is contribution from 
	the interpolated background.} \label{fig:fig4}
\end{figure}
%%%-----------------------------------

We note that the $K^\ast(892)$ contributions arise entirely from
background. The fitted fractions and phases are given in 
Table~\ref{tab:tab1}. Both $\eta_c\to K\bar K\pi$ decay modes are 
dominated by the $(K\pi \mathcal{S}$-wave) $\bar K$ amplitude, with 
significant interference effects.
%%%-----------------------------------
\begin{table}[h]
\caption{Results from the $\eta_c\to K_S^0K^\pm\pi^\mp$ and $\eta_c\to 
	K^+K^-\pi^0$ MIPWA. Phases are determined relative to the $(K\pi\ 
	\mathcal{S}$-wave) $\bar K$ amplitude which is fixed to $\pi/2$ 
	at 1.45~GeV/c$^2$.}
\label{tab:tab1}
{\small
\begin{center}
\begin{tabular}{|l | r@{}c@{}r | r@{}c@{}r | r@{}c@{}r | r@{}c@{}r|}
\hline \\ [-2.3ex]
	  & \multicolumn{6}{c|} {$\eta_c\to K_S^0K^\pm\pi^\mp$} & \multicolumn{6}{c|}{$\eta_c\to K^+K^-\pi^0$}  \cr
\hline \\ [-2.3ex]
Amplitude & \multicolumn{3}{c|} {Fraction (\%)} & \multicolumn{3}{c|}{Phase (rad)} & \multicolumn{3}{c|}{Fraction (\%)}& \multicolumn{3}{c|}{Phase (rad)}\cr
\hline \\ [-2.3ex]
$(K\pi \ \mathcal{S}$-wave) $\bar K$ & 107.3 $\pm$  & \, 2.6 $\pm$  & \, 17.9 & &  fixed  &  & 125.5 $\pm$  & \, 2.4 $\pm$ & \, 4.2 & & fixed  &  \cr
$a_0(980)\pi$  & 0.8 $\pm$  & \, 0.5 $\pm$  & \, 0.8 & 1.08 $\pm$  & \, 0.18  $\pm$  & \, 0.18 & 0.0 $\pm$  & \, 0.1 $\pm$ & \, 1.7 &  & - & \cr
$a_0(1450)\pi$ & 0.7 $\pm$  & \, 0.2 $\pm$  & \, 1.4 & 2.63 $\pm$  & \, 0.13 $\pm$  & \, 0.17 & 1.2 $\pm$  & \, 0.4 $\pm$ & \, 0.7 & 2.90 $\pm$  & \, 0.12 $\pm$  & \, 0.25\cr
$a_0(1950)\pi$ & 3.1 $\pm$  & \, 0.4 $\pm$  & \, 1.2 & $-$1.04 $\pm$  & \, 0.08 $\pm$  & \, 0.77& 4.4 $\pm$  & \, 0.8 $\pm$ & \, 0.8& $-$1.45 $\pm$  & \, 0.08 $\pm$  & \, 0.27\cr
$a_2(1320)\pi$ & 0.2 $\pm$  & \, 0.1 $\pm$  & \, 0.1 & 1.85 $\pm$  & \, 0.20 $\pm$  & \, 0.20 & 0.6 $\pm$  & \, 0.2 $\pm$ & \, 0.3& 1.75 $\pm$  & \, 0.23 $\pm$  & \, 0.42\cr
$K^\ast_2(1430) \bar K$ & 4.7 $\pm$  & \, 0.9 $\pm$  & \, 1.4 & 4.92 $\pm$  & \, 0.05 $\pm$  & \, 0.10 & 3.0 $\pm$  & \, 0.8 $\pm$  & \, 4.4 & 5.07 $\pm$  & \, 0.09 $\pm$  & \, 0.30\cr
\hline \\ [-2.3ex]
Total & 116.8 $\pm$  & \, 2.8  $\pm$ & \, 18.1 &   & &    & 134.8 $\pm$  & \, 2.7 $\pm$ & \, 
6.4 & & &   \cr $\chi^2/N_{\rm cells}$ & \multicolumn{3}{c|} {301/254=1.17} & & & &  \multicolumn{3}{c|}{283.2/233=1.22} & & & \cr
\hline
\end{tabular}
\end{center}
}
\end{table}
%%%-----------------------------------

We use as figure of merit describing the fit quality the 2-D 
$\chi^2/N_{cells}$ on the Dalitz plot and obtain a good description of the 
data with $\chi^2/N_{\rm cells}=1.17$ and  $\chi^2/N_{\rm cells}=1.22$ for 
the  two $\eta_c$ decay modes.

In comparison, the isobar model gives a worse description of the data, with 
$\chi^2/N_{cells}=457/254=1.82$ and $\chi^2/N_{cells}=383/233=1.63$, 
respectively for the two $\eta_c$ decay modes. The resulting $K \pi$ 
$\mathcal{S}$-wave amplitude and phase for the two $\eta_c$ decay modes 
is shown in Fig.~\ref{fig:fig5}. We observe a clear $K^\ast_0(1430)$ 
resonance signal with the corresponding expected phase motion. At high 
mass we observe the presence of the broad $K^\ast_0(1950)$ 
contribution with good agreement between the two $\eta_c$ decay modes.
%----------------------------------
\begin{figure}[tbp]
    \centering
    \includegraphics[width=0.8\textwidth]{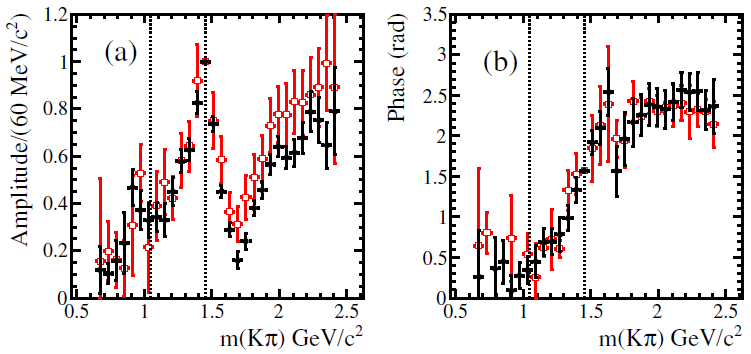}

    \caption{The $I=1/2$ $K \pi$ $\mathcal{S}$-wave amplitude (a) and
	phase (b) from $\eta_c\to K_S^0K^\pm\pi^\mp$ (solid (black)
	points) and $\eta_c\to K^+K^-\pi^0$ (open (red) points); only 
	statistical uncertainties are shown. The dotted lines indicate 
	the $K \eta$ and $K \eta'$ thresholds.} \label{fig:fig5}
\end{figure}
%%%-----------------------------------

Comparing with LASS~\cite{Aston:1987irf} and E791~\cite{Aitala:2005yhf} 
experiments we note that phases before the $K \eta'$ threshold are similar, 
as expected from Watson theorem~\cite{Watson:1952jif} but amplitudes are 
very different.

A preliminary K-matrix fit which include $K^-\pi^-\to K^-\pi^-$ 
$S$-wave data~\cite{lass2f}, LASS data and $\eta_c$ decays has been 
performed~\cite{ppf}, obtaining  a description of the data in  terms of 
three-poles:
\bea
{\rm Pole~1}\quad E_{P1}&=& \;\,659\, -\, i 302\,{\rm MeV\quad on\, 
	Sheet\, II},\\
	{\rm Pole~2}\quad E_{P2}&=& 1409\,-\,i 128 \,{\rm MeV\quad on\, Sheet\, 
	III},\\
	{\rm Pole~3}\quad E_{P3}&=& 1768\, -\, i 107\, {\rm MeV\quad on\, Sheet\, 
	III}.
\eea
Pole~1 is identified with the $\kappa$, the pole position of which was 
found to be at $[(658\pm 7)\,-\,i\ (278\pm 13)]$~MeV, in the dispersive 
analysis of Ref.~\cite{dispf}. Pole~2 is identified with $K^\ast_0(1430)$, 
to be compared with $[(1438 \pm 8 \pm 4)\,-\, i\ (105 \pm 20 \pm 12)]$~MeV 
using the Breit-Wigner form (Eq.~(3)). Pole~3 may be identified with the 
$K^\ast_0(1950)$ with a pole mass closer to that of the reanalysis of the 
LASS data from Ref.~\cite{anisovichf} with a pole at $E\,=\, (1820\pm 
20)\,-\,i (125\pm 50)$~MeV. For pole~2, the $K_0^\ast(1430)$, a ratio of 
$K\eta$/$K\pi$ decay rate of 0.05 is obtained, consistent with that 
reported in the present analysis (Eq.~(2)).

%%%-----------------------------------
\item \textbf{Dalitz Plot Analysis of $J/\psi\to K_S^0K^-\pi^\mp$}

We study the following reaction:
\begin{center}
	$e^+e^-\to\gamma_{\rm ISR} K_S^0K^\pm\pi^\mp$
\end{center}
where $\gamma_{\rm ISR}$ indicate the ISR photon~\cite{Lees:2017ndyf}.
Candidate events for this reaction are selected from the sample of events 
having exactly four charged tracks including the $K_S^0$ candidate. We 
compute $M_{rec} \equiv (p_{e-}+p_{e+}-p_{K_S^0}-p_K-p_\pi)^2$, which peaks 
near zero for ISR events. We select events in the ISR region by requiring 
$|M_{rec}^2|<1.5~GeV^2/c^4$ and obtain the $K_S^0K^\pm\pi^\mp$ mass spectrum 
shown in Fig.~\ref{fig:fig6} where a clean $J/\psi$ signal can be 
observed.
%----------------------------------
\begin{figure}[tbp]
    \centering
    \includegraphics[width=0.6\textwidth]{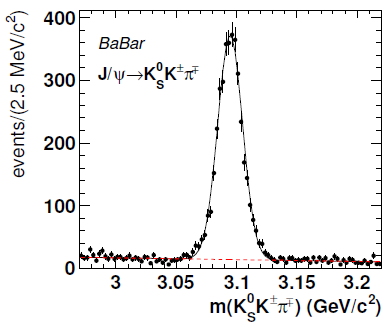}

    \caption{$K_S^0K^\pm\pi^\mp$ mass spectrum from ISR events.} 
	\label{fig:fig6}
\end{figure}
%%%-----------------------------------
We fit the $K_S^0K^\pm\pi^\mp$ mass spectrum using the Monte Carlo resolution 
functions described by a Crystal Ball+Gaussian function and obtain 3694 $\pm$ 64 
events with 93.1 $\pm$ 0.4 purity.
%----------------------------------
\begin{figure}[tbp]
    \centering
    \includegraphics[width=1\textwidth]{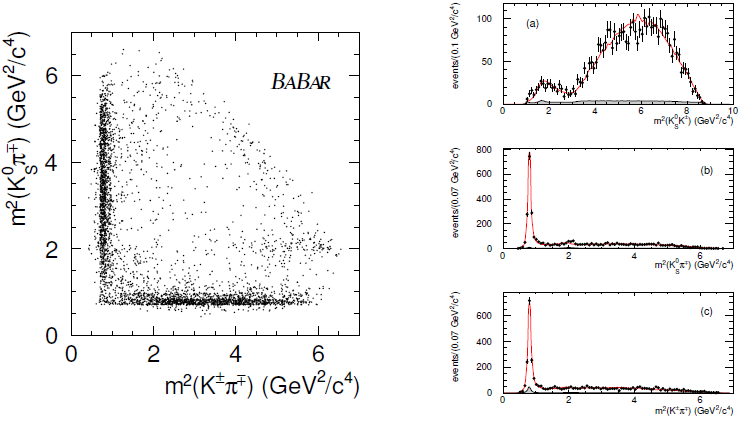}

    \caption{(Left) $J\psi\to K_S^0K^\pm\pi^\mp$ Dalitz plot.
	(Right) Dalitz plot projections with fit results for
	$J\psi\to K_S^0K^\pm\pi^\mp$. Shaded is the background 
	interpolated from $J\psi$ sidebands.} \label{fig:fig7}
\end{figure}
%%%-----------------------------------

%%%-----------------------------------

Figure~\ref{fig:fig7}(Left) shows the Dalitz plot for the $J/\psi$ signal region and 
Fig.~\ref{fig:fig7}(Right) shows the Dalitz plot projections. We perform the Dalitz 
plot analysis of $J\psi\to K_S^0K^|pm\pi^MP$ using the isobar model and express the 
amplitudes in terms of Zemach tensors~\cite{zemachf,dionisif}. We observe the 
following features:
\begin{itemize}
\item{} The decay is dominated by the $K^\ast(892) \bar K$, $K^\ast_2(1430) \bar K$, and 
	$\rho(1450)^{\pm}\pi^{\mp}$ amplitudes with a smaller contribution from the
	$K^\ast_1(1410) \bar K$ amplitude.
\item{} We obtain a significant improvement of the description of the data by leaving 
	free the $K^\ast(892)$ mass and width parameters and obtain
\begin{eqnarray}
	m(K^\ast(892)^+) = 895.6\pm 0.8~MeV/c^2, ~~~~~~ \Gamma(K^\ast(892)^+) = 43.6\pm 1.3~MeV,
\nonumber\\
	m(K^\ast(892)^0) = 898.1\pm 1.0~MeV/c^2, ~~~~~~ \Gamma(K^\ast(892)^0) = 52.6\pm 1.7~MeV.
\end{eqnarray}
\end{itemize}
The measured parameters for the charged $K^\ast(892)^+$ are in good agreement with those 
measured in $\tau$ lepton decays~\cite{pdgf}.

%%%-----------------------------------
\item \textbf{Acknowledgments}

This work was supported (in part) by the U.S. Department of Energy, Office
of Science, Office of Nuclear Physics under contract DE--AC05--06OR23177.
\end{enumerate}

%%%-----------------------------------

%%%%%%%%%%%%%%%%%%%%%%%%%%%%%%%%%%%%%%%%%%%%%%%%%%%%%%%%%%%%%%%%%%%%
\newpage
\subsection{Kaon and Light-Meson Resonances at COMPASS}
\addtocontents{toc}{\hspace{2cm}{\sl Boris Grube}\par}
\setcounter{figure}{0}  
\setcounter{table}{0}   
\setcounter{equation}{0}
\setcounter{footnote}{0}
\halign{#\hfil&\quad#\hfil\cr
\large{Boris Grube (for the COMPASS Collaboration)}\cr
\textit{Technical University Munich}\cr
\textit{85748 Garching, Germany}\cr}

%%%%-----------------------------------
\begin{abstract}
COMPASS is a multi-purpose fixed-target experiment at the CERN Super
Proton Synchrotron aimed at studying the structure and spectrum of
hadrons.  One of the main goals of the experiment is the study of
the light-meson spectrum.  In diffractive reactions with a 190~GeV
negative secondary hadron beam consisting mainly of pions and kaons,
a rich spectrum of isovector and strange mesons is produced.  The
resonances decay typically into multi-body final states and are
extracted from the data using partial-wave analysis techniques.

We present selected results of a partial-wave analysis of the
$K^-\pi^-\pi^+$ final state based on a data set of diffractive
dissociation of a 190~GeV $K^-$ beam impinging on a proton target.
This reaction allows us to study the spectrum of strange mesons up
to masses of about 2.5~GeV.  We also discuss a possible future
high-intensity kaon-beam experiment at CERN.
\end{abstract}

%%%-----------------------------------
\begin{enumerate}
\item \textbf{Introduction}

The excitation spectrum of light mesons is studied since many
decades but is still not quantitatively understood.  For higher
excited meson states experimental information is often scarce or
non-existent.  This is in particular true for the strange-meson
sector as illustrated by Fig.~\ref{fig:grube:kaon_spectrum}.  The
PDG~\cite{Patrignani:2016xqpe} lists only 25~kaon states below
3.1~GeV: 12~states that are considered well-known and established
and in addition 13~states that need confirmation.  Many higher
excited states that are predicted by quark-model calculation
(Fig.~\ref{fig:grube:kaon_spectrum} shows as an example the one from
Ref.~\cite{Ebert:2009ube}) have not yet been found by experiments.
In addition, for certain combinations of spin~$J$ and parity~$P$ the
quark model does not describe the experimental data well.  This is
most notably the case for the scalar kaon states with $J^P = 0^+$,
where the $K_0^\ast(800)$ seems to be a supernumerous state.
%----------------------------------
  \begin{figure}[tbp]
    \centering
    \includegraphics[width=0.75\textwidth]{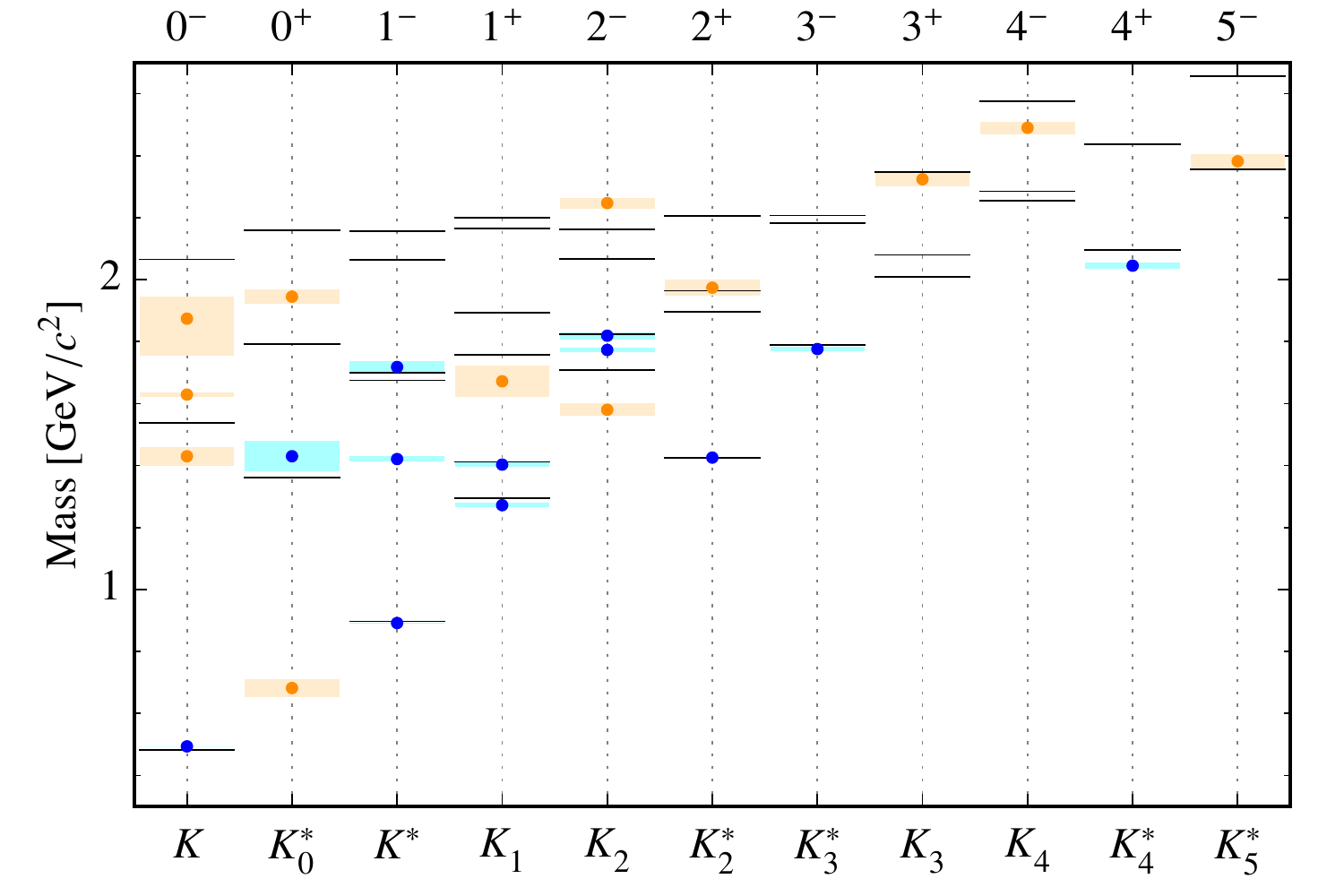}

    \caption{Strange-meson spectrum: Comparison of the measured masses
      of kaon resonances (colored points with boxes representing the
      uncertainty) with the result of a quark-model
      calculation~\protect\cite{Ebert:2009ube} (black lines).  States 
      that are considered well established by the 
      PDG~\protect\cite{Patrignani:2016xqpe}
      are shown in blue, states that need confirmation in orange.}
    \label{fig:grube:kaon_spectrum}
  \end{figure}
%-----------------------------------

In the last 30~years, little progress has been made on the
exploration of the kaon spectrum.  Since 1990, only four kaon states
were added to the PDG and only one of them to the summary table.
However, precise knowledge of the kaon spectrum is crucial to
understand the light-meson spectrum.  In particular, the
identification of supernumerous states that could be related to new
forms of matter beyond conventional quark-antiquark states---like
multi-quark states, hybrids, or glueballs---requires the observation
of complete SU(3)$_\text{flavor}$ multiplets.  The kaon spectrum
also enters in analyses that search for CP violation in multi-body
decays of $D$ and $B$ mesons, where kaon resonances appear in the
subsystems of various final states.

%%%-----------------------------------
\item \textbf{Diffractive Production of Kaon Resonances}

A suitable reaction to produce excited kaon states is diffractive
dissociation of a high-energy kaon beam, as it was already measured
in the past by the WA3 experiment at CERN (see,
e.g., Ref.~\cite{Daum:1981hbe}) and the LASS experiment at SLAC (see,
e.g., Refs.~\cite{Aston:1986jbe,Aston:1990yse}).  In these peripheral
reactions, the beam kaon scatters softly off the target particle and
is thereby excited into intermediate states, which decay into the
measured $n$-body hadronic final state.

These reactions were also measured by the COMPASS experiment using a
secondary hadron beam provided by the M2 beam line of the CERN SPS.
The beam was tuned to deliver negatively charged hadrons of 190~GeV
momentum passing through a pair of beam Cherenkov detectors (CEDARs)
for beam particle identification.  The beam impinged on a 40~cm long
liquid-hydrogen target with an intensity of $5 \times 10^7$
particles per SPS spill (10~s extraction with a repetition time of
about 45~s).  At the target, the hadronic component of the beam
consisted of 96.8\%~$\pi^-$, 2.4\%~$K^-$, and 0.8\%~$\bar{p}$.  At
the beam energy of 190~GeV, the reaction is dominated by Pomeron
exchange.  Elastic scattering at the target vertex was ensured by
measuring the slow recoil proton.  This leads to a minimum
detectable reduced squared four-momentum transfer $t'$ of about
0.07~$(\text{GeV})^2$.  By selecting the $K^-$ component of the beam
using the CEDAR information, we have studied diffractive production
of kaon resonances.  Charged kaons that appear in some of the
produced forward-going final states were separated from pions by a
ring-imaging Cherenkov detector (RICH) in the momentum range between
2.5~and 50~GeV.

%%%-----------------------------------
\item \textbf{Partial-Wave Analysis of $\bm{K^-\pi^-\pi^+}$ Final State}

In the 2008 and 2009 data-taking campaigns, COMPASS acquired a large
data sample on the diffractive dissociation reaction
\begin{equation}
    \label{eq:grube:reaction}
    K^- + p \to K^-\pi^-\pi^+ + p.
\end{equation}
The measurement is exclusive, i.e., all four final-state particles
are measured and energy-momentum conservation constraints are
applied in the event selection.  In
reaction~\eqref{eq:grube:reaction}, intermediate kaon resonances are
produced that decay into the 3-body $K^-\pi^-\pi^+$ final state.  A
first analysis of this reaction was performed based on a data sample
that corresponds to a fraction of the available data and consists of
about 270\,000~events with $K^-\pi^-\pi^+$ mass below 2.5~GeV and in
the range
$0.07 < t' <
0.7~(\text{GeV})^2$~\cite{phd_jasinskie,jasinski_hadron2011e}.  This
data sample is similar in size to the one of the WA3 experiment.
Fig.~\ref{fig:grube:kinematics} shows selected kinematic
distributions.  The distribution of the energy sum of the
forward-going particles peaks at the nominal beam energy.  The
non-exclusive background below the peak is small.  The invariant
mass distribution of the $K^-\pi^-\pi^+$ system and that of the
$K^-\pi^+$ and $\pi^-\pi^+$ subsystems exhibit peaks that correspond
to known resonances.  All three distribution are similar to the ones
obtained by WA3~\cite{Daum:1981hbe}.
%-------------------------------
  \begin{figure}[tbp]
    \centering
    \includegraphics[width=0.49\textwidth]{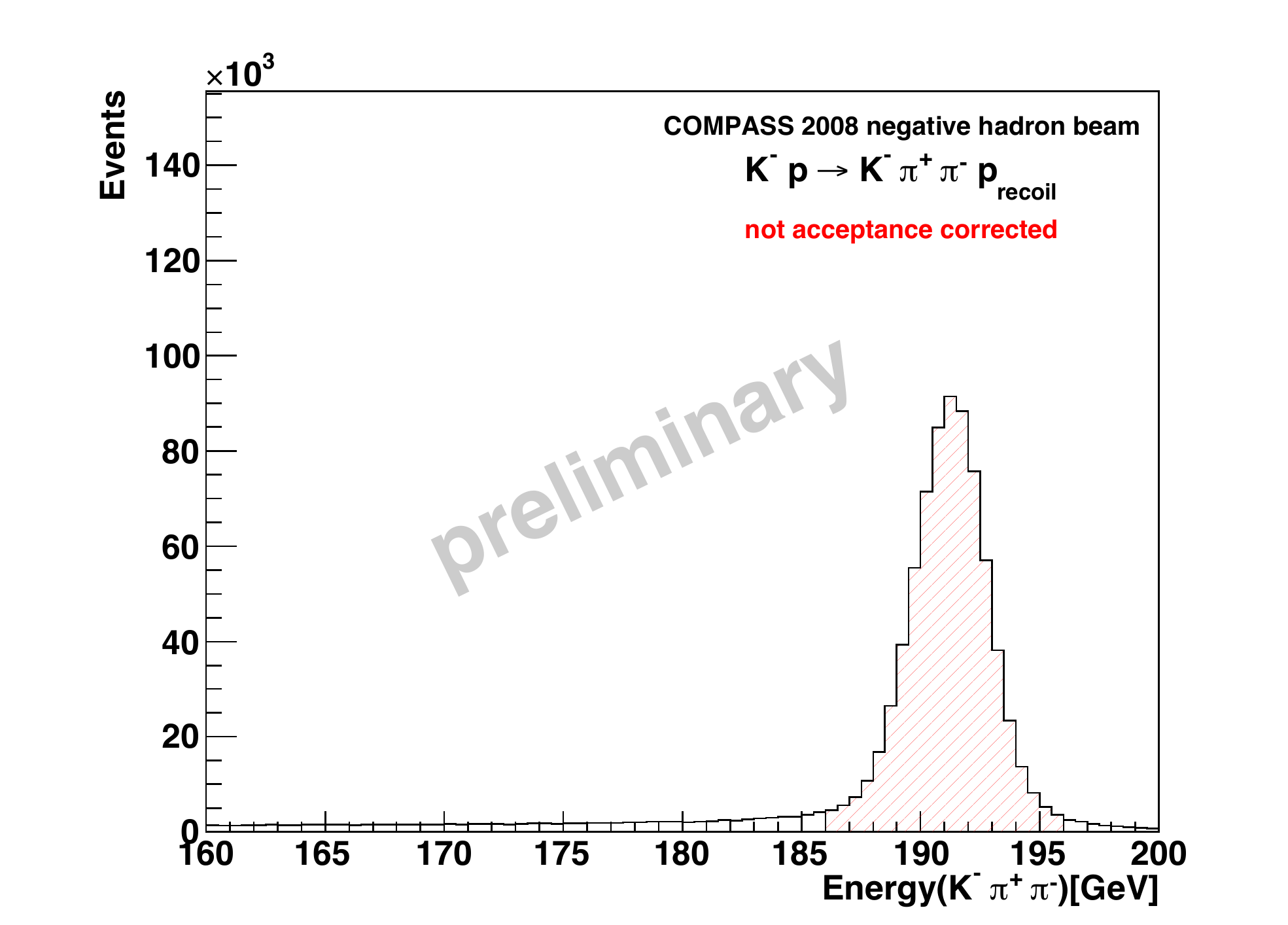}
    \includegraphics[width=0.49\textwidth]{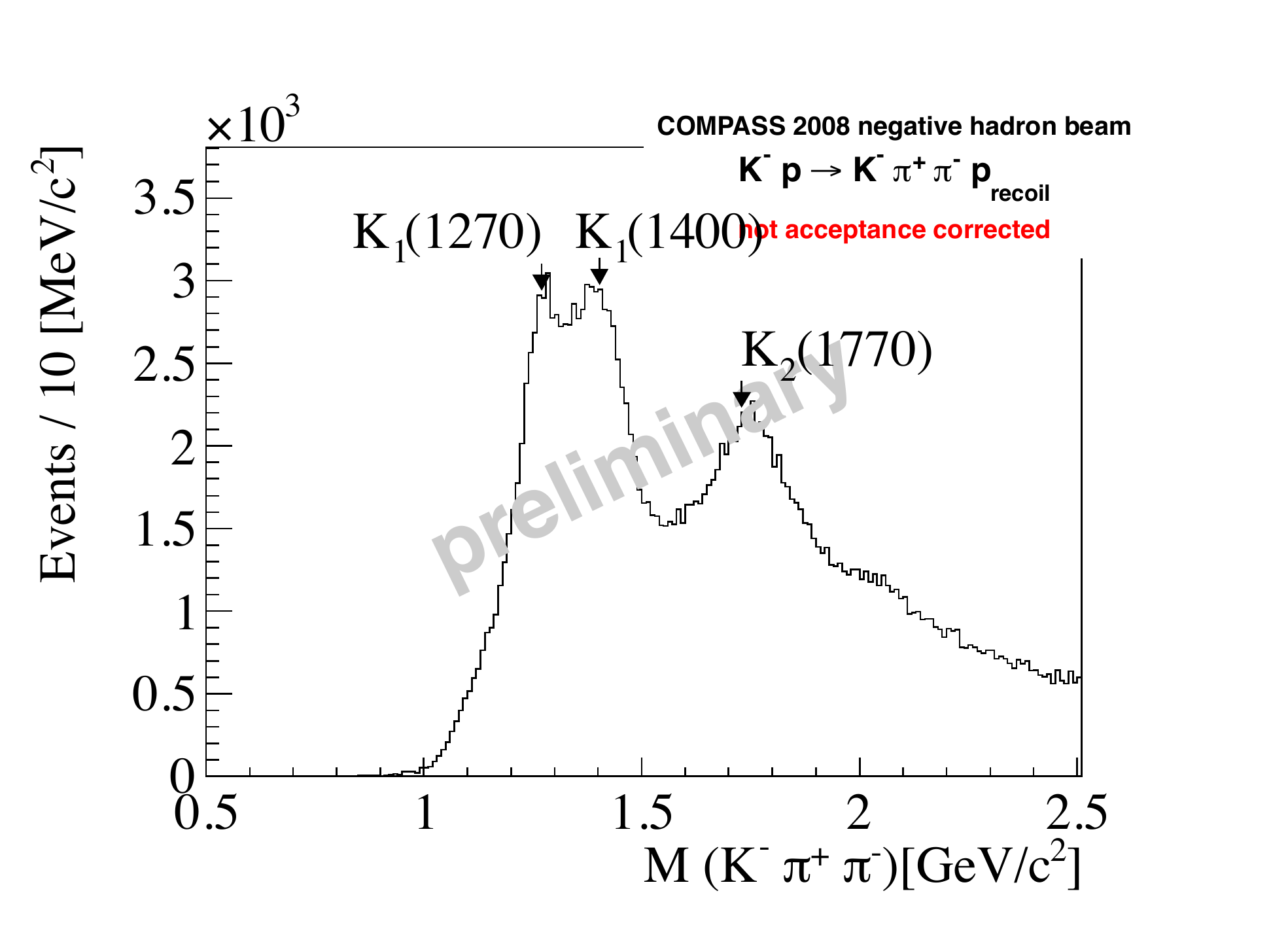} \\
    \includegraphics[width=0.49\textwidth]{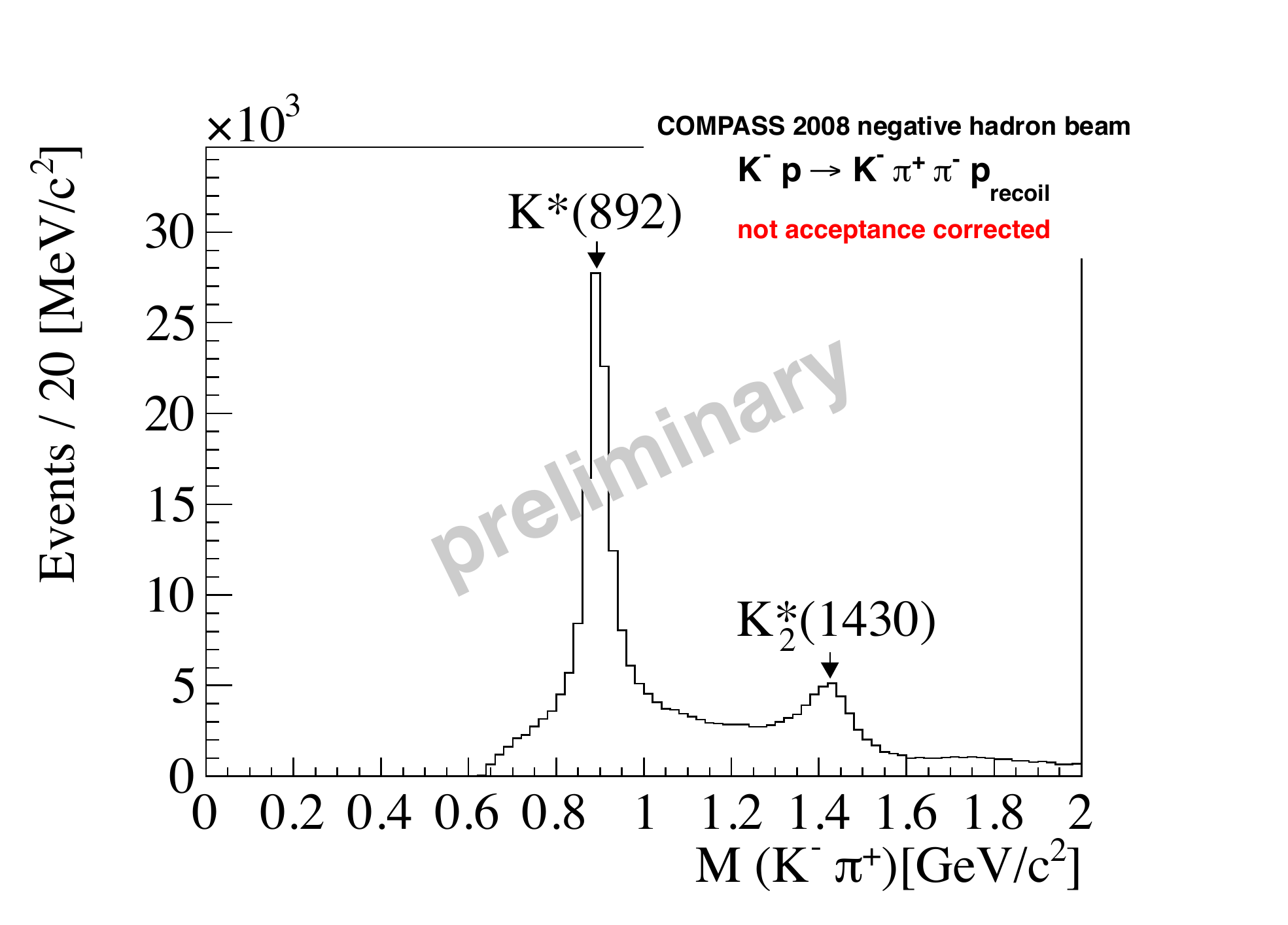}
    \includegraphics[width=0.49\textwidth]{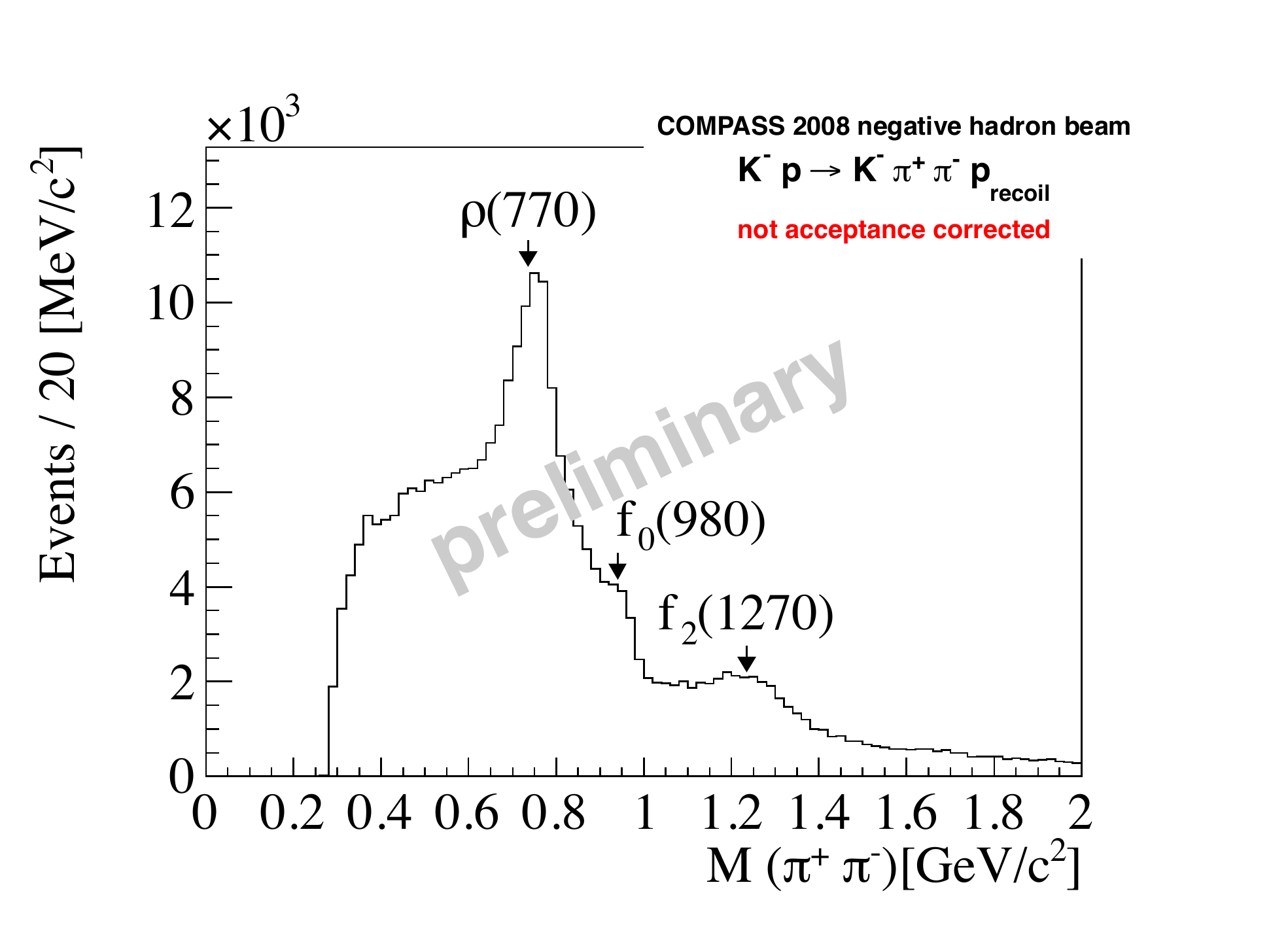}

    \caption{Kinematic distributions~\protect\cite{jasinski_hadron2011e}: 
      (Top left) distribution of the energy sum of the forward-going
      particles in reaction~\protect\eqref{eq:grube:reaction}, 
      (Top right) $K^-\pi^-\pi^+$ invariant mass distribution, 
      (Bottom left) $K^-\pi^+$ invariant mass distribution, and 
      (Bottom right) $\pi^-\pi^+$ invariant mass distribution.}
      \label{fig:grube:kinematics}
  \end{figure}
%-----------------------------------

As a first step towards a description of the measured
$K^-\pi^-\pi^+$ mass spectrum in terms of kaon resonances, we
performed a partial-wave analysis (PWA) using a model similar to the
one used by the ACCMOR collaboration in their analysis of the WA3
data~\cite{Daum:1981hbe}.  The PWA formalism is based on the isobar
model and is described in detail in Ref.~\cite{Adolph:2015tqae}.  The
PWA model takes into account three $K^-\pi^+$ isobars [$K_0^\ast(800)$,
$K^\ast(892)$, and $K_2^\ast(1430)$] and three $\pi^-\pi^+$ isobars
[$f_0(500)$, $\rho(770)$, and $f_2(1270)$].  Based on these isobars,
a wave set is constructed that consists of 19~waves plus an
incoherent isotropic wave, which absorbs intensity from events with
uncorrelated $K^-\pi^-\pi^+$, e.g., non-exclusive background.  A
partial-wave amplitude is completely defined by the spin~$J$,
parity~$P$, spin projection~$M^\varepsilon$, and the decay path of
the intermediate state.  The spin projection is expressed in the
reflectivity basis~\cite{Chung:1974fqe}, where $M \geq 0$ and
$\varepsilon = \pm 1$ is the naturality of the exchange particle.
Since the reaction is dominated by Pomeron exchange, all waves have
$\varepsilon = +1$.  We use the partial-wave notation
\wave{J}{P}{M}{\varepsilon}{\text{[isobar]}}{L}, where $L$ is the
orbital angular momentum between the isobar and the third
final-state particle.

Fig.~\ref{fig:grube:intensities} shows the intensity distributions
of selected waves.  The \wave{1}{+}{0}{+}{K^\ast(892)}{S}\ wave
intensity exhibits two clear peaks at the positions of the
$K_1(1270)$ and $K_1(1400)$.  We also see a clear peak of the
$K_2^\ast(1430)$ in the \wave{2}{+}{1}{+}{K^\ast(892)}{D}\ wave intensity.
However, there is no clear signal from the $K_2^\ast(1980)$.  The
\wave{2}{-}{0}{+}{K_2^\ast(1430)}{S}\ wave shows a broad bump in the
intensity distribution peaking slightly below 1.8~GeV that could be
due to the $K_2(1770)$ and/or $K_2(1820)$.  But also contributions
from $K_2(1580)$ and/or $K_2(2250)$ are not excluded.
%-----------------------------------
  \begin{figure}[tbp]
    \centering
    \includegraphics[width=0.49\textwidth]{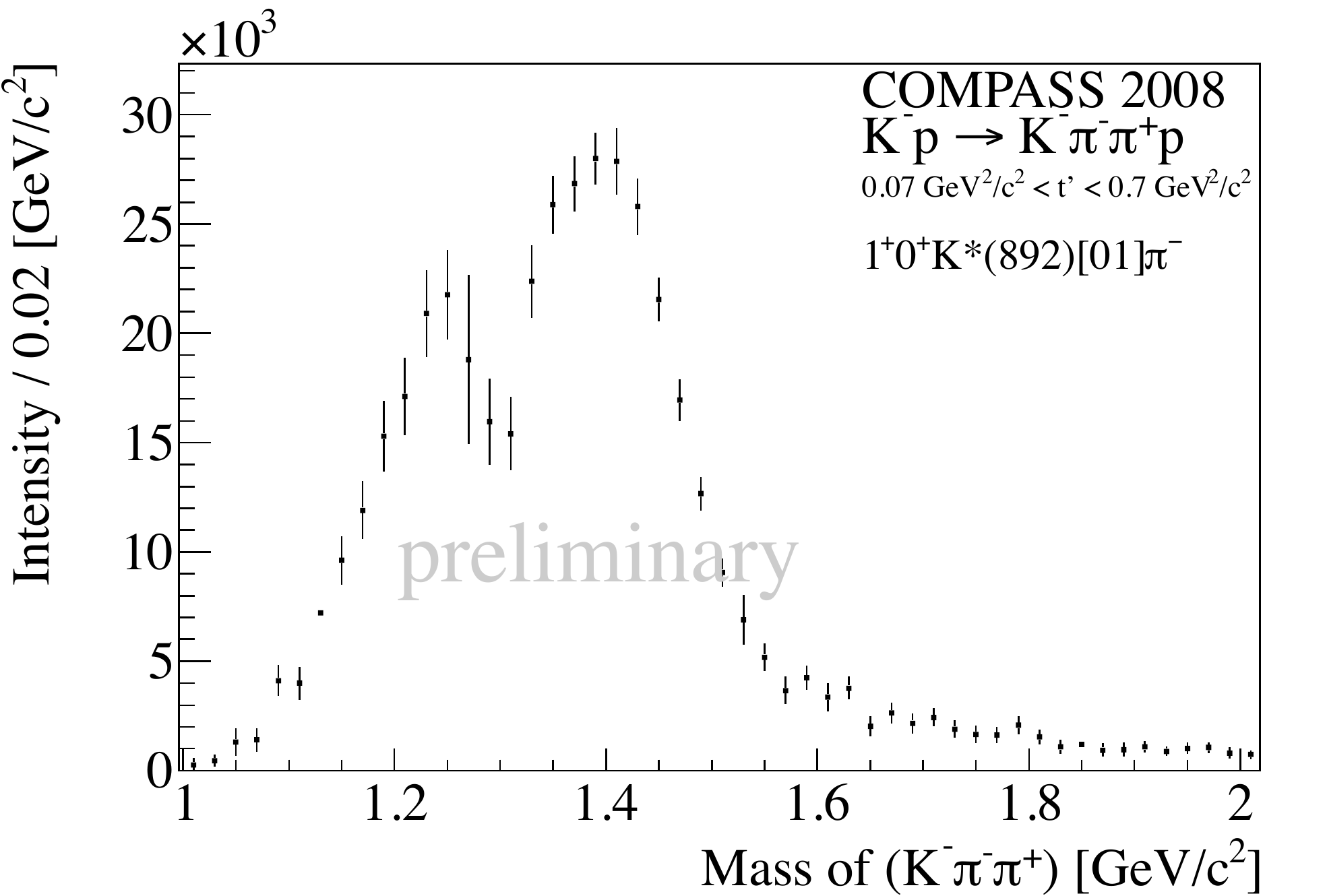}
    \includegraphics[width=0.49\textwidth]{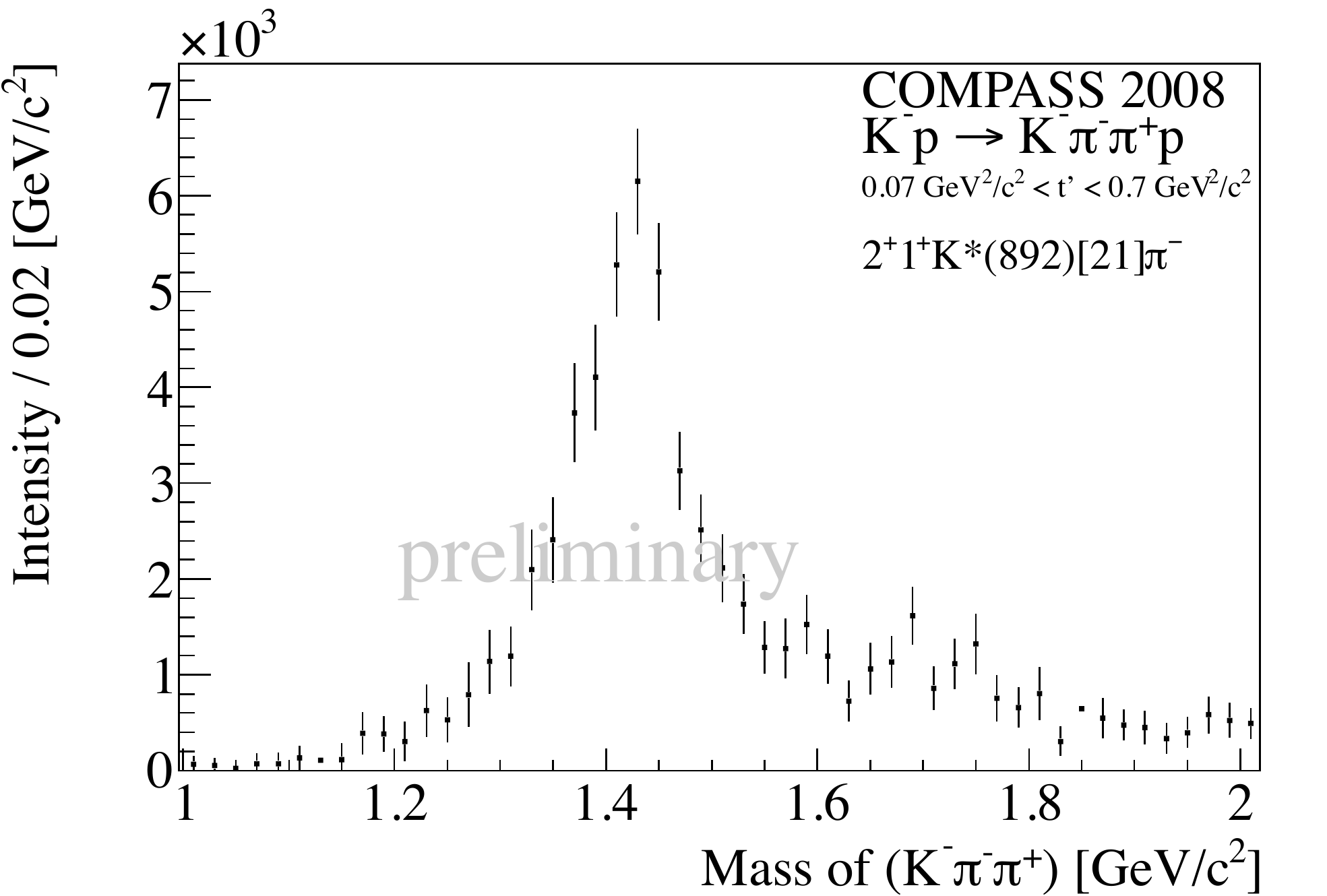} \\
    \includegraphics[width=0.49\textwidth]{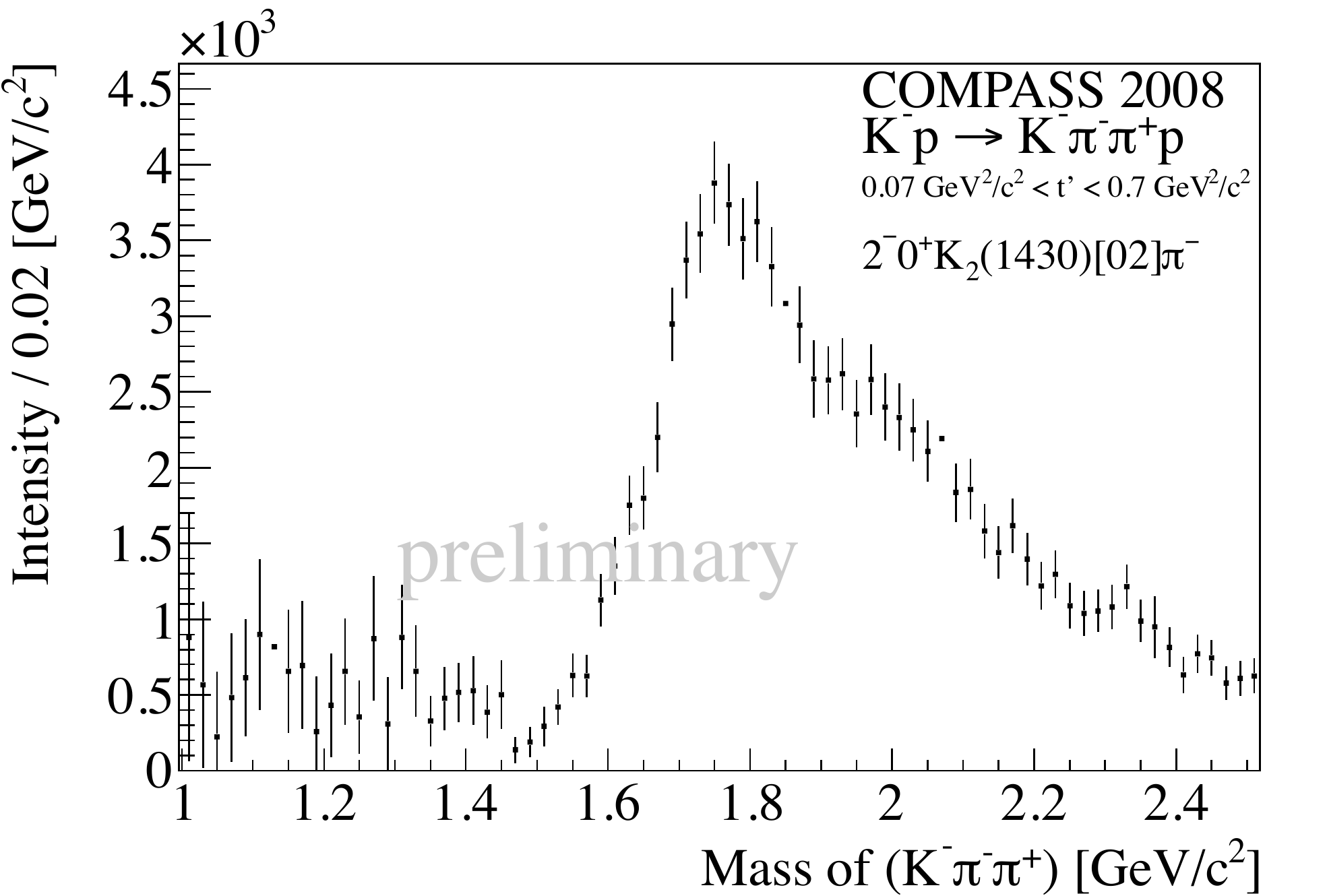}

    \caption{Intensities of selected waves as a function of the
      $K^-\pi^-\pi^+$ mass~\cite{jasinski_hadron2011e}: (Top left)
      \wave{1}{+}{0}{+}{K^\ast(892)}{S}\ wave, (Top right)
      \wave{2}{+}{1}{+}{K^\ast(892)}{D}\ wave, and (Bottom)
      \wave{2}{-}{0}{+}{K_2^\ast(1430)}{S}\ wave.}
    \label{fig:grube:intensities}
  \end{figure}
%--------------------------------

The analysis is currently work in progress.  With an improved beam
particle identification and event selection the full data sample
consists of about 800\,000~exclusive events, making it the world's
larges data set of this kind.  Also the PWA model will be improved
by using more realistic isobar parametrizations and parameters and
by including the $f_0(980)$ as an additional $\pi^-\pi^+$ isobar.
In order to extract kaon resonances and their parameters, we will
also perform a resonance-model fit similar to the one of the
$\pi^-\pi^-\pi^+$ final state in Ref.~\cite{Akhunzyanov:2018pnre}.

%%%-----------------------------------
\item \textbf{Possible Future Measurements with Kaon Beam}

The COMPASS collaboration has submitted a proposal for a future
fixed-target experiment within in the framework of CERN's ``Physics
beyond Colliders'' initiative.  Among other things, we propose to
perform a high-precision measurement of the kaon spectrum using a
high-energy kaon beam similar to the COMPASS measurements in 2008
and 2009.  The goal of this experiment would be to acquire a
high-precision data set that is at least 10~times larger than that
of COMPASS.  Such a large data set would allow us not only to search
for small signals but also to apply the novel analysis techniques
that we developed for the analysis of the COMPASS $\pi^-\pi^-\pi^+$
data sample, which consists of $46 \times 10^6$
events~\cite{Adolph:2015tqae,Akhunzyanov:2018pnre,Krinner:2017dbae}.
In particular, we could study in detail the amplitude of the scalar
$K^-\pi^+$ subsystem with $J^P = 0^+$ in the the $K^-\pi^-\pi^+$
final state as a function of the $K^-\pi^+$ mass, the
$K^-\pi^-\pi^+$ mass, and the quantum numbers of the $K^-\pi^-\pi^+$
system, similar to the analysis of the scalar $\pi^-\pi^+$ subsystem
in the $\pi^-\pi^-\pi^+$ final state in Ref.~\cite{Adolph:2015tqae}.
With these data, one could learn more about the scalar kaon states.
The PWA could also be performed in bins of the reduced squared
four-momentum $t'$ in order to extract the $t'$~dependence of the
resonant and non-resonant partial-wave components like it was done
for the $\pi^-\pi^-\pi^+$ final state in
Ref.~\cite{Akhunzyanov:2018pnre}.  This gives information about the
production processes.

To obtain such a high-precision data set for kaon spectroscopy, the
rate of beam kaons on the target must be increased with respect to
the COMPASS measurements in 2008 and 2009.  This can be achieved by
increasing the kaon fraction in the beam using RF-separation
techniques~\cite{Panofsky:1956jge,Bernard:1968gge} similar to the ones
that have already been used in the past at CERN~\cite{Citron:1978sre}
(see Fig.~\ref{fig:grube:rf_separation}).  First preliminary
estimates for the M2~beam line at CERN show that a high-energy kaon
beam with a kaon rate of about $3.7 \times 10^6$~s$^{-1}$ seems to
be feasible~\cite{pbc_gatignone,pbc_bernharde}.  This would correspond
to about 10~to $20 \times 10^6$ $K^-\pi^-\pi^+$ events within a year
of running.  However, more detailed feasibility studies are still
needed.
%----------------------------------------
\begin{figure}[tbp]
    \centering
    \includegraphics[width=\textwidth]{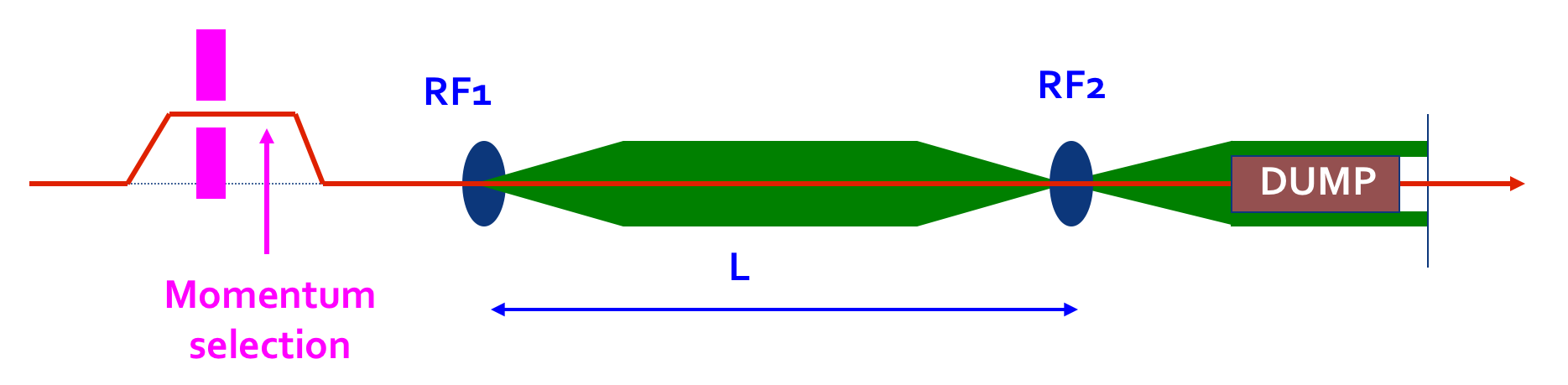}

    \caption{Principle of an RF-separated beam~\protect\cite{pbc_gatignone}: A
      momentum-selected beam passes through two RF-cavities that act
      as a time-of-flight selector.  The first RF-cavity deflects the
      beam.  This deflection is cancelled or enhanced by the second
      RF-cavity, depending on the particle type.  Hence particles can
      be selected in the plane transverse to the beam by putting
      absorbers or collimators.}
    \label{fig:grube:rf_separation}
\end{figure}

%%%-----------------------------------
\newpage
\item \textbf{Acknowledgments}

This work was supported by the BMBF, the Maier-Leibnitz-Laboratorium 
(MLL), the DFG Cluster of Excellence Exc153 ``Origin and Structure of 
the Universe,'' and the computing facilities of the Computational 
Center for Particle and Astrophysics (C2PAP). 
\end{enumerate}

%%%-----------------------------------

%%%%%%%%%%%%%%%%%%%%%%%%%%%%%%%%%%%%%%%%%%%%%%%%%%%%%%%%%%%%%%%%%%%%%%%%%
\newpage
\subsection{Recent Belle Results Related to $\pi-K$ Interactions}
\addtocontents{toc}{\hspace{2cm}{\sl Bilas Pal}\par}
\setcounter{figure}{0}
\setcounter{table}{0}
\setcounter{equation}{0}
\setcounter{footnote}{0}
\halign{#\hfil&\quad#\hfil\cr
\large{Bilas Pal (for the Belle Collaboration)}\cr
\textit{Brookhaven National Laboratory}\cr
\textit{Upton, NY 11973, U.S.A.~\&}\cr
\textit{University of Cincinnati}\cr
\textit{Cincinnati, OH, U.S.A}\cr}

%%%-----------------------------------
\begin{abstract}
We report the recent results related to  $\pi-K$ interactions
based on the data collected by the Belle experiment at the KEKB
collider. This includes the branching fraction and $CP$
asymmetry measurements of $B^+\to K^+K^-\pi^+$ decay, search for
the $\Lambda_c^+\to\phi p\pi^0$, $\Lambda_c^+\to P_s^+\pi^0$
decays, branching fraction measurement of $\Lambda_c^+\to
K^-\pi^+p\pi^0$, first observation of  doubly Cabibbo-suppressed
decay $\Lambda_c^+\to K^+\pi^-p$, and the measurement of CKM
angle $\phi_3$ ($\gamma$) with a model-independent Dalitz plot
analysis of $B^{\pm}\to DK^{\pm},D\to K_S^0\pi^+\pi^-$ decay.
\end{abstract}

%%%-----------------------------------
\begin{enumerate}
\item \textbf{Introduction}

In this report, we present some recent results related to $\pi-K$
interactions based on the  data, collected by the Belle experiment
at the KEKB $e^+e^-$ asymmetric-energy collider~\cite{KEKBb}.
(Throughout this paper charge-conjugate modes are implied.)
The experiment took data at center-of-mass energies corresponding
to several $\Upsilon(nS)$ resonances; the total data sample
recorded exceeds $1~{\rm ab}^{-1}$.

The Belle detector is a large-solid-angle magnetic spectrometer that 
consists of a silicon vertex detector (SVD), a 50-layer central drift 
chamber (CDC), an array of aerogel threshold Cherenkov counters
(ACC), a barrel-like arrangement of time-of-flight scintillation 
counters (TOF), and an electromagnetic calorimeter comprised of 
CsI(Tl) crystals (ECL) located inside a super-conducting solenoid
coil that provides a 1.5~T magnetic field. An iron flux-return 
located outside of the coil is instrumented to detect $K^0_L$
mesons and to identify muons (KLM). The detector is described in 
detail elsewhere~\cite{Belleb,svd2b}.

%%%-----------------------------------
\item \textbf{$CP$ Asymmetry in $B^+\to K^+K^-\pi^+$ Decays}

In the recent years, an unidentified structure has been observed by
BaBar~\cite{Aubert:2007xb} and LHCb 
experiments~\cite{Aaij:2013blab,Aaij:2014ivab} in the low $K^+K^-$ 
invariant mass spectrum of the $B^+\to K^+K^-\pi^+$ decays. The 
LHCb reported a nonzero inclusive $CP$ asymmetry of $-0.123\pm0.017
\pm0.012\pm0.007$ and a large unquantified local $CP$ asymmetry in 
the same mass region. These results suggest that final-state 
interactions may contribute to $CP$ 
violation~\cite{Bhattacharya:2013boab, Bediaga:2013elab}. In 
this analysis, we attempt to quantify the $CP$ asymmetry and 
branching fraction as a function of the $K^+K^-$ invariant mass, 
using $711~{\rm fb^{-1}}$ of data, collected at $\Upsilon(4S)$ 
resonance~\cite{Hsu:2017kirb}.

The signal yield is extracted by performing a two-dimensional unbinned 
maximum likelihood fit to the variables: the beam-energy constrained 
mass $M_{\rm bc}$ and the energy difference $\Delta E$. The resulting 
branching fraction and $CP$ asymmetry are
\begin{eqnarray*}
	\mathcal{B}(B^+\to K^+K^-\pi^+)&=&(5.38\pm0.40\pm0.35)\times10^{-6},\\
	A_{CP}&=&-0.170\pm0.073\pm0.017,
\end{eqnarray*}
where the quoted uncertainties are statistical and systematic, 
respectively.

To investigate the localized $CP$ asymmetry in the low $K^+K^-$
invariant mass region, we perform the 2D fit (described above) to 
extract the signal yield and $A_{CP}$ in bins of $M_{K^+K^-}$. The 
fitted results are shown in Fig.~\ref{fig:cp_all} and 
Table~\ref{tab:cp_all}. We confirm the excess and local $A_{CP}$ in the 
low $M_{K^+K^-}$ region, as reported by the LHCb, and quantify the 
differential branching fraction in each $K^+K^-$ invariant mass bin. We 
find a 4.8$\sigma$ evidence for a negative $CP$ asymmetry in the region 
$M_{K^+K^-}<1.1$ GeV/$c^2$. To understand the origin of the low-mass 
dynamics, a full Dalitz analysis from experiments with a sizeable
data set, such as LHCb and Belle II, will be needed in the future.
%------------------------------------
\begin{figure}[htb]
%\vskip -0.3cm
\centering
	\includegraphics[width=0.7\textwidth]{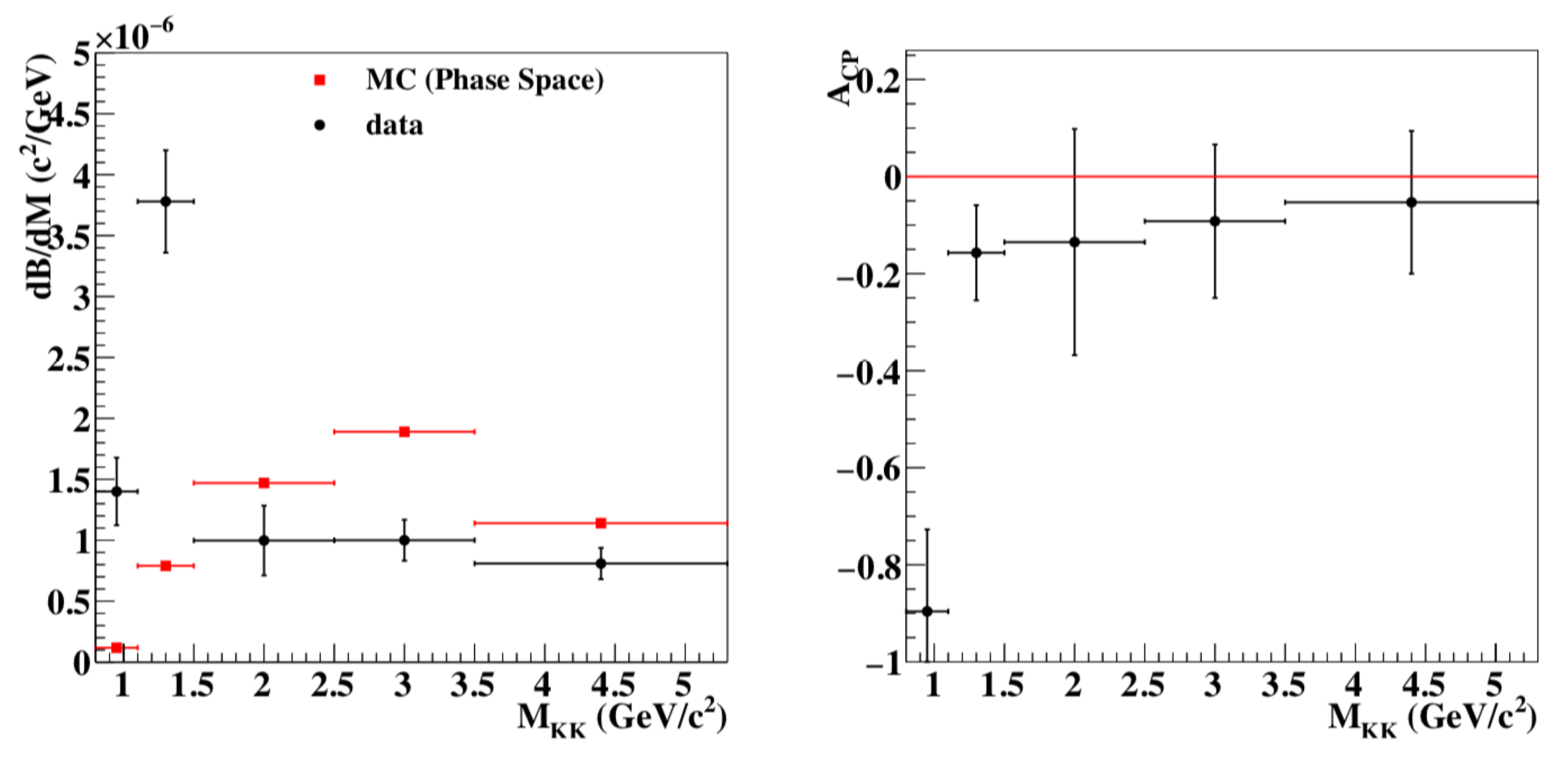}
	\vskip -0.35cm

	\caption{\small  Differential branching fractions~(left) and measured 
	$A_{CP}$~(right) as a function of $M_{K^+K^-}$. Each point is 
	obtained from a two-dimensional fit with systematic uncertainty 
	included. Red squares with error bars in the left figure show 
	the expected signal distribution in a three-body phase space 
	MC. Note that the phase space hypothesis is rescaled to the 
	total observed $K^+K^-\pi^+$ signal yield.} \label{fig:cp_all}
\end{figure}
%------------------------------------
%------------------------------------
\begin{table}[h!t!p!]
\centering

	\caption{\small Differential branching fraction, and $A_{CP}$ for 
	individual $M_{K^+K^-}$ bins. The first uncertainties are 
	statistical and the second systematic.}
\vspace{0.2cm}
\begin{tabular}{c|cc}
\hline\hline
$M_{K^+K^-}$ & $d\mathcal{B}/dM (\times 10^{-7})$ & $A_{CP}$\\
\hline
0.8--1.1 & $14.0\pm2.7\pm0.8$ & $-0.90\pm0.17\pm0.04$\\
1.1--1.5 & $37.8\pm3.8\pm1.9$ & $-0.16\pm0.10\pm0.01$\\
1.5--2.5 & $10.0\pm2.3\pm1.7$ & $-0.15\pm0.23\pm0.03$\\
2.5--3.5 & $10.0\pm1.6\pm0.6$ & $-0.09\pm0.16\pm0.01$\\
3.5--5.3 & $8.1\pm1.2\pm0.5$   & $-0.05\pm0.15\pm0.01$\\
\hline    \hline
\end{tabular}
\label{tab:cp_all}
\end{table}

%------------------------------------
\item \textbf{Search for \mbox{\boldmath$\Lambda_c^+\to\phi p \pi^0$} and 
	Branching Fraction Measurement of  \mbox{\boldmath$\Lambda_c^+\to 
	K^-\pi^+ p \pi^0$} }

The story of exotic hadron spectroscopy begins with the discovery  
of the $X(3872)$ by the Belle collaboration in 2003~\cite{Choi:2003ueb}. 
Since then, many exotic $X\!Y\!Z$ states have been reported by Belle 
and other experiments~\cite{Agashe:2014kdab}. Recent observations of 
two hidden-charm pentaquark  states $P_c^+(4380)$ and $P_c^+(4450)$ 
by the LHCb collaboration in the $J/\psi p$ invariant mass spectrum 
of the $\Lambda_b^0\to J/\psi pK^-$ process~\cite{Aaij:2015tgab} raises 
the question of whether a hidden-strangeness pentaquark $P_s^+$, where 
the $c\bar{c}$ pair in $P_c^+$ is replaced by an $s\bar{s}$ pair,
exists~\cite{Kopeliovich:2015vqab,Zhu:2015bbab,Lebed:2015dcab}. The 
strange-flavor analogue of the $P_c^+$ discovery channel is the decay 
$\Lambda_c^+\to\phi p\pi^0$~\cite{Kopeliovich:2015vqab,Lebed:2015dcab}, 
shown in Fig.~\ref{fig:Feynman} (a). The detection of a 
hidden-strangeness pentaquark could be possible through  the $\phi p$ 
invariant mass spectrum within this channel [see Fig.~\ref{fig:Feynman} 
(b)] if the underlying mechanism creating the $P_c^+$ states also 
holds for $P_s^+$, independent of the flavor~\cite{Lebed:2015dcab}, and 
only if the mass of $P_s^+$ is less than $M_{\Lambda_c^+}-M_{\pi^0}$.
In an analogous $s\bar{s}$ process of $\phi$ photoproduction $(\gamma 
p\to\phi p)$,  a forward-angle  bump structure at $\sqrt{s}\approx2.0$~GeV
has been observed by  the LEPS~\cite{Mibe:2005erb} and CLAS 
collaborations~\cite{Dey:2014tfab}.  However, this structure appears only 
at the most forward angles, which is  not   expected for  the decay of a 
resonance~\cite{Lebed:2015fpab}.
%%%-----------------------------------
\begin{figure}[htb]
%\vskip -0.3cm
\centering
	\includegraphics[width=0.30\textwidth]{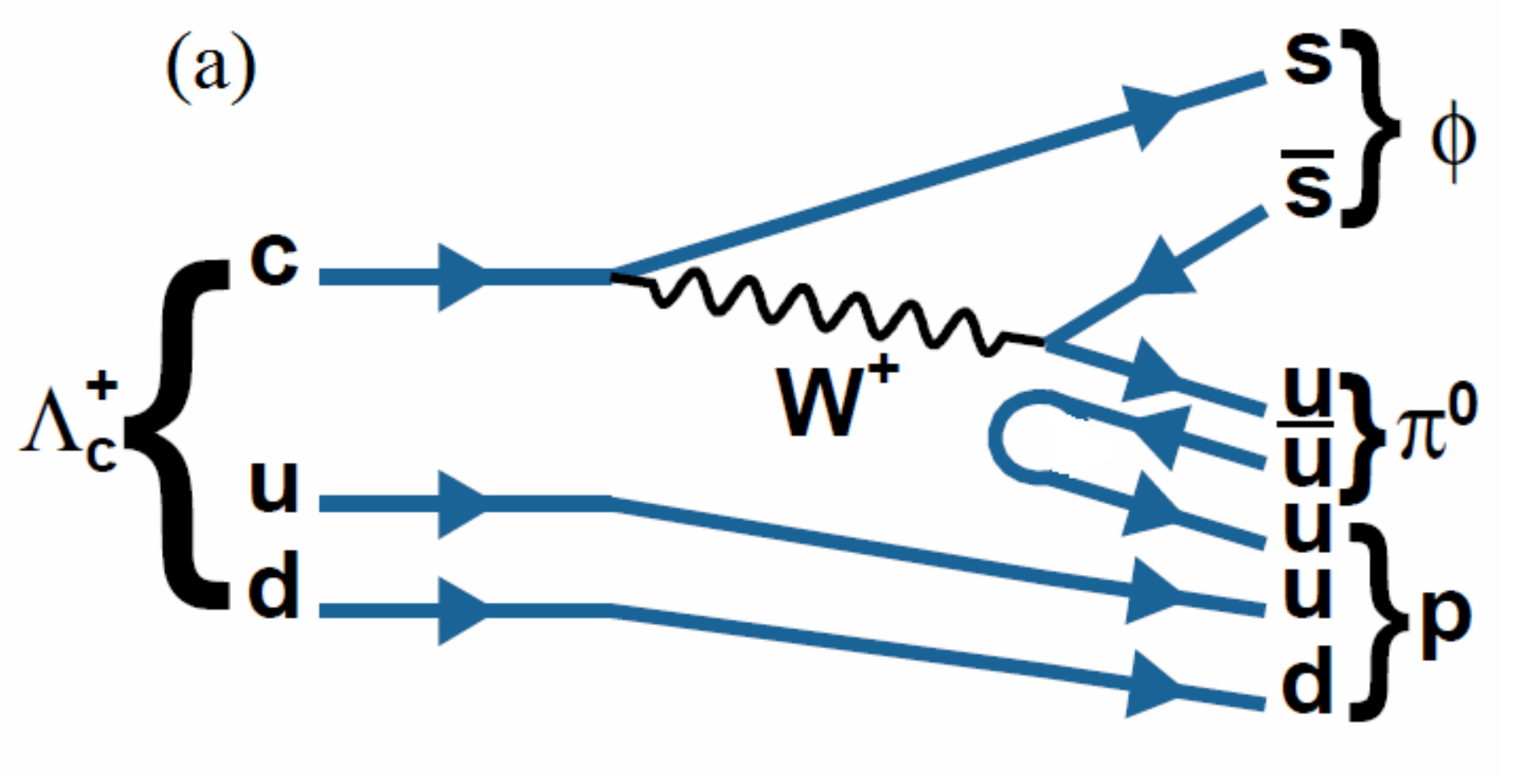}%	
	\includegraphics[width=0.30\textwidth]{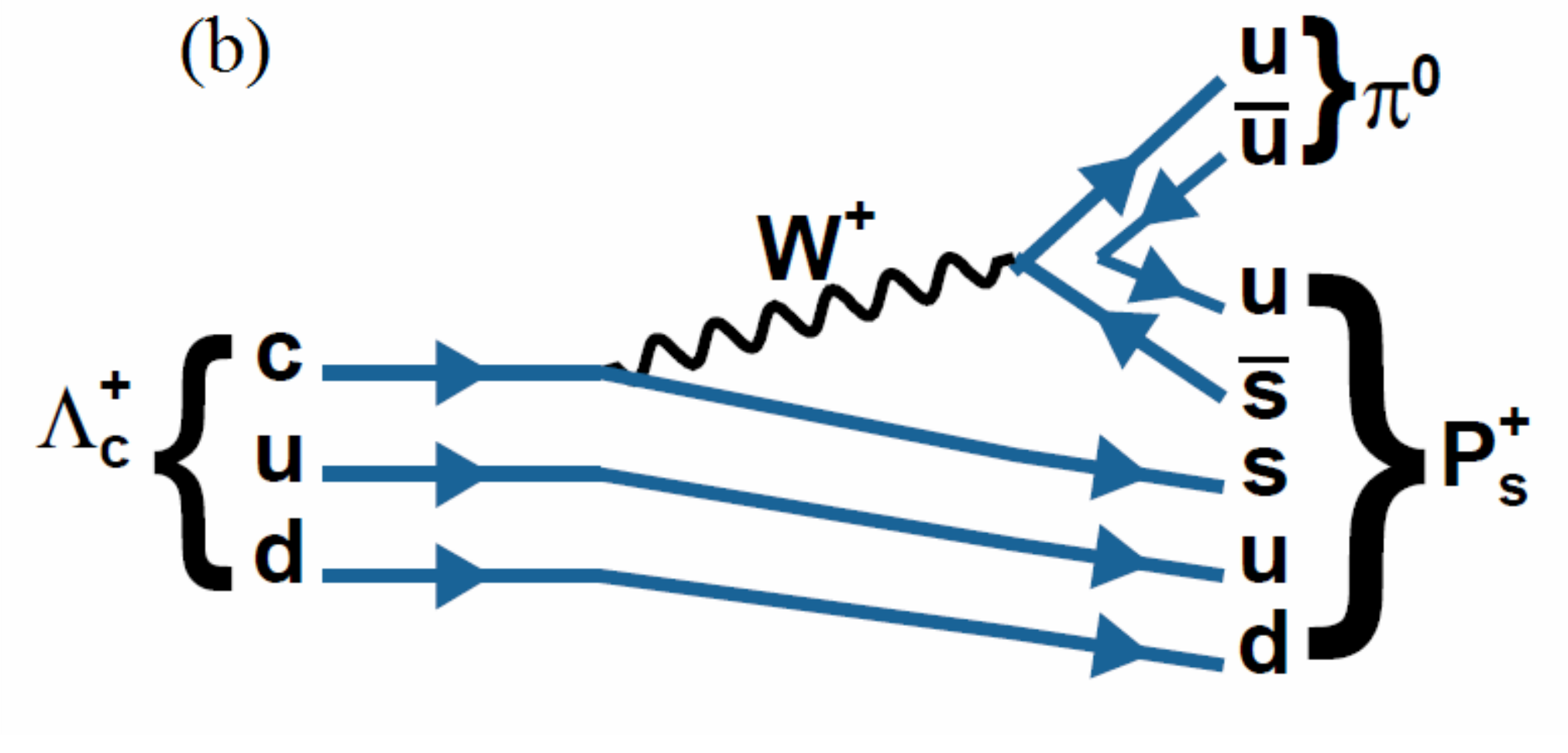}
\vskip -0.3cm

	\caption{\small Feynman diagram for the decay (a) $\Lambda_c^+\to\phi 
	p\pi^0$ and (b) $\Lambda_c^+\to P_s^+\pi^0$.}
	\label{fig:Feynman}
\end{figure}
%%%-----------------------------------

Previously, the decay $\Lambda_c^+\to\phi p\pi^0$ has not been studied 
by any experiment. Here, we report a search for this decay, using 915 
$\rm fb^{-1}$ of data~\cite{Pal:2017yppb}. In addition, we search for 
the nonresonant decay $\Lambda_c^+\to K^+K^-p\pi^0$ and measure the 
branching fraction of the  Cabibbo-favored decay $\Lambda_c^+\to 
K^-\pi^+p\pi^0$.

In order to extract the signal yield, we perform a two-dimensional 
(2D) unbinned extended maximum likelihood fit to the variables $m 
(K^+K^-p\pi^0)$ and $m(K^+K^-)$.  Projections of the fit result are 
shown in Fig.~\ref{fig:2dfit}.  From the fit, we extract $148.4\pm61.8$ 
signal events, $75.9\pm84.8$ nonresonant events, and $7158.4\pm36.4$ 
combinatorial background events.  The statistical significances are 
found to be 2.4 and 1.0 standard deviations for $\Lambda_c^+\to\phi p 
\pi^0$ and nonresonant $\Lambda_c^+\to K^+K^- p \pi^0$ decays, 
respectively. We use the well-established decay $\Lambda_c^+\to p 
K^-\pi^+$~\cite{Agashe:2014kdab} as the normalization channel for the 
branching fraction measurements.
%%%-----------------------------------
\begin{figure}[h!tp]
\begin{center}
    \includegraphics[width=0.44\textwidth]{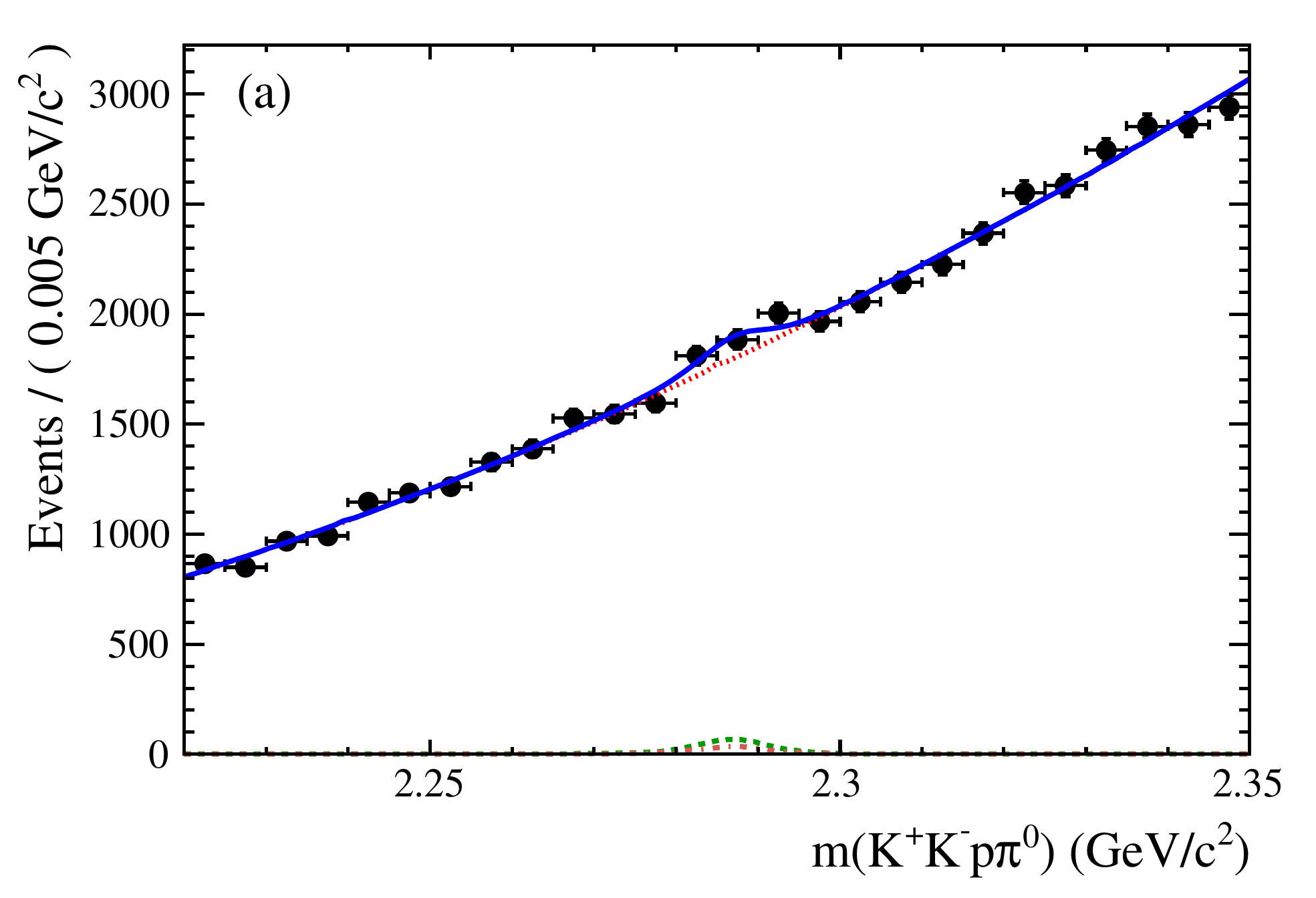}%
    \includegraphics[width=0.44\textwidth]{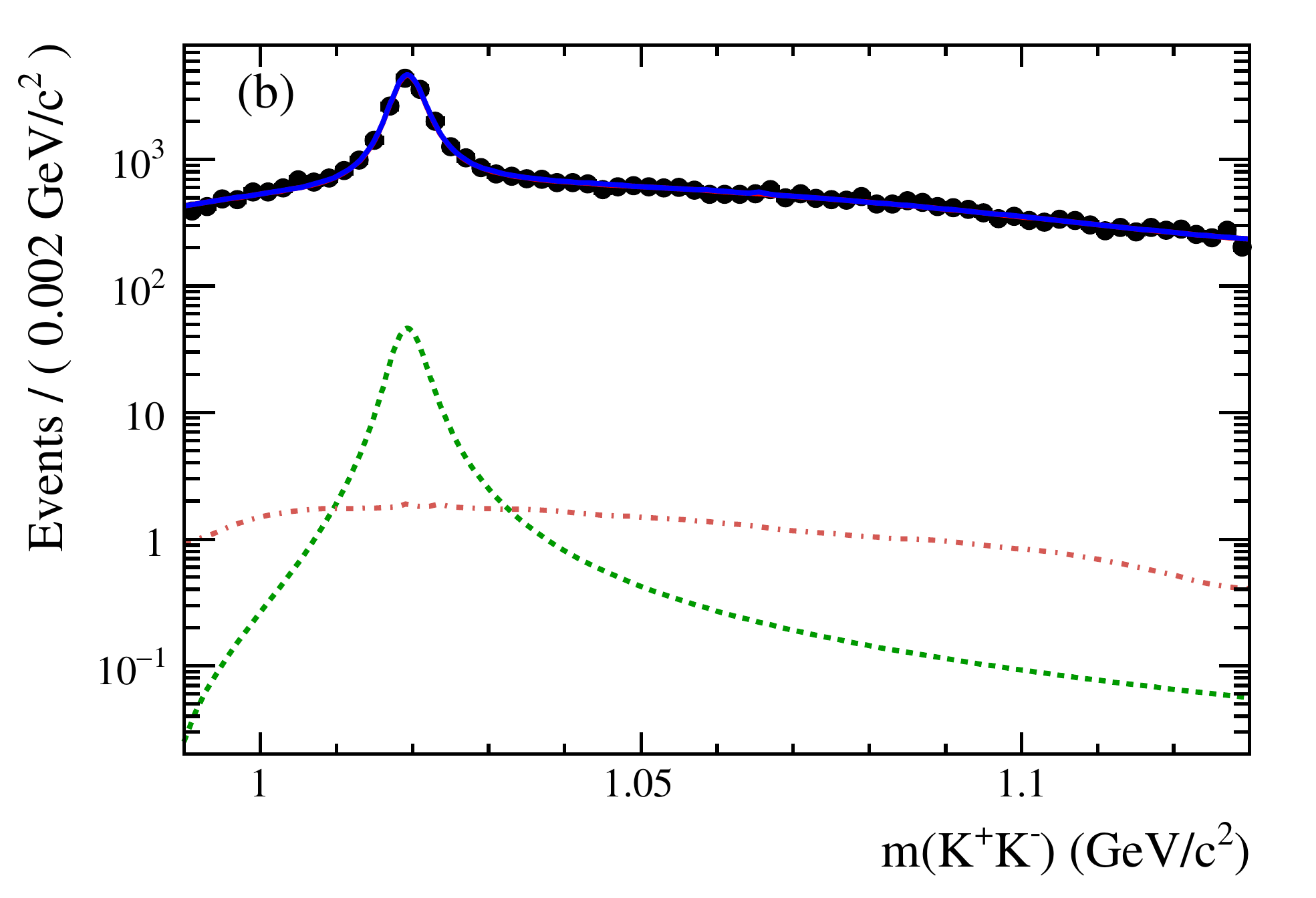}
\end{center}
\vskip -0.75cm

\caption{\small Projections of the 2D fit: (a) $m(K^+K^- p\pi^0)$ and 
	(b) $m(K^+K^-)$.  The points with the error bars are the  data, 
	and the (red) dotted, (green) dashed and (brown) dot-dashed 
	curves represent the combinatorial,  signal and nonresonant 
	candidates, respectively, and (blue) solid curves represent 
	the total PDF. The solid curve in (b) completely overlaps the 
	curve for the combinatorial background.} \label{fig:2dfit}
\end{figure}
%%%-----------------------------------

Since the significances are below 3.0 standard deviations both for $\phi 
p\pi^0$ signal and $K^+K^-p\pi^0$ nonresonant decays, we set upper limits 
on their branching fractions  at 90\% confidence level (CL) using a 
Bayesian approach. The results are
\begin{eqnarray*}
	\mathcal{B}(\Lambda_c^+\to \phi p\pi^0) &<& 15.3\times10^{-5} ,\\
	\mathcal{B}(\Lambda_c^+\to K^+K^-p\pi^0)_{\rm NR} &<&6.3\times10^{-5} ,
\end{eqnarray*}
which are the first limits on these branching fractions.

To search for a putative $P_s^+\to\phi p$ decay, we select $\Lambda_c^+\to 
K^+K^-p\pi^0$ candidates in which $m(K^+K^-)$ is within 0.020~GeV/$c^2$ of 
the $\phi$ meson mass~\cite{Agashe:2014kdab} and plot the  
background-subtracted $m(\phi p)$ distribution (Fig.~\ref{fig:bkg-sub_dis}). 
This distribution is obtained by performing 2D fits as discussed above in 
bins of $m(\phi p)$.  The data shows no clear evidence for a $P_s^+$ state.
We set an upper limit on the product branching fraction 
$\mathcal{B}(\Lambda_c^+\to P_s^+\pi^0) \times \mathcal{B}(P_s^+\to \phi p)$ 
by fitting the distribution of Fig.~\ref{fig:bkg-sub_dis} to the sum of a RBW 
function and a phase space distribution determined from a sample of simulated 
$\Lambda^+_c\to\phi p\pi^0$ decays. We obtain $77.6\pm28.1$ $P_s^+$ events 
from the fit, which gives an upper limit of
\begin{eqnarray*}
	\mathcal{B}(\Lambda_c^+\to P_s^+\pi^0) \times
	\mathcal{B}(P_s^+\to \phi p)  <  8.3\times 10^{-5}
\end{eqnarray*}
at 90\% CL. %This limit is calculated using the same procedure as that used 
for our limit on ${\cal B}(\Lambda_c^+\rightarrow \phi p \pi^0)$. From the 
fit, we also obtain, $M_{P_s^+}=(2.025\pm 0.005)$~GeV/$c^2$ and
$\Gamma_{P_s^+}=(0.022\pm 0.012)$~GeV, where the uncertainties are statistical 
only.
%--------------------------------------------------
\begin{figure}[htb]
%\vskip -0.3cm
\centering
	\includegraphics[width=0.5\textwidth]{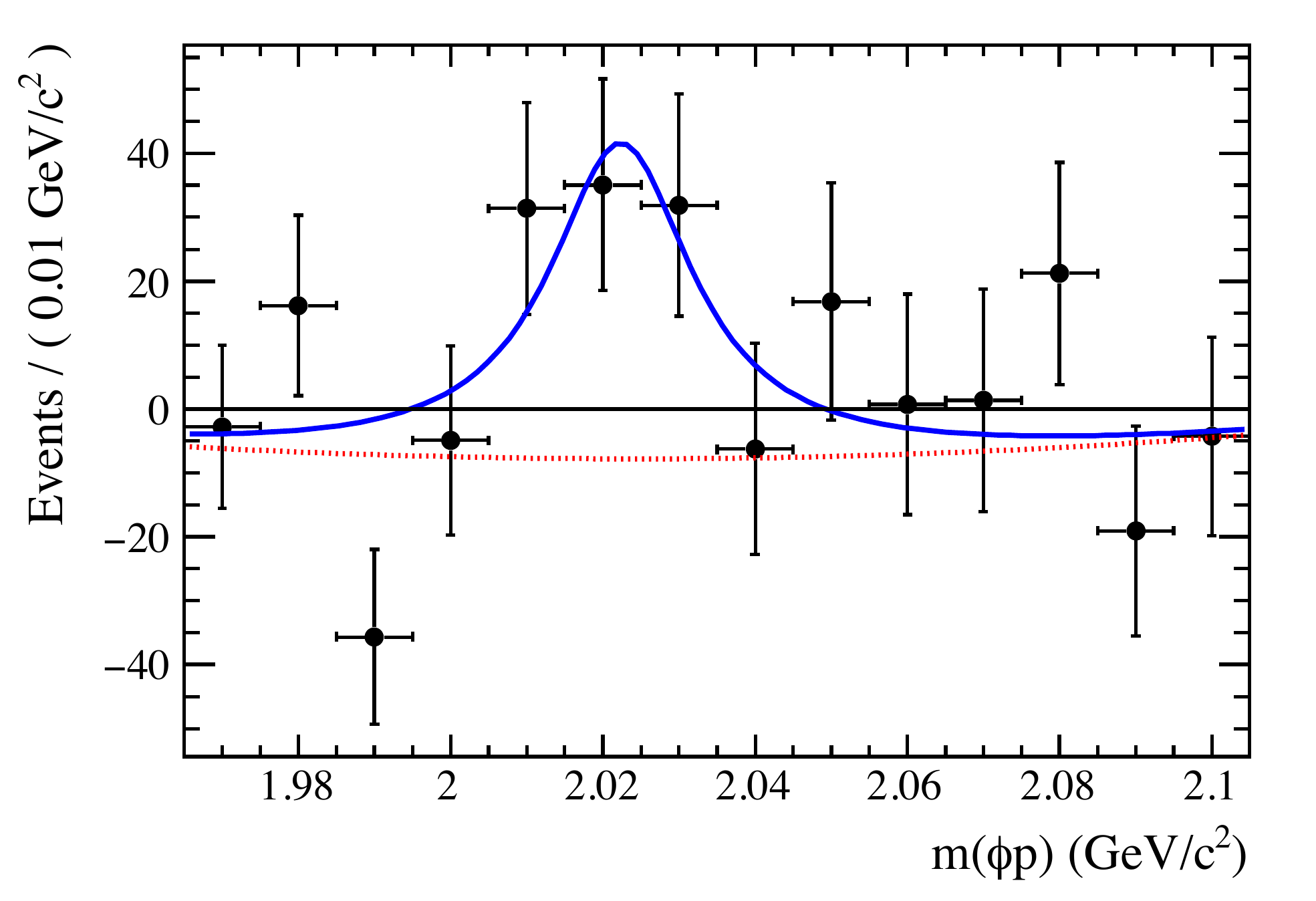}
\vskip -0.3cm
	\caption{\small The background-subtracted distribution of 
	$m(\phi p)$ in the $\phi p\pi^0$ final state. The points with 
	error bars are data, and   the (blue) solid line shows the 
	total PDF. The (red) dotted curve shows the fitted phase space 
	component (which has fluctuated negative).}
	\label{fig:bkg-sub_dis}
\end{figure}
%--------------------------------------------------

The high statistics decay  $\Lambda_c^+\to K^-\pi^+p\pi^0$ is used to 
adjust the data-MC differences in the $\phi p\pi^0$ signal and $K^+K^-p\pi^0$ 
nonresonant decays. For the $\Lambda_c^+\to K^-\pi^+p\pi^0$ sample, the mass 
distribution is plotted in Fig.~\ref{fig:invmass_control2_data}. We fit this
distribution to obtain the signal yield. We find $242\,039\pm \,2342$ signal 
candidates and $472\,729\pm\,467$ background candidates. We measure the ratio 
of branching fractions,
\begin{eqnarray*}
	\frac{\mathcal{B}(\Lambda_c^+\to K^-\pi^+p\pi^0)}{\mathcal{B}
	(\Lambda_c^+\to K^-\pi^+p)}=(0.685\pm0.007\pm 0.018),
\end{eqnarray*}
where the first uncertainty is statistical and the second is systematic. 
Multiplying this ratio by the world average value of 
$\mathcal{B}(\Lambda_c^+\to K^-\pi^+p)=(6.46\pm0.24)\%$~\cite{Amhis:2016xyhb}, 
we obtain 
\begin{eqnarray*}
	\mathcal{B}(\Lambda_c^+\to K^-\pi^+p\pi^0)=(4.42\pm0.05\pm 0.12\pm0.16)\%,
\end{eqnarray*}
where the first uncertainty is statistical, the second is systematic, and 
the third reflects the uncertainty due to the branching fraction of the 
normalization decay mode.  This is the most precise measurement of 
$\mathcal{B}(\Lambda_c^+\to K^-\pi^+p\pi^0)$ to date and is consistent with 
the recently measured value $\mathcal{B}(\Lambda_c^+\to K^-\pi^+p\pi^0)=
(4.53\pm0.23\pm0.30)\%$ by the BESIII collaboration~\cite{Ablikim:2015flgb}.
%-------------------------------------
\begin{figure}[h!tb]
\centering
	\includegraphics[width=0.5\textwidth]{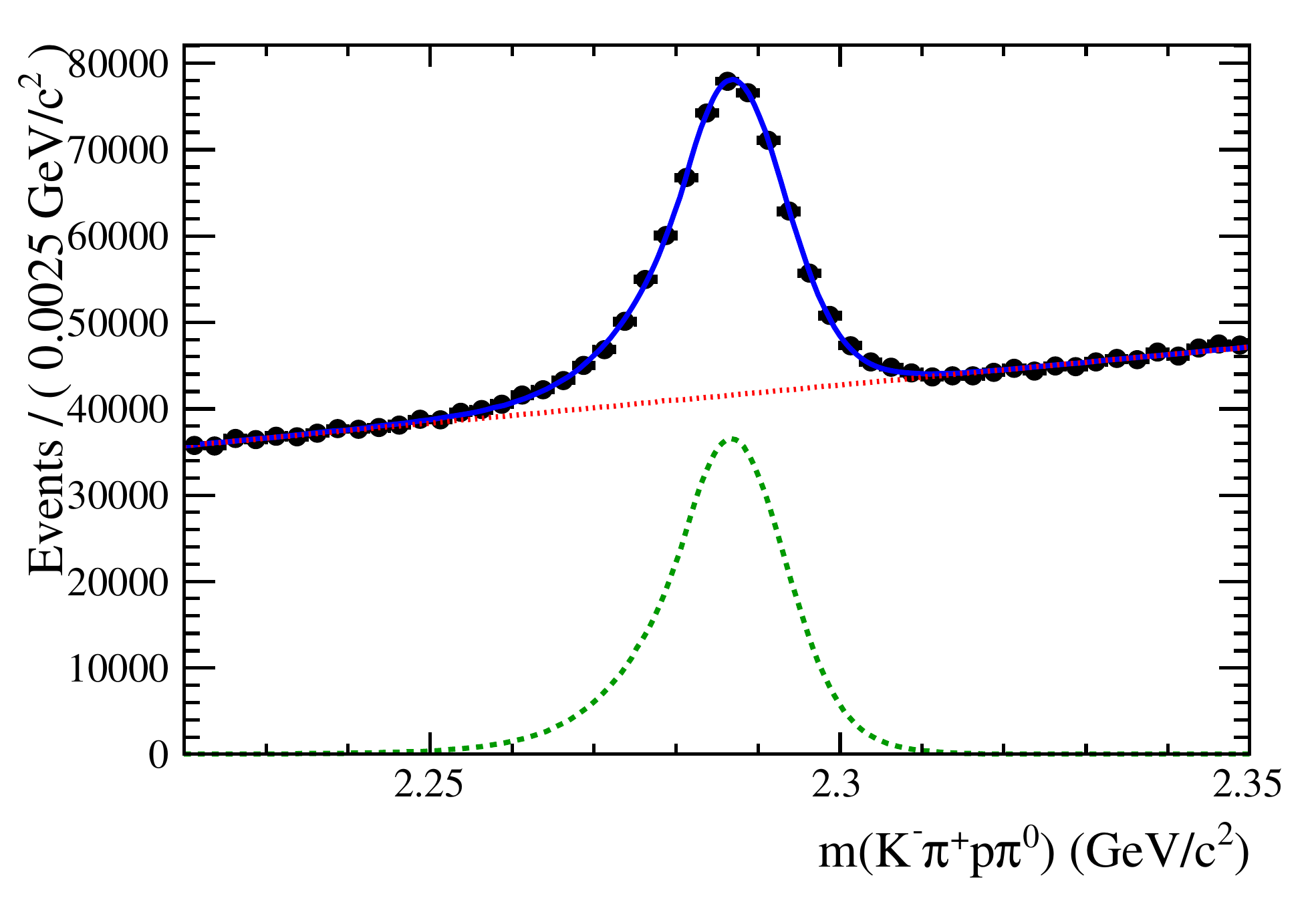}
\vskip -0.3cm
	\caption{\small  Fit to the invariant mass distribution of 
	$m(K^-\pi^+p\pi^0)$. The points with the error bars are the  
	data, the (red) dotted and (green) dashed curves represent 
	the combinatorial and signal candidates, respectively, and 
	(blue) curve represents the total PDF.  %The $\chi^2/$ (number 
	of bins) of the fit is 1.43, which indicate that the fit gives 
	a good description of the data.}
	\label{fig:invmass_control2_data}
\end{figure}
%-------------------------------------

%-------------------------------------
\item \textbf{Observation of the Doubly Cabibbo-Suppressed 
	\mbox{\boldmath$\Lambda_c^+$} Decay }

Several doubly Cabibbo-suppressed (DCS) decays of charmed mesons have been
observed~\cite{Agashe:2014kdab}. Their measured branching ratios with respect 
to the corresponding Cabibbo-favored (CF) decays play an important role in
constraining models of the decay of charmed hadrons and in the study of 
flavor- $SU(3)$ symmetry~\cite{Lipkin:2002zab,Gao:2006nbb}. On the other 
hand, because of the smaller production cross-sections for charmed baryons, 
DCS decays of charmed baryons have not yet been observed, and only an upper 
limit, $\frac{\mathcal{B}(\Lambda_c^+\to pK^+\pi^-)}{\mathcal{B}(\Lambda_c^+
\to pK^-\pi^+)}<0.46\%$ at 90\% CL, has been reported by the FOCUS 
Collaboration~\cite{Link:2005ymb}. Here we present the first observation of 
the DCS decay $\Lambda_c^+\to pK^+\pi^-$ and the measurement of its 
branching ratio with respect to the CF decay $\Lambda_c^+\to pK^-\pi^+$, 
using $980~{\rm fb}^{-1}$ of data~\cite{Yang:2015ytmb}.

Figure~\ref{fig:dcs} shows the invariant mass distributions of (a) $pK^-\pi^+$ 
(CF) and (b) $pK^+\pi^-$ (DCS) combinations. DCS decay events are clearly 
observed in $M(pK^+\pi^-)$. In order to obtain the signal yield, a binned 
least-$\chi^2$ fit is performed. From the mass fit, we extract 
$(1.452\pm0.015)\times10^6$ $\Lambda_c^+\to pK^-\pi^+$ events and $3587\pm380$ 
$\Lambda_c^+\to pK^+\pi^-$ events. The latter has a peaking background from
the single Cabibbo-suppressed (SCS) decay $\Lambda_c^+\to\Lambda(\to 
p\pi^-)K^+$, which has the same final-state topology. After subtracting the 
SCS contribution, we have $3379\pm380\pm78$ DCS events, where the first 
uncertainty is statistical and the second is the systematic due to SCS 
subtraction. The corresponding statistical significance is 9.4 standard 
deviations. We measure the branching ratio,
\begin{eqnarray*}
	\frac{\mathcal{B}(\Lambda_c^+\to pK^+\pi^-)}{\mathcal{B}(\Lambda_c^
	+\to pK^-\pi^+)}=(2.35\pm0.27\pm0.21)\times10^{-3},
\end{eqnarray*}
where the uncertainties are statistical and systematic, respectively.  This 
measured branching ratio corresponds to $(0.82\pm0.21)\tan^4\theta_c$, where 
the uncertainty is the total, which is consistent  within 1.5 standard 
deviations with the na{\"i}ve expectation 
($\sim\tan^4\theta_c$~\cite{Link:2005ymb}). LHCb's recent measurement of 
$\frac{\mathcal{B}(\Lambda_c^+\to pK^+\pi^-)}{\mathcal{B}(\Lambda_c^+\to
pK^-\pi^+)}=(1.65\pm0.15\pm0.05)\times10^{-3}$~\cite{Aaij:2017rinb} is 
lower than our ratio at the 2.0$\sigma$ level.  Multiplying this ratio with 
the previously measured $\mathcal{B} (\Lambda_c^+\to pK^-\pi^+)
=(6.84\pm0.24^{+0.21}_{-0.27})\%$ by the Belle 
Collaboration~\cite{Zupanc:2013ikib}, we obtain the the absolute branching 
fraction of the DCS decay,
\begin{eqnarray*}
	\mathcal{B}(\Lambda_c^+\to pK^+\pi^-)
	=(1.61\pm0.23^{+0.07}_{-0.08})\times10^{-4},
\end{eqnarray*}
where the first uncertainty is due to the total uncertainty of the branching 
ratio and the second is uncertainty due to the branching fraction of the CF 
decay.  After subtracting the contributions of $\Lambda^\ast(1520)$ and 
$\Delta$ isobar intermediates, which contribute only to the CF decay, the 
revised ratio, $\frac{\mathcal{B}(\Lambda_c^+\to pK^+\pi^-)}{\mathcal{B}
(\Lambda_c^+\to pK^-\pi^+)}=(1.10\pm0.17)\tan^4\theta_c$ is consistent with 
the na{\"i}ve expectation within 1.0 standard deviation.
%------------------------------
\begin{figure}[htb!]
\begin{center}
	\includegraphics[width=0.7\textwidth, height=5.0cm]{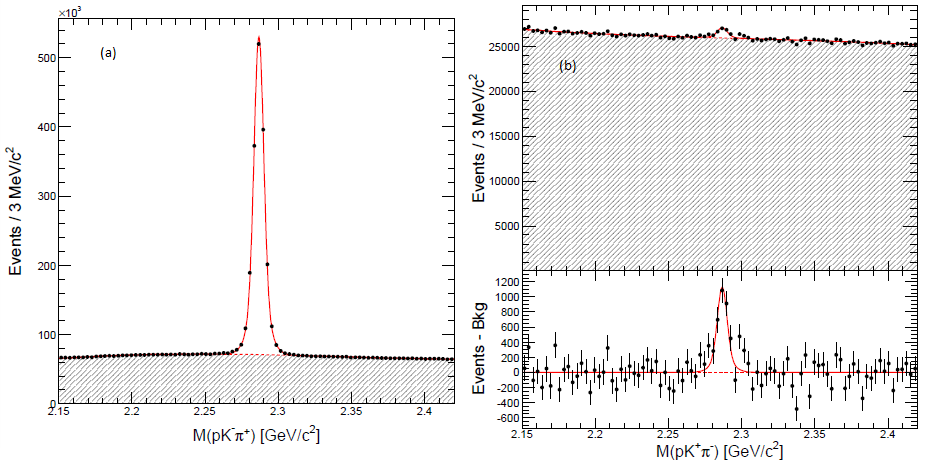}
\vskip -0.3cm

\caption{\small Distributions of (a) $M(pK^-\pi^+)$ and (b) $M(pK^+\pi^-)$ 
	and residuals of data with respect to the fitted combinatorial 
	background. The solid curves indicate the full fit model and the 
	dashed curves the combinatorial background.} \label{fig:dcs}
\end{center}
\end{figure}
%------------------------------

%------------------------------
\item \textbf{$\phi_3$ Measurement with a Model-independent Dalitz Plot 
	Analysis of $B^{\pm}\to DK^{\pm},D\to K_S^0\pi^+\pi^-$ Decay}

The CKM angle $\phi_3$ (also denoted as $\gamma$) is one of the least 
constrained parameters of the CKM Unitary Triangle. Its determination is 
however theoretically clean due to absence of loop contributions; $\phi_3$ 
can be determined using tree-level processes only, exploiting the 
interference between $b\to u\bar{c}s$ and $b\to c\bar{u}s$ transitions 
that occurs when a process involves a neutral $D$ meson reconstructed in 
a final state accessible to both $D^0$ and $\bar{D}^0$  decays (see 
Fig.~\ref{fig:phi3_fn}). Therefore, the angle $\phi_3$ provides a SM 
benchmark, and its precise measurement is crucial in order to disentangle 
non-SM contributions to other processes, via global CKM fits. The size of 
the interference also depends on the ratio ($r_B$) of the magnitudes of 
the two tree diagrams involved and $\delta_B$, the strong phase difference 
between them. Those hadronic parameters will be extracted from data together 
with the angle $\phi_3$.
%%%-----------------------------------
\begin{figure}[htb!]
\begin{center}
	\includegraphics[width=0.5\textwidth]{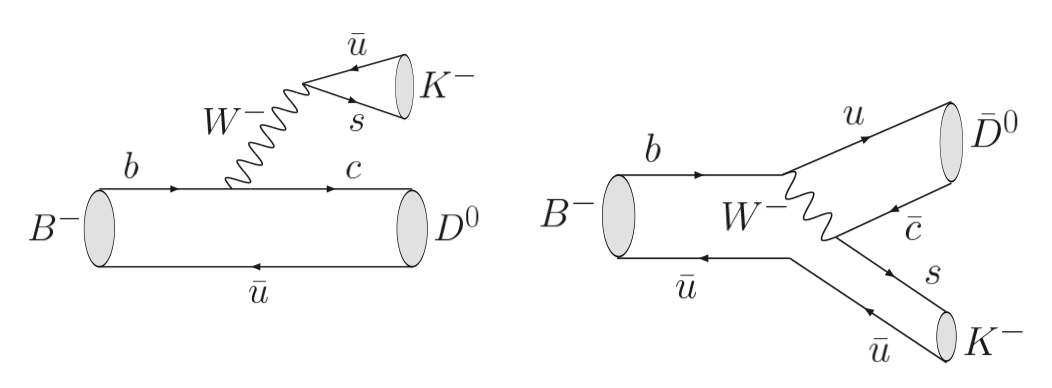}
\vskip -0.3cm

	\caption{\small Feynman diagram for $B^-\to D^0K^-$ and 
	$B^-\to\bar{D}^0K^-$ decays.} \label{fig:phi3_fn}
\end{center}
\end{figure}
%%%-----------------------------------

The measurement are performed in three different ways: (a) by utilizing decays 
of $D$ mesons to $CP$ eigenstates, such as $\pi^+\pi^-$, $K^+K^-$ ($CP$ even) 
or $K_S^0\pi^0$, $\phi K_S^0$ ($CP$ odd), proposed by Gronau, London, and Wyler 
(and called the GLW method~\cite{Gronau:1990rab,Gronau:1991dpb}) (b) by making 
use of DCS decays of $D$ mesons, $e.g.$, $D^0\to K^+\pi^-$, proposed by Atwood, 
Dunietz, and Soni (and called the ADS method~\cite{Atwood:2000ckb}) and (c) by 
exploiting the interference pattern in the Dalitz plot of the $D$ decays such 
as $D^0\to K_S^0\pi^+\pi^-$, proposed by Giri, Grossman, Soffer, and Zupanc 
(and called the GGSZ method~\cite{Giri:2003tyb}).

Using a model-dependent Dalitz plot method, Belle's earlier 
measurement~\cite{Poluektov:2010wzb} based on a data sample of $605~{\rm 
fb^{-1}}$ integrated luminosity yielded $\phi_3
=(78.4^{+10.8}_{-11.6}\pm3.6\pm8.9)^{\circ}$ and 
$r_B=0.160^{+0.040}_{-0.038}\pm0.011^{+0.050}_{-0.010}$, where the uncertainties 
are statistical, systematic and  Dalitz model dependence, respectively. Although 
with more data one can squeeze on the statistical part, the result will still 
remain limited by the model uncertainty.

In a bid to circumvent this problem, Belle has carried out a model-independent
analysis~\cite{Aihara:2012awb}, using GGSZ method~\cite{Giri:2003tyb}, that is 
further extended in a latter work~\cite{Bondar:2008hhb}. The analysis is based 
on the $711~{\rm fb^{-1}}$ of data, collected at the $\Upsilon(4S)$ resonance. 
In contrast to the conventional Dalitz method, where the $D^0\to K_S^0\pi^+
\pi^-$ amplitudes are parameterized as a coherent sum of several quasi two-body 
amplitudes as well as a nonresonant term, the model-independent approach invokes 
study of a binned Dalitz plot. In this approach, the expected number of events 
in the $i^{th}$ bin of the Dalitz plan for the $D$ mesons from $B^{\pm}\to 
DK^{\pm}$ is given by
%\begin{equation}
%	N^{\pm}_i=h_B\big[K_{\pm i}+ r^2_BK_{\mp i}
%	+2\sqrt{K_iK_{-i}}(x_{\pm}c_i\pm y_{\pm}s_i)\big],
%\end{equation}
where $h_B$ is the overall normalization and $K_i$ is the number of events in 
the $i^{th}$ Dalitz bin of the flavor-tagged (whether $D^0$ or $\bar{D}^0$) 
$D^0\to K_S^0\pi^+\pi^-$ decays, accessible via the charge of the slow pion 
in $D^{\ast\pm}\to D\pi^{\pm}$. The terms $c_i$  and $s_i$  contain information 
about the strong-phase difference between the symmetric Dalitz points 
[$m^2(K_S^0\pi^+),~m^2(K_S^0\pi^-)$] and [$m^2(K_S^0\pi^-),~m^2(K_S^0\pi^+)$];
they are the external inputs obtained from quantum correlated $D^0\bar{D^0}$ 
decays at the $\psi(3770)$ resonance in CLEO~\cite{Briere:2009aab,Libby:2010nub}. 
Finally $x_{\pm} = r_B \cos(\delta_B \pm \phi_3)$ and $y_{\pm} = r_B 
\sin(\delta_B \pm \phi_3)$, where $\delta_B$ is the strong-phase difference 
between $B^{\pm}\to \bar{D}^0K^{\pm}$ and $B^{\pm}\to D^0K^{\pm}$.

We perform a combined likelihood fit to four signal selection variables in 
all Dalitz bins (16 bins in our case) for the $B^{\pm}\to DK^{\pm}$ signal 
and Cabibbo-favored $B^{\pm}\to D\pi^{\pm}$ control samples; the free 
parameters of the fit are $x_{\pm}$, $y_{\pm}$, overall normalization (see 
Eq. 1) and background fraction. Table~\ref{tab:xy} summarizes the results 
obtained for $B^{\pm}\to DK^{\pm}$ decays. From these results, we obtain 
$\phi_3 = (77.3^{+15.1}_{-14.9}\pm 4.1 \pm 4.3)^{\circ}$ and $r_B = 
0.145\pm0.030\pm0.010\pm0.011$, where the first error is statistical, the 
second is systematic, and the last error is due to limited precision on 
$c_i$ and $s_i$. Although $\phi_3$ has a mirror solution at $\phi_3 + 
180^{\circ}$, we retain the value consistent with $0^{\circ} < \phi_3 
<180^{\circ}$.
%---------------------------------------
\begin{table}
\centering

	\caption{\small Results of  the $x, y$ parameters and their 
	statistical correlation for $B^{\pm}\to DK^{\pm}$ decays. The 
	quoted uncertainties are statistical, systematic, and
	precision on $c_i$, $s_i$, respectively.}
\begin{tabular}{c|c}
\hline\hline
   Parameter & \\
\hline
$x_{+}$ &  $+0.095\pm0.045\pm0.014\pm0.010$\\
$y_{+}$  &  $+0.135^{+0.053}_{-0.057}\pm0.015\pm0.023$\\
corr($x_{+}$, $y_{+}$) & -0.315\\
$x_{-}$ & $-0.110\pm0.043\pm0.014\pm0.007$\\
$y_{-}$ & $-0.050^{+0.052}_{-0.055}\pm0.011\pm0.017$\\
corr($x_{-}$, $y_{-}$) & +0.059\\
\hline    \hline
\end{tabular}
\label{tab:xy}
\end{table}
%----------------------------------------
We report evidence for direct $CP$ violation, the fact that $\phi_3$ is 
nonzero, at the 2.7 standard deviations  level. Compared to results of 
the model-dependent Dalitz method, this measurement has somewhat poorer 
statistical precision despite a larger data sample used. There are two 
factors responsible for lower statistical sensitivity: 1) the statistical
error for the same statistics is inversely proportional to the $r_B$ 
value, and the central value of $r_B$ in this analysis is smaller, and 
2) the binned approach is expected to have the statistical precision that 
is, on average, 10--20\% poorer than the unbinned one. On the positive 
side, however, the large model uncertainty for the model-dependent study
($8.9^{\circ}$) is now replaced by a purely statistical uncertainty due 
to limited size of the $\psi(3770)$ data sample available at CLEO 
($4.3^{\circ}$). With the use of BES-III data, this error will decrease 
to $1^{\circ}$ or less.

The model-independent approach therefore offers an ideal avenue for 
Belle~II and LHCb in their pursuits of $\phi_3$. We expect that the 
statistical error of the $\phi_3$ measurement using the statistics of a 
$50~{\rm ab^{-1}}$ data sample that will be available at Belle~II will 
reach $1-2^{\circ}$. We also expect that the experimental systematic error 
can be kept at the level below $1^{\circ}$, since most of its sources are 
limited by the statistics of the control channels.

%%%-----------------------------------
\item \textbf{Acknowledgments}

The author thanks the organizers of PKI2018 for excellent hospitality
and for assembling a nice scientific program. This work is supported
by the U.S. Department of Energy.

\end{enumerate}

%%%-----------------------------------

%%%%%%%%%%%%%%%%%%%%%%%%%%%%%%%%%%%%%%%%%%%%%%%%%%%%%%%%%%%%%%%%%%%%%%%%%
\newpage
\subsection{Study of $\tau\to K\pi\nu$ Decay at the B Factories}
\addtocontents{toc}{\hspace{2cm}{\sl Denis Epifanov}\par}
\setcounter{figure}{0}
\setcounter{table}{0}
\setcounter{footnote}{0}
\setcounter{equation}{0}
\halign{#\hfil&\quad#\hfil\cr
\large{Denis Epifanov}\cr
\textit{Budker Institute of Nuclear Physics SB RAS}\cr
\textit{Novosibirsk, 630090 Russia}\cr}

%%%-----------------------------------
\begin{abstract}
Recent results of high-statistics studies of the $\tau\to K\pi\nu$ decays
at $B$ factories are reviewed. We discuss precision measurements of the 
branching fractions of the $\tau^-\to K^0_S\pi^-\nu_\tau$ and
$\tau^-\to K^-\pi^0\nu_\tau$ decays, and a study of the $K^0_S\pi^-$
invariant mass spectrum in the $\tau^-\to K^0_S\pi^-\nu_\tau$ decay.  
Searches for CP symmetry violation in the $\tau^-\to K^0_S\pi^-(\geq 
0\pi^0)\nu_\tau$ decays are also briefly reviewed. We emphasize the 
necessity of the further studies of the $\tau\to K\pi\nu$ decays
at $B$ and Super Flavour factories.
\end{abstract}

%%%-----------------------------------
\begin{enumerate}
\item \textbf{Introduction}

The record statistics of $\tau$ leptons collected at the $e^+ e^-$ $B$
factories~\cite{Bevan:2014igaq} provide unique opportunities of the
precision tests of the Standard Model (SM). In the SM, $\tau$ decays due 
to the charged weak interaction described by the exchange of $W$ boson. 
Hence, there are two main classes of $\tau$ decays, leptonic and hadronic 
$\tau$ decays. Leptonic decays provide very clean laboratory to probe 
electroweak couplings~\cite{Fetscher:1993kiq}, while hadronic $\tau$ 
decays offer unique tools for the precision study of low energy 
QCD~\cite{Pich:1997ymq}. The hadronic system is produced from the QCD 
vacuum via decay of the $W^-$\footnote{Unless specified otherwise, charge 
conjugate decays are implied throughout the paper.} boson into $\bar{u}$ 
and $d$ quarks (Cabibbo-allowed decays) or $\bar{u}$ and $s$ quarks 
(Cabibbo-suppressed decays). As a result the decay amplitude can be 
factorized into a purely leptonic part including the $\tau^-$ and 
$\nu_\tau$ and a hadronic spectral function.

Of particular interest are strangeness changing Cabibbo-suppressed
hadronic $\tau$ decays. The decays $\tau^-\to\bar{K^0}\pi^-\nu_\tau$ and
$\tau^-\to K^-\pi^0\nu_\tau$ (or, shortly, $\tau\to K\pi\nu$) provide
the dominant contribution to the inclusive strange hadronic spectral
function, which is used to evaluate $s$-quark mass and $V_{us}$ element
of Cabibbo-Kobayashi-Maskawa (CKM) quark flavor-mixing 
matrix~\cite{Gamiz:2004arq}. In the $\tau\to K\pi\nu$ decays the $K\pi$ 
system is produced in the clean experimental conditions without 
disturbance from the final state interactions. Hence, $\tau\to K\pi\nu$ 
decays provide complementary information about $K$-$\pi$ interaction to 
the experiments with kaon beams~\cite{lassq,Dobbs:2016malq}. In the 
leptonic sector, CP symmetry violation (CPV) is strongly suppressed in the 
SM ($A_{\rm SM}^{\rm CP}\lesssim 10^{-12}$) leaving enough room
to search for the effects of New Physics~\cite{Delepineq}. Of particular
interest are strangeness changing Cabibbo-suppressed hadronic $\tau$ 
decays, in which large CPV could appear from a charged scalar boson 
exchange in some Multi-Higgs-Doublet models (MHDM)~\cite{Kuhnq}.

Recently, Belle and BaBar performed an extended study of the
$\tau\to K\pi\nu$ decays and searches for CPV in these 
decays~\cite{Epifanov:2007rfq,Bischofberger:2011pwq,Ryu:2014vpcq,
Aubert:2007jhq,Aubert:2008anq,BABAR:2011aaq}.

%%%-----------------------------------
\item \textbf{Study of $\tau\to K\pi\nu$ Decays at Belle and BaBar}

The first analysis of $\tau^-\to K^0_S\pi^-\nu_\tau$ decay at Belle
was done with a $351~{\rm fb}^{-1}$ data sample that contains
$323\times 10^6\ \tau^+\tau^-$ pairs~\cite{Epifanov:2007rfq}.
So called lepton-tagged events were selected, in which $\tau^+$ decays to 
leptons, $\tau^+\to\ell^+\nu_\ell\bar{\nu_\tau},~\ell=e,\,\mu$, while the 
other one decays to the signal $K^0_S \pi^-\nu_\tau$ final state.
Events where both $\tau$'s decay to leptons were used for the 
normalization. In the calculation of the $\tau^- \to K^0_S \pi^- 
\nu_{\tau}$ branching fraction the detection efficiencies for the signal 
and two-lepton events were determined from Monte Carlo (MC) simulation 
with the corrections from the experimental data. The obtained branching 
fraction: \[{\cal B}(\tau^-\to K^0_S\pi^-\nu_\tau)=(4.04\pm 0.02({\rm 
stat.})\pm 0.13 ({\rm syst.}))\times 10^{-3}\] is consistent with the 
other measurements.
%%%-----------------------------------
\begin{figure}[htbp]
\includegraphics[width=0.40\textwidth]{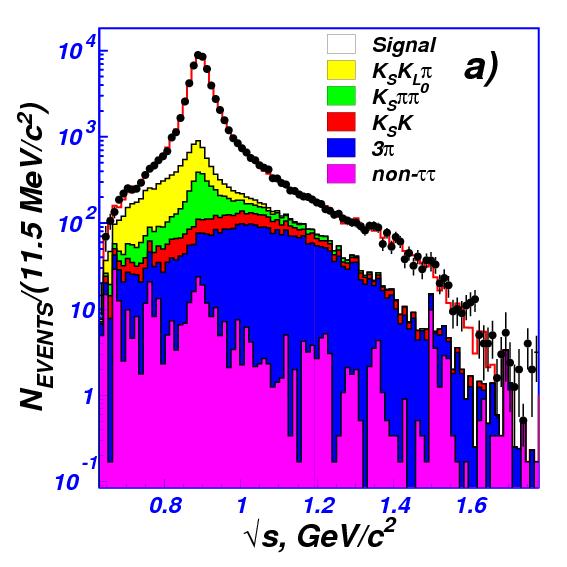}\hfill
\includegraphics[width=0.40\textwidth]{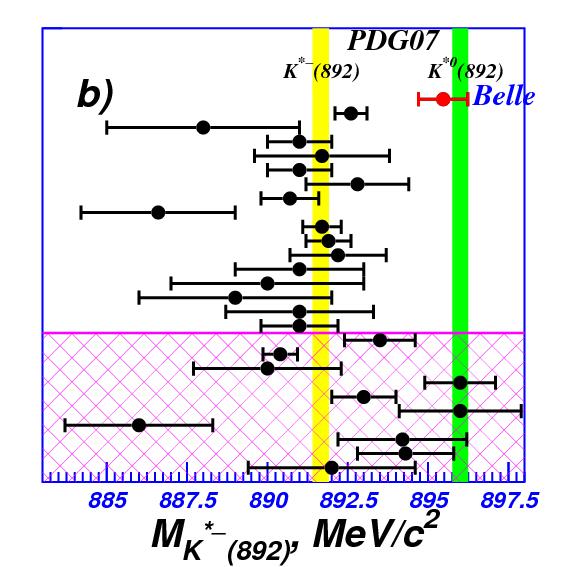}\\

	\parbox[t]{0.95\textwidth}{\caption{ (a) The $K^0_S\pi$ mass 
	distribution, points are experimental data, the histogram is the 
	$K_0^\ast(800)^-+K^\ast(892)^-+K^\ast(1410)^-$ model; (b) The 
	$K^\ast(892)^-$ mass measured in different experiments.
	\label{figone}}}
\end{figure}
%%%-----------------------------------

The $K^0_S \pi^-$ invariant mass distribution shown in
Fig.~\ref{figone} (a) is described in terms of the vector ($F_V$)
and the scalar ($F_S$) form factors according to Ref.~\cite{finkq}:
\begin{equation}
	\frac{d\Gamma}{d\sqrt{s}}\sim\frac{1}{s}\biggl(1-\frac{s}{m^2_{\tau}}\biggr)^2 \biggl(1+2\frac{s}{m^2_{\tau}}\biggr)P\biggl\{P^2|F_V|^2+\frac{3(m^2_K-m^2_{\pi})^2}{4s(1+2\frac{s}{m^2_{\tau}})}|F_S|^2 \biggr\},
\end{equation}
where $s$ is squared $K^0_S\pi^-$ invariant mass, $P$ is $K^0_S$
momentum in the $K^0_S\pi^-$ rest frame. The vector form factor is 
parametrized by the $K^\ast(892)^-$, $K^\ast(1410)^-$ and $K^\ast(1680)^-$ 
meson amplitudes, while the scalar form factor includes the 
$K_0^\ast(800)^-$ and $K_0^\ast(1430)^-$ contributions. The 
$K^\ast(892)^-$ alone is not sufficient to describe the $K^0_S \pi^-$ 
invariant mass spectrum. To describe the enhancement near threshold, we 
introduce a $K_0^\ast(800)^-$ amplitude, while for the description of the 
distribution at higher invariant masses we try to include the 
$K^\ast(1410)^-$, $K^\ast(1680)^-$ vector resonances or the scalar 
$K_0^\ast(1430)^-$. The best description is achieved with the 
$K_0^\ast(800)^-+K^\ast(892)^-+K^\ast(1410)^-$ and $K_0^\ast(800)^-+
K^\ast(892)^-+K_0^\ast(1430)^-$ models. The parameterization of $F_S$ 
suggested by the LASS experiment~\cite{lassq} was also tested:
\begin{equation}
	F_S=\lambda\frac{\sqrt{s}}{P}(\sin\delta_Be^{i\delta_B}+e^{2i\delta_B}BW_{K_0^\ast(1430)}(s)),
\end{equation}
where $\lambda$ is a real constant, $P$ is $K^0_S$ momentum in the
$K^0_S\pi^-$ rest frame, and the phase  $\delta_B$ is determined from the
equation $\cot\delta_B=\frac{1}{aP}+\frac{bP}{2}$, where $a$, $b$ are the 
model (fit) parameters. In this parameterization the non-resonant mechanism 
is given by the effective range term $\sin\delta_Be^{i\delta_B}$,
while the resonant structure is described by the $K_0^\ast(1430)$ amplitude.
%%%-----------------------------------
\begin{figure}[htbp]
\includegraphics[width=0.45\textwidth]{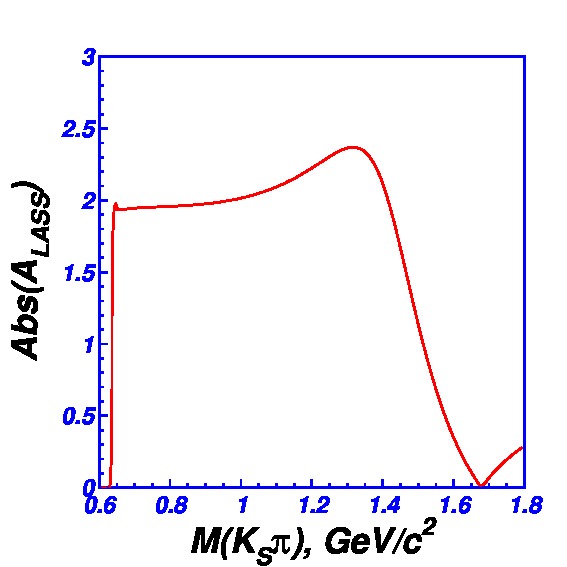} \hfill
\includegraphics[width=0.45\textwidth]{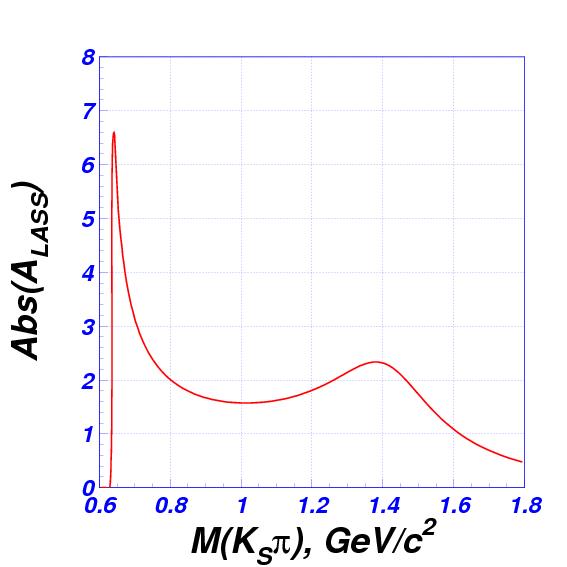} \\

\parbox[t]{0.95\textwidth}{\caption{The absolute value of $F_S$ from 
	LASS experiment (left) and from the $\tau^-\to K^0_S\pi^-\nu_\tau$ 
	study at Belle (right).\label{figtwo}}}
\end{figure}
%%%-----------------------------------

The shape of the optimal scalar form factor in the LASS experiment strongly
differs (especially, near the threshold of the $K^0_S\pi^-$ production) from
that obtained in the fit of the $K^0_S\pi^-$ mass distribution in the study
of $\tau^-\to K^0_S\pi^-\nu_\tau$ decay at Belle, see Fig.~\ref{figtwo}.
There is large systematic uncertainty in the near $K^0_S\pi^-$ production 
threshold part of the spectrum due to the large background from the 
$\tau^-\to K^0_S\pi^- K^0_L\nu_\tau$ decay, whose dynamics is not precisely 
known. In the new study at $B$ factories it will be possible to suppress 
this background essentially applying special kinematical constraints.

A fit to the $K^0_S\pi^-$ invariant mass spectrum also provides a high 
precision measurement of the $K^\ast(892)^-$ mass and width:
$M(K^\ast(892)^-)=(895.47\pm 0.20({\rm stat.})\pm 0.44({\rm syst.})\pm
0.59({\rm mod.}))\ {\rm MeV}/c^2$, $\Gamma(K^\ast(892)^-)=(46.2\pm 0.6({\rm 
stat.})\pm 1.0({\rm syst.})\pm 0.7({\rm mod.}))\ {\rm MeV}$. While our 
determination of the width is compatible with most of the previous
measurements within experimental errors, our mass value, see
Fig.~\ref{figone} (b), is considerably higher than those before and is 
consistent with the world average value of the neutral $K^\ast(892)^0$ mass, 
which is $(896.00 \pm 0.25)$~MeV/$c^2$~\cite{pdg06q}.

The second analysis of the $\tau^-\to K^0_S\pi^-\nu_\tau$ decay at Belle
was based on the data sample with the luminosity integral of
${\cal L}=669$~fb$^{-1}$ which comprises 615 million $\tau^+\tau^-$ 
events~\cite{Ryu:2014vpcq}. One inclusive decay mode $\tau^-\to K^0_S 
X^-\nu_{\tau}$ and 6 exclusive hadronic $\tau$ decay modes with $K^0_S$
($\tau^-\to K^0_S\pi^-\nu_\tau$, $\tau^-\to K^0_SK^-\nu_\tau$, $\tau^-\to
K^0_SK^0_S\pi^-\nu_\tau$, $\tau^-\to K^0_S\pi^-\pi^0\nu_\tau$, $\tau^-\to 
K^0_SK^-\pi^0\nu_\tau$, $\tau^-\to K^0_SK^0_S\pi^-\pi^0\nu_\tau$)
were studied in Ref.~\cite{Ryu:2014vpcq}. In this study signal events
were tagged by one-prong tau decays (into $e\nu\nu$, $\mu\nu\nu$ or
$\pi/K(n\geq 0)\pi^0\nu$ final states) with the branching fraction 
${\cal B}_{1-\rm prong}=(85.35\pm 0.07)\%$. In total, $157836$ events of 
the $\tau^-\to K^0_S\pi^-\nu_\tau$ decay were selected with the fraction 
of the non-cross-feed background of $(8.86\pm 0.05)\%$ and the detection 
efficiency $\varepsilon_{\rm  det} = (7.09\pm 0.12)\%$. The obtained
branching fraction: \[ {\cal B}(\tau^-\to K^0_S\pi^-\nu_\tau)=(4.16\pm 
0.01({\rm stat.})\pm 0.08({\rm syst.}))) \times 10^{-3}\] supersedes the 
previous Belle result and has the best accuracy.
%%%-----------------------------------
\begin{figure}[htbp]
\includegraphics[width=0.40\textwidth]{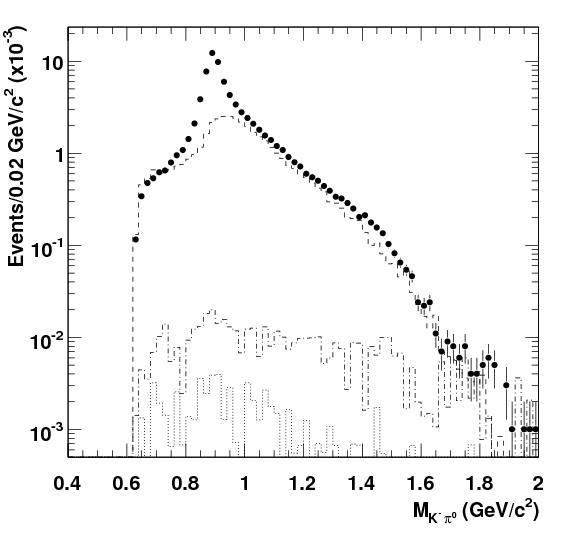}\hfill
\includegraphics[width=0.45\textwidth]{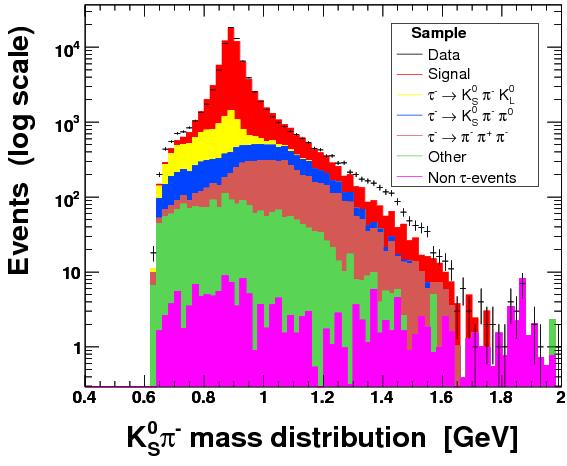} \\

\parbox[t]{0.95\textwidth}{\caption{ The $K^-\pi^0$ invariant 
	mass distribution (left) from Ref.~\protect\cite{Aubert:2007jhq}. 
	The dots are experimental data, histograms are background
	MC events with selection and efficiency corrections: 
	background from $\tau\tau$ (dashed line), $q\bar{q}$ (dash-dotted 
	line), $\mu^+\mu^-$ (dotted line). The $K^0_S\pi^-$ invariant mass
	distribution (right) from Ref.~\protect\cite{Aubert:2008anq}. The 
	dots are experimental data, histograms are signal and background 
	MC: signal events (red), dominant background from the
	$\tau^-\to K^0_SK^0_L\pi^-\nu_\tau$ decay (yellow), non-$\tau\tau$
	events (magenta).\label{figthr}}}
\end{figure}
%%%-----------------------------------

Precision measurement of the branching fraction of the $\tau^-\to
K^-\pi^0\nu_\tau$ decay with a $230.2$~fb$^{-1}$ data sample collected 
at the $\Upsilon(4S)$ resonance has been carried out by 
BaBar~\cite{Aubert:2007jhq}. The result: \[ {\cal B}(\tau^-\to 
K^-\pi^0\nu_\tau)=(4.16\pm 0.03({\rm stat.})\pm
0.18({\rm syst.}))) \times 10^{-3}\]
is consistent with the previous measurements and ${\cal B}(\tau^-\to
K^0_S\pi^-\nu_\tau)$ and has the best accuracy. The $K^-\pi^0$ 
invariant mass distribution is shown in Fig.~\ref{figthr}.

The preliminary result on the ${\cal B}(\tau^-\to K^0_S\pi^-\nu_\tau)$
using a $384.6$~fb$^{-1}$ data sample was also published by 
BaBar~\cite{Aubert:2008anq}:
\[ {\cal B}(\tau^-\to K^0_S\pi^-\nu_\tau)=(4.20\pm 0.02({\rm stat.})\pm
0.12({\rm syst.}))) \times 10^{-3}.\]
It is consistent with the other measurements. The distribution of the
invariant mass of the $K^0_S\pi^-$ system is shown in Fig.~\ref{figthr}, 
experimental data exhibit additional contribution around the invariant 
mass of 1.4~GeV/$c^2$, which is not included in the signal MC simulation.

%%%-----------------------------------
\item \textbf{Search for CPV in $\tau\to K\pi\nu$}

Recent studies of CPV in the $\tau^-\to\pi^- K_S(\geq 0\pi^0)\nu_{\tau}$
decays at BaBar~\cite{BABAR:2011aaq} as well as in the $\tau^-\to 
K_S\pi^-\nu_{\tau}$ decay at Belle~\cite{Bischofberger:2011pwq} provide 
complementary information about sources of CPV in these hadronic decays.

The decay-rate asymmetry $A_{\rm CP}=\frac{\Gamma(\tau^+\to\pi^+
K_S(\geq 0\pi^0)\nu_{\tau})-\Gamma(\tau^-\to\pi^- 
K_S(\geq 0\pi^0)\nu_{\tau})}{\Gamma(\tau^+\to\pi^+ K_S(\geq 0\pi^0)\nu_{\tau})
+\Gamma(\tau^-\to\pi^- K_S(\geq 0\pi^0)\nu_{\tau})}$
was studied at BaBar with the $\tau^+\tau^-$ data sample of
$\int Ldt=476$~fb$^{-1}$.  The obtained result $A_{\rm CP}=(-0.36\pm 
0.23\pm 0.11)\%$ is about $2.8$ standard deviations from the SM 
expectation $A^{K^0}_{\rm CP}=(+0.36\pm 
0.01)\%$.

At Belle, CPV search was performed as a blinded analysis based on
a $699~{\rm fb}^{-1}$ data sample. Specially constructed asymmetry, 
which is a difference between the mean values of the $\cos\beta\cos\psi$
for $\tau^-$ and $\tau^+$ events, was measured in bins of $K^0_S\pi^-$
mass squared ($Q^2=M^2(K^0_S\pi)$):
\[ A_i^{CP}(Q^2_i) = \frac{\int\limits_{\Delta Q^2_i}\cos{\beta}\cos{\psi}  \left( \frac{d\Gamma_{\tau^-}}{d\omega} -
\frac{d\Gamma_{\tau^+}}{d\omega} \right) d\omega}{\frac{1}{2}
\int\limits_{\Delta Q^2_i}\left(\frac{d\Gamma_{\tau^-}}{d\omega}
+\frac{d\Gamma_{\tau^+}}{d\omega} \right)d\omega} \simeq \langle
\cos\beta\cos\psi \rangle_{\tau^-} - \langle \cos\beta\cos\psi
\rangle_{\tau^+}, \]
where $\beta$, $\theta$ and $\psi$ are the angles, evaluated from
the measured parameters of the final hadrons, $d\omega=dQ^2 d\!\cos\theta 
d\!\cos\beta$. In contrary to the decay-rate asymmetry the introduced
$A_i^{CP}(Q^2_i)$ is already sensitive to the CPV effects from the
charged scalar boson exchange~\cite{Kuhn:1996dvq}.  No CP violation 
was observed and the upper limit on the CPV parameter $\eta_S$ was 
extracted to be: $|{\rm Im}(\eta_S)|<0.026$~at~90\% CL.
Using this limit parameters of the Multi-Higgs-Doublet 
models~\cite{Grossman:1994jbq,Choi:1994chq} can be constrained as $|{\rm 
Im}(XZ^\ast)|<0.15~M_{H^\pm}^2/(1\,\mathrm{GeV}^{2}/c^4)$, where
$M_{H^\pm}$ is the mass of the lightest charged Higgs boson, the
complex constants Z and X describe the coupling of the Higgs boson to 
leptons and quarks respectively.

%%%-----------------------------------
\item \textbf{Further Studies of $\tau\to K\pi\nu$ Decays}

In the analysis of the $\tau^-\to K^0_S\pi^-\nu_\tau$ decay, it is
very desirable to measure separately vector ($F_V$), scalar ($F_S$) 
form factors and their interference. The $K^\ast(892)^-$ mass and width are 
measured in the vector form factor taking into account the effect of the 
interference of the $K^\ast(892)^-$ amplitude with the contributions from 
the radial exitations, $K^\ast(1410)^-$ and $K^\ast(1680)^-$. The scalar 
form factor is important to unveil the problem of the $K_0^\ast(800)^-$ 
state as well as for the tests of the various fenomenological models and 
search for CPV. The interference between vector and scalar form factors 
is necessary in the search for CPV in $\tau^-\to K^0_S\pi^-\nu_\tau$ decay.

To elucidate the nature of the $K^\ast(892)^- - K^\ast(892)^0$ mass 
difference it is important to study the following modes: $\tau^-\to 
K^0_S\pi^-\nu_\tau$, $\tau^-\to K^0_S\pi^-\pi^0\nu_\tau$, $\tau^-\to 
K^0_S K^-\pi^0\nu_\tau$.  $K^\ast(892)^-$ mass and width can be measured 
in the clean experimental conditions without disturbance from the final 
state interactions in the $\tau^-\to K^0_S\pi^-\nu_\tau$ decay. While a 
study of the $\tau^-\to K^0_S\pi^-\pi^0\nu_\tau$ mode allows one to 
measure simultaneously in one mode the $K^\ast(892)^-$($K^0_S\pi^-$) and 
the $K^\ast(892)^0$($K^0_S\pi^0$) masses. The effect of the pure hadronic
interaction of the $K^\ast(892)^-$ ($K^\ast(892)^0$) and $\pi^0$($\pi^-$) 
on the $K^\ast(892)^-$($K^\ast(892)^0$) mass can be precisely measured as 
well. It is also important to investigate precisely the effect of the pure 
hadronic interaction of the $K^\ast(892)^-$($K^\ast(892)^0$) and 
$K^0_S$($K^-$) on the $K^\ast(892)^-$($K^\ast(892)^0$) mass in the 
$\tau^-\to K^0_S K^-\pi^0\nu_\tau$ decay.

%%%-----------------------------------
\item \textbf{Summary}

Belle and BaBar essentially improved the accuracy of the branching
fractions of the $\tau^-\to (K\pi)^-\nu_\tau$ decays. At Belle the 
$K^0_S\pi$ invariant mass spectrum was studied and the $K^\ast(892)^-$ 
alone is not sufficient to describe the $K^0_S\pi$ mass spectrum.
The best description is achieved with the 
$K_0^\ast(800)^-+K^\ast(892)^-+K^\ast(1410)^-$ and
$K_0^\ast(800)^-+K^\ast(892)^-+K_0^\ast(1430)^-$ models. For the first 
time the the $K^\ast(892)^-$ mass and width have been measured in $\tau$ 
decay at $B$ factories. The $K^\ast(892)^-$ mass is significantly 
different from the current world average value, it agrees with the 
$K^\ast(892)^0$ mass. Precision study of the $\tau^-\to 
K^0_S\pi^-\nu_\tau$, $\tau^-\to K^0_S\pi^-\pi^0\nu_\tau$ and $\tau^-\to 
K^0_S K^-\pi^0\nu_\tau$ decays at the $B$ factories as well as the $e^+ 
e^- \to K^0_SK^\pm\pi^\mp$ reaction at the 
VEPP-2000~\cite{Shatunov:2000zcq,Berkaev:2012qeq} and 
$K-\pi$ scattering amplitude at the coming GlueX 
experiment~\cite{Dobbs:2016malq} could provide additional valuable 
information about the $K^\ast(892)^-$ mass, namely unveil an  impact of 
the hadronic and electromagnetic interactions in the final state.

%%%-----------------------------------
\item \textbf{Acknowledgments}

I am grateful to Prof. Igor Strakovsky for inviting me to this very interesting
workshop. This work was supported by Russian Foundation for Basic Research
(Grant No.~17--02--00897--a).
\end{enumerate}

%%%-----------------------------------

%%%%%%%%%%%%%%%%%%%%%%%%%%%%%%%%%%%%%%%%%%%%%%%%%%%%%%%%%%%%%%%%%%%%%%%%%
\newpage
\subsection{From $\pi{K}$ Amplitudes to $\pi{K}$ Form Factors (and Back)}
\addtocontents{toc}{\hspace{2cm}{\sl Bachir Moussallam}\par}
\setcounter{figure}{0}
\setcounter{table}{0}
\setcounter{equation}{0}
\setcounter{footnote}{0}
\halign{#\hfil&\quad#\hfil\cr
\large{Bachir Moussallam}\cr
\textit{IPN, Universit\'e Paris-Sud 11}\cr
\textit{91406 Orsay, France}\cr}

%%%%-----------------------------------
\begin{abstract}
The dispersive construction of the scalar $\pi\pi$ and the scalar and vector 
$\pi K$ form factors are reviewed. The experimental properties of the $\pi\pi$ 
and $\pi K$ scattering $J=0,1$ amplitudes are recalled, which allow for an 
application of final-state interaction theory in a much larger range than the 
exactly elastic energy region. Comparisons are made with recent lattice QCD 
results and with experimental $\tau$ decay results. The latter indicate that 
some corrections to the $\pi K$ $P$-wave phase-shifts from LASS may be needed.
\end{abstract}

%%%-----------------------------------
\begin{enumerate}
\item \textbf{Introduction}

Pions and kaons are the lightest hadrons in QCD. The $\pi\pi$ interaction
plays an important role in generating stable nuclei. Studying the $\pi\pi$ and
the $\pi K$ interactions allows to probe the chiral symmetry aspects of QCD at
low energy and the resonance structure at higher energy. Sufficiently
precise measurements also associated with theoretical, analyticity properties
of QCD  have allowed to establish the existence of  light, exotic,
resonances~\cite{Caprini:2005zrt} which had remained elusive for many decades.

In recent years a considerable amount of data on decays of heavy mesons or
leptons into light pseudo-scalar mesons have accumulated. The energy
dependencies of these decay amplitudes are  essentially controlled by the
final-state interactions (FSI) among the light mesons. FSI theory relies on
analyticity and the existence of a right-hand cut, in each energy variable,
which can be associated with unitarity~\cite{Omnes:1958hvt}.

Two-mesons form factors are the simplest functions to which FSI theory can be
applied in the sense that they depend on a single variable and that the
unitarity cut is the only one.  We review below some aspects of the $\pi\pi$
scalar and the $\pi K$ scalar and vector form factors. We will recall, in
particular, the experimental properties of the interactions which allow for an
application of FSI theory in a much larger energy range than the range of
purely elastic scattering. We will also consider the possibility, given
sufficiently precise determination of the form factors, to improve the
precision of the determination of the light meson amplitudes.

%%%-----------------------------------
\item \textbf{The $\pi\pi$ Scalar Form Factors}

$\pi\pi$ scattering is exactly elastic in QCD in the energy range $s\le
16m^2_\pi$. However, it has long been known that it can be considered as
effectively elastic in a larger range. The two main assumptions which underlie
the dispersive construction of the form factors in Ref.~\cite{Donoghue:1990xht}
is that, in the $S$-wave, this effectively elastic region extends up to the
$K\bar{K}$ threshold and that beyond this, there exists an energy region in
which two-channel unitarity effectively holds. These assumptions are supported
by the experimental measurements of the $\pi\pi$ phase-shifts and
inelasticities. For instance Fig.~5a of Ref.~\cite{Hyams:1973zft} showing the
inelasticity $\eta_0^0$ indicates clearly that the $\eta_0^0$ is driven away 
from 1 by the $f_0(980)$ resonance and the cusp-like shape of the curve 
suggests that $K\bar{K}$ is the leading inelastic channel, while $4\pi$ or 
$6\pi$ channels play a negligible role up to 1.4~GeV.

This can be further probed using experimental measurements of the $\pi\pi\to
K\bar{K}$ amplitudes. Indeed, if two-channel unitarity holds then the modulus
of this amplitude is simply related to the inelasticity. This is illustrated
from fig~\ref{fig:comparinel} which shows fits to the inelasticity as deduced
from the $\pi\pi\to K\bar{K}$ measurements of Refs.~\cite{Cohen:1980cqt,Etkin:1981sgt} 
compared with the inelasticity measure in $\pi\pi$. The two experiments are 
in good agreement in the energy energy region $E \ge 1.2$~GeV and the figure 
indicates that the two-channel unitarity picture could hold up to $E \simeq 
1.6$~GeV.
%%%-----------------------------------
\begin{figure}[hbt]
\centering
	\includegraphics[width=0.5\linewidth]{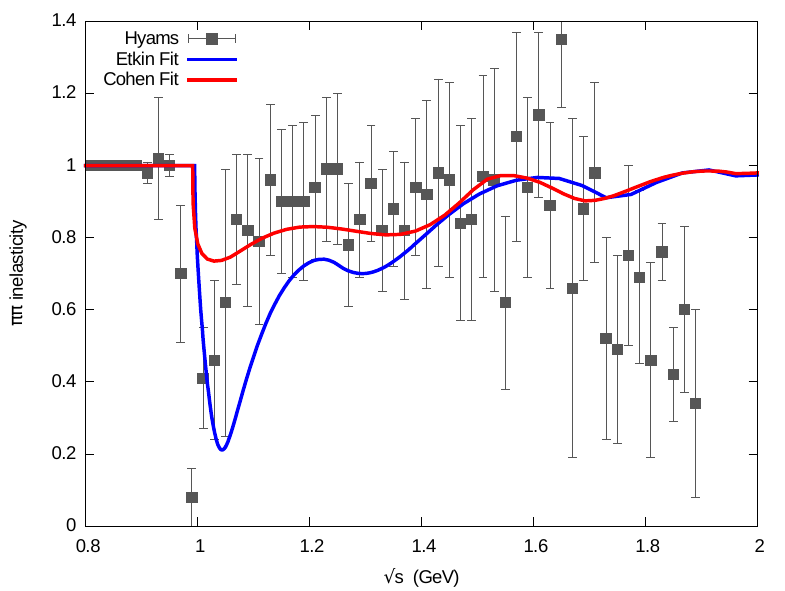}

	\caption{\small Inelasticity parameter of the $\pi\pi$ $I=0$
  	$S$-wave compared to that deduced from $\pi\pi\to K\bar{K}$
  	measurements assuming two-channel unitarity.}
	\label{fig:comparinel}
\end{figure}
%%%-----------------------------------

Scalar form factors, for instance $\Gamma_\pi(t)=\langle{\pi(p)\pi(q)\vert 
m_u\bar{u}u+m_d\bar{d}d\vert 0\rangle}$, with $t=(p+q)^2$ are analytic 
functions of $t$ in the whole complex plane except for a right-hand cut on 
the real axis. Assuming that it has no zeroes, it can be expressed as a phase
dispersive representation
\begin{equation}
	\Gamma_\pi(t)=\Gamma_\pi(0)\exp\left[\frac{t}{\pi}\int_{4\mpid}^\infty
	\frac{dt'}{t'(t'-t)}\,\phi_\pi(t')\ .
	\right]
\end{equation}
From Watson's theorem the form factor phase $\phi_\pi$ is equal to the
$\pi\pi$ scattering phase in the elastic scattering region, that is,
according to the preceding discussion
\begin{equation}
	\phi_\pi(t)=\delta_0^0(t),\ 4\mpid \le t \le 4m^2_K\ .
\end{equation}
Furthermore, when $t\to \infty$ one must have $\phi_\pi(t)\to \pi$
which ensures compatibility with a Brodsky-Lepage type
behavior~\cite{Lepage:1980fjt}. Based on these arguments, a simple
interpolating model for the phase in the range $[4m_K^2,\infty]$ was
proposed in ref.~\cite{Yndurain:2003vkt}. As was pointed out
in~\cite{Ananthanarayan:2004xyt} this simple interpolation leads to an
overestimate of the pion scalar form factor $\langle
r^2\rangle^\pi_S=\dot{\Gamma}_\pi(0)/(6 \Gamma_\pi(0))$ because the
$f_0(980)$ resonance effect is not properly accounted for. A more
accurate interpolation can be performed by exploiting the existence of
a two-channel unitarity region and generating the form
factor from a solution of the corresponding coupled-channel
Muskhelishvili equations
\begin{equation}\label{eq:pionFFSMO}
	\vec{F}(t)=\frac{1}{\pi}\int _{4\mpid}^\infty dt' \frac{
	\bm{T}(t')\Sigma(t') \vec{F}^\ast(t')}{t'-t} ,\
\end{equation}
where $\bm{T}$ is the $\pi\pi-K\bar{K}$ coupled-channel matrix and\\
$\Sigma(t')=diag\left(\sqrt{1-4\mpid/t'},
\sqrt{1-4\mkd/t'}\theta(t'-4\mkd)\right)$.
Beyond the region where two-channel unitarity applies, a proper
asymptotic phase interpolation is performed in this model by imposing
asymptotic conditions on the $T$-matrix. It is also convenient to
choose these conditions such that the so-called Noether
index~\cite{FritzNoethert} is equal to ${\cal N}=2$ which ensures that
Eq.~\ref{eq:pionFFSMO} has a unique solution once two conditions are
imposed, e.g., $\Gamma_\pi(0)$, $\Gamma_K(0)$.

Fig.~\ref{fig:FFSphase} illustrates the result for the phase of the form
factor, $\phi_\pi$, obtained from solving these equations. The $f_0(980)$
resonance is predicted to generate a sharp drop in this phase at 1
GeV. Nevertheless, it is noteworthy that the asymptotic phase is
reached from above in this approach as is required in
QCD~\cite{Yndurain:2003vkt} (see right plot in Fig.~\ref{fig:FFSphase}).  The
result for the pion scalar radius~\cite{Ananthanarayan:2004xyt}: $\langle r^2
\rangle^\pi_S=(0.61\pm 0.04)$ $\hbox{fm}^2$ is in good agreement with lattice
QCD simulations, e.g.,~\cite{Gulpers:2013ucat}.
%%%-----------------------------------
\begin{figure}[hbt]
\centering
	\includegraphics[width=0.49\linewidth]{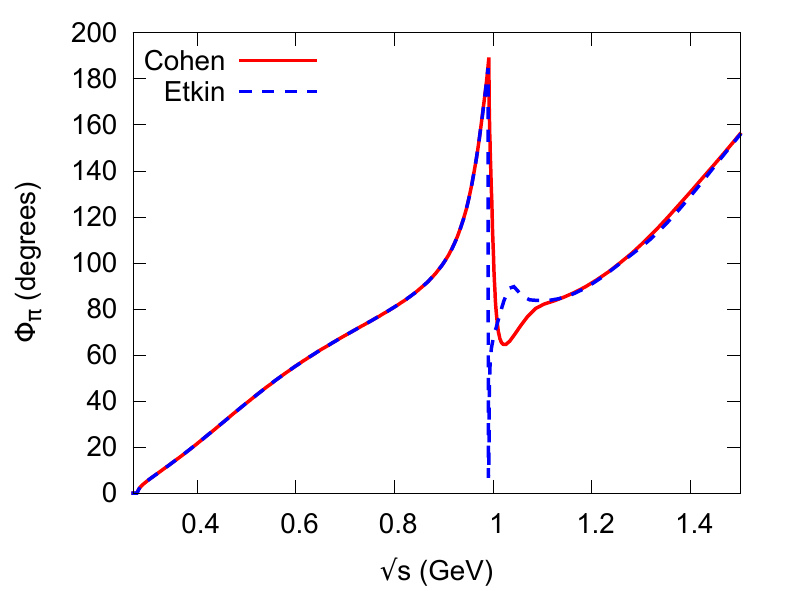}
	\includegraphics[width=0.49\linewidth]{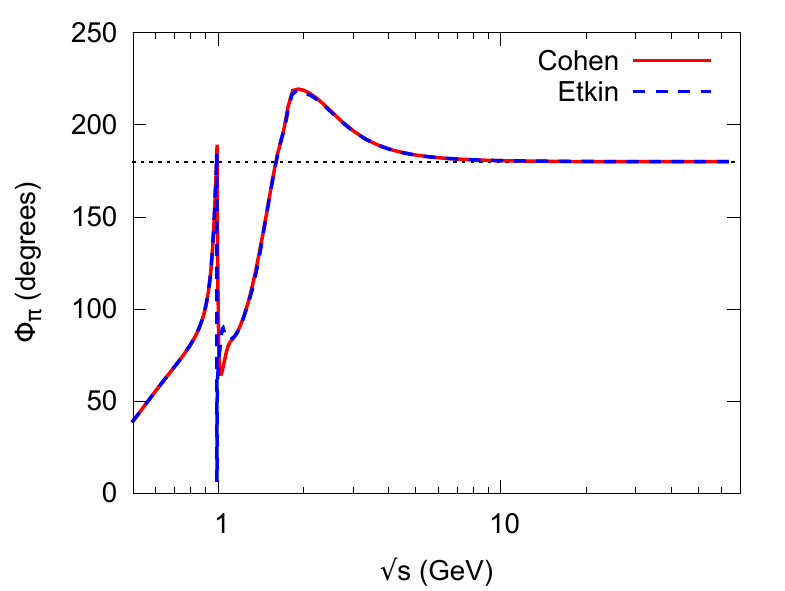}

	\caption{\small Phase of the pion scalar form factor from solving 
	the Muskhelishvili equations~\protect\ref{eq:pionFFSMO}. The right 
	plot illustrates how the asymptotic value is reached.}
	\label{fig:FFSphase}
\end{figure}
%%%-----------------------------------

%%%-----------------------------------
\item \textbf{$\pi K$ Scalar Form Factor}

$\pi K$ scattering in the $S$-wave is exactly elastic in QCD in the
region $t \le (m_K+3m_\pi)^2$. By analogy with $\pi\pi$ one would
expect the amplitude to be effectively elastic in a substantially
larger range. This seems to be confirmed by experiment. 
Fig.~\ref{fig:piKSinel} shows the inelasticity parameter $\eta_0^{1/2}$ 
obtained from a fit to the two most recent amplitudes
determinations~\cite{Estabrooks:1977xet,Aston:1987irt}. One sees that
$\eta_0^{1/2}$ remains close to 1 up to $\sqrt{t}\simeq 1.9$ GeV and
it is driven away from 1 very sharply by the $K^\ast_0(1950)$ resonance.
The detailed properties of this resonance are yet unknown. It is often
assumed that its main decay channel is $\eta' K$ which would imply the
existence of a two-channel unitarity region for the $\pi K$
$S$-wave. A dispersive derivation of the $\pi K$ scalar form factor
exactly analogous to that of $\pi\pi$ can then be
performed~\cite{Jamin:2001zqt}.

An important difference with $\pi\pi$, though, is that the $\pi K$ scalar 
form factor is measurable. It can be determined from semi-leptonic decay 
amplitudes $K\to \pi l\nu_l$ and $\tau \to K \pi \nu_\tau$ since the 
matrix element of the vector current involves both the vector and the 
scalar form factors, $f_+^{K\pi}$, $f_0^{K\pi}$
\begin{equation}
	\sqrt2\langle K^+(p_K)\vert
	\bar{u}\gamma^\mu{s}\vert\pi^0(p_\pi)\rangle=
	f_+^{K\pi}(t)\,\left(p_K+p_\pi-\frac{\Delta_{K\pi}}{t}(p_K-p_\pi)\right)^\mu+
	f_0^{K\pi}(t)\,\frac{\Delta_{K\pi}}{t}\left(p_K-p_\pi\right)^\mu
\end{equation}
The prediction from the two-channel dispersive form factor model for
the slope parameter $\lambda_0=m_{\pi^+}^2 \dot{f}_0^{K\pi}(0)/
f_0^{K\pi}(0)$ is~\cite{Jamin:2006tjt} $\lambda_0=(14.7\pm0.4)\cdot10^{-3}$ 
which is in rather good agreement with the most recent determination by NA48/2, 
NA62 experiments~\cite{Piccini:2018wfpt}: $\lambda_0=(14.90\pm0.55\pm0.80)
\cdot10^{-3}$. Further verifications of this form factor, in particular in the 
region of the $K^\ast_0(1430)$ resonance, would be necessary. This is possible, 
in principle, from $\tau \to K \pi\nu_\tau$ decays if one measures both the 
energy and the angular distributions in order to disentangle the
contributions from the two form factors. At low energy $f_0^{K\pi}(t)$
displays a strong $\kappa$ meson induced enhancement. As can be seem from
fig.~\ref{fig:fitscombined} (left plot) below, this feature is in accord with
the data already existing on $\tau$ decay.

%%%-----------------------------------
\item \textbf{$\pi K$ Vector Form Factor and $J=1$ Amplitude}

For the $P$-wave amplitude, fits to the experimental data indicate that the
quasi-elastic range extends up to the $K^\ast\pi$ threshold (see
fig.~\ref{fig:piKSinel}).  The inelasticity is driven by the $K^\ast(1410)$ 
and by the $K^\ast(1680)$ resonance. The decay properties of these two 
resonances have been studied in detail in Ref.~\cite{Aston:1986jbt}: they 
essentially involve the two quasi two-body channels $K^\ast\pi$ and $K\rho$. 
This suggests that a plausible model based on three-channel unitarity can 
be developed for describing the $\pi K$ $P$-wave scattering which can be 
applied to the dispersive construction of the vector form factor. Models of 
this type were considered~\cite{Moussallam:2007qct,Bernard:2013jxat}. We 
present below slightly updated results from~\cite{Moussallam:2007qct}.
%%%-----------------------------------
\begin{figure}[hbt]
\centering
	\includegraphics[width=0.45\linewidth]{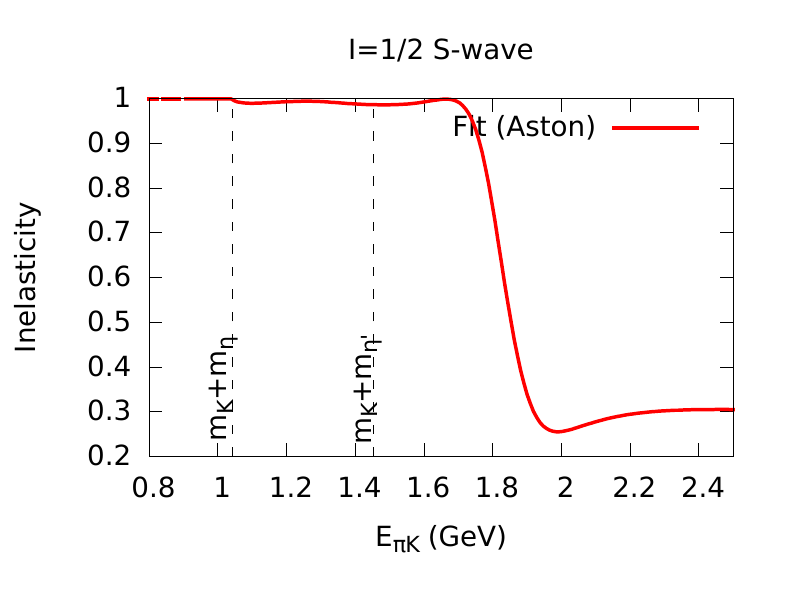}
	\includegraphics[width=0.45\linewidth]{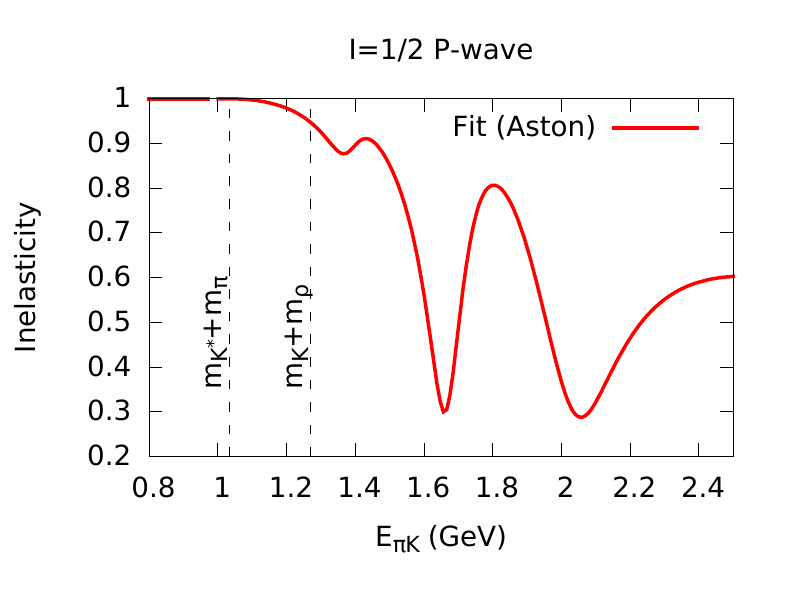}

	\caption{\small $\pi K$ inelasticity from fits to the
	data of~\protect\cite{Estabrooks:1977xet,Aston:1987irt}. 
	Left plot: $S$-wave, right plot: $P$-wave.} 
	\label{fig:piKSinel}
\end{figure}
%%%-----------------------------------

At first, a reasonable fit of the experimental data can be achieved
(with a  $\chi^2/{d.o.f}=1.8$) in the whole energy range $0.8 \le E \le2.5$
GeV within a model which implements exact three-channel
unitarity via a simple standard $K$-matrix approach,
\begin{equation}
	K_{ij}=\sum_R\frac{ g_R^i\,g_R^j}{m_R^2-s} + K_{ij}^{back}\ .
\end{equation}
The model has four $P$-wave resonances and contains 15 parameters
which are determined from a fit to the $\pi K \to \pi K$ experimental data.
The parameters are also constrained with respect to the inelastic channels
$K^\ast\pi$ and $K\rho$ by the measured branching fractions of the 
$K^\ast(1410)$, $K^\ast(1680)$ resonances.

The three vector form factors associated with $K\pi$, $K^\ast\pi$, $K\rho$ 
can then be determined by solving the coupled integral dispersive equations
analogous to eq.~\ref{eq:pionFFSMO}. Having chosen ${\cal N}=3$ for the
Noether index, it is necessary to provide three constraints in order to
uniquely specify the solution, e.g. the values of the three form factors 
$H_i$ at $t=0$. The value of $H_1(0)=f_+^{K\pi}(0)$ is rather precisely 
known from chiral symmetry and from lattice QCD simulations
(see~\cite{Aoki:2016frlt}). Concerning $H_2(0)$, $H_3(0)$ we can only obtain 
a qualitative order of magnitude in the limit of exact three-flavor chiral
symmetry and assuming exact vector-meson dominance. This gives
\begin{equation}
	H_2(0)=-H_3(0)=\frac{\sqrt2 N_c M_V}{16\pi^2 F_V F_\pi}\simeq
	1.50\ \hbox{GeV}^{-1}
\end{equation}
We can thus parametrize the exact values as $H_2(0)=1.50(1+a)$,
$H_3(0)=1.50(1+b)$ and we expect $\vert{a}\vert, \vert{b}\vert < 30\%$.
Fitting the Belle data~\cite{Epifanov:2007rft} on $\tau\to K\pi \nu_\tau$ with
these two parameters $a$, $b$ gives a rather poor $\chi^2 =8.6$.  While this
could simply be blamed on the inadequacy of the model, we will argue instead
that the $P$-wave phase-shifts of LASS~\cite{Aston:1986jbt} may need some
updating. At first, one sees that much of the large $\chi^2$ value originates
from the $K^\ast(892)$ resonance region and simply reflects a significant
difference in the $K^\ast(892)$ resonance width between the LASS data and the
Belle data. Indeed, in the recent re-analysis of the LASS
data constrained by dispersions relations~\cite{Pelaez:2016klvt}  a value for
the width, $\Gamma_{K^\ast}=58\pm2$ MeV is found, which is much larger 
than the result of Belle: $\Gamma_{K^\ast}= 46.2\pm1.4$ MeV.

In the present approach one can go further and determine how the $P$-wave
results from LASS need to be modified in order for the vector form factor to
be in better agreement with the Belle results. This is done by refitting the
parameters of the three-channel $T$ matrix including both LASS and Belle data
sets. Doing this, a reasonable agreement with the $\tau$ data can be achieved
with a $\chi^2/N({Belle})=2.4$ while the $\chi^2$ for the LASS
data is $\chi^2/N({LASS})=3.5$. This is illustrated in
fig.~\ref{fig:fitscombined}. The left plot shows the $\pi K$ energy
distribution in the $\tau\to \pi K \nu_\tau$ mode and the right plot shows the
phase of the $\pi^+ K^-\to \pi^+ K^-$ $J=1$ amplitude as a function of energy.
Beyond the region of the $K^\ast(892)$ a visible modification of the
phase in the region of the $K^\ast(1410)$ seems to be also needed.

%%%-----------------------------------
\begin{figure}[hbt]
\centering
	\includegraphics[width=0.45\linewidth]{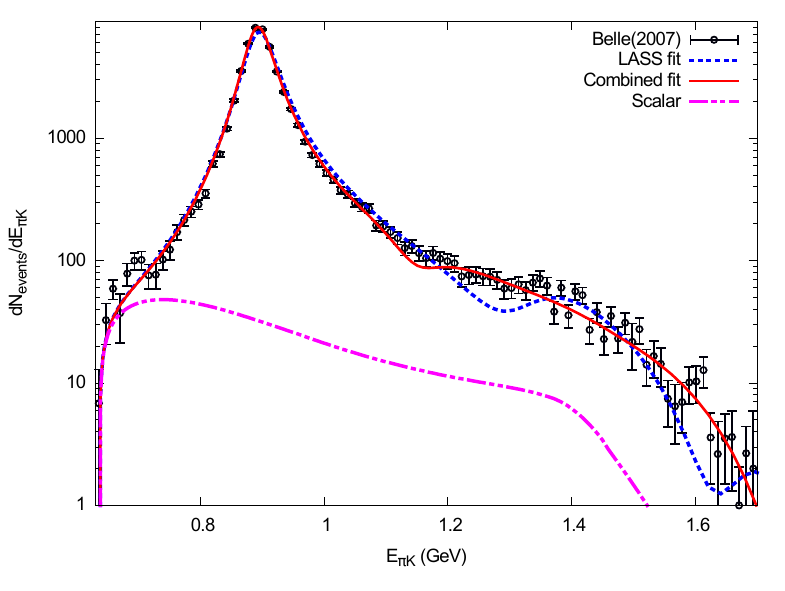}
	\includegraphics[width=0.45\linewidth]{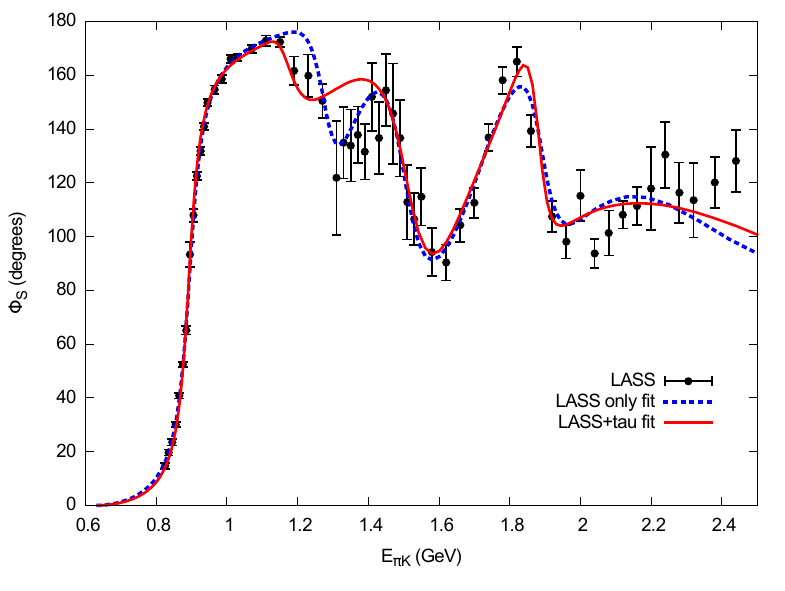}

	\caption{\small Left plot: energy distribution in the $\tau\to \pi 
	K \nu_\tau$ mode. Right plot: Phase of the $\pi^+ K^-\to \pi^+ K^-$ 
	$J=1$ amplitude. The dashed blue curves show the results when the 
	$T$-matrix parameters are fitted to the LASS data only, while the 
	solid red curves when the fit includes both LASS and Belle data. 
	The dash-dot magenta curve shows the contribution associated with 
	the scalar form factor.}
\label{fig:fitscombined}
\end{figure}
%%%-----------------------------------
\end{enumerate}

%%%-----------------------------------

%%%%%%%%%%%%%%%%%%%%%%%%%%%%%%%%%%%%%%%%%%%%%%%%%%%%%%%%%%%%%%%%%%%%
\newpage
\subsection{Three-Body Interaction in Unitary Isobar Formalism}
\addtocontents{toc}{\hspace{2cm}{\sl Maxim Mai}\par}
\setcounter{figure}{0}
\setcounter{table}{0}
\setcounter{footnote}{0}
\setcounter{equation}{0}
\halign{#\hfil&\quad#\hfil\cr
\large{Maxim Mai}\cr
\textit{Institute for Nuclear Studies and}\cr
\textit{Department of Physics}\cr
\textit{The George Washington University}\cr
\textit{Washington, DC 20052, U.S.A.}\cr}

%%%-----------------------------------
\begin{abstract}
In this talk, I present our recent results on the three-to-three
scattering amplitude constructed from the compositeness principle of
the $S$-matrix and constrained by two- and three-body unitarity. The
resulting amplitude has important applications in the infinite volume, 
but can also be used to derive the finite volume quantization condition
for the determination of energy eigenvalues obtained from ab-initio
Lattice QCD calculations of three-body systems.
\end{abstract}

%%%-----------------------------------
\begin{enumerate}
\item \textbf{Introduction}
\label{sec:intro}

Interest in the description of three-hadron systems has re-sparked in recent
years due to two aspects of modern nuclear physics. \underline{First}, there
are substantial advances of experimental facilities such as
COMPASS~\cite{Abbon:2007pqk}, GlueX~\cite{Ghoul:2015ifwk} and
CLAS12~\cite{Glazier:2015cpak} experiments, aiming for the study of meson
resonances with mass above 1~GeV including light hybrids. The large branching
ratio of such resonances to three pions is expected to generate important
features via the final state interaction. Furthermore, effects such as the
log-like behavior of the irreducible three-body interaction, associated with
the $a_1(1420)$~\cite{Ketzer:2015tqak} can be studied in detail with a full
three-body amplitude. Similarly, the study of the $XYZ$
sector~\cite{Lebed:2016hpik,Guo:2017jvck} currently explored by LHCb, BESIII,
Belle and BaBar~\cite{Alves:2008zzk,Fang:2016gshk} can be conducted in more
detal  when the three-body interactions are taken into account. The prominent
Roper puzzle can be re-addressed when the features of the $\pi\pi N$ amplitude
are sufficiently under control, see, e.g., Ref.~\cite{Aitchison:1978pwk}. Finally,
the proposed Klong beam experiment at
JLab~\cite{Amaryan:2017ldwk,Alba:2017cbrk,Horn:2017vkzk,Albrow:2016ibsk} can
give new insights into properties of, e.g., $\kappa$-resonance from the
$K\pi\pi$ channel. \underline{Second}, the algorithmic and computational
advances in \textit{ab-intio} Lattice QCD calculations make the analysis
of such interesting systems as the Roper-resonance ($N(1440)1/2^+$) or the
$a_1(1260)$ possible. Some first studies in these systems have already been
conducted, using gauge configurations with unphysically heavy quark
masses~\cite{Lang:2014tiak,Lang:2016hnnk,Kiratidis:2016hdak}, but no 3-body
operators have been included there yet. Many other groups are working on
conducting similar studies, such as, e.g., the $\pi\rho$ scattering in the
$I=2$ sector~\cite{Woss:2018irjk}. Also in these systems the pion mass is
very heavy, such that the $\rho$ is stable and the infinite-volume
extrapolation can effectively be carried out using the two-body L\"uscher
formalism. In the future, it has to be expected that these and similar
studies will be carried out at lower pion masses with an unstable $\rho$
decaying into two pions. For these cases the full understanding of the
infinite volume extrapolations including three-body dynamics is desired.
Important progress has been achieved in the last 
years~\cite{Sharpe:2017jejk,Hammer:2017kmsk,Hammer:2017uqmk,Hansen:2017mndk,
Guo:2017ismk,Agadjanov:2016maok,Hansen:2016ynck,Hansen:2016fzjk,Hansen:2015ztak,
Hansen:2015zgak,Meissner:2014deak,Polejaeva:2012utk,Mai:2017bgek}, and first 
numerical case studies have now been conducted with
three different approaches~\cite{Briceno:2018mlhk,Doring:2018xxxk,Mai:2017bgek,
Hammer:2017kmsk}. While still exploratory, they mark an important step in the
development of the three-body quantization condition.

In this work, we show theoretical developments of one of these
approaches~\cite{Mai:2017bgek} as well as numerical studies based on it.
This framework is based on the general formulation of the infinite volume
three-to-three scattering amplitude which respects two- and three-body
unitarity. At its core, the two-body sub-amplitudes are parametrized by a
tower of functions of invariant mass with correct right-hand singularities
of the corresponding partial-waves. The truncation of such a series of
functions in a practical calculations is the only approximation of such
an approach and takes account of the sparsity of the lattice data. The
imaginary parts of such an amplitude are fixed by unitarity, giving rise
to a power-law (in $ML$ -- a dimensionless product of pion mass and size
of cubic lattice volume) finite volume dependence, when replacing continuous
momenta in such an amplitude by discretized ones due to boundary conditions
imposed in Lattice QCD studies. Therefore, such a framework gives a natural
way to study three-body systems in the infinite and in finite volume
simultaneously.

%%%-----------------------------------
\item \textbf{Three-Body Scattering in the Infinite Volume}

%%%-----------------------------------
\begin{figure}
\begin{center}
\includegraphics[width=0.8\linewidth]{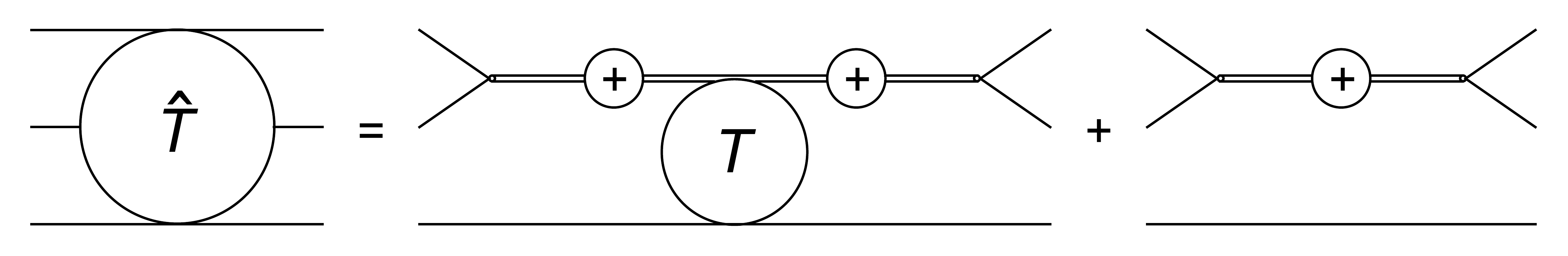}
\end{center}

\caption{Total scattering amplitude $\hat T$ consisting of a connected
	($\hat T_c$) and a disconnected contribution ($\hat T_d$), represented
	by the first and second term on the right-hand side of 
	Eq.~\protect\eqref{eq:t33full}, respectively. Single lines indicate 
	the elementary particle, double lines represent the isobar, empty dots 
	stand for isobar dissociation vertex $v$, while time runs from right 
	to left. $T$ and $"+"$ denote the isobar-spectator scattering amplitude 
	and isobar propagator $\tau$, respectively.} \label{fig:T_total}
\end{figure}
%%%-----------------------------------
The interaction of three-to-three asymptotic states is described by the scattering
amplitude $T$. We assume here that the particles in question are stable, spinless
and identical (of mass $M$) for simplicity. The connectedness-structure of matrix
elements dictates that the scattering amplitude consists of two parts: the fully
connected one and one-time\footnote{two particle sub-amplitude are still fully
connected, while the third particle takes the role of what we refer to as the
``spectator".} disconnected, denoted by the subscript $c$ and $d$ in the following.
As discussed in Refs.~\cite{Bedaque:1999vbk,Hammer:2017kmsk}, the full two-body
amplitude can be re-parametrized by a tower of  ``isobars", which for given quantum
numbers of the two-body sub-system describe the correct right-hand singularities
of each partial wave in that system. In this sense the isobar formulation is not
an approximation but a re-parametrization of the full two-body amplitude, see the
discussion in the original work~\cite{Mai:2017votk}. In summary, the three-to-three
scattering amplitude (depicted in Fig.~\ref{fig:T_total}) reads
\begin{align}\label{eq:t33full}
 	\langle q_1,q_2,q_3&|\hat{T}| p_1,p_2,p_3\rangle
    	=
    	\langle q_1,q_2,q_3|\hat{T}_c |p_1,p_2,p_3\rangle
    	+\langle q_1,q_2,q_3|\hat{T}_d| p_1,p_2,p_3\rangle\\
    	&=
    	\frac{1}{3!}\sum_{n=1}^3\sum_{m=1}^3\,
	v(q_{\bar{n}},q_{\barr{n}})v(p_{\bar{m}},p_{\barr{m}})\nonumber\\
	%\underbrace{
    	&~~~~~~~~~~~\Bigg(
    	\tau(\sigma(q_n))\,
    	T(q_n,p_m;s)\,
    	\tau(\sigma(p_m))\,
    	-2E({\boldsymbol q}_n)\tau(\sigma(q_n))(2\pi)^3\delta^3(\boldsymbol{q}_n-\boldsymbol{p}_m)
    	\Bigg)\,,\nonumber
\end{align}
where $P$ is the total four-momentum of the system, $s=W^2=P^2$ and
$E({\boldsymbol p})=\sqrt{{\boldsymbol p}^2+M^2}$. All four-momenta
$p_1,q_1,...$ are on-mass-shell, and the square of the invariant mass
of the isobar reads $\sigma(q):=(P-q)^2=s+M^2-2WE(\boldsymbol q)$ for
the spectator momentum $q$. We work in the total center-of-mass frame
where ${\boldsymbol P}=\boldsymbol{0}$ and denote throughout the
manuscript three-momenta by bold symbols. The dissociation vertex
$v(p,q)$ of the isobar decaying in asymptotically stable particles,
e.g., $\rho\,(p+q)\to \pi(p)\pi(q)$, is chosen to be cut-free in the
relevant energy region, which is always possible. The notation is such
that, e.g., for a spectator momentum $q_n$ the isobar decays into two
particles with momenta $q_{\bar{n}}$ and $q_{\barr{n}}$. Finally, $T$
describes the isobar-spectator interaction and is the function of eight
kinematic variables allowed by momentum and energy conservation, such
as the full three-to-three scattering amplitude.

The above equation contains three unknown functions $\tau(\sigma)$, $T$
and $v$. Since real and imaginary parts of the scattering amplitude are
related by unitarity, the latter building blocks of the scattering
amplitude are related between each other as well. Specifically, including
a complete set of three-particle intermediate states (inclusion of
higher-particle states will be studied in a future work) the three-body
unitarity condition can be written as
\begin{align}\label{eq:unit0}
    	\langle q_1,q_2,q_3|(\hat T-\hat T^\dagger)| &p_1,p_2,p_3\rangle=\\
	&i\int\prod_{\ell=1}^3\frac{\mathrm{d}^4k_\ell}{(2\pi)^{4}}\,(2\pi)\delta^+(k_\ell^2-m^2)
	\,(2\pi)^4\delta^4\left(P-\sum_{\ell=1}^3\,k_\ell\right)\nonumber\\
    	&~~~~~~~~~~~~~~~~~~~~~~\langle q_1,q_2,q_3|\hat T^\dagger|k_1,k_2,k_3\rangle\,
    	\langle k_1,k_2,k_3|\hat T| p_1,p_2,p_3\rangle \,,\nonumber
\end{align}
where $\delta^+(k^2-m^2):=\theta(k_0)\delta(k^2-m^2)$. To reveal the relations
between $T$, $\tau$ and $v$ we make the Bethe-Salpeter Ansatz for the
isobar-spectator amplitude
\begin{align}\label{eq:bse1}
	T(q_n,p_m;s)=
	B(q_n,p_m;s)
	+
	\int\frac{\mathrm{d}^4 k}{(2\pi)^4}
	B(q_n,k;s)\hat\tau(\sigma(k))T(k,p_m;s) \,,
\end{align}
which holds for any in/outgoing spectator momenta $p_m/q_n$ (not necessarily
on-shell), and effectively re-formulates the unknown function $T$ by yet
another two unknown functions: the isobar-spectator interaction kernel $B$
and the isobar-spectator Green's function $\hat\tau$. In particular, we can
rewrite the left-hand-side of the unitarity relation~\eqref{eq:unit0} into
eight different topologies, which symbolically read
\begin{align}\label{eq:tstep}
    \hat T-\hat T^\dagger=&~
    v(\tau-\tau^\dagger) v
    +v\left(\tau-\tau^\dagger \right)T\tau v
    +v\tau^\dagger T^{\dagger}\left(\tau-\tau^\dagger\right)v\nonumber\\
    &
    +v\tau^\dagger (B-B^\dagger)\tau v+v\tau^\dagger(B-B^\dagger)\hat\tau T\tau v
    +v\tau^\dagger T^{\dagger}\hat\tau^{\dagger}(B-B^\dagger)\tau v\\
    &
    +v\tau^\dagger T^{\dagger}(\hat\tau -\hat\tau^{\dagger})T\tau v
    +v\tau^\dagger T^{\dagger}\hat\tau^{\dagger}(B-B^\dagger)\hat\tau  T\tau v \,.\nonumber
\end{align}
The main goal of such a decomposition is to express the discontinuity of $\hat T$ as
a sum of discontinuities of simpler building blocks in corresponding variables.

The same kind of decomposition can be achieved for the right-hand-side of unitarity
relation~\eqref{eq:unit0}, inserting $\hat T=\hat T_d+\hat T_c$ there, while keeping
in mind the permutations of particle indices as in Eq.~\eqref{eq:t33full}. The exact
formulation of this is discussed in detail in the original
publication~\cite{Mai:2017votk}. Here we restrict ourselves of mentioning that
piecewise comparison of all structures to those of Eq.~\eqref{eq:tstep} leads to
eight independent matching relations. These can be fulfilled simultaneously imposing
\begin{align}\label{eq:match_s}
	&\hat \tau(\sigma(k)) &=&~-(2\pi)\delta^+(k^2-m^2)\tau(\sigma(k))\,,&\\[6pt]
	&B(q,p;s) - B^\dagger(q,p;s) &=&
	~iv(Q,q)(2\pi)\delta^+(Q^2-m^2)v(Q,p) \,,&\\[4pt]
	&\Big(\tau^\dagger(\sigma(k))\Big)^{-1}-
	\Big(\tau (\sigma(k))\Big)^{-1}&=&~
	i\int \frac{\mathrm{d}^4\bar K}{2(2\pi)^2}\,
	\delta^+\left(\left(\tilde P+\bar K  \right)^2-m^2\right)
	\delta^+\left(\left(\tilde P-\bar K  \right)^2-m^2\right)&\\
&&&~~~~~~~~~~~~~~~
	\left(v\left(\tilde P+\bar K,\tilde P-\bar K\right)\right)^2\,,&\nonumber
\end{align}
where $\tilde P:= (P-k)/2$ and $Q:=P-p-q$. The direct consequence of the above
relations is that the number of unknown in the integral equation defining the
three-to-three scattering amplitude~\eqref{eq:t33full} is naturally reduced from
three ($\hat\tau$, $B$, $\tau$) to two ($\tau$, $B$). The remaining two can be
determined using twice subtracted dispersion relation w.r.t the invariant mass
of the two-body system ($\sigma$) as well an un-subtracted dispersion relation
in $Q^2$, respectively. The exact integral representation of $B$ and $\tau$ is
given in the original publication~\cite{Mai:2017votk} and resembles the
one-particle exchange as well as a fully dressed isobar propagator. In both
cases a real-valued function can be added, without altering the discontinuity
relations and, thus, being allowed by the unitarity constraint discussed here.

The above considerations finalize the form of the three-to-three scattering
amplitude as demanded by three-body unitarity and compositeness principle. It
is a fully relativistic three-dimensional integral equation, which becomes a
coupled-channel equation when more than one isobar is considered for the
parametrization of the two-to-two scattering. The parameters of these isobars
(subtraction constants) can be fixed from the two-body scattering data, while
the real part of $B(q,p;s)$ has to be fixed from the three-body data. The
application of this approach to systems like the $a_1(1260)$ with two isobars,
i.e., isovector and isoscalar $\pi\pi$ channels, is work in progress.

%%%-----------------------------------
\item \textbf{Three-Body Scattering in Finite Volume}

%%%-----------------------------------
\begin{figure}
\begin{center}
\includegraphics[width=0.49\linewidth,trim=0.5cm 0.75cm 0cm 0cm]{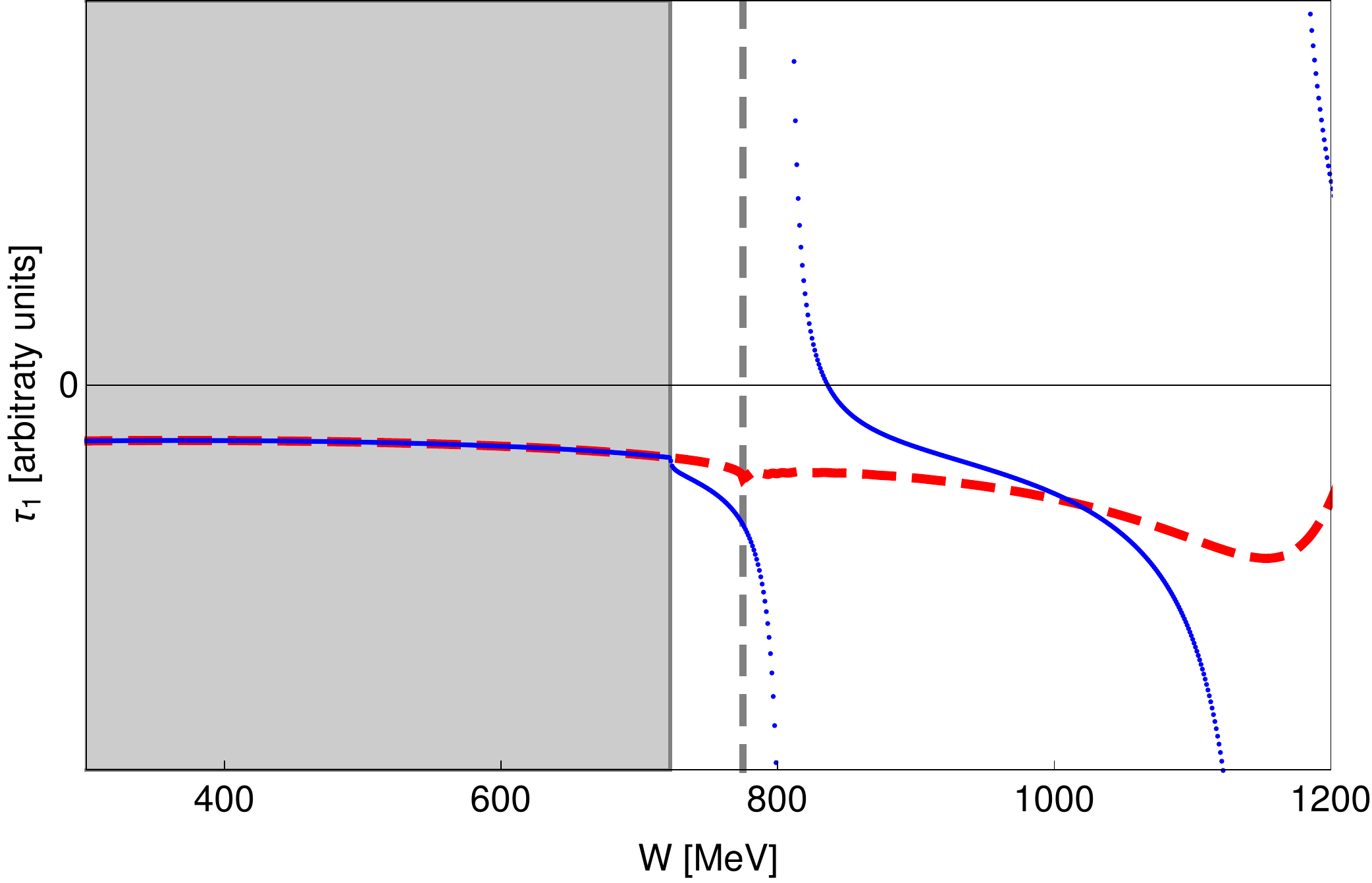}
\includegraphics[width=0.49\linewidth,trim=0.5cm 0.75cm 0cm 0cm]{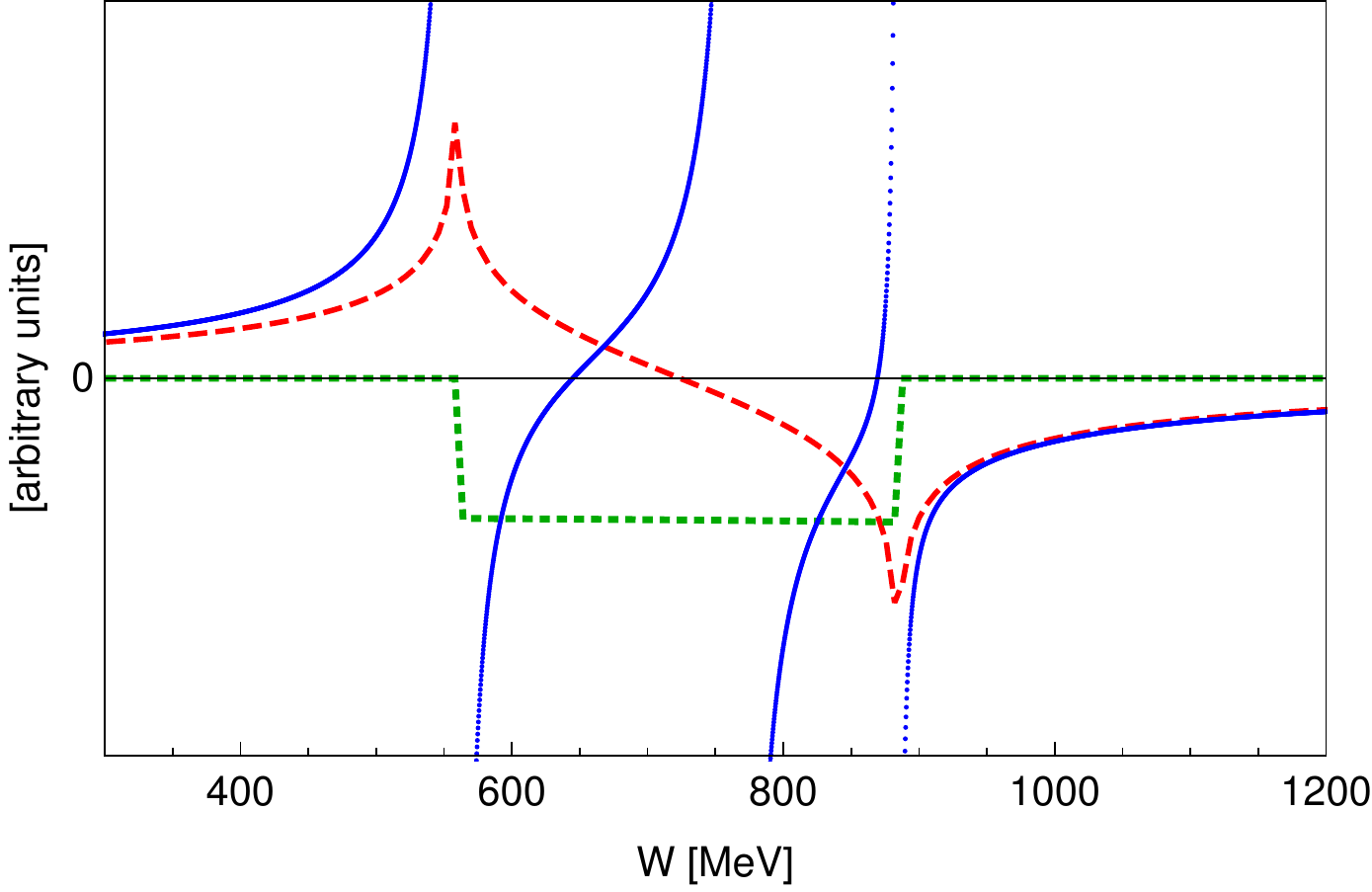}
\end{center}

\caption{Left: isobar propagator $\tau$ in the finite volume (blue line)
	and the real part of the infinite volume one (red dashed line) for a given
	boost $\boldsymbol{l}=(0,0,2\pi/L)$ and $L=3.5$~fm. The gray dashed line
	denotes the onshell-condition of two particles ($\sigma(\boldsymbol{l})=(2M)^2$),
	whereas the gray area represents the energy range for which
	$\sigma(\boldsymbol{l})\le0$.
	Right: Red dashed line (green dotted line) represent real (imaginary) part 
	of the infinite-volume $S$-wave projection of the isobar-spectator interaction
	kernel $B$. In comparison, the finite-volume projection $B^{A_1^+}$ for the
	transition from shell 1 to shell 1 at $L=6$~fm (blue dots).} \label{fig:comp}
\end{figure}
%%%-----------------------------------

As discussed in the introduction, one of the main goals for the present
investigations on the unitarity constraints on three-body systems is the
derivation of the finite-volume spectrum in such systems. To re-iterate,
the main idea is that unitarity fixes the imaginary parts of the amplitude,
which themselves lead to the power-law dependence of finite-volume corrections.

To begin, we note that since $v$ (the isobar dissociation vertex into
asymptotically stable particles) is a cut-free function in momenta, it
will not lead to any power-law finite-volume effects. Thus, the desired
quantization condition will be entirely derived from the part in brackets
of the second line of Eq.~\eqref{eq:t33full}. The non-trivial part of this
expression is the isobar-spectator scattering amplitude $T$, which in
integral form reads
\begin{align}\label{eq:T}
	T(q,p;s)=
	B(q,p;s)-
	\int
	\frac{\mathrm{d}^3\boldsymbol{l}}{(2\pi)^3}
	B(q,l;s)
	\frac{\tau(\sigma(l))}{2E({\boldsymbol l})}
	T(l,p;s)\,,
\end{align}
where the parameters of the isobar-propagator $\tau$ can be fixed from
two-body scattering data, defining the two-to-two scattering amplitude
via $T_{22}:=v\tau v$.

In the finite cubic volume with periodic boundary conditions the momenta
are discretized. In particular, in a cube of a size $L$ only the following
three-momenta are allowed (organized by ``shells'')
\begin{align}\label{eq:discrete}
	{\boldsymbol q}_{ni}=\frac{2\pi}{L}\,{\boldsymbol r}_i
	\text{~for~}
	\{{\boldsymbol r}_i\in \mathds{Z}^3 | {\boldsymbol r}_i^2=n,i=1,\dots,\vartheta(n)\} \ .
\end{align}
where $\vartheta(n)=1,6,12,\dots$ indicates the multiplicity (number of points
in shell  $n=0,1,2,\dots$) that can be calculated as described, e.g., in
Refs.~\cite{Doring:2011ipk,Doring:2018xxxk}. In principle, replacing all momenta
in Eqs.~(\ref{eq:t33full}, \ref{eq:T}, \ref{eq:B}) including the replacement
of the appearing integrals over solid angle as
${\int d\Omega_{\boldsymbol p_n}\to \frac{4\pi}{\vartheta(n)}\sum_{i=1}^{\vartheta(n)}}$ 
leads to a generic three-body quantization condition -- an equation which determines the 
positions of singularities of such an amplitude in energy. However, several subtleties arise 
from the breakdown of the spherical symmetry on the lattice, which we wish to discuss in the 
following.

First of all, the isobar propagator in the second line of Eq.~(\ref{eq:B})
is evaluated in the isobar center-of-mass frame. In the finite volume, however,
the allowed momenta given by Eq.~(\ref{eq:discrete}) are defined in the three-body
rest frame at $\boldsymbol{P}=\boldsymbol{0}$. For the calculation of the
finite-volume self-energy one, therefore, has to boost the momenta to the
isobar rest frame. In this context it is important to recall that the two-body
sub-system can become singular when the invariant mass of the system becomes
real, see, e.g., Fig.~\ref{fig:comp}. The tower (for all spectator momenta in
question) of two-body singularities has to cancel such that only genuine
three-body singularities remain in the final expression. As it is shown
analytically in the original publication~\cite{Mai:2017bgek}, such cancellation
occur in the full quantization condition when all terms (including disconnected
topology $\hat T_d$) and boosts are taken into account accordingly.

Another important observation w.r.t the breakdown of spherical symmetry is
that the isobar-spectator interaction kernel is singular for specific
combination of momenta and energies. Expressed differently, when projecting
to a partial wave in infinite volume, this term develops an imaginary part
below threshold as presented for the S-wave projection of $B$ in
Fig.~\ref{fig:comp}. Thus, in finite volume the same term has to have a series
of singularities in this region. This indeed happens when projecting $B$ to the
corresponding irreducible representation ($A_1^+$ for the depicted case) of the
cubic symmetry group $O_h$. Furthermore, the projection to irreps of $O_h$ of
the three-body scattering amplitude has two additional advantages. For once,
the results of Lattice QCD calculation are usually projected to these irreps.
Additionally, the projection to different irreps reduces the dimensionality of
the three-body scattering amplitude and therefore also that of the quantization
condition in the finite volume.

There are various ways of projecting to definite irreps in the finite volume.
In Ref.~\cite{Doring:2018xxxk} a method has been developed, which has a form
very similar to the usual partial-wave projection in infinite volume. We refer
the reader for more details on the construction techniques of this method to
the original publication~\cite{Doring:2018xxxk}, and quote here only the
corresponding result. A given function $f^s(\bf\hat p_j)$ acting on momenta
of the shell $s$ can be expanded as
\begin{align}\label{eq:exp-f-finvol_NEW}
	f^s({\bf\hat p}_j)=\sqrt{4\pi}\sum_{\Gamma\alpha}\sum_u
	f^{\Gamma \alpha s}_u \chi_u^{\Gamma \alpha s}({\bf\hat p}_j)\,
	\text{~~for~~}
	f^{\Gamma \alpha s}_u&=\frac{\sqrt{4\pi}}{\vartheta(s)}\,\sum_{j=1}^{\vartheta(s)}f^s({\bf\hat p}_j)\chi^{\Gamma \alpha s}_u({\bf\hat p}_j)\,,
\end{align}
where $\Gamma$, $\alpha$, and $u$ denote the irrep, basis vector of the irrep
and the corresponding index, respectively. In these indices the functions
$\chi_u^{\Gamma \alpha s}$ build an orthonormal basis of functions acting on
momenta on the shell $s$.

Using the projection method presented above our final result for three-body
quantization condition for the irrep $\Gamma$ reads
\begin{align}\label{eq:qcond2}
	{\rm Det}\left[B^{\Gamma ss'}_{uu'}(W^2)
	+
	\frac{2E_s \,L^3}{\vartheta(s)}
	\tau_s(W^2)^{-1}\delta_{ss'}\delta_{uu'}\right]=0\,,
\end{align}
where $W$ is the total energy of three-body system, $E_s=\sqrt{{\bf p}_i^2+M^2}$
and $\tau_s(W^2):=\tau(\sigma(p_i))$ for any three-momentum on the shell $s$.
Note that the determinant is taken with respect to a matrix in shell-indices
($s,s'$), as well as basis indices $(u,u')$ of the functional basis to the irrep
$\Gamma$, whereas the dependence on $\alpha$ drops off naturally.
To demonstrate the usefulness of the derived quantization condition we fix
$m=139$~MeV and $L=3.5$~fm. Furthermore, we assume one S-wave isobar with
$v(p,q):=\lambda f((p-q)^2)$ with $f$ such that it is 1 for $(p-q)^2=0$ and
decreases sufficiently fast for large momentum difference, e.g.,
$f(Q^2)=\beta^2/(\beta^2+Q^2)$ to regularize integrals of the scattering
equation. Note that one is not obliged to use form factors but can instead
formulate the dispersive amplitude through multiple subtractions rendering
it automatically convergent, see Eq.~(14) in Ref.~\cite{Mai:2017votk}. This
leads to
\begin{align}\label{eq:B}
	B(q,p;s)
	=
	&\frac{-\lambda^2 f(( P-q-2p)^2)f((P-2q-p)^2)}
	{2E({\boldsymbol q}+{\boldsymbol p})\left(W-E({\boldsymbol q})-E({\boldsymbol p})-E({\boldsymbol q}+{\boldsymbol p})+i\epsilon\right)}+C(q,p;s)\,,\\[8pt]
	\frac{1}{\tau(\sigma(l))}=
	&\sigma(l)-M_0^2
	-
	\int \frac{d^3\boldsymbol{k}}{(2\pi)^3}
	\frac{\lambda^2 (f(4\boldsymbol{k}^2))^2}{2E({\boldsymbol k})(\sigma(l)-4E({\boldsymbol k})^2+i\epsilon)} \,,\nonumber
\end{align}
where $C$ is a real-valued function of total energy and both spectator
momenta, while $M_0$ is a free parameter that is fixed (together with
$\lambda$ and $\beta$) to reproduce some realistic two-body scattering
data. In the present case we take the experimental phase-shifts for the
$\pi\pi$ scattering in the isovector channel for demonstration, and fix
$C=0$ and $\Gamma=A_1^+$ for simplicity.
%%%-----------------------------------
\begin{figure}
\includegraphics[width=0.99\linewidth,trim=0cm 1cm 0cm 0cm]{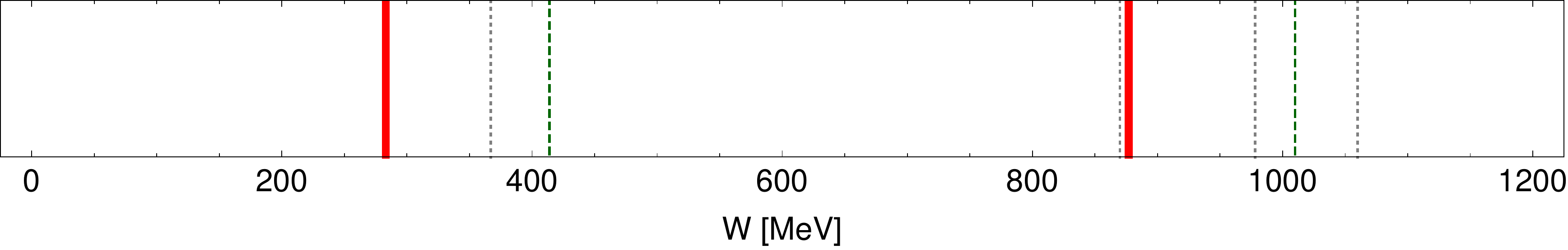}

\caption{The red lines show finite-volume energy eigenvalues for
	$L=3.5$~fm and $M=139$~MeV. The doted gray vertical lines show the
	positions of singularities of $\tau$ for all used boost momenta,
	while the dashed green vertical lines show the position of
	non-interacting energy eigenvalues of three particles.}
	\label{fig:final1}
\end{figure}
%%%-----------------------------------

The result of the numerical investigation is depicted in
Fig.~\ref{fig:final1}. It shows the non-interacting levels
of the three-body system (green, dashed lines) along with
the solutions of the quantization condition~\eqref{eq:qcond2}
(full, red lines) representing the interacting energy eigenvalues.
Since $\chi^{A_1^+}=1/\sqrt{4\pi}$ the dimensionality of the matrix
is given entirely by the set of considered shells. We have checked
that using more than 8 first shells does not lead to visible change
of the position of the interacting levels. Constraining ourselves
to these shells is equivalent to a momentum cutoff of $\sim 1$~GeV
for the given lattice volume. Note also that the range of applicability
of the quantization condition is, in principle, also restricted by
construction, due to missing intermediate higher-particle states.
The later is common to all present studies of the three-body in
finite volume, and has to be eased at some point in future. However,
this study demonstrates clearly the usefulness and practical
applicability of the derived quantization condition. Further
studies, such as volume dependence and inclusion of multiple
isobars are work in progress.

%%%-----------------------------------
\item \textbf{Acknowledgments}

The speaker thanks the organizers of the workshop for the invitation and
productive discussions during the workshop. Furthermore, he is grateful
for the financial support by the German Research Foundation (DFG), under
the fellowship MA 7156/1-1, as well as for the George Washington University
for the hospitality and inspiring environment.
\end{enumerate}

%%%-----------------------------------

%%%%%%%%%%%%%%%%%%%%%%%%%%%%%%%%%%%%%%%%%%%%%%%%%%%%%%%%%%%%%%%%%%%%%%%%%
\newpage
\subsection{Study of the Processes $e^+e^-\to K\bar{K}n\pi$ with the CMD-3 
	Detector at VEPP-2000 Collider}
\addtocontents{toc}{\hspace{2cm}{\sl Vyacheslav Ivanov}\par}
\setcounter{figure}{0}
\setcounter{table}{0}
\setcounter{footnote}{0}
\setcounter{equation}{0}
\halign{#\hfil&\quad#\hfil\cr
\large{Vyacheslav Ivanov (on behalf of CMD-3 Collaboration)}\cr
\textit{Budker Institute of Nuclear Physics}\cr
\textit{Novosibirsk, 630090 Russia}\cr}

%%%-----------------------------------
\begin{abstract}
This paper describes the preliminary results of the study of
processes of $e^{+}e^{-}$ annihilation in the final states with
kaons and pions with the CMD-3 detector at VEPP-2000 collider.
The collider allows the c.m. energy scanning in the range from
0.32 to 2.0~GeV, and about $100$ pb$^{-1}$ of data has been taken
by CMD-3 up to now. The results on the $K^{+}K^{-}\pi^{+}\pi^{-}$,
$K^{+}K^{-}\eta$, $K^{+}K^{-}\omega(782)$ and $K^{+}K^{-}\pi^{0}$
final states are considered.
\end{abstract}

%%%-----------------------------------
\begin{enumerate}
\item \textbf{Introduction}

A high-precision measurement of the inclusive $e^{+}e^{-}{\to}~hadrons$
cross section is required for a calculation of the hadronic contribution
to the muon anomalous magnetic moment $(g-2)_\mu$ in the frame of the
Standard Model. To confirm or deny the observed difference between the
calculated $(g-2)_\mu$ value~\cite{hagiwarar} and the measured 
one~\cite{bnlr}, more precise measurements of the exclusive channels of
$e^{+}e^{-}{\to}~hadrons$ are necessary. The exclusive $KK(n)\pi$ final
states are of special interest, since their producltion involves rich
intermediate dynamics which allows the test of isotopic relations and
measurement of intermediate vector mesons parameters.

In this paper we describe the current status of the study of
$e^{+}e^{-}{\to}K^{+}K^{-}\pi^{+}\pi^{-}$, $K^{+}K^{-}\eta$,
$K^{+}K^{-}\omega(782)$ and $K^{+}K^{-}\pi^{0}$ processes with
the CMD-3 detector at VEPP-2000 collider (Novosibirsk, Russia), based
on about $20$ pb$^{-1}$ of data, collected in the runs of 2011-2012
years, and about $50$ pb$^{-1}$ in the runs of 2017 year. The preliminary
results for $\phi(1680)$ meson parameters were obtained from
$K^{+}K^{-}\eta$ cross section fitting. We see also the indication on
non trivial behavior of $e^{+}e^{-}{\to}K^{+}K^{-}\pi^{+}\pi^{-}$
process cross section at $p\bar{p}$ threshold.

%%%-----------------------------------
\item \textbf{VEPP-2000 collider and CMD-3 detector}

The VEPP-2000 $e^+e^-$ collider~\cite{vepp2000r} at Budker Institute of
Nuclear Physics covers the $E_{\rm c.m.}$ range from 0.32 to 2.0\,GeV and
employs a technique of round beams to reach luminosity up to
10$^{32}$\,cm$^{-2}$s$^{-1}$ at $E_{\rm c.m.}$=2.0\,GeV. The Cryogenic
Magnetic Detector (CMD-3) described in~\cite{cmd3r} is installed in one
of the two beam interaction regions. The tracking system of the \mbox{CMD-3} 
detector consists of a cylindrical drift chamber (DC) and a double-layer 
cylindrical multiwire proportional Z-chamber, installed inside a superconducting 
solenoid with a 1.0--1.3~T magnetic field (see \mbox{CMD-3} layout in 
Fig.~\ref{fig:CMD3}). Amplitude information from the DC wires is used to 
measure the specific ionization losses ($dE/dx_{\rm DC}$) of charged particles. 
Bismuth germanate crystals of 13.4 $\rm X_{0}$ thickness are used in the endcap 
calorimeter. The barrel calorimeter, placed outside the solenoid, consists of two 
parts: internal (based on liquid Xenon (LXe) of 5.4 $\rm X_{0}$ thickness) and
external (based on CsI crystals of 8.1 $\rm X_{0}$ thickness)~\cite{CsIr}.
%%%-----------------------------------
\begin{figure}[ht]
  \begin{center}
  \includegraphics[width=6cm]{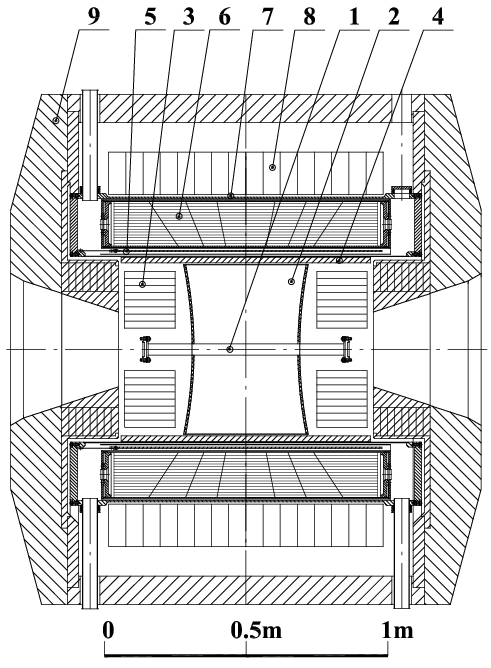}

    \caption{The \mbox{CMD-3} detector layout: 1 - beam pipe, 2 - drift
	chamber, 3 - BGO endcap calorimeter, 4 - Z-chamber, 5 - 
	superconducting solenoid, 6 - LXe calorimeter, 7 - time-of-flight 
	system, 8 - CsI calorimeter, 9 - yoke. \label{fig:CMD3}}
  \end{center}
\end{figure}
%%%-----------------------------------

The physics program of CMD-3 includes:
\begin{itemize}
\item precise measurement of the $R=\sigma(e^{+}e^{-}{\to}hadrons)/
	\sigma(e^{+}e^{-}{\to}\mu^{+}\mu^{-})$, neccessary for clarification
	of $(g-2)_{\mu}$ puzzle;
\item study of the exclusive hadronic channels of $e^{+}e^{-}$ annihilation,
	test of isotopic relations;
\item study of the $\rho$, $\omega$, $\phi$ vector mesons and their
	excitations;
\item CVC tests: comparison of isovector part of $\sigma(e^{+}e^{-}
	{\to}hadrons)$ with $\tau$ lepton decay spectra;
\item study of $G_{E}/G_{M}$ of nucleons near threshold;
\item diphoton physics (e.g. $\eta^{\prime}$ production).
\end{itemize}

%%%-----------------------------------
\item \textbf{Study of $e^{+}e^{-}{\to}KK(n)\pi$ Processes}
%%%-----------------------------------
\begin{enumerate}
\item \textbf{Charged Kaon/Pion Separation}

The starting point of the analysis of the final state with charged kaons
and pions is the kaon/pion separation, and to perform it we use the
measurement of the specific ionization losses $dE/dx$ of particles in
the DC. For the event with $n_{\rm tr}$ DC-tracks the log-likelihood
function (LLF) for the hypothesis that for $i=1,2,...,n_{\rm tr}$ the
particle with the momentum $p_{i}$ and energy losses $(dE/dx)_{i}$ is
the particle of $\alpha_{i}$ type ($\alpha_{i}=K$ or $\pi$) is defined as
\begin{eqnarray}
	L(\alpha_{1},\alpha_{2},...,\alpha_{n_{\rm tr}})=\sum_{i=1}^{n_{\rm tr}}ln\Biggl(\frac{f_{\alpha_{i}}(p_{i},(dE/dx)_{i})}{f_{K}(p_{i},(dE/dx)_{i})+f_{\pi}(p_{i},(dE/dx)_{i})}\Biggr),
\end{eqnarray}
where the functions $f_{K/\pi}(p,dE/dx)$ represent the probability density
for charged kaon/pion with the momentum $p$ to produce the energy losses
$dE/dx$ in the DC. To perform the particle identification (PID) we search
for the $(\alpha_{1},\alpha_{2},...,\alpha_{n_{\rm tr}})$ combination (with
two oppositely charged kaons and zero net charge) that delivers the maximum
to LLF. In what follows we use $L_{2K(n_{\rm tr}-2)\pi}$ designation for the
LLF maximum value. The cut on $L_{2K(n_{\rm tr}-2)\pi}$ value is used to
avoid misidentification.

Unfortunatelly, the described $(dE/dx)_{\rm DC}$-based separation for single
kaons and pions works reliably only up to the momenta $p<450$~MeV/c, see
Fig.~\ref{fig:dEdx_k_pi_DC}. For the $K^{+}K^{-}$, $K^{+}K^{-}\pi^{0}$,
$K^{+}K^{-}\pi^{0}\pi^{0}$ and $K_{S}K^{\pm}\pi^{\mp}$ final states studies
we are developing other technique based on the $dE/dx$ in 14 layers of
LXe-calorimeter, see detailed description in~\cite{Ivanov:2017oobr}.

%%%-----------------------------------
\item \textbf{Study of the $e^+e^-{\to}K^+K^-\pi^{+}\pi^{-}$ Process}

The study of $e^+e^-{\to}K^+K^-\pi^{+}\pi^{-}$ process has been performed
on the base of $23$ pb$^{-1}$ of data, collected in 2011-2012. The events
with 3 and 4 DC-tracks were considered with the kaon/pion separation using
the LLF maximization and cuts on $L_{2K2\pi}$ and $L_{2K\pi}$, see
Fig.~\ref{fig:L_KKpipi}. For the class of events with 4 tracks the pure
sample of signal events was selected using energy-momentum conservation
law (see Fig.~\ref{fig:dE_dP}). For the 3-tracks class the signal/background
separation was performed by the fitting of energy disbalance
${\Delta}E_{3+1}{\equiv}E_{K^+}+E_{K^-}+E_{\pi}+\sqrt{m^{2}_{\pi}
+(\vec{p}_{K^{+}}+\vec{p}_{K^{-}}+\vec{p}_{\pi})^{2}}-\sqrt{s}$
distribution, see Fig.~\ref{fig:kkpipi_selection}. In total we selected
about 24000 of signal events.
%%%-----------------------------------
\begin{figure}[hbtp]
  \begin{minipage}[t]{0.5\textwidth}
    \centerline{\includegraphics[width=1.0\textwidth]{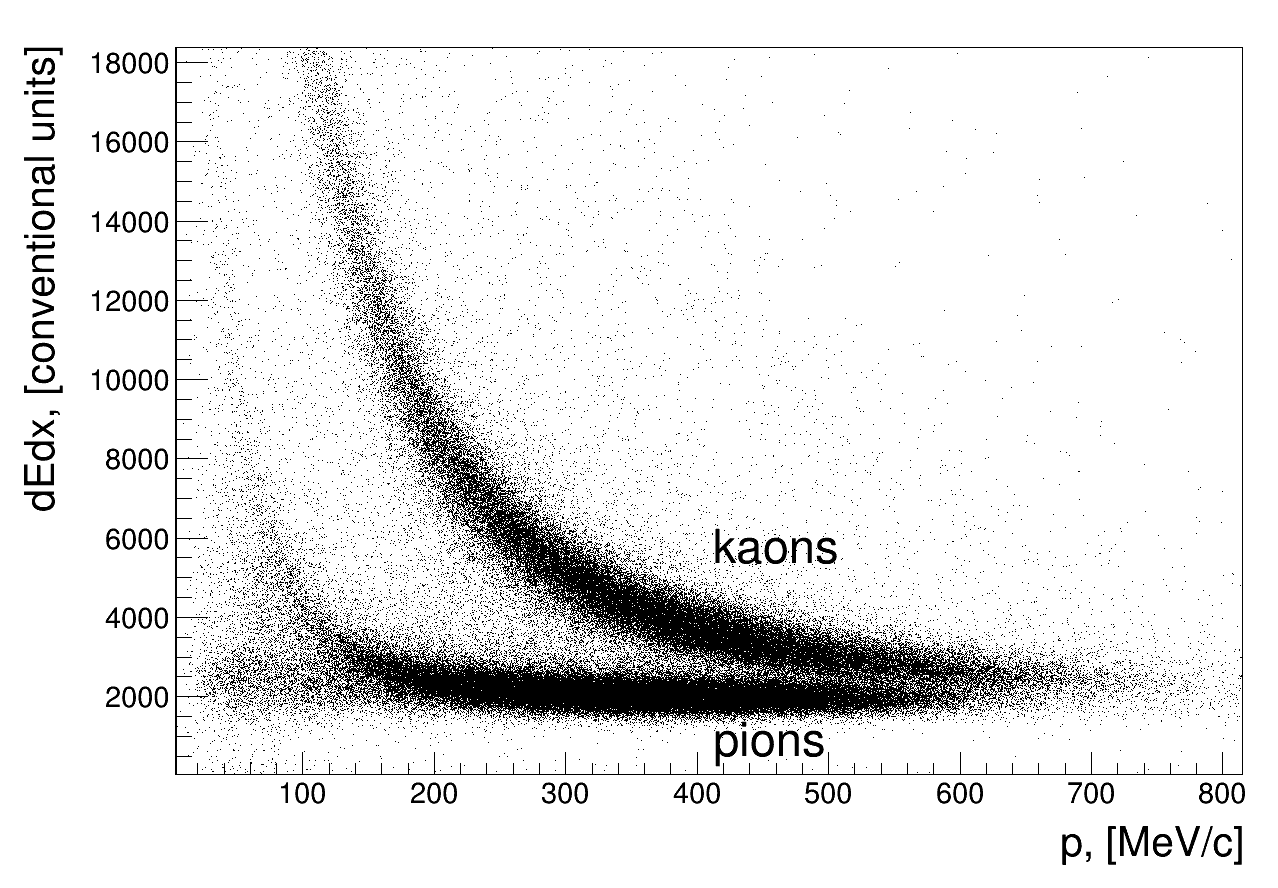}}

    \caption{The $(dE/dx)_{\rm DC}$ distribution for the simulated kaons 
	and pions in $K^{+}K^{-}\pi^{+}\pi^{-}$ final state.
    \label{fig:dEdx_k_pi_DC}}
  \end{minipage}\hfill\hfill
  \begin{minipage}[t]{0.45\textwidth}
    \centerline{\includegraphics[width=1.0\textwidth]{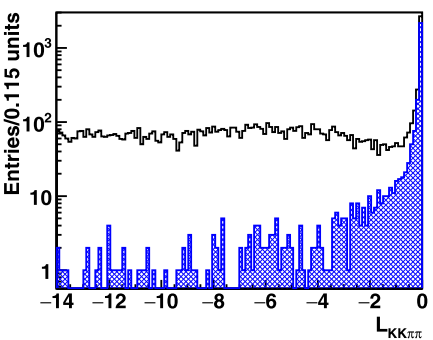}}

    \caption{The $L_{2K2\pi}$ distribution for the 4-tracks events for
	data (open histogram) and $K^{+}K^{-}\pi^{+}\pi^{-}$ simulation 
	(blue histogram). The cut $L_{2K2\pi}>-3.0$ is applied to avoid 
	misidentification. All c.m. energy points are combined.
    \label{fig:L_KKpipi}}
  \end{minipage}\hfill\hfill
\end{figure}
%%%-----------------------------------
%%%-----------------------------------
\begin{figure}[hbtp]
  \begin{minipage}[t]{0.5\textwidth}
    \centerline{\includegraphics[width=1.0\textwidth]{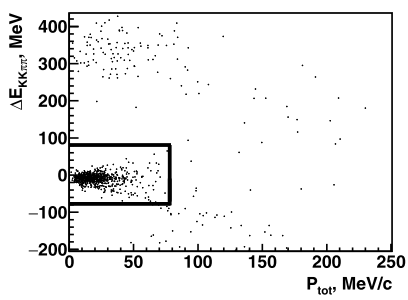}}

    \caption{The energy disbalance vs. total momentum for the 4-tracks
	events after kaon/pion separation (data, $E_{\rm c.m.}=1.98{\,}\rm GeV$).
	The events inside the frame are considered to be signal.
    \label{fig:dE_dP}}
  \end{minipage}\hfill\hfill
  \begin{minipage}[t]{0.45\textwidth}
    \centerline{\includegraphics[width=1.0\textwidth]{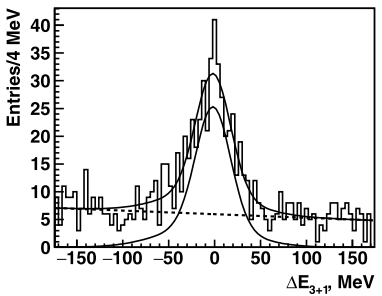}}

    \caption{The signal/background separation by fitting the energy
	disbalance distribution for 3-tracks events (data, $E_{\rm c.m.}=
	1.98{\,}\rm GeV$).
        \label{fig:kkpipi_selection}}
  \end{minipage}\hfill\hfill
\end{figure}
%%%-----------------------------------

The cross section measurement for $e^+e^-{\to}K^+K^-\pi^{+}\pi^{-}$
process requires the amplitude analysis of the final state production.
The major intermediate mechanisms were found to be:
\begin{itemize}
\item $f_{0}(500,980)\phi(1020)$;
\item $\rho(770)(KK)_{\rm S-wave}$;
\item $(K_{1}(1270,1400)K)_{\rm S-wave}{\to}(K^{\ast}(892)\pi)_{\rm S-wave}K$;
\item $(K_{1}(1400)K)_{\rm S-wave}{\to}(\rho(770)K)_{\rm S-wave}K$.
\end{itemize}

The relative amplitudes of these mechanisms at each c.m. energy point were
found using the unbinned fit of the data, see the Monte-Carlo-data comparison
after the fit in
Figs.~\ref{fig:unbinned_fit_results}a--~\ref{fig:unbinned_fit_results}d.
%%%-----------------------------------
\begin{figure}[hbtp]
  \begin{minipage}[t]{0.5\textwidth}
    \centerline{\includegraphics[width=1.0\textwidth]{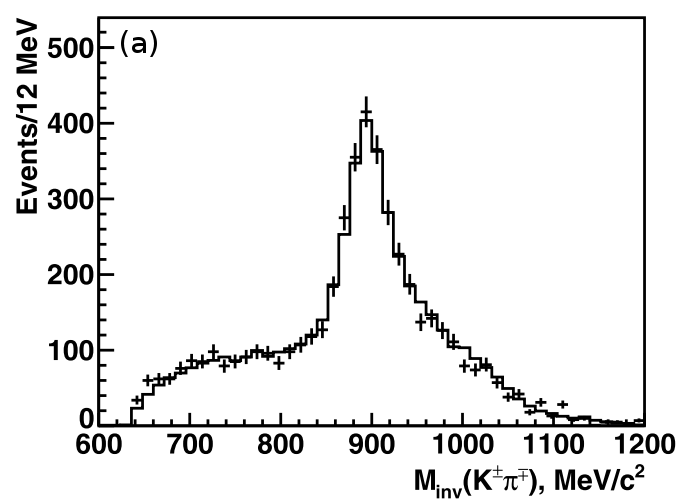}}
  \end{minipage}\hfill\hfill
  \begin{minipage}[t]{0.5\textwidth}
    \centerline{\includegraphics[width=1.0\textwidth]{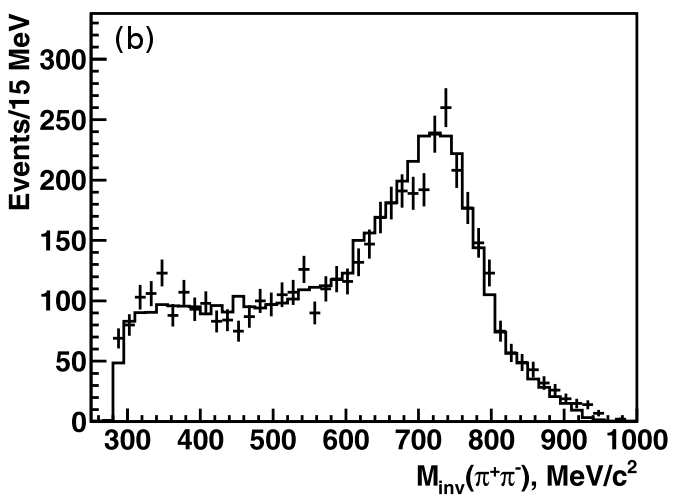}}
  \end{minipage}\hfill\hfill
  \begin{minipage}[t]{0.5\textwidth}
    \centerline{\includegraphics[width=1.0\textwidth]{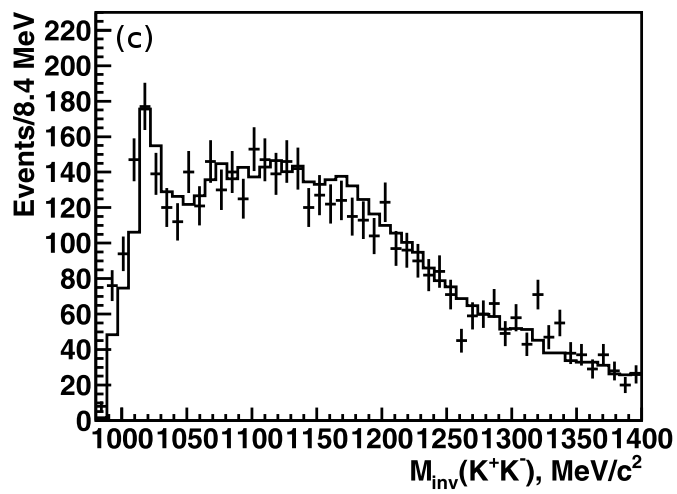}}
  \end{minipage}\hfill\hfill
  \begin{minipage}[t]{0.5\textwidth}
    \centerline{\includegraphics[width=1.0\textwidth]{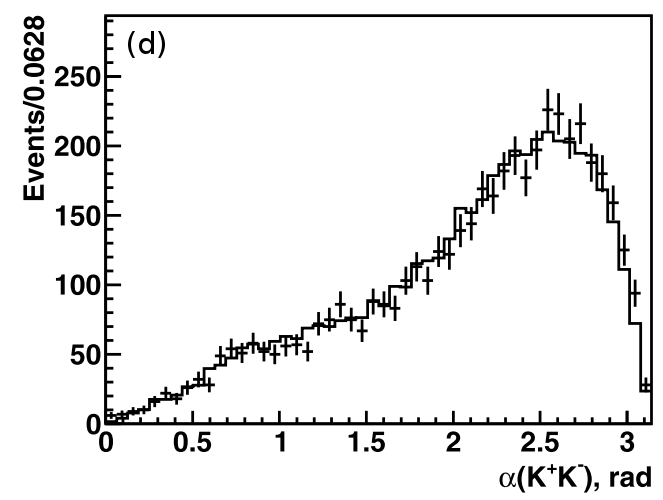}}
  \end{minipage}\hfill\hfill

  \caption{The distributions of the $K^{\pm}\pi^{\mp}$ (a), $\pi^{+}\pi^{-}$
	(b), $K^{+}K^{-}$ (c) invariant masses and the angle between momenta 
	of kaons (d). Data (points) and simulation (open histogram) for
	$E_{\rm c.m.}=1.95{\,}\rm GeV$. \label{fig:unbinned_fit_results}}
\end{figure}
%%%-----------------------------------

The results for the cross section measurement, based on the runs of 2011-2012
(published in~\cite{shemyakin_kkpipir}), are shown in Fig.~\ref{fig:cs_kkpipi}.
The preliminary results of the analyzis of new data of 2017 show a drop of
about $10\%$ in the visible cross section at $p\bar{p}$ threshold
(see Fig.~\ref{fig:cs_kkpipi_drop}), similar to that in
$e^{+}e^{-}{\to}3\pi^{+}3\pi^{-}$ cross section~\cite{cmd3_6pir}. Such a drop
at $p\bar{p}$ threshold is firstly observed in the final state with kaons,
and, being confirmed, will require theoretical explanation.
%%%-----------------------------------
\begin{figure}[ht]
  \begin{center}
    \includegraphics[width=7cm]{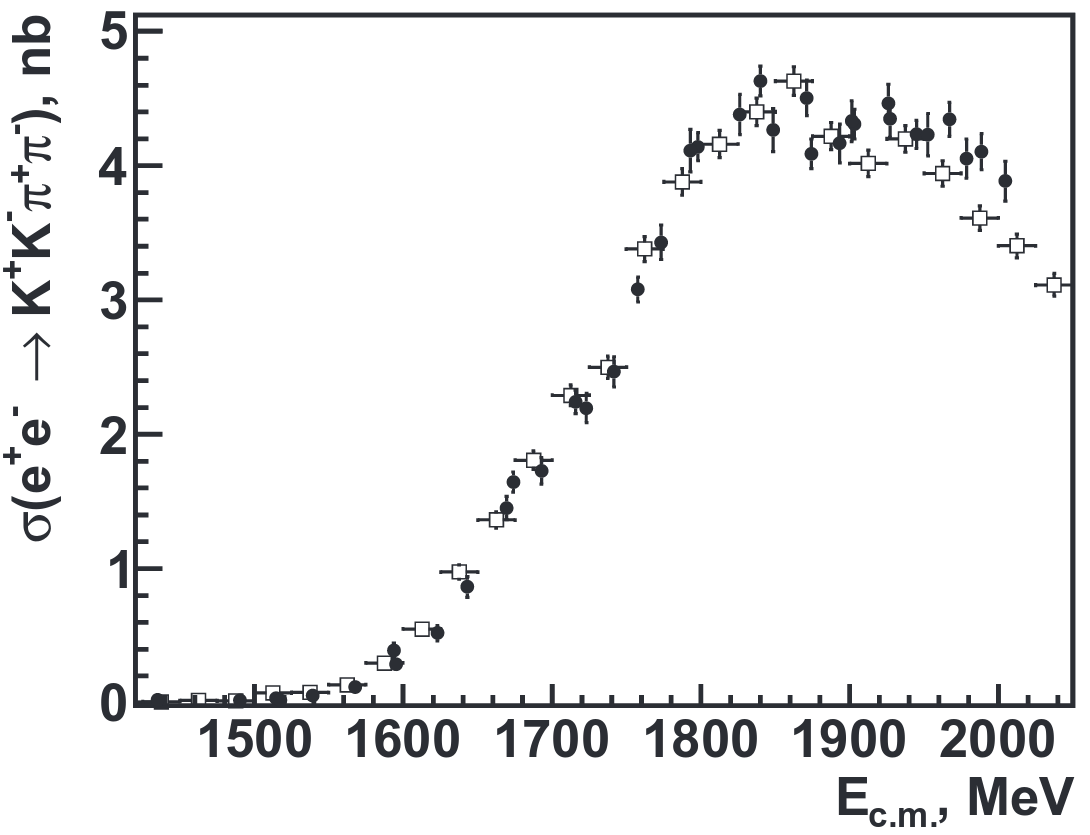}

    \caption{The $e^{+}e^{-}{\to}K^{+}K^{-}\pi^{+}\pi^{-}$ cross section,
	measured on the base of the runs of 2011-2012 years (black circles), 
	along with the BaBar results (open 
	bars)~\protect\cite{babar_kkpipir}.\label{fig:cs_kkpipi}}
  \end{center}
\end{figure}
%%%-----------------------------------
%%%-----------------------------------
\begin{figure}[ht]
  \begin{center}
    \includegraphics[width=10cm]{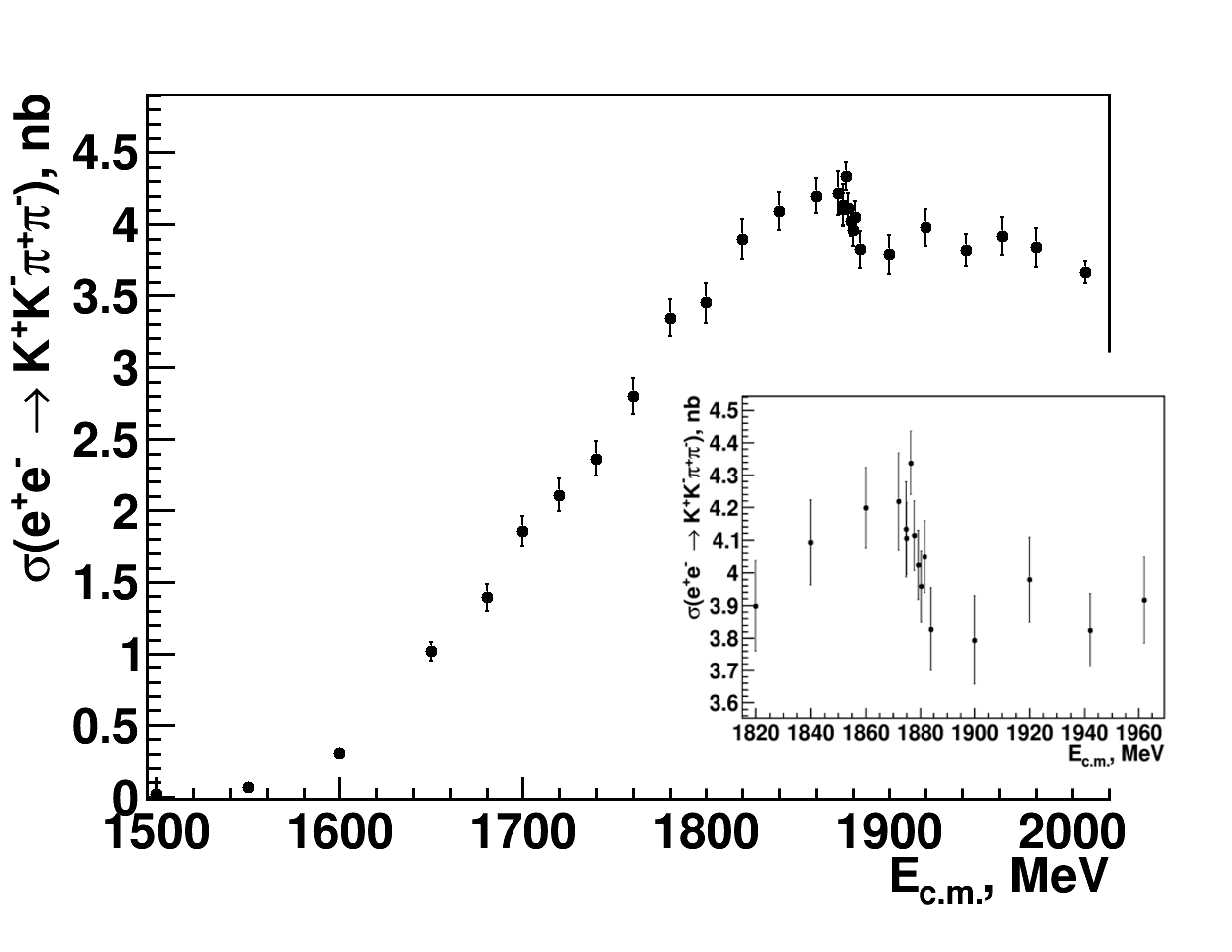}

    \caption{The visible cross section of $e^{+}e^{-}{\to}K^{+}K^{-}\pi^{+}
	\pi^{-}$ process with the drop at $p\bar{p}$ threshold (on the 
	base of 2017 year runs). \label{fig:cs_kkpipi_drop}}
  \end{center}
\end{figure}
%%%-----------------------------------

%%%-----------------------------------
\item \textbf{Study of the $e^+e^-{\to}K^+K^-\eta,{\,}K^+K^-\omega(782)$
	processes}

The study of $e^+e^-{\to}K^+K^-\eta,{\,}K^+K^-\omega(782)$ processes has
been performed on the base of $19$ pb$^{-1}$ of data, collected in 2011-2012.
In these two analyzes the $\eta$ and $\omega(782)$ were treated as the recoil
particles. The events with 2, 3 and 4 DC-tracks were considered. The kaon/pion
separation was performed using the LLF maximization and cuts on $L_{2K}$,
$L_{2K\pi}$ and $L_{2K2\pi}$, see Fig.~\ref{fig:L_KKpi_kketa}. The
$K^{+}K^{-}\pi^{+}\pi^{-}$ final state dominates the background in 3 and
4-tracks classes, but we suppress it's contribution by the cuts on the
$2K\pi$ and $2K2\pi$ missing masses, see Fig.~\ref{fig:Mmiss_2k2pi_kketa}.
Since we observed only $\phi(1020)\eta$ intermediate mechanism of
$K^{+}K^{-}\eta$ production, we applied the cut on the $K^{+}K^{-}$
invariant mass to select the $\phi(1020)$ meson region, see
Fig.~\ref{fig:Minv_KK}. The signal/background separation in both processes
is performed by approximation of the $K^{+}K^{-}$ missing mass distribution,
see Fig.~\ref{fig:Mmiss_KK}. The preliminary results for the cross sections
are shown in Figs.~\ref{fig:cs_kketa}--\ref{fig:cs_kkomega}.
%%%-----------------------------------
\begin{figure}[hbtp]
  \begin{minipage}[t]{0.48\textwidth}
    \centerline{\includegraphics[width=1.0\textwidth]{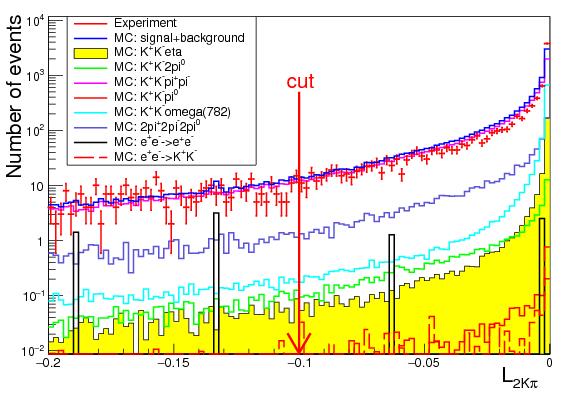}}

    \caption{The distribution of $L_{2K\pi}$ parameter in the experiment
	and simulation of signal and major background processes. All c.m. 
	energy points are combined.
      \label{fig:L_KKpi_kketa}}
  \end{minipage}\hfill\hfill
  \begin{minipage}[t]{0.48\textwidth}
    \centerline{\includegraphics[width=1.0\textwidth]{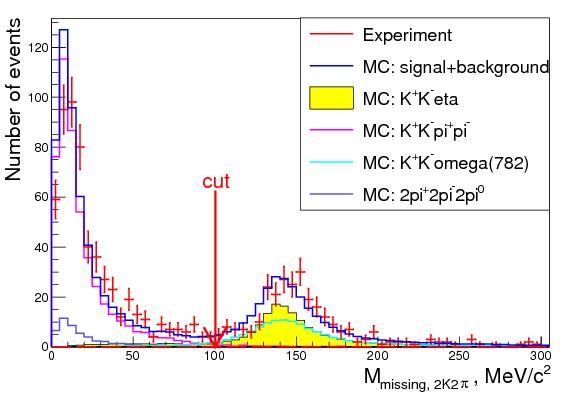}}

    \caption{The distribution of the $2K2\pi$ missing mass in the 
	experiment and simulation of signal and major background processes. 
	All c.m. energy points are combined.
      \label{fig:Mmiss_2k2pi_kketa}}
  \end{minipage}\hfill\hfill
\end{figure}
%%%-----------------------------------
%%%-----------------------------------
\begin{figure}[hbtp]
  \begin{minipage}[t]{0.48\textwidth}
    \centerline{\includegraphics[width=1.0\textwidth]{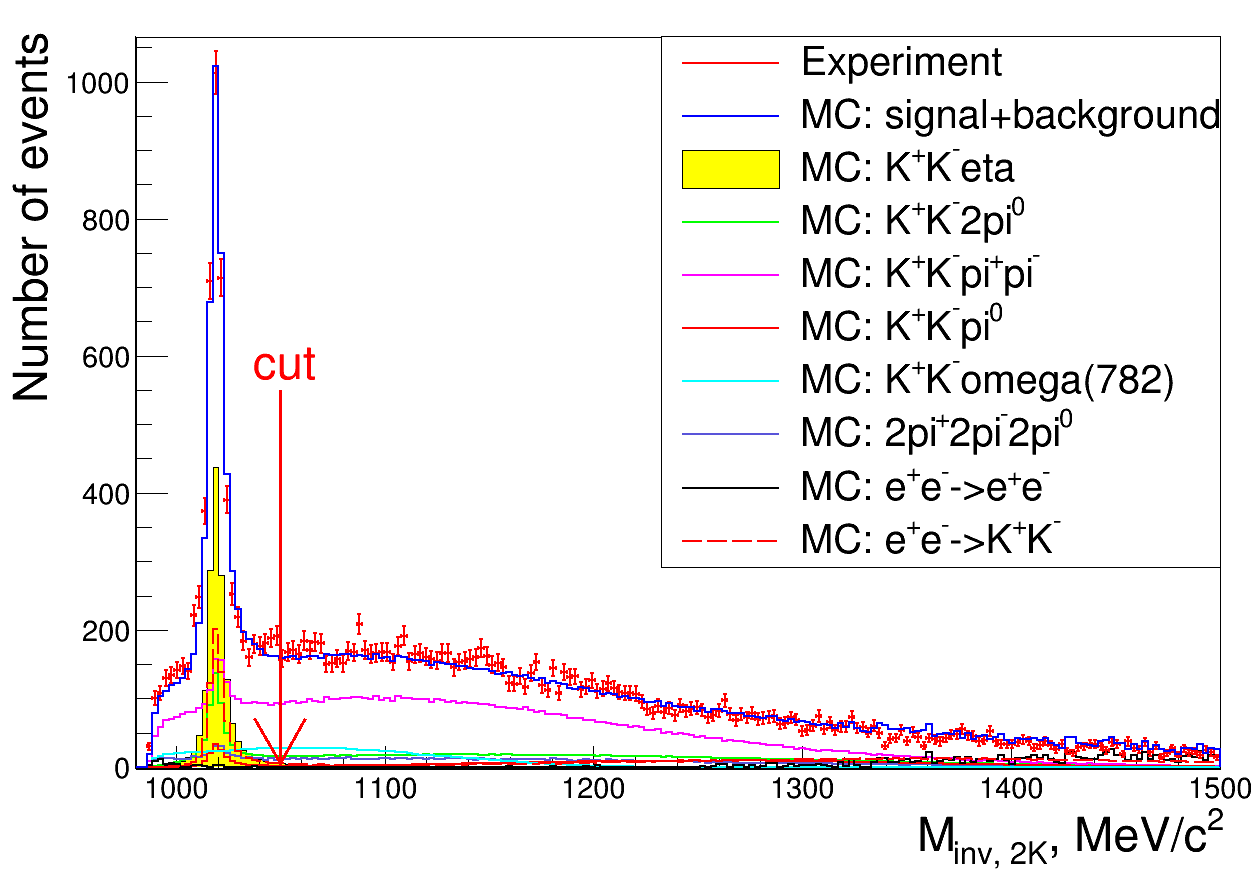}}

    \caption{The distribution of $K^{+}K^{-}$ invariant mass in the 
	experiment and simulation of signal and major background 
	processes. All c.m. energy points are combined.
        \label{fig:Minv_KK}}
  \end{minipage}\hfill\hfill
  \begin{minipage}[t]{0.48\textwidth}
    \centerline{\includegraphics[width=1.0\textwidth]{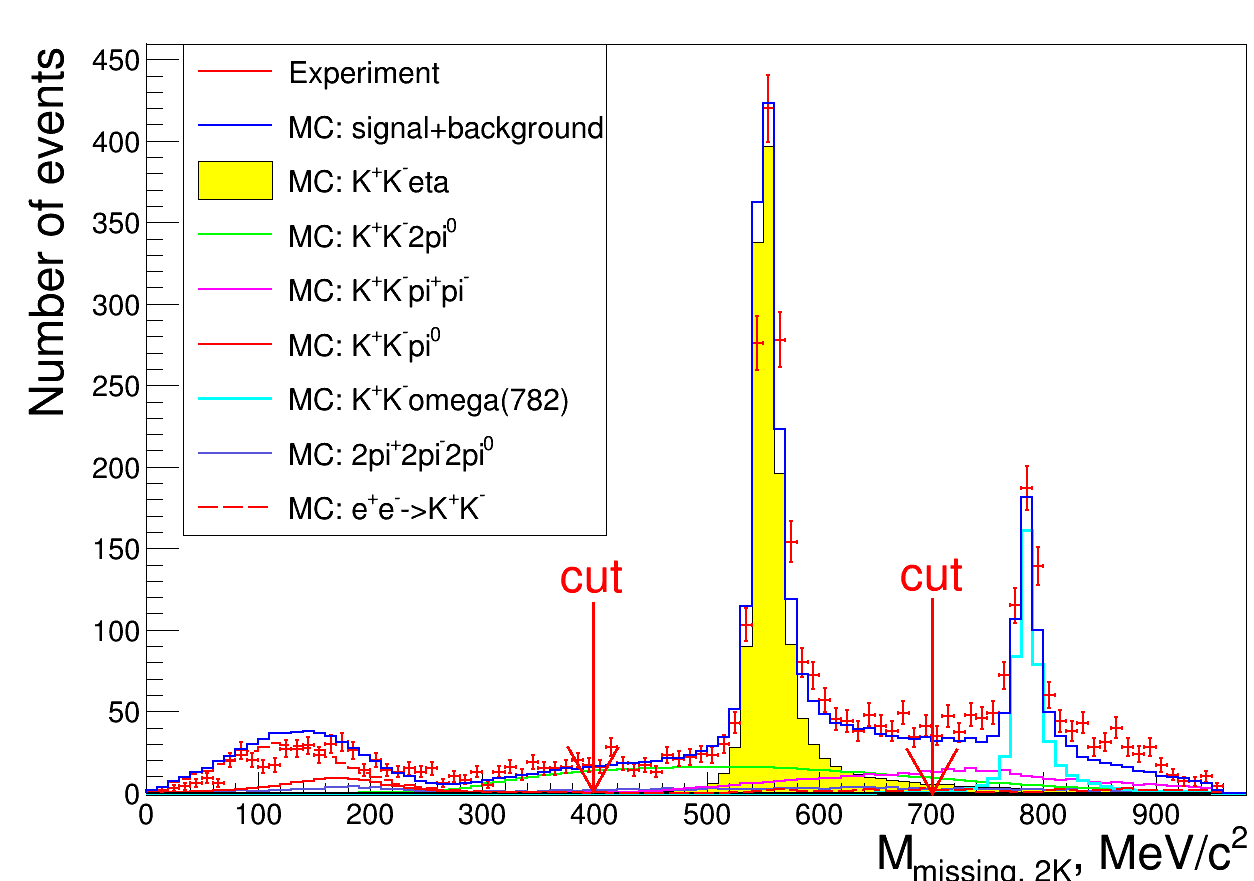}}

\caption{The distribution of $K^{+}K^{-}$ missing mass in the 
	experiment and simulation of signal and major background 
	processes. All c.m. energy points are combined.
        \label{fig:Mmiss_KK}}
  \end{minipage}\hfill\hfill
\end{figure}
%%%-----------------------------------
%%%-----------------------------------
\begin{figure}[hbtp]
  \begin{minipage}[t]{0.48\textwidth}
    \centerline{\includegraphics[width=1.0\textwidth]{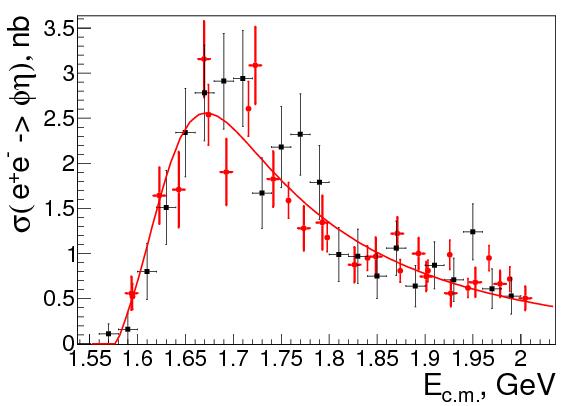}}

    \caption{The results for the $e^{+}e^{-}{\to}K^{+}K^{-}\eta$ cross
	section (red - CMD-3, preliminary; black - 
	BaBar~\protect\cite{babar_kpkmeta_2gammar}).
        \label{fig:cs_kketa}}
  \end{minipage}\hfill\hfill
  \begin{minipage}[t]{0.48\textwidth}
    \centerline{\includegraphics[width=1.0\textwidth]{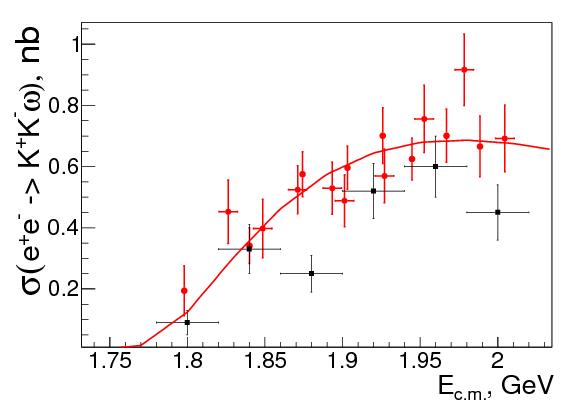}}

    \caption{The results for the $e^{+}e^{-}{\to}K^{+}K^{-}\omega(782)$
	cross section (red - CMD-3, preliminary; black -
	BaBar~\protect\cite{babar_kpkmomegar}).
        \label{fig:cs_kkomega}}
  \end{minipage}\hfill\hfill
\end{figure}
%%%-----------------------------------

Since the $\phi(1020)\eta$ production is dominated by $\phi(1680)$ meson
decay, the approximation of the $e^{+}e^{-}{\to}\phi(1020)\eta$ cross
section allows us to measure the $\phi(1680)$ parameters. We perform the
approximation using the following formula:
\begin{eqnarray}
	F(s)={\Biggl|}A_{\rm non-\phi^{\prime}}(s)e^{i\Psi}+\sqrt{\frac{(\Gamma^{\phi^{\prime}}_{ee}\mathcal{B}(\phi^{\prime}{\to}\phi\eta))\Gamma_{\phi^{\prime}}m^{3}_{\phi^{\prime}}}{|\vec{p}_{\phi}(m_{\phi^{\prime}})|^3}}D_{\phi^{\prime}}(s){\Biggr|}^{2}.
\end{eqnarray}

In these formulae $D_{\phi^{\prime}}(s)=1/(s-m^{2}_{\phi^{\prime}}
+i\sqrt{s}\Gamma_{\phi^{\prime}}(s))$ and $D_{\phi}(p^2_{\phi})=
1/(p^{2}_{\phi}-m^{2}_{\phi}+i\sqrt{p^{2}_{\phi}}\Gamma_{\phi}
(p^{2}_{\phi}))$ are the inverse denominators of the $\phi$ and
$\phi^{\prime}$ propagators, $|\vec{p}_{\phi}(\sqrt{s})|$ is the
momentum of the $\phi$ in the $\phi^{\prime}{\to}\phi\eta$ decay
in $\phi^{\prime}$ rest frame, $|\vec{p}_{K}(\sqrt{p^{2}_{\phi}})|$
is the momentum of the kaon in the $\phi{\to}K^{+}K^{-}$ decay in
$\phi$ rest frame, $\theta_{\rm normal}$ is the polar angle of the
normal to the plane, formed by the $\vec{p}_{K^{+}}$ and $\vec{p}_{K^{-}}$
vectors, $d\Phi_{K^{+}K^{-}\eta}$ is the element of three-body phase space,
the function $A_{\rm non-\phi^{\prime}}(s)=a/s$ is introduced to describe
the possible contribution of the resonances apart from $\phi^{\prime}$
($a$ is a constant), $\Psi$ is the relative phase between two amplitudes.
The results of the fit, shown in the Table~1, are in good agreement with
those in BaBar study~\cite{babar_kpkmeta_2gammar}.

For the $\sigma(e^{+}e^{-}{\to}K^{+}K^{-}\omega(782))$ approximation
we used the shape, obtained from the integration of the squared matrix
element of $e^{+}e^{-}{\to}\phi^{\prime}{\to}K^{+}K^{-}\omega(782)$,\\ 
$\omega(782){\to}\rho^{\pm,0}\pi^{\mp,0}{\to}\pi^{+}\pi^{-}\pi^{0}$
decay chain over the 5-body phase space.
%%%-----------------------------------
\begin{table}[h]
  \begin{center}
    \caption{The $\phi^{\prime}$ parameters obtained from the fit.
      \label{fit_results}}
    \begin{tabular}{ccc}
      \hline
      Parameter & Value \\
      \hline
      $\chi^{2}/{\rm n.d.f}$ & $46.3/33{\approx}1.4$ \\
      $\Gamma^{\phi^{\prime}}_{ee}\mathcal{B}(\phi^{\prime}{\to}\phi\eta),'\rm eV$ &$163{\pm}37_{\rm stat}{\pm}6_{\rm mod}$ \\
      $m_{\phi^{\prime}},\rm MeV$ & $1690{\pm}12_{\rm stat}{\pm}3_{\rm mod}$ \\
      $\Gamma_{\phi^{\prime}},\rm MeV$ & $327{\pm}88_{\rm stat}{\pm}14_{\rm mod}$ \\
      \hline
    \end{tabular}
  \end{center}
\end{table}
%%%-----------------------------------

%%%-----------------------------------
\item \textbf{Study of the $e^+e^-{\to}K^+K^-\pi^{0}$ Process}

The study of $e^+e^-{\to}K^+K^-\pi^{0}$ process has been performed on the base of
$30$ pb$^{-1}$ of data, collected in 2011-2012. The events with two oppositely
charged DC-tracks and no less than 2 photons with the $E_{\gamma}>40{\,}\rm MeV$
were considered. Then, assimung energy-momentum conservation, the 4C-kinematic fit
was performed with the $\chi^{2}_{4C}<35$ cut, see Fig.~\ref{fig:chi2}. The major
background sources were found to be $K^{+}K^{-}\gamma$, $K^{+}K^{-}2\pi^{0}$,
$K_{S,L}K^{\pm}\pi^{\mp}$, $\pi^{+}\pi^{-}\pi^{0}$, $\pi^{+}\pi^{-}2\pi^{0}$.
The suppression of these backgrounds was done using the training of BDT classifiers
(see Fig.~\ref{fig:BDT}) with the following input variables: 1) $(dE/dx)_{\rm DC}$; 2)
momenta and angles of charged particles and photons; 3) missing mass of $K^{+}K^{-}$
system. The preliminary results for the $e^{+}e^{-}{\to}K^{+}K^{-}\pi^{0}$
cross section are shown in Fig.~\ref{fig:cs_kkpi0}.
%%%-----------------------------------
\begin{figure}[hbtp]
  \begin{minipage}[t]{0.48\textwidth}
    \centerline{\includegraphics[width=1.0\textwidth]{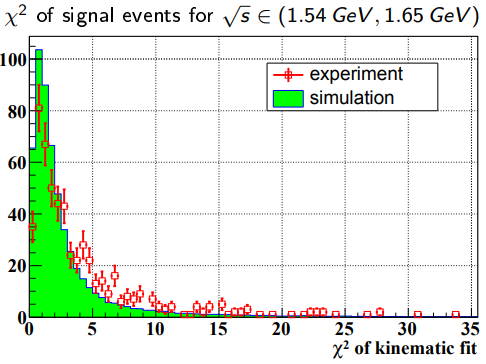}}

    \caption{The distribution of the $\chi^2$ of the 4C-kinematic fit of 
	the events (red - experiment, green - simulation of signal process). 
	The energy points in range $1.54{\,}\rm GeV<E_{\rm c.m.}<1.65{\,}\rm 
	GeV$ are combined.
        \label{fig:chi2}}
  \end{minipage}\hfill\hfill
  \begin{minipage}[t]{0.48\textwidth}
    \centerline{\includegraphics[width=1.0\textwidth]{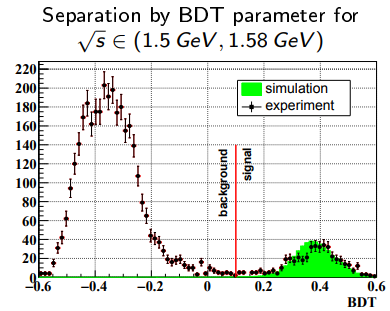}}

    \caption{The distribution of the BDT response of the events (red - 
	experiment, green - simulation of signal process). The energy 
	points in range $1.5{\,}\rm GeV<E_{\rm c.m.}<1.58{\,}\rm GeV$ 
	are combined.
        \label{fig:BDT}}
  \end{minipage}\hfill\hfill
\end{figure}
%%%-----------------------------------
%%%-----------------------------------
\begin{figure}[ht]
  \begin{center}
    \includegraphics[width=8cm]{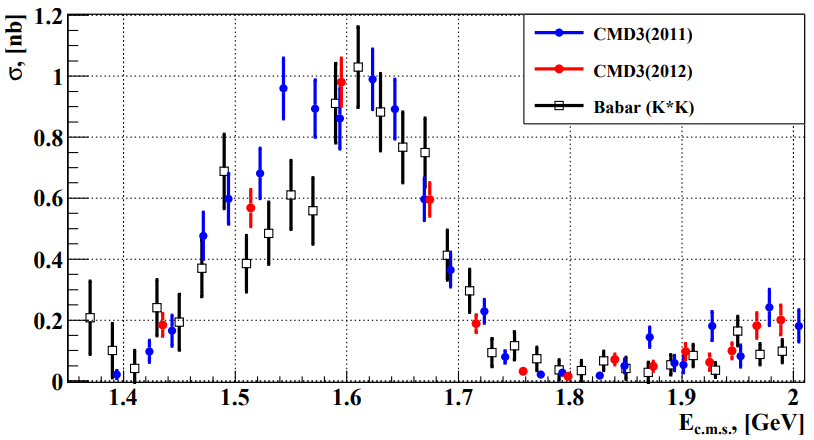}

    \caption{The results for the $e^{+}e^{-}{\to}K^{+}K^{-}\pi^{0}$ 
	cross section (blue and red - CMD-3 2011-2012, preliminary; black - 
	BaBar~\cite{babar_kpkmeta_2gammar}).
        \label{fig:cs_kkpi0}}
  \end{center}
\end{figure}
%%%-----------------------------------
\end{enumerate}

%%%-----------------------------------
\item \textbf{Conclusion}

The current status of the study of processes of $e^{+}e^{-}{\to}KK(n)\pi$ with the
CMD-3 detector was considered. The CMD-3 has alredy collected about $100$~pb$^{-1}$
of data and now is continuing datataking to collect about 1~fb$^{-1}$ in the next
few years. We are in good disposition to provide the best precision for the 
$\phi(1680)$ vector meson parameters and to perform the study of $KK(n)\pi$ final 
states in all charge modes to test the isotopic relations between them. The drop in 
the $e^{+}e^{-}{\to}K^{+} K^{-}\pi^{+}\pi^{-}$ cross section, seen in the preliminary 
analysis of the data of 2017 year runs, is firstly observed in the final state with 
kaons, and, being confirmed, will require theoretical explanation.

%%%-----------------------------------
\item \textbf{Acknowledgments}

We thank the VEPP-2000 personnel for the excellent machine operation.
The work was supported by the Russian Fund for Basic Research grants
RFBR 15--02--05674--a, RFBR 14--02--00580--a, RFBR 16--02--00160--a,
RFBR 17--02--00897--a. Part of this work related to the photon
reconstruction algorithm in the electromagnetic calorimeter is
supported by the Russian Science Foundation (project No. 14--50--00080).
\end{enumerate}

%%%-----------------------------------

%%%%%%%%%%%%%%%%%%%%%%%%%%%%%%%%%%%%%%%%%%%%%%%%%%%%%%%%%%%%%%%%%%%%%%%%%
\newpage
\subsection{The GlueX Meson Program}
\addtocontents{toc}{\hspace{2cm}{\sl Justin Stevens}\par}
\setcounter{figure}{0}
\setcounter{table}{0}
\setcounter{equation}{0}
\setcounter{footnote}{0}
\halign{#\hfil&\quad#\hfil\cr
\large{Justin Stevens (for the GlueX Collaboration}\cr
\textit{Department of Physics}\cr
\textit{College of William \& Mary}\cr
\textit{Williamsburg, VA 23187, U.S.A.}\cr}

%%%-----------------------------------
\begin{abstract}
The GlueX experiment is located in Jefferson Lab's Hall~D, and
provides a unique capability to study high-energy photoproduction,
utilizing a 9~GeV linearly polarized photon beam. Commissioning of the
Hall~D beamline and GlueX detector was recently completed and the data
collected in 2017 officially began the GlueX physics program.
\end{abstract}

%%%-----------------------------------
\begin{enumerate}
\item \textbf{Meson Photoproduction}

The GlueX experiment, shown schematically in Fig.~\ref{fig:dirc},
utilizes a tagged photon beam derived from Jefferson Lab's 12~GeV
electron beam.  Coherent bremsstrahlung radiation from a thin diamond
wafer, yields a linearly polarized photon beam with a maximum intensity
near 9~GeV.  The primary goal of the experiment is to search for and
ultimately study an unconventional class of mesons, known as exotic
hybrid mesons which are predicted by Lattice QCD
calculations~\cite{Dudek:2013yjau}.
%%%-----------------------------------
\begin{figure}[htb!]
    \centering
    \includegraphics[width=0.85\textwidth]{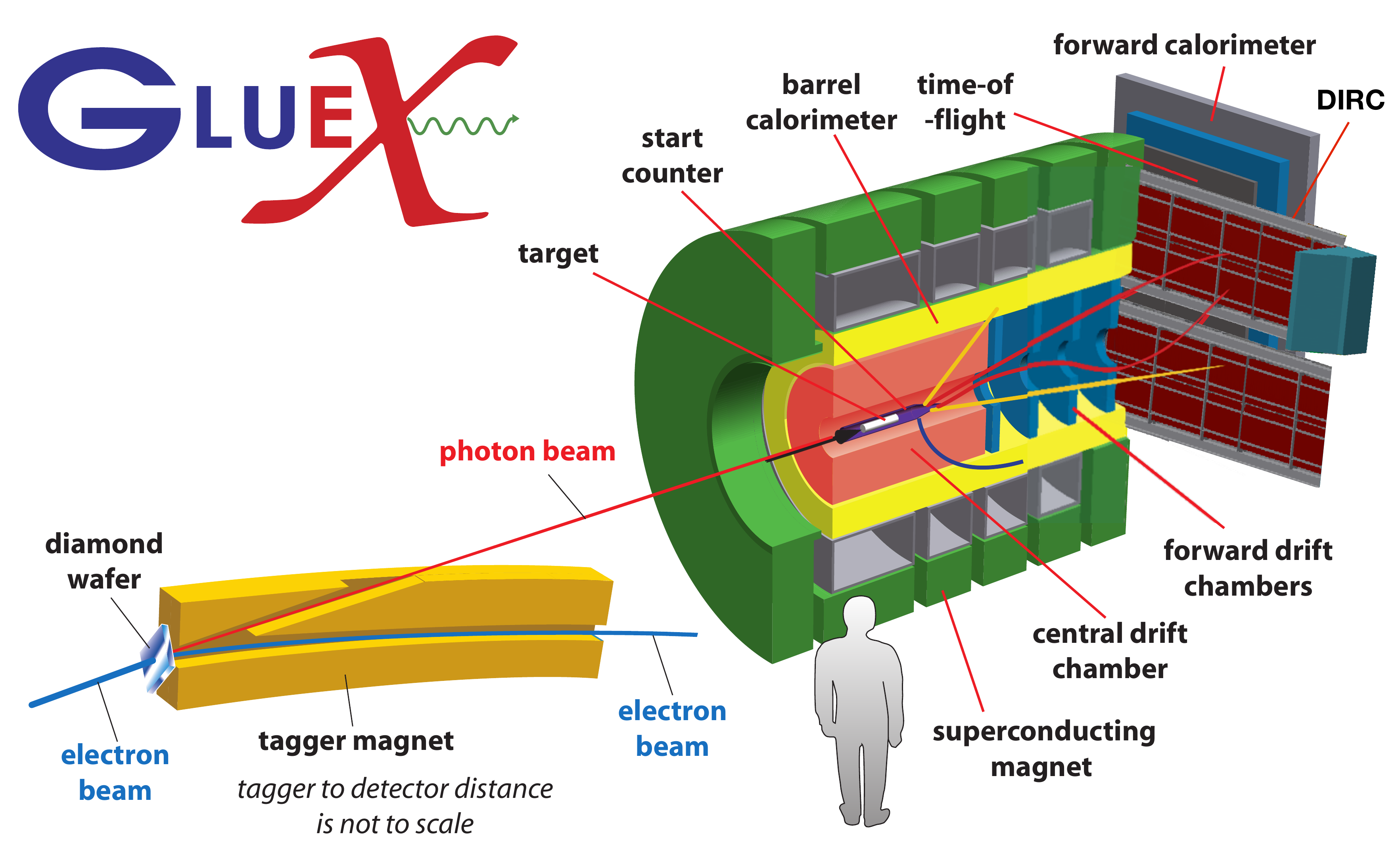}

        \caption{A schematic of the Hall~D beamline and GlueX detector at
        Jefferson Laboratory.  The DIRC detector upgrade will be installed
        directly upstream of the time-of-flight detector in the forward
        region.} \label{fig:dirc}
\end{figure}
%%%-----------------------------------
%%%-----------------------------------
\begin{figure}[htb!]
    \centering
    \includegraphics[width=0.4\textwidth]{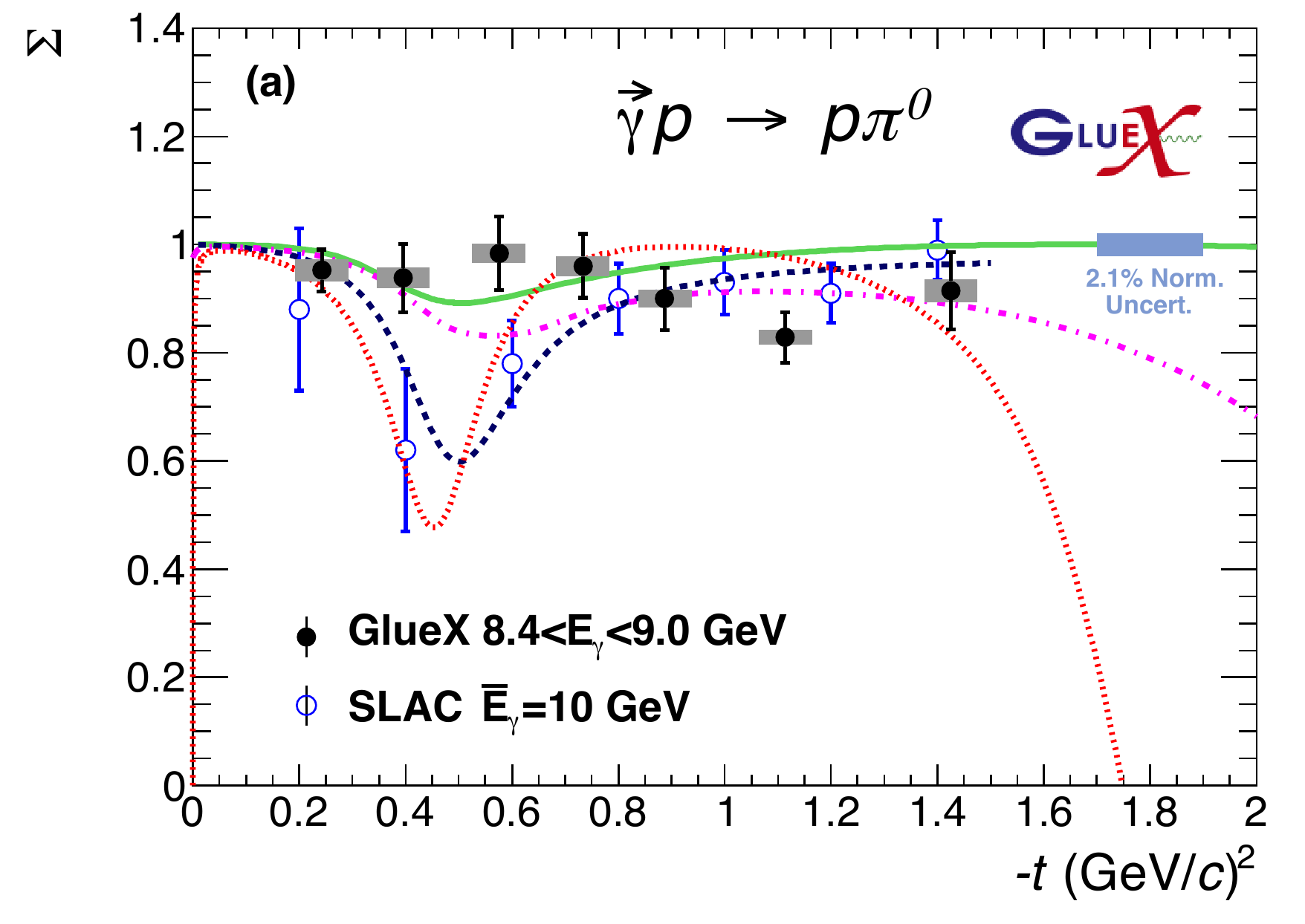}
    \includegraphics[width=0.4\textwidth]{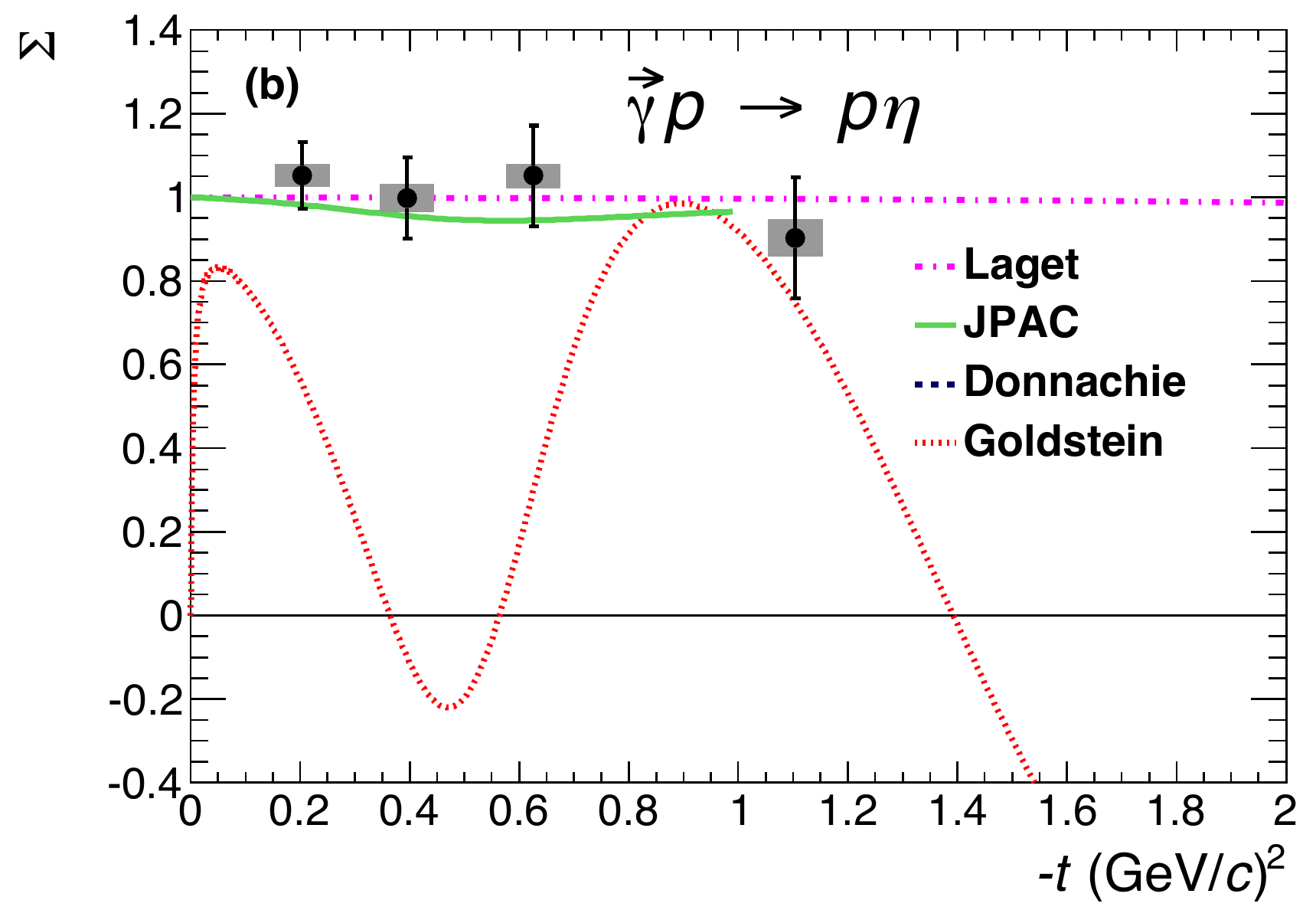}

    	\caption{Beam asymmetry $\Sigma$ for (a) $\pi^0$ and
	(b) $\eta$ photoproduction from 
	Ref.~\protect\cite{AlGhoul:2017nbpu}.}
    	\label{fig:asym}
\end{figure}
%%%-----------------------------------

To pursue the search for exotic hybrid mesons in photoproduction the
production of conventional states, such as pseudoscalar and vector
mesons, must first be understood.  Previous measurements of pseudoscalar
photoproduction at these energies are limited, especially for the linearly
polarized beam asymmetry $\Sigma$.  This asymmetry provides direct
information on the quantum numbers of the $t-$channel Reggeon exchange,
with $\Sigma=1$ for purely ``natural" exchange (\textit{e.g.} $J^P = 1^-$)
and $\Sigma=-1$ for ``unnatural" exchange (\textit{e.g.} $J^P = 1^+$).
First measurements from GlueX~\cite{AlGhoul:2017nbpu} indicate that the
natural vector meson exchange is dominant in this regime for $\pi^0$
and $\eta$ meson production, as seen in Fig.~\ref{fig:asym}.  Further
studies of the beam asymmetry and polarization observables for other
reactions will provide critical insights into the meson photoproduction
mechanisms in this energy regime.

%%%-----------------------------------
\item \textbf{DIRC Upgrade}

As described above, an initial physics program to search for and study
hybrid mesons which decay to non-strange final state particles is well
underway.  However, an upgrade to the particle identification capabilities
of the GlueX experiment is needed to fully exploit its discovery 
potential, by studying the quark flavor content of the potential hybrid 
states.

This particle identification upgrade for GlueX will utilize fused silica
radiators from the BaBar DIRC (Detection of Internally Reflected Cherenkov
light) detector~\cite{Adam:2004fqu}, with new, compact expansion volumes 
to detect the produced Cherenkov light.  The GlueX DIRC will provide 
$\pi/K$ separation for momenta up to 4~GeV, significantly extending the  
discovery potential of the GlueX program~\cite{Stevens:2016ciau}.
The charged kaon identification provided by the DIRC may also yield useful
identification of charged kaons for the $K_L$ beam facility (KLF) proposed
for Hall~D~\cite{Amaryan:2017ldwu}.  This may be particularly relevant in
the production of strange mesons from the high momentum component of the
$K_L$ beam, where the charged kaons are produced at forward angles.

%%%-----------------------------------
\item \textbf{Acknowledgments}

This work is supported by the Department of Energy Early Career Award
contract DE--SC0018224.
\end{enumerate}

%%%-----------------------------------

%%%%%%%%%%%%%%%%%%%%%%%%%%%%%%%%%%%%%%%%%%%%%%%%%%%%%%%%%%%%%%%%%%%%
\newpage
\subsection{Strange Meson Spectroscopy at CLAS and CLAS12}
\addtocontents{toc}{\hspace{2cm}{\sl Alessandra Filippi (for the CLAS Collaboration)}\par}
\setcounter{figure}{0}
\setcounter{table}{0}
\setcounter{equation}{0}
\setcounter{footnote}{0}
\halign{#\hfil&\quad#\hfil\cr
\large{Alessandra Filippi}\cr
\textit{I.N.F.N. Sezione di Torino}\cr
\textit{10125 Torino, Italy}\cr}

%%%-----------------------------------
\begin{abstract}
The CLAS Experiment, that had been operating at JLAB for about one decade,
recently obtained the first high statistics results in meson spectroscopy
exploiting photon-induced reactions. Some selected results involving 
production of strangeness are reported, together with a description of 
the potentialities of the new CLAS12 apparatus for studies of reactions 
induced by quasi-virtual photons at higher energies.
\end{abstract}

%%%-----------------------------------
\begin{enumerate}
\item \textbf{Introduction}

The identification of states containing open and hidden strangeness is
still an open issue in light meson spectroscopy investigations.
Apart form a handful of confirmed states, still little is known, for 
instance, about the radial excitations of the $\phi(1020)$ meson, and even 
less about strangeonia with quantum numbers other than $1^{--}$. Below 
2.1~GeV just about half the open strangeness kaonia states (composed by a 
strange and a light quark)  expected by the Constituent Quark Model 
(CQM)~\cite{re:CQMl} have been observed so far, while less than ten  
strangeonia states, mesons of $\bar ss$ structure, have been steadily 
observed out of a total of at least 20 expected states.

Nonetheless, the knowledge of the properties of such states can provide 
important inputs for hadron spectroscopy. Strangeonia, in fact, feature an 
intermediate mass between the heavier systems where the quark model is 
approximately valid and the lighter meson sector. Unfortunately, their 
experimental signatures are less clear as compared to charmonium states, 
as they are broader and lie in a mass range where the overlap probability 
with other light quark states is very strong, and moreover most of the
decay channels modes are shared among all of them.

To further complicate the problem, in the same mass region some other
structures of exotic composition ($\bar q qg$ states, known as hybrids, 
$ggg$ states, the so-called glueballs, or even molecular states of 
multi-quark composition) are expected by QCD. Recent QCD calculations on 
the lattice are able to predict most of the conventional meson spectrum in 
good agreement with experimental findings~\cite{re:lqcdl}; these 
confirmations of course strengthen the confidence in their predictive 
power for the searches of new states. According to these calculations, the 
lightest hybrids and glueballs are predicted in the 1.4--3~GeV mass 
range: namely, at 2~GeV for the lightest $J^{PC} = 0^{+-}$ state,
and at 1.6~GeV for the  $1^{-+}$ one. This is actually the mass region 
where signatures of still unobserved strangeonia or kaonia are expected to 
show up, and indeed a number of candidates for exotics has been suggested 
over the years, still all awaiting for confirmation, fitting the same 
slots where strangeonia would be expected.

The use of photons as probes to study strangeness was not used extensively 
in the past due to the small production rates, and the lack of beams of 
suitable intensity and momentum resolution. In fact, the production of 
strange quarks in a non-strange environment involves disconnected quark 
diagrams whose occurrence is suppressed as a consequence of the OZI rule. 
However, a good step forward is expected from the results obtained with
the photon beam produced by brehmsstrahlung from the continuous electron 
beam at the CEBAF machine at JLAB.

Smoking guns for the existence of open and hidden strangeness
states would by their observations in the $\phi\eta$ or $\phi\pi$ 
invariant mass systems~\cite{re:photoproductionl}; especially in
the first case, due to the strange content of the $\eta$ meson, the 
production of strangeness should be eased \cite{re:smokingunl}. A typical 
diagram for the photoproduction of the $\phi\eta$ final state is shown in
Fig.~\ref{fig:phieta}.
%%%-----------------------------------
\begin{figure}[h]
\centering
\includegraphics[height=5.5truecm,clip=true]{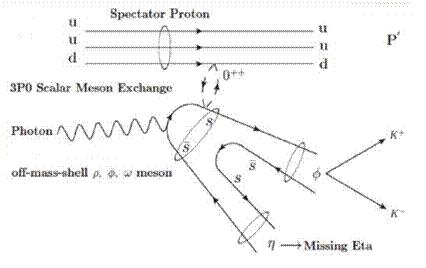}

\caption{Quark line diagram for the photoproduction of the 
	$\phi\eta$ final state.} \label{fig:phieta}
\end{figure}
%%%-----------------------------------

Very few  events in these channels have been observed so far; however, 
there is a good possibility, as will be shown in 
Sec.~\ref{clas12_strangeness}, that in the upcoming meson spectroscopy 
experiment at CLAS12 good samples of such reactions could be collected,
opening therefore new opportunities to widen the knowledge of this sector 
of the meson spectrum.

%%%-----------------------------------
\item \textbf{Meson Spectroscopy Studies in Photon Induced Reactions}

Several reactions and beams have been exploited so far for meson 
spectroscopy searches: among most important, high-energy meson (mainly 
pion) and proton beams, based on the peripheral and central production 
mechanisms, antinucleon annihilation at rest and in flight, which convey 
the formation of a gluon-rich environment suitable for glueball 
production, and $e^+e^-$ annihilation. The latter reaction has been 
studied extensively since the LEP era, and is an environment where also
$\gamma\gamma$ collisions can be measured, which provide quite useful
information as they are a natural anti-glueball filter. The $e^+e^-$ 
annihilation differs from hadronic reactions for the fact that only 
$1^{--}$ systems can be formed, so these reactions provide naturally a 
powerful quantum number selection.

On the other hand, the use of electromagnetic probes, and in particular of
photons, in fixed target reactions was not used very extensively for meson 
spectroscopy purposes, as meantioned earlier. Nevertheless, 
electromagnetic reactions could deliver important complementary 
information. In fact, first of all, electromagnetic processes can be 
exactly reproduced to a high level of precision through QED diagrams 
thanks to the smallness of the electromagnetic coupling, which is 
prevented in the case of strong interactions. Moreover, photons can
excite with larger probability the production of spin-1 mesons as compared
to pion or kaon induced reactions, since in the latter case a spin-flip
is required. Therefore, the production of vector hybrid mesons could occur 
in photoproduction reactions with a rate comparable to those of 
conventional vector mesons~\cite{re:photoproductionl,re:smokingunl}.
This feature applies not only to hybrids, but also to spin-1 $\bar ss$ 
excitations.

%%%-----------------------------------
\item \textbf{Selected Results from CLAS}

The CLAS apparatus, which had operated up to 2010, is described in detail
elsewhere~\cite{re:clas6l}. In the following some selected results from
recent meson spectroscopy papers involving strangeness production
will be summarized, in connection with some still open issues.

%%%-----------------------------------
\begin{enumerate}
\item \textbf{The Scalar Glueball Search Case}

In spite of the efforts by the experiments in the early Nineties, in
particular those studying antinucleon annihilations, like Crystal Barrel 
and OBELIX, the full composition of the scalar meson sector is not 
completely clear yet. Observations have been made of several mesons whose 
existence was not foreseen by the CQM; one of them, the $f_0(1500)$, 
seemed to have the right features as lightest scalar glueball 
candidate~\cite{re:f01500l}. Among these, the most important is the fact 
that it was observed to decay in several channels, a clear hint to its
flavor-blindness. However, more data are still desirable to confirm
these properties as several other structures tentatively identified as 
scalars as well. All the existing observations make the interpretation of 
the scalar sector difficult, since it is populated more than expected by
broad and overlapping states. In this mass region one should also recall 
the existence of the $\sigma$ state (also known as $f_0(600)$),
corresponding to a very broad $\pi\pi$ non-resonant iso-scalar $S$-wave 
interaction, whose nature and properties are still unclear. Also scarcely 
known are the features of the $\kappa$, the analogous of $\sigma$ observed 
in the $K\overline K$ channel.

A search was carried on in CLAS exploiting the $\gamma p\rightarrow p
K^0_SK^0_S$ reaction~\cite{re:KSKSl}, with real photons of energy in the 
ranges $(2.7-3)$ and $(3.1-5.1)$~GeV. The $K^0_S$ were fully reconstructed 
through their decay in two pions, while the proton was identified from the 
event missing mass. A strong correlation between the two $K^0_S$ was 
found, which allowed to collect a clean sample of good statistics, enough 
to perform selections in momentum transfer $t (= Q^2)$. The selection in 
$t$ is useful to understand the production mechanism of the intermediate 
states: small values of momentum transfers are correlated to a dominant 
production from the $t$-channel, while $s$-channel production is 
characterized by a wider range of transferred momenta.
Fig.~\ref{fig:ksks_IM} shows the distributions of the $(K^0_SK^0_S)$ 
invariant mass system, after proper background subtraction, in two 
momentum transfer ranges: for $|t|<1$ GeV$^2$ on the left side, and for 
$|t|>1$ GeV$^2$ on the right. A clean peak at about 1500~MeV appears in 
the first case, while at larger momentum transfer no evidence for it can 
be observed. This means that the structure at 1500~MeV, a possible 
indication for the $f_0(1500)$ observed in its $K^0_SK^0_S$ decay mode, is 
predominantly produced via $t$-channel. This observation could support 
its possible interpretation as glueball.
%%%-----------------------------------
\begin{figure}[h]
\centering
\includegraphics[width=0.8\linewidth,clip=true]{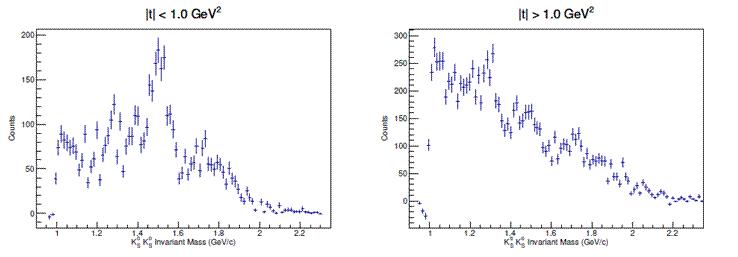}

\caption{Invariant mass of the $(K^0_SK^0_S)$ system for events selected 
	in the $\gamma p\rightarrow p K^0_SK^0_S$ reaction, in two 
	momentum transfer ranges: left, for $|t|<1$ GeV$^2$, right for 
	$|t|>1$ GeV$^2$.} \label{fig:ksks_IM}
\end{figure}
%%%-----------------------------------

In order to fully characterize the features of this state a complete
spin-parity analysis of the sample is required. Unfortunately, the task is 
not straightforward due to the limited apparatus acceptance at small 
forward and backwards angles. Nonetheless, an angular analysis was 
attempted to test which of the spin-parity hypotheses for the events in 
the peak region would provide a better fit to the Gottried-Jackson angular 
distributions.

We recall that a $(K^0_SK^0_S)$ bound system may just have $J^{PC} =
(even)^{++}$ quantum numbers, so the spin 0 and 2 hypotheses need to be 
tested. An example of angular distribution is reported in 
Fig.~\ref{fig:ksks_ANG}, in a $(K^0_SK^0_S)$ mass window centered at 1525 
MeV. The curves superimposed to the experimental data show the 
contributions of $S$ and $D$ waves (blue and red, respectively) to the 
total fit (green). The $D$-wave contribution plays a marginal r\^ole, 
larger for masses higher than 1550~MeV, from which one can deduce that
the scalar hypothesis for the observed resonance is mostly supported;
therefore, it can be more easily identified as the $f_0(1500)$.
%%%-----------------------------------
\begin{figure}[h]
\centering
\includegraphics[width=7truecm,clip=true]{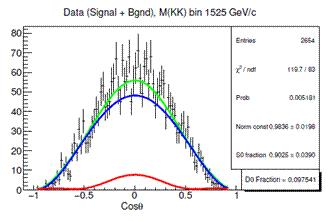}

\caption{Gottfried-Jackson angular distribution for $K^0_SK^0_S$ events
	selected in the 1525 MeV invariant mass slice, with superimposed the
	expected trend, on the basis of Monte Carlo simulations, for 
	production from $S$ wave (blue), $D$ wave (red), and the global 
	fit with the two components (green).} \label{fig:ksks_ANG}
\end{figure}
%%%-----------------------------------

%%%-----------------------------------
\item \textbf{The Axial/Pseudoscalar Sector at 1.4~GeV Case}

Kaonia radial excitations were widely studied in the past together with
$\eta$ excitations in the same mass range, to search for possible exotic
states. Many observation of $\eta$'s and $f_1$ mesons have been reported
since the Sixties, when the issue of the overlap of many axial and
pseudoscalar states and the difficulty of their identification posed
the so-called $E/\iota$ puzzle~\cite{re:masonil}. While annihilation 
experiments, and in particular OBELIX, could provide a solution to this 
puzzle addressing the production of several pseudoscalar and axial states, 
high statistics photoproduction reactions are expected to deliver new 
complementary information which will be able to improve the knowledge in 
this sector.

A systematic study was performed by CLAS to study the photoproduction of
states, decaying into $\eta\pi^+\pi^-$ and $K^0 K^\pm \pi^\mp$ and 
recoiling against a proton, in the $\gamma p$ reaction  with photons in 
the energy range $(3-3.8)$~GeV~\cite{re:dicksonl}. 
Fig.~\ref{fig:missing_p} shows the missing mass plot of the system 
recoiling against a proton for $\gamma p\rightarrow p \eta\pi^+\pi^-$ 
selected events, where a clean signal due to the $\eta^\prime(958)$ 
appears, together with a structure at about 1280~MeV, that can tentatively 
be identified as the $f_1(1285)$.
%%%-----------------------------------
\begin{figure}[h]
\centering
\includegraphics[height=5.5truecm,clip=true]{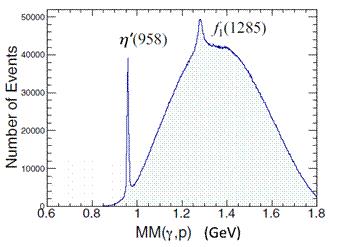}

\caption{Missing mass recoling against a proton for events selected in the
	$\gamma p\rightarrow p\eta\pi^+\pi^-$ reaction.}
	\label{fig:missing_p}
\end{figure}
%%%-----------------------------------

Concerning the reaction witk kaons, the $K^\pm$ were identified by CLAS 
through time of flight techniques, while the $K^0$ via the missing mass 
information. The missing mass plots for the system recoiling against the 
proton for events selected in the two channels $\gamma p\rightarrow 
p\overline{K^0}K^+\pi^-$ and $\gamma p\rightarrow p{K^0}K^-\pi^+$ are 
shown, respectively, in Fig.~\ref{fig:missing_kkp}(a) and (b).
%%%-----------------------------------
\begin{figure}[h]
\centering
\includegraphics[width=0.8\linewidth,clip=true]{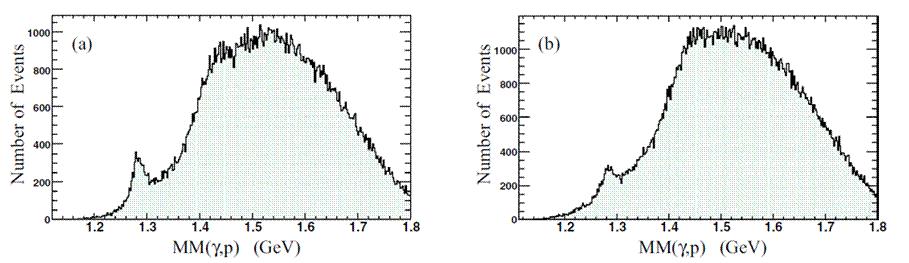}

\caption{Missing mass recoling against a proton for events selected in the
	a) $\gamma p\rightarrow p\overline{K^0}K^+\pi^-$ and
	b) $\gamma p\rightarrow p{K^0}K^-\pi^+$ reactions.}
	\label{fig:missing_kkp}
\end{figure}
%%%-----------------------------------

In both of them a clear peak appears at about 1.3 GeV, but no further
evidence for higher mass states that could be addressed to additional
pseudoscalar states, like the $\eta(1405)$ and $\eta(1470)$, or to axial 
states, as the $f_1(1420)$ of $f_1(1510)$, is present.

From the richest $\eta\pi^+\pi^-$ sample, one can get information for the
identification of the observed state. There are a few hints that support 
the identification as axial $f_1(1285)$ against the pseudoscalar
$\eta(1295)$. The first one is given by the values obtained for the mass 
and width of the signal, $M=(1281.0\pm 0.8)$~MeV and $\Gamma = (18.4\pm 
1.4)$~MeV, which are closer to those already observed for 
$f_1(1285)$~\cite{re:pdgl}. The ratio itself of the decay rates in 
$\eta\pi^+\pi^-$ versus $K\overline K\pi$, that amounts to about five,
is consistent with the value quoted by PDG  for the $f_1(1285)$ decays, 
while no such ratio was ever measured  for $\eta(1295)$~\cite{re:pdgl}.

The second hint is given by the trend of the differential cross section, 
for events selected in the band of $\eta^\prime(958)$ and around 1280~MeV: 
as shown in Fig.~\ref{fig:etaprime_xsec}, the two are remarkably different 
as a function of the center-of-mass angle, indicating a different 
production mechanism and possibly a different spin configuration of the 
produced state.
%%%-----------------------------------
\begin{figure}[h]
\centering
\includegraphics[height=5.5truecm,clip=true]{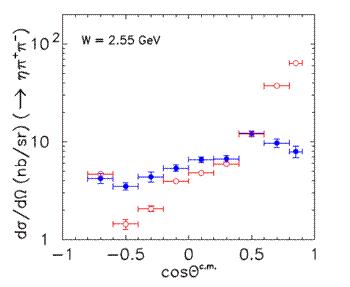}

\caption{Differential cross section of the $\gamma p\rightarrow 
	\eta\pi^+\pi^- p$ reaction for events selected in the
	$\eta^\prime(958)$ mass band (red open points) and the $f_1(1285)$ 
	one (blue points). The energy in the center of mass is fixed at 
	2.55~GeV.} \label{fig:etaprime_xsec}
\end{figure}
%%%-----------------------------------

The comparison of the cross sections for events in the 1280 MeV band
with the expectations from $t$-channel based 
models~\cite{re:t_model1l,re:t_model2l,re:t_model3l}, shown in 
Fig.~\ref{fig:xsec_t_models}, indicates clearly a poor match with
this production hypothesis: a substantial contribution from $s$-wave could
be needed, or a mechanism different from meson exchange involving
$N^\ast$ excitations or $KK^\ast$ molecular interactions. In both cases 
the identification of the structure as $f_1(1285)$ would get larger 
support.
%%%-----------------------------------
\begin{figure}[h]
\centering
\includegraphics[width=0.8\linewidth,clip=true]{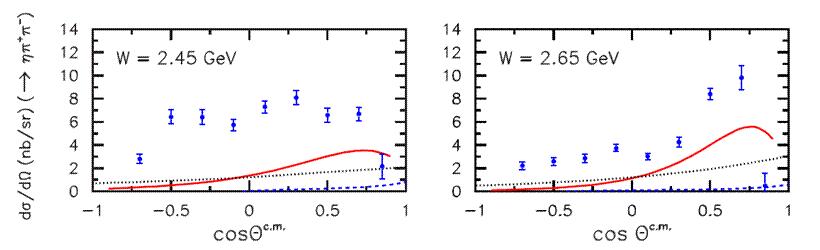}

\caption{Differential cross sections of the $\gamma p\rightarrow 
	\eta\pi^+\pi^- p$ reaction for events selected in the
	$f_1(1285)$ mass band in two center-of-mass energy bins (left: 
	2.45~GeV, right: 2.85~GeV), compared with a few $t$-channel 
	exchange based models (none of which matches the experimental 
	points): solid red line from 
	Ref.~{\protect\cite{re:t_model1l}}, blue dashed line from 
	Ref.~{\protect\cite{re:t_model2l}}, black dotted line from 
	Ref.~{\protect\cite{re:t_model3l}}.} \label{fig:xsec_t_models}
\end{figure}
%%%-----------------------------------
\end{enumerate}

%%%-----------------------------------
\item \textbf{Prospects for CLAS12}

The CLAS12 spectrometer, an upgraded version of CLAS, is a multipurpose
facility dedicated to hadron physics studies, from nuclear properties and 
structure to meson spectroscopy investigations. A full description of 
CLAS12 is given elsewhere~\cite{re:clas12l}. A second experiment installed 
on the CEBAF machine, GlueX, is operating in parallel with a program fully 
committed to meson spectroscopy~\cite{re:gluexl}.

Due to the increased beam energy, as compared to the previous 
installation, the CLAS12 spectrometer has a more compact structure, which
limits its geometric acceptance, although allowing an acceptable 
hermeticity. However, its detectors feature better momentum
resolution and particle identification capabilities as compared to GlueX,
therefore the two experimental setups offer complementary qualities.

While GlueX will use a real photon bremsstrahlung beam, this is not be 
possible for CLAS12 as the bending dipole magnet available in Hall~B will
not be powerful enough to steer the 11~GeV electron beam into the tagger 
beam dump. A new technique was therefore conceived to produce a photon 
beam for the study of photoproduction reactions, based on electron 
scattering at very low transferred momentum ($Q^2 < 10^{-1}$~GeV$^2$). In 
this situation the electrons are scattered at very small polar angles, and 
the reaction on protons is induced by quasi-real photons. A dedicated part 
of the CLAS12 apparatus, the {\it Forward Tagger}, was devised to detect 
these forward scattered electrons. For momentum transfers in the interval 
$0.01 < Q^2 < 0.3$~GeV$^2$, the electrons scattered between $2.5^\circ$ 
and $4.5^\circ$ in polar angle in the lab have an energy in the
0.5--4.5~GeV range, being produced by the interaction of 6.5--10.5~GeV 
quasi-real photons on protons. Measuring the electron momentum, each 
virtual photon can be tagged. The use of the CLAS12 spectrometer to 
measure in coincidence the particles produced in the photoproduction
reaction will allow to perform a complete event recontruction, necessary 
for meson spectroscopy studies. With this tagging technique one can 
measure also the virtual photon polarization, that is linear and can be 
deduced, event by event, from the energy and the angle of the scattered 
electron. The systematic uncertainty affecting the polarization depends 
only on the electron momentum resolution. High electron currents may be 
used, therefore a good luminosity can be obtained even with thin targets, 
that are not operable with real photon bremsstrahlung beams. For instance, 
using a 5~cm long LH$_2$ target, the resulting hadronic rate will be 
equivalent to that achievable by a real photon flux of about $5\times 
10^8\gamma$/s.

The Forward Tagger equipment, described in detail in Ref.~\cite{re:FTl}, 
is located about 190~cm away from the target and fits within a 5$^\circ$ 
cone around the beam axis. It is made up of:

\begin{itemize}
\item an electromagnetic calorimeter (FT-Cal): composed by 332 PbWO$_4$
	crystals, 20 cm long and with square $15\times 15$~mm$^2$ 
	cross-section. It is used to identify the scattered electron and 
	measure its energy, from which the photon energy and its 
	polarization can be deduced (the polarization being given
	by $\epsilon^{-1} \sim 1+\nu^2/(EE^{\prime})$, where $\nu = 
	E-E^{\prime}$ is the photon energy, and $E$ and $E^{\prime}$ the 
	energies of the incident and of the scattered electron, 
	respectively). It is also used to provide a fast trigger signal. 
	Its expected design resolution is $\sigma_E/E \sim
	(2\%/\sqrt{E(\mathrm{GeV})} \oplus 1\%)$;
 \item a scintillator hodoscope made of plastic scintillator tiles:
	located in front  of the calorimeter, is used to veto photons; its
	spatial and timing resolution is required to be comparable with 
	FT-Cal's;
\item a tracker: located in front of the hodoscope and composed by
	Micromegas detectors, is used to measure the angle of the scattered
	electron and the photon polarization plane.
\end{itemize}

A sketch of the Forward Tagger region is shown in Fig.~\ref{fig:ft}.
%%%-----------------------------------
\begin{figure}[htb]
\centering
\includegraphics[width=0.48\textwidth,clip]{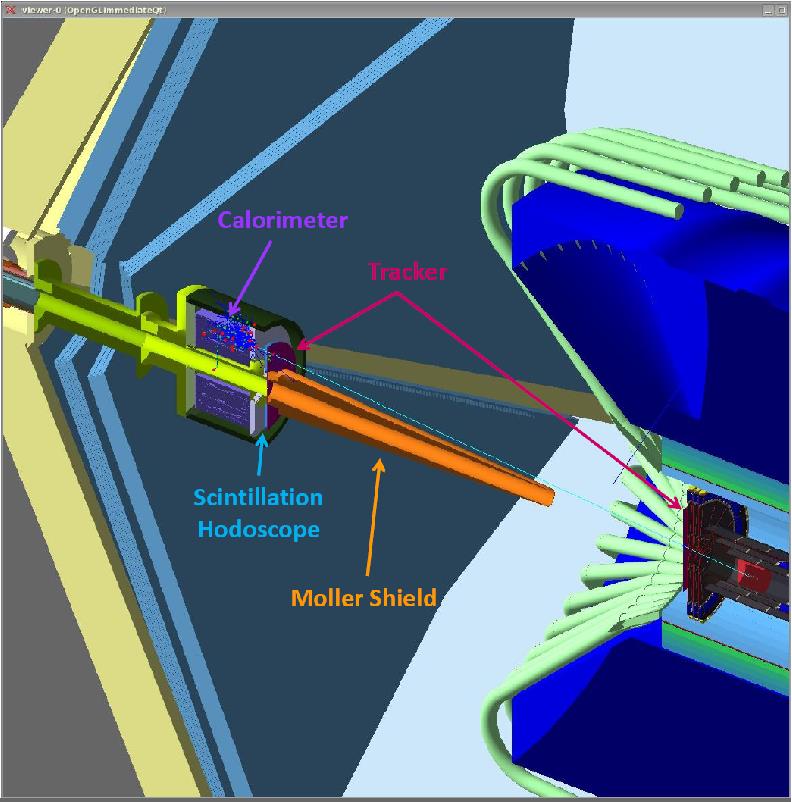}

\caption{Schematic view of the Forward Tagger equipment, to be hosted
	in CLAS12.} \label{fig:ft}
\end{figure}
%%%-----------------------------------

%%%-----------------------------------
\item \textbf{Strange Meson Spectroscopy with CLAS12}
\label{clas12_strangeness}

The Meson-EX experiment at CLAS12 (Exp-11-005~\cite{re:mesonexl})
was proposed to study of the meson spectrum in the 1--3 GeV
mass range through quasi-real photon induced reactions, for the 
identification of gluonic excitations of mesons and other exotic quark 
configurations beyond CQM. The use of the Forward Tagger will allow to 
identify the photoproduction reaction on protons through the tagging of 
the forward scattered electron, while the full CLAS12 apparatus will 
perform a complete reconstruction and identification of the charged and 
neutral particles produced in the interaction. Some golden channels have 
been selected as particularly suitable for the search of exotic or still 
unknown particles, whose production, as mentioned before, will be favored 
in photoproduction especially in the case of spin 1 particles. In 
particular, for the search of hybrids with open strangeness and
strangeonia, the following channels are expected to provide interesting 
new indications: $\gamma p\rightarrow \phi\pi^0 p$, $\gamma p\rightarrow 
\phi\eta p$ and $\gamma p\rightarrow K\overline{K}\pi p$. For the first 
two, extensive simulations were carried over to test the feasibility of 
such measurements and the expected collectable statistics, in a 80-days 
long data taking at the full CEBAF luminosity ($\sim 
10^{35}\;\mathrm{s}^{-1}\mathrm{cm}^{-2}$) with a total expected
trigger rate for photoproduction reactions less than 10~kHz.
Assuming an apparatus acceptance for four track events of the order of 
15\%, it will be possible to collect as many as 3000 events per 10~MeV 
mass bin, for reactions with a cross section as small as 10~nb, expected 
for instance for strangeonia production. This statistics is considered to 
be enough to perform detailed partial-wave analysis studies.

More in detail, concerning the $\phi\eta$ channel and the possible 
production of the $\phi(1850)$ excitation, simulations showed that an 
overall acceptance of about 10\% can be expected, due to the coverage at 
small angles provided by the Forward Tagger electromagnetic calorimeter in 
which the photons from the $\eta$ decay can be detected. The full event 
would be reconstructed identifying, in addition, at least one of the kaons 
from the $\phi$ decay and the recoiling proton, the $K^-$ being 
identifiable through the reaction missing mass. A cross section
on the order of 10 nb is tentatively expected, given an existing  
estimation by the CERN Omega Collaboration~\cite{re:omegal} of about 6~nb 
at higher energies, for the production of $\phi(1850)$ and its decay in 
the $K^+K^-$ mode, whose branching ratio is expected to be about twice as 
large as compared to $\phi\eta$.

On the other hand, concerning the $\phi\pi^0$ intermediate state which 
could be a possible source of exotic systems, again a resonable detection 
efficiency could be achieved requiring a full identification of the 
positive kaon from the $\phi$ and the reconstruction of the $\pi^0$ 
through two photons in the Forward Tagged calorimeter. The negative kaon 
acceptance, largely reduced as it bends inwards at small angles, therefore 
in a region of scarce detector coverage, can be recovered identifying the 
particle via the missing mass information; this of course demands again a 
good resolution on the photon energy. In these conditions, an acceptance 
similar to those of more ``conventional'' intermediate states (composed 
for instance by pions only) can be expected, with basically no impact by 
the chosen intensity of the magnetic field.

%%%-----------------------------------
\item \textbf{Conclusion}

Nowadays many communities are involved all over the world in
hadron spectroscopy studies, both from experimental and from the 
theoretical point of view. Experimental facilities are presently 
delivering, and will deliver in the near future, large amounts of  good 
quality data that will allow to extract for the first time, from different 
reactions and experimental environments, new information on many still 
unsolved issues.

Photoproduction experiments at JLab will play a big role in the coming
future. At JLab, GlueX and CLAS12 will provide data samples of
unprecedented quality and richness. Their contribution is expected to grow 
to sizeable relevance and have a big impact in the present and future 
meson spectroscopy experimental scenario, complementing the information 
that will be provided by $e^+e^-$ collisions (by BESIII and BelleII), by 
pion induced interactions (by COMPASS), by proton-proton interactions (at 
fixed target by LHCb, and in high-energy $pp$ collisions by ATLAS and CMS)
and by antiproton-proton annihilations (by PANDA at FAIR).
\end{enumerate}

%%%-----------------------------------

%%%%%%%%%%%%%%%%%%%%%%%%%%%%%%%%%%%%%%%%%%%%%%%%%%%%%%%%%%%%%%%%%%%%%%%%
%\newpage
%\subsection{Non-leptonic Charmless Three Body Decays at LHCb}
%\addtocontents{toc}{\hspace{2cm}{\sl Rafael Silva Coutinho}\par}
%\setcounter{figure}{0}
%\setcounter{table}{0}
%\setcounter{equation}{0}
%\setcounter{footnote}{0}
%\halign{#\hfil&\quad#\hfil\cr
%\large{Rafael Silva Coutinho}\cr
%\textit{Institut f\"ur Kernphysik}\cr
%\textit{Goethe-Universit\"at Frankfurt}\cr
%\textit{D-60438 Frankfurt am Main, Germany}\cr}
%
%%%-----------------------------------
%\begin{abstract}
%xxxxxxxxxxxxxxxxxxxxxxxxxxxxxxxxxxxxxxxx
%\end{abstract}
%
%%%-----------------------------------
%\begin{enumerate}
%\item \textbf{Introduction}
%
%%%-----------------------------------
%\item \textbf{Conclusion}
%
%%%-----------------------------------
%\item \textbf{Acknowledgments}
%
%xxxxxxxxxxxxxxxxxxxxxxxxxxx
%\end{enumerate}
%
%%%-----------------------------------
%\begin{thebibliography}{99}
%
%\end{thebibliography}

%%%%%%%%%%%%%%%%%%%%%%%%%%%%%%%%%%%%%%%%%%%%%%%%%%%%%%%%%%%%%%%%%%%%
\newpage
\subsection{Dispersive Determination of the $\pi-K$ Scattering Lengths}
\addtocontents{toc}{\hspace{2cm}{\sl Jacobo Ruiz de Elvira, Gilberto 
	Colangelo, and Stefano Maurizio}\par}
\setcounter{figure}{0}
\setcounter{table}{0}
\setcounter{equation}{0}
\setcounter{footnote}{0}
\halign{#\hfil&\quad#\hfil\cr
\large{Jacobo Ruiz de Elvira, Gilberto Colangelo, and Stefano Maurizio}\cr
\textit{Albert Einstein Center for Fundamental Physics}\cr
\textit{Institute for Theoretical Physics, University of Bern}\cr
\textit{3012 Bern, Switzerland}\cr}

%%%-----------------------------------
\begin{abstract}
The pion-kaon scattering lengths are one of the most relevant
quantities to study the dynamical constraints imposed by chiral 
symmetry in the strange-quark sector
and hence, they are a key quantity for understanding the interaction 
of hadrons at low energies.
In this talk we review the current status of their determination.
After discussing the predictions expected from chiral symmetry at
different orders in the chiral expansion,
we review current experimental and lattice determinations. We then
focus on the dispersive determination, based on a Roy-Steiner 
equation analysis of pion-kaon scattering, and discuss in detail 
the current tension between the chiral symmetry and dispersive 
solutions.  We finish this talk providing an explanation of this 
disagreement.
\end{abstract}

%%%-----------------------------------
\begin{enumerate}
\item \textbf{Introduction}

Pion-kaon scattering is one of the simplest processes to test our
understanding of the chiral symmetry-breaking pattern in the presence
of the strange quark. In particular, its low-energy parameters, most 
notably the scattering lengths, encode relevant information about the 
spontaneous and explicit chiral symmetry breaking in this sector.
Being low-energy observables, their properties can be efficiently studied
using the effective field theory of Quantum Chromodynamics (QCD) at low
energies,  Chiral Perturbation Theory
(ChPT)~\cite{Weinberg:1978kzj,Gasser:1983ygj,Gasser:1984ggj}, constructed
as a systematic expansion around the chiral limit of QCD in terms of
momenta and quark masses.

Pion-kaon scattering can be expressed in terms of two independent 
invariant amplitudes with well defined isospin $I=1/2$ and $I=3/2$ in 
the $\pi K\to\pi K$ channel, namely $T^{1/2}$ and $T^{3/2}$.
Nevertheless, for convenience, it is useful to combine them in terms of
isospin-even and -odd amplitudes $I=\pm$, which are defined as
\begin{equation}
	T_{ab}=\delta_{ab}T^{+}+\frac{1}{2}\left[\tau_a,\tau_b\right]T^{-},
\end{equation}
where $a$ and $b$ denote pion isospin indices and $\tau_{a}$ stand for the
Pauli matrices.
Both basis are related by simple isospin transformations:
\begin{equation}
	T^{1/2}=T^{+}+2T^{-},\qquad T^{3/2}=T^{+}-T^{-}.
\end{equation}

At leading order (LO) in the chiral expansion, i.e., in the expansion in
pion and kaon masses and momenta, the scattering amplitude is given by
the Feynman diagram shown in Fig.~\ref{fig:Op1+2}a,
%%%-----------------------------------
\begin{figure}[t]
\centering
\includegraphics[width=0.4\linewidth]{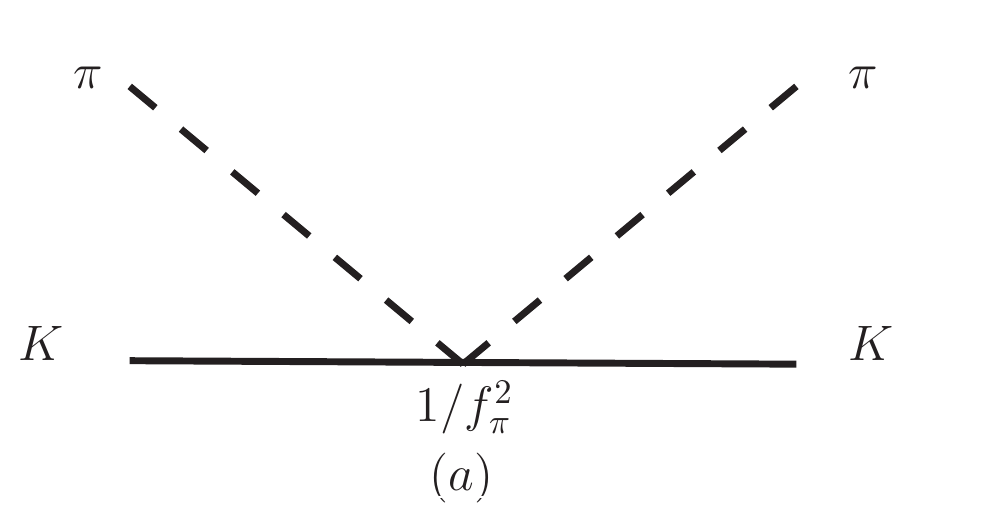}
\includegraphics[width=0.4\linewidth]{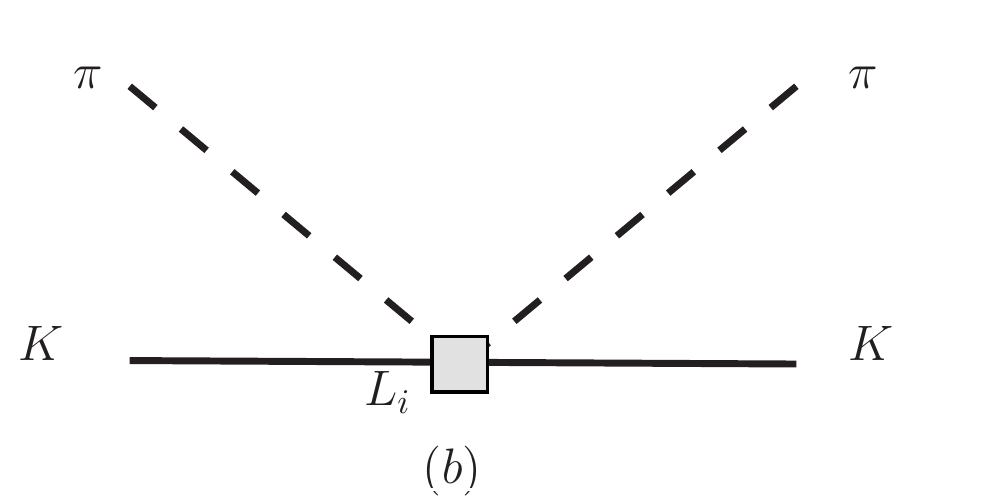}

\caption{(a) Leading-order diagrams for $\pi K$ scattering in chiral 
	perturbation theory. Kaon are denoted by full, pions by dashed 
	lines.
	(b) Next-to-leading-order diagrams depending on low-energy constants 
	$L_{1-8}$.}\label{fig:Op1+2}
\end{figure}
%%%-----------------------------------
resulting in the well-known low-energy theorems for the $S$-wave
scattering lengths, the amplitudes evaluated at
threshold~\cite{Gasser:1984ggj,Weinberg:1966kfj}:
\begin{equation}
	a^-_{0} = \frac{m_\pi m_K}{8\pi(m_\pi+m_K) f_\pi^2}+ \Order\big({m_i^4}\big) , \qquad
	a^+_{0} = \Order \big({m_i^4}\big),
\end{equation}
where $m_i$ denotes the pion ($\mpi$) or kaon ($\mK$) mass.
The isospin-odd scattering length is hence predicted solely in terms
of the pion and kaon masses as well as the pion decay constant $\Fpi$,
while the isospin-even one is suppressed at low energies.
The LO ChPT value for the pion-kaon scattering lengths is denoted by
a star in Fig.~\ref{fig:SL}, where the pion-kaon scattering length
plane is plotted in the $I=1/2$ and $I=3/2$ basis.
%%%-----------------------------------
\begin{figure}[!t]
\centering
\includegraphics[width=\textwidth]{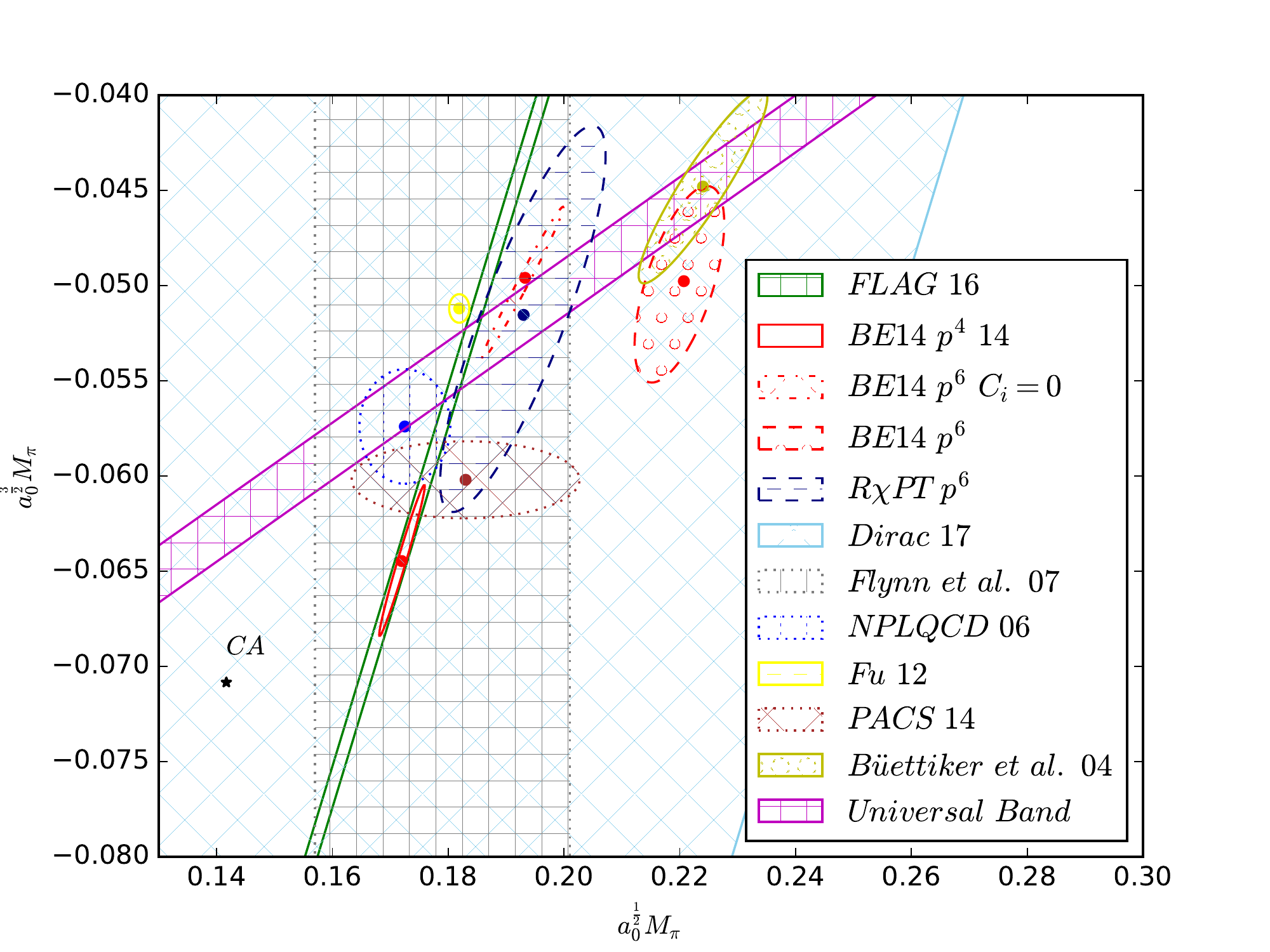}

\caption{Different determinations of the pion-kaon scattering lengths
	in the $I=1/2$, $I=3/2$ basis. The LO ChPT value, just a result 
	of current algebra, is denoted by a star. The NLO prediction of 
	the isospin-odd scattering length is given by the dark-green band 
	labelled as FLAG16. The inclusion of the isospin-even scattering 
	length, using the LECs provided 
	in Ref.~\protect\cite{Bijnens:2014leaj}, leads to the solid-red 
	ellipse BE14 $p^4$. The red dot-dashed ellipse, BE14 $p^6$ 
	$C_i=0$, corresponds to the NNLO chiral prediction when the 
	$\Order\big(p^6\big)$ LECs are set to zero.
	The full NNLO result is represented by the dashed-red ellipse, 
	albeit, as explained in the text, this result is biased to 
	reproduce the RS dispersive values given  
	in Ref.~\protect\cite{Buettiker:2003ppj}, solid-green ellipse.
	The NNLO results with $\Order\big(p^6\big)$ LECs estimated by 
	resonance saturation is denoted by the dashed-blue ellipse, $R\chi 
	PT$ $p^6$. The universal band obtained from the RS analysis 
	performed in this work is given by the violet band.  The remaining 
	experimental and lattice results are explained in the main text.} 
	\label{fig:SL}
\end{figure}
%%%-----------------------------------

The pion-kaon scattering amplitude at next-to-leading order (NLO) was
derived and studied first in~\cite{Bernard:1990kxj,Bernard:1990kwj}.
It involves one-loop diagrams, which might generate large contributions.
Nevertheless, they are suppressed at threshold and hence their role for
the pion-kaon scattering lengths is relatively small. In addition, the 
$\pi K$ scattering amplitude depends at NLO on a list of low-energy 
constants (LECs), conventionally denoted by $L_{1-8}$, which, encoding 
information about heavier degrees of freedom, can not be constrained from 
chiral symmetry solely, Fig.~\ref{fig:Op1+2}b. Once determined in one 
process, these LECs can subsequently be used to predict others. The NLO 
pion-kaon scattering lengths were analyzed in detail 
in~\cite{Bernard:1990kxj}. The contribution from the LECs
reads~\cite{Bernard:1990kxj}:
\begin{align}
	a^-_{0}\big\vert_{\mathrm{LECs}} =&\frac{m_K  m_\pi^3}{\pi(m_\pi+m_K)f_\pi^4}L_5+\Order\big(m_i^6\big),\\
	a^+_{0}\big\vert_{\mathrm{LECs}} =&\frac{m_K^2m_\pi^2}{\pi(m_\pi+m_K)f_\pi^4}\left(4(L_1+L_2-L_4)+2L_3-L_5+2(2L_6+2L_8)\right)+\Order\big(m_i^6\big).
	\nonumber
\end{align}

On the one hand, whereas the NLO LECs contribution to the isospin-odd
scattering length is suppressed by a $\mpi^3$ factor, the isospin-even
scattering length is proportional to $m_K^2$.  Thus, one should expect a 
small higher-order correction in the chiral expansion for $a^-_{0}$ but a 
much larger shift for $a^+_{0}$. On the other hand, while the isovector 
scattering length only depends on one LEC, $L_5$, which is indeed well 
constrained from the pion and kaon mass and decay constant values,
the isoscalar scattering length involve seven LECs and hence sizable
uncertainties for this quantity associated with LECs errors are expected.
The constraint imposed by the isovector scattering length at NLO in
ChPT is depicted by a dark-green band in Fig.~\ref{fig:SL}, where the
central value and error for $L_5$ is taken from the FLAG group
estimate~\cite{Aoki:2016frlj}. Note that the small width of this band is 
completely determined by the $L_5$ uncertainty. Furthermore, using the 
LECs collected in Table~1 in Ref.~\cite{Bijnens:2014leaj} (first column), 
one can also include the NLO prediction for the isoscalar scattering 
length, which leads to the solid-red ellipse in Fig.~\ref{fig:SL}.
As we have already anticipated, this ellipse is stretched out in the
isoscalar direction but shrunk in the isovector one.

One might wonder how stable are the NLO predictions against higher-order
corrections. A pion-kaon low-energy theorem~\cite{Bernard:1990kxj} imposes 
higher $\Order\big({m_i^{2\,n}}\big)$ contributions to the isospin-odd 
scattering length arising from contact terms to be at most:
\begin{equation}\label{LET}
	a^{-}_0\big\vert_{N^{n}LO}\propto a^{-}_0\big\vert_{LO}\,\left(\frac{\mpi}
	{4\pi f_\pi}\right)^2\,\left(\frac{\mK}{4\pi f_\pi}\right)^{2\,n}\quad\textrm{with}\quad n \ge 2.
\end{equation}

Whereas the factor $(\mK/4\pi f_\pi)\sim 0.2$ is relatively large, the
prefactor $\mpi^2/(4\pi f_\pi)^2\sim 0.015$ suppresses higher order
corrections by roughly two orders of magnitude, i.e., the isospin-odd 
scattering length is protected from higher-order correction and hence one 
should expect small deviations from the NLO ChPT prediction for $a_0^-$.

The pion-kaon scattering amplitude at next-to-next-to-leading order
(NNLO) in the chiral expansion was derived in~\cite{Bijnens:2004buj}.
It involves a set of 32 new $\Order\big(p^6\big)$ LECs, the so called
$C_{1-32}$, which unfortunately are still not well constrained from 
experiment. As a first step, one could estimate the size of the NNLO 
chiral corrections by setting all the $C_i$ to zero. Using for the
$\Order\big(p^4\big)$ $L_i$ the corresponding fit 
in~\cite{Bijnens:2014leaj}, the outcome is the red dot-dashed ellipse 
plotted in Fig.~\ref{fig:SL}. This result is consistent with our previous 
statement, i.e., whereas the shift between the NLO and NNLO ellipsis is 
small in the isovector direction, it is much larger in the isoscalar one.

The $C_i$ entering in pion-kaon scattering were also estimated
in~\cite{Bijnens:2014leaj} by performing a global fit to different
$\pi\pi$ and $\pi K$ observables. Nevertheless, among them, the 
dispersive determination of the $\pi K$ scattering lengths 
in~\cite{Buettiker:2003ppj} was used as constraint. Consequently, the full 
$\Order\big(p^6\big)$ results in~\cite{Bijnens:2014leaj} are not a genuine 
ChPT prediction but they are biased to satisfy the results given 
in~\cite{Buettiker:2003ppj}. The scattering length results 
in~\cite{Bijnens:2014leaj} and~\cite{Buettiker:2003ppj} are denoted in 
Fig.~\ref{fig:SL} by the dashed-red and dashed-green circle-filled 
ellipse, respectively. As we will discuss in detail below, the large 
difference one finds between the NLO and NNLO chiral estimates in the 
isovector direction is just a consequence of the large discrepancy between 
the dispersive result in~\cite{Buettiker:2003ppj} and chiral expectations.
Alternatively, one can estimate the value of the $C_i$ by using
resonance saturation. The contribution from vector and scalar
resonances to the saturation of the $\Order\big(p^6\big)$ LECs was
also studied in~\cite{Bijnens:2004buj}. Using the vector and scalar 
resonance parameter values extracted in~\cite{Ledwig:2014claj} from a 
global $\pi\pi$ and $\pi K$ fit, one obtains the dashed-blue ellipse in 
Fig.~\ref{fig:SL}, which is now consistent with the NLO prediction for the 
isovector scattering length.

The only direct experimental information about the pion-kaon scattering
lengths comes from the DIRAC experiment at CERN~\cite{Adeva:2017ocoj},
where the lifetime of hydrogen-like $\pi K$ atoms was measured.
They are a electromagnetically bound state of charged pions and kaons,
$\pi^+K^-$ and $\pi^-K^+$, which decay predominantly by strong
interactions to the neutral pairs $\pi^0 \bar K^0$ and $\pi^0K^0$.
The $\pi K$ atom lifetime and the scattering length are related
through the so-called modified Deser formula~\cite{Bilenky:1969zdj,
Schweizer:2004irj,Schweizer:2004qej}, namely
\begin{equation}
	\Gamma_{1S}=8\alpha^3\mu^2 p\,a_{0}^{-\,2}\left(1+\delta_K\right),
\end{equation}
where $\alpha$ is the fine structure constants, $\mu$ is the reduced
mass of the $\pi^\pm K^\mp$ system, $p$ is the outgoing momentum in
the centre-of-mass frame and $\delta_K$ accounts for isospin breaking
corrections~\cite{Schweizer:2004irj,Schweizer:2004qej,Kubis:2001ijj}. 
The experimental determination of $\Gamma_{1S}$ obtained at CERN
yields~\cite{Adeva:2017ocoj}
\begin{equation}
	a_0^- = \left(0.072^{+0.031}_{-0.020}\right)\,\mpi^{-1},
\end{equation}
which is denoted in Fig.~\ref{fig:SL} by a light-blue squared-filled
band. Unfortunately, the experimental errors are still too large to
provide useful information about the pion-kaon scattering lengths.
Nevertheless, there is still room for improvement, the statistical
precision is expected to improve by a factor 20 if the DIRAC 
Collaboration manages to run its experiment using the LHC 450~GeV 
proton beam. {\bf The proposed kaon beam experiment at JLab could 
certainly help to improve the current experimental information about 
the pion-kaon scattering lengths.}

On the lattice side, there is a plethora of results and we will only
consider unquenched analyses. From a lattice analysis of the $\pi K$ 
scalar form factor in semileptonic $K_{l3}$ decays, the value 
$a_0^{1/2}=0.179\left(17\right)\left(14\right)\,m_\pi^{-1}$ was
reported in~\cite{Flynn:2007kij} for the pion-kaon scattering length 
in the $I=1/2$ channel. This value corresponds to the gray squared-filled 
band in Fig.~\ref{fig:SL}. The first fully dynamical calculation with 
$N_f=2+1$ flavors was performed by the NPLQCD collaboration, leading 
to~\cite{Beane:2006gjj}
\begin{equation}
	a_0^{1/2}=\left(0.173^{+0.003}_{-0.016}\right)\,m_\pi^{-1},\quad  a_0^{3/2}
	=\left(-0.057^{+0.003}_{-0.006}\right)\,m_\pi^{-1},
\end{equation}
which is denoted in Fig.~\ref{fig:SL} by a dotted-blue ellipse.
Further dynamical results for $N_f=2+1$ flavors were reported
in~\cite{Fu:2011wcj} using a staggered-fermion formulation,
$a_0^{1/2}=0.182\left(4\right)\,m_\pi^{-1},\quad$ \\
$a_0^{3/2}=-0.051\left(2\right)\,m_\pi^{-1}$,
and by the PACS collaboration considering an improved Wilson
action~\cite{Sasaki:2013vxaj}, $a_0^{1/2}=0.183\left(18\right)
\left(35\right)\,m_\pi^{-1},\quad  
a_0^{3/2}=-0.060\left(3\right) \left(3\right)\,m_\pi^{-1}$. These 
results are depicted in Fig.~\ref{fig:SL} by a solid-yellow and a 
dotted-brown ellipse, respectively.

As we have seen, all the previous results are consistent with chiral
predictions, i.e., all of them are consistent within one standard
deviation for the isospin-odd direction, whereas much larger differences
are found in the isospin-even component. Nevertheless, the most precise 
up-to-date result was reported in~\cite{Buettiker:2003ppj} by solving a 
complete system of Roy-Steiner equations, corresponding to
\begin{equation}\label{eq:bachir}
	a_0^{1/2}=0.224(22)m_\pi^{-1},\quad a_0^{3/2}=-0.045(8)m_\pi^{-1}.
\end{equation}
This result is denoted in Fig.~\ref{fig:SL} by a light-green circle-filled
ellipse and it lies more than 3.5 standard deviations away from the NLO
ChPT result. This disagreement is particularly puzzling in the isospin-odd 
direction, where the ChPT prediction is protected by the low-energy 
theorem given in~\eqref{LET} and one should expect NLO ChPT to provide a 
reasonably precise value for the pion-kaon scattering lengths. In fact, 
previous dispersive analyses for $\pi\pi$ scattering  provided results for 
the scattering lengths only within a universal 
band~\cite{Ananthanarayan:2000htj,Colangelo:2001dfj}. High accuracy values 
were reached only after constraining dispersive results with chiral 
symmetry. Thus, one might wonder why should things be different in 
pion-kaon scattering. In the remaining part of this talk we will try to 
answer that question.

%%%-----------------------------------
\item \textbf{Roy--Steiner Equations for $\pi K$ Scattering}
\label{rs-eqs}

Dispersion relations have repeatedly proven to be a powerful tool for
studying processes at low energies with high precision. They are built 
upon very general principles such as Lorentz invariance, unitarity, 
crossing symmetry, and analyticity.

For $\pi\pi$ scattering, Roy equations (RE)~\cite{Roy:1971tcj} are 
obtained from a twice-subtracted fixed-$t$ dispersion relation, where 
the $t$-dependent subtraction constants are determined by means of 
$s\leftrightarrow t$ crossing symmetry, and performing a partial-wave 
expansion. This leads to a coupled system of partial-wave dispersion 
relations (PWDRs) for the $\pi\pi$ partial waves where the scattering 
lengths---the only free parameters---appear as subtraction constants.
The use of RE for $\pi\pi$ scattering has led to a determination of the
low-energy $\pi\pi$ scattering amplitude with unprecedented
accuracy~\cite{Ananthanarayan:2000htj,Colangelo:2001dfj,
GarciaMartin:2011cnj}, which, for the first time, allowed for a precise 
determination of the $f_0(500)$ pole 
parameters~\cite{Caprini:2005zrj,GarciaMartin:2011jxj}.

In the case of $\pi K$ scattering, a full system of PWDRs has to include
dispersion relations for two distinct physical processes, $\pi 
K\rightarrow\pi K$ ($s$-channel) and $\pi\pi\rightarrow \bar KK$ 
($t$-channel), and the use of $s\leftrightarrow t$ crossing symmetry will 
intertwine $s$- and $t$-channel equations. Roy-Steiner (RS) 
equations~\cite{Hite:1973pmj} are a set of PWDR that 
combine the $s$- and $t$- channel physical region by means of hyperbolic 
dispersion relations. The construction and solution of a complete system 
of RS equations for $\pi K$ scattering has been presented 
in~\cite{Buettiker:2003ppj}.

In more detail, the starting point for the work in~\cite{Buettiker:2003ppj}
is a set of fixed-$t$ dispersion for the pion-kaon isospin-even and -odd
scattering amplitudes, where the $t$-dependent subtraction constants are 
expressed in terms of hyperbolic dispersion relations passing through the 
threshold, i.e., where the internal and external Mandelstam variables $s$ 
and $u$ satisfy the condition $s\cdot u=\mK^2-\mpi^2$. A twice- and 
once-subtracted version was considered for the isospin-even -odd 
amplitude, respectively, where the subtraction constants are the
$a_0^{\pm}$ scattering lengths and the slope of the hyperbola in the
$t$-direction for the isospin-even amplitude, $b^+$, which, in the end,
is written in terms of a sum rule involving the $a_0^{-}$ scattering
length. Finally, the solution of the RS equations is achieved by 
minimizing the $\chi^2$-like function
\begin{equation}\label{eq:chi2}
	\chi_{\rm phys}^2=\sum_{l,I_s}\sum_{j=1}^N\left(\Re f_{l}^{I_s}(s_j)-F[{f_{l}^{I_s}](s_j)}\right)^2,
\end{equation}
where $f_{l}^{I_s}$ denotes pion-kaon partial-waves with angular 
momentum $l$ and isospin $I_s$, $F[f_{l}^{I_s}]$ stands for the functional 
form of the RS equations for the $f_{l}^{I_s}$ partial wave, and the 
minimizing parameters are the partial waves and the pion-kaon
scattering lengths. In this way, the minimum of~\eqref{eq:chi2} provides 
as an output the pion-kaon scattering length values given 
in~\eqref{eq:bachir}. A relevant question is whether this solution is 
unique. In principle, the subtracted version built 
in~\cite{Buettiker:2003ppj} is constructed in such a way that it matches 
the conditions ensuring a unique RS equation solution investigated 
in~\cite{Gasser:1999hzj}. In the $\pi\pi$ RE case studied  in 
Ref.~\cite{Ananthanarayan:2000htj}, it was observed that the $\pi\pi$ 
scattering lengths were determined only within a universal band.
Something similar was observed in the RS solution for $\pi N$ presented
in Ref.~\cite{Hoferichter:2015hvaj}, where precise results were obtained 
once the $\pi N$ scattering lengths were imposed as constraints.
More precisely, the problem is connected with the number of no-cusp
conditions required in order to ensure a smooth matching in the three
partial waves between the dynamical solution of the RS equations and the 
input considered at higher energies. In~\cite{Buettiker:2003ppj}, no-cusp 
conditions for the $f_0^{1/2}$ and $f_1^{1/2}$ partial waves were imposed, 
matching precisely the number of free subtraction constants, the two 
pion-kaon scattering lengths $a_0^{\pm}$. However, in 
Ref.~\cite{Ananthanarayan:2000htj} it was found for $\pi\pi$
scattering that only one no-cusp condition was enough to ensure a smooth
matching, leading to a $\pi\pi$ scattering length universal band.
%%%-----------------------------------
\begin{figure}[!t]
\centering
\includegraphics[width=\textwidth]{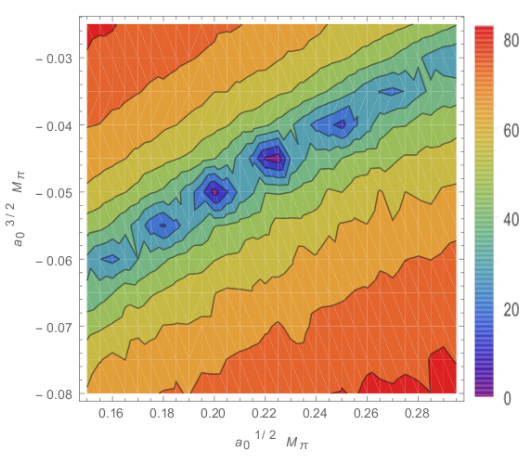}

\caption{Value of the $\pi K$ RS $\chi$-like function defined in 
	(2.1) for a grid of points on the  $I=1/2$, $I=3/2$ 
	scattering length plane. This result suggests that one can 
	achieved an exact solution of the pion-kaon RS equations 
	on the universal band.} \label{fig:universal}
\end{figure}
%%%-----------------------------------

In order to analyze whether something similar might happen in the $\pi K$
case, we have studied further possible cusp-free RS solutions in a grid of
points in the pion-kaon scattering length plane. The results are plotted 
in Fig.~\ref{fig:universal}, where one can see that RS equation solutions 
for $\pi K$ scattering can be achieved within a universal band. Although 
the solution presented in~\cite{Buettiker:2003ppj} lies perfectly within 
this universal band, it is clearly not enough to fully constrain the values 
of the pion-kaon scattering lengths. As we can see in Fig.~\ref{fig:SL}, 
this universal band is indeed consistent with both chiral predictions and 
the different lattice results studied above. The next step of this project 
will be to study whether the combination of RS equations with sum rules 
for subtraction constants allows one to obtain a unique and consistent 
solution of the $\pi K$ scattering lengths.

%%%-----------------------------------
\item \textbf{Acknowledgments}

We would like to thank the organizers for a wonderful workshop, and for 
the invitation to talk about our work on pion--kaon scattering. We are 
grateful to B.~Moussallam for helful dicussions and to J.~Bijnens for 
providing us the code for the NNLO pion-kaon scattering amplitudes. We 
also acknowledge useful discussions with M.~Hoferichter, B.~Kubis and
U.-G.~Mei\ss ner. Financial support by the Swiss National Science 
Foundation is gratefully acknowledged.
\end{enumerate}

%%%-----------------------------------

%%%%%%%%%%%%%%%%%%%%%%%%%%%%%%%%%%%%%%%%%%%%%%%%%%%%%%%%%%%%%%%%%%%%%%%%
\newpage
\subsection{Dispersive Analysis of Pion-kaon Scattering}
\addtocontents{toc}{\hspace{2cm}{\sl Jos\'e R. Pelaez, Arkaitz Rodas, and Jacobo Ruiz de Elvira}\par}
\setcounter{figure}{0}
\setcounter{table}{0}
\setcounter{equation}{0}
\setcounter{footnote}{0}
\halign{#\hfil&\quad#\hfil\cr
\large{Jos\'e R. Pelaez and Arkaitz Rodas}\cr
\textit{Departamento de F\'{\i}sica Te\'orica}\cr
\textit{Universidad Complutense}\cr
\textit{28040, Madrid, Spain}\cr\cr
\large{Jacobo Ruiz de Elvira}\cr
\textit{Albert Einstein Center for Fundamental Physics}\cr
\textit{Institute for Theoretical Physics, University of Bern}\cr
\textit{3012 Bern, Switzerland}\cr}

%%%-----------------------------------
\begin{abstract}
After briefly motivating the interest of $\pi K$ scattering and light strange
resonances, we discuss the relevance of dispersive methods to constrain the 
amplitude analysis and for the determination of resonances parameters. Then 
we review our recent results on a precise determination of $\pi K$ amplitudes 
constrained with Forward Dispersion Relations, which are later used together with
model-independent methods based on analyticity to extract the parameters of
the lightest strange resonances. In particular we comment on our most recent
determinations of the $\kappa/K_0^\ast(800)$ pole using dispersive and/or
techniques based on analytic properties of amplitudes.  We also comment
on the relevance that a new kaon beam at JLab may have for a precise
knowledge of these amplitudes and the light strange resonances.
\end{abstract}

%%%-----------------------------------
\begin{enumerate}
\item \textbf{Motivation to Study $\pi K$ Scattering}

Pion and kaons are the Goldstone bosons of the spontaneous $SU(3)$ chiral
symmetry breaking in SU(3) and their masses are due to the small explicit
breaking due to non-vanishing quark masses. Thus, by studying their 
interactions we are testing our understanding of this spontaneous symmetry 
breaking, which is rigorously formulated in terms of the low-energy effective 
theory of QCD, namely Chiral Perturbation Theory, as well as the role of 
quark masses and the breaking of the flavor SU(3) symmetry. In addition, pion 
and kaons appear in the final state of almost all hadronic
interactions involving strangeness and a precise understanding of $\pi K$
scattering is therefore of relevance to describe the strong final-state 
interactions of many hadronic processes, including those presently under an 
intense experimental and theoretical study: $B$ decays, $D$ decays, CP 
violation, etc...

Finally, most of our knowledge of strange resonances below 2 GeV comes from
$\pi K$ experiments. Strange resonances are very helpful in order to determine
how many flavor multiplets exist, which in turn helps to determine how many 
non-strange resonances are needed to complete these multiplets.  Any additional 
flavorless state would then clearly suggest the existence of glueball states. 
Moreover, the mass hierarchy between the strange and non-strange members of a 
multiplet can also reveal the internal nature of meson resonances (ordinary $q 
\bar q$ states, tetraquarks, molecules, etc...). In particular, as we will see
below, the strong theoretical constraints on $\pi K$ scattering provide
the most reliable method to determine the existence of the lightest strange 
scalar meson, the controversial $\kappa$ or $K_0^\ast(800)$ meson, which still 
``Needs Confirmation'' according to the Review of Particle Properties 
(RPP)~\cite{PDGm}.

%%%-----------------------------------
\item \textbf{Analyticity, Dispersion Relations, and Resonance Poles}

Analyticity constraints in the $s,t,u$ Mandelstam variables are the 
mathematical expression of causality. For example,  as illustrated on the left 
side of Figure~\ref{fig1}, causality implies that two-body scattering 
amplitudes $T(s,t)$ have no singularities in the first Riemann sheet of complex 
$s$-plane for fixed $t$, except for cuts in the real axis. In particular, there 
is one ``physical'' cut on the real axis that reflects the existence of a 
threshold, which extends from that threshold to $+\infty$. In addition, due to 
crossing symmetry, the physical cuts in the $u$-channel are seen  as cuts in the
complex $s$-plane from $-\infty$ to $-t$. By using Cauchy Theorem one can now 
calculate the value of the amplitude anywhere in the complex plane as an 
integral of the amplitude over the contour $C$. If the amplitude decreases 
sufficiently fast when $s\rightarrow\infty$ then, by sending the curved
part of contour $C$ to infinity, one is left only with integrals over the right 
and left cuts in the real axis. Since the amplitude in the upper half plane is 
conjugate to the amplitude in the lower plane, these are integrals over the 
imaginary part of the amplitude.  If the amplitude does not decrease 
sufficiently fast at infinity, one repeats the argument but for the amplitude 
divided by a polynomial of sufficiently high degree. Then  the amplitude is
determined up to a polynomial of that degree, whose coefficients are called 
``subtraction constants''. The resulting relation is called a subtracted 
Dispersion Relation.
%%%-----------------------------------
  \begin{figure}[!htbp] 
    \centering
    \includegraphics[width=0.3\columnwidth]{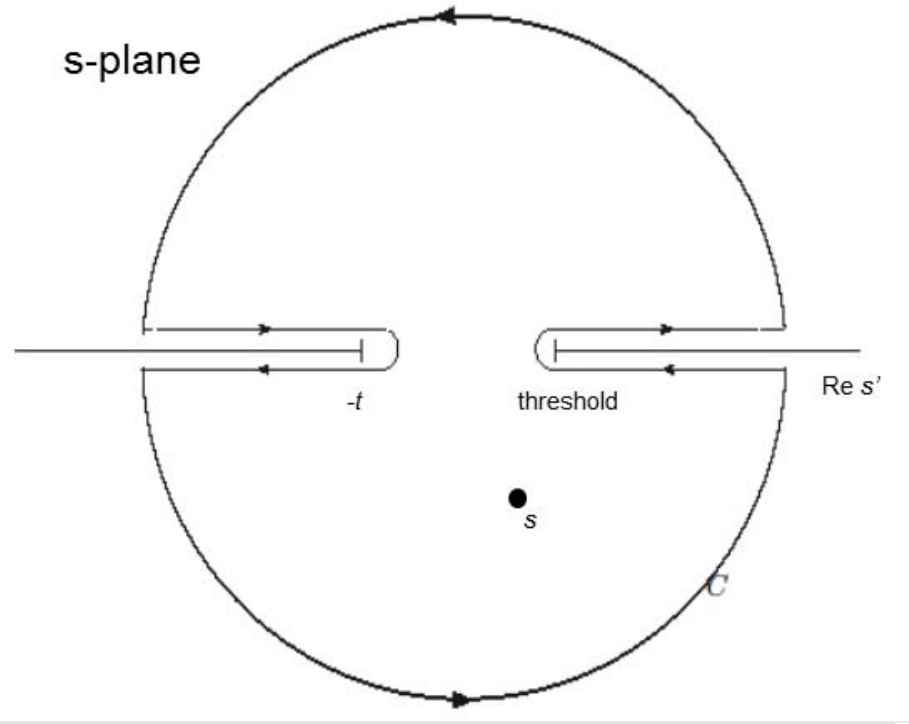}
    \includegraphics[width=0.55\columnwidth]{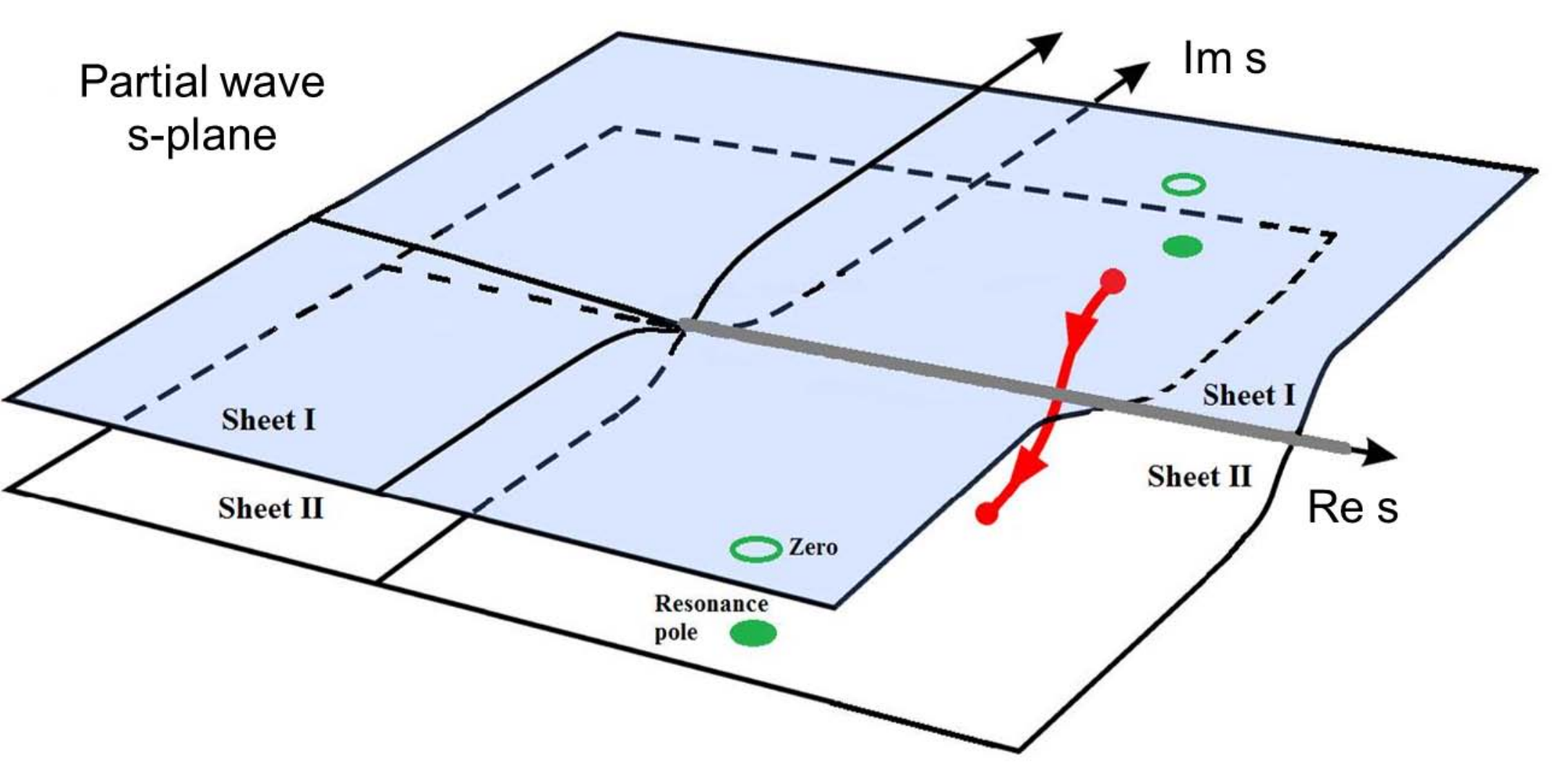}

     \caption{Left: Analytic structure in the $s$-plane for fixed-$t$.
	Right: Two-sheet structure for elastic partial waves.
	Poles appear in the second sheet. In the
	same location a zero appears in the $S$-matrix.
\label{fig1}}
 \end{figure}
%%%-----------------------------------

Dispersion relations are useful for calculating the amplitude where there is no 
data, to constrain data analyses or to look for resonance poles in the complex 
plane. Actually, the rigorous mathematical definition of a resonance involves 
the existence of a pair of conjugated poles in the second Riemann sheet of a 
partial wave amplitude.  This sheet is reached by crossing continuously the 
physical cut, as seen on the right side of Figure~\ref{fig1}. The position of 
the pole in the lower half plane is related to the mass $M$ and with $\Gamma$
of the resonance as $\sqrt{s_{pole}}=M-i\Gamma/2$.

So far we have discussed the physical cut and its sheet structure, but
in Figure~\ref{fig:kappaplane} we show the complicated analytic structure of
$\pi K$ scattering partial waves, which includes the "unphysical" left cut and 
an additional circular cut which appears due to the different pion and kaon 
masses. Now, when the pole of a resonance is far from other singularities and  
close to the real axis, namely $\Gamma <<M $, as it is the case of the 
$K^\ast(892)$, whose width is $\sim 50\,$MeV, it is seen in the real axis, i.e., 
experimentally, as a peak in the amplitude. Simple parameterizations of data 
just around its nominal mass, like a Breit-Wigner formula, can describe such 
resonances rather well. However, for very wide resonances like the 
$\kappa/K^\ast_0(800)$, the energy region around its nominal mass is more 
distant than other energy regions and particularly the threshold region. This 
threshold will distort dramatically the peak naively expected for a resonance. 
In addition, chiral symmetry implies the existence of the so-called Adler zero 
below threshold, which once again is close to the pole ans distorts its simple 
shape when seen at physical energies. Finally, the nominal mass region is also 
at a comparable distance to the pole as the circular cut and the left cut and 
their contributions should not be discarded a priori but should taken into 
account for precise pole determinations.

Therefore, the $\kappa/K_0^\ast(800)$ resonance cannot be described with 
precision just by the knowledge of data around its nominal mass, and the pole 
cannot be extracted with simple formulas that do not posses the correct 
analytic structures or that do not have the correct low-energy constraints of 
the QCD spontaneous chiral symmetry breaking. This is the reason why dispersion 
relations and analyticity properties are relevant tools to analyze data and the 
use of simple models can lead to considerable confusion or artifacts. In 
addition, data in the near threshold region impose strong constraints to the 
pole position, even if the resonance is nominally at a higher mass. This partly 
explains the situation of the $\kappa/K^\ast_0(800)$ in the RPP where, as we 
will see in the final section, the poles obtained with Breit-Wigner formulas are 
quite spread and do not coincide with those obtained using more rigorous 
formalisms.
%%%-----------------------------------
  \begin{figure}[!htbp] %  figure placement: here, top, bottom, or page
    \centering
    \includegraphics[width=0.7\columnwidth]{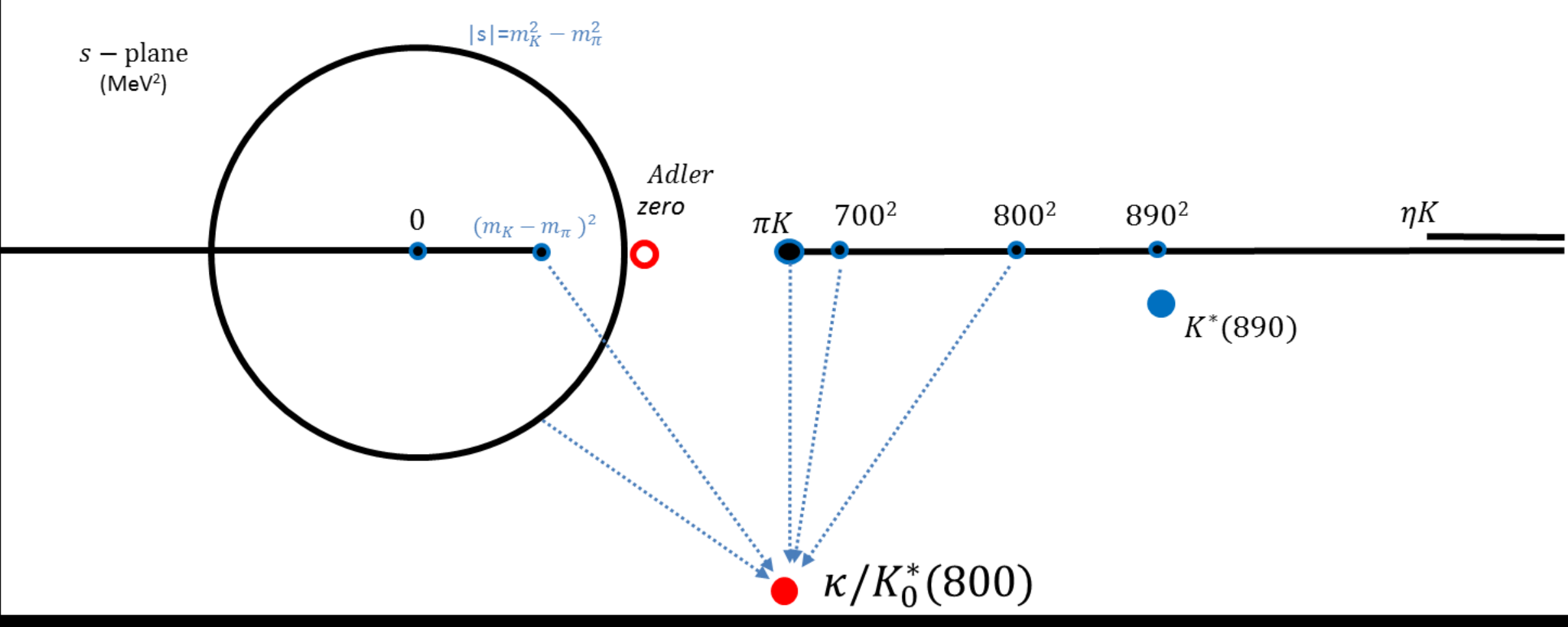}

    \caption{Analytic structure of the $\pi K$ partial waves in the complex 
	$s$-plane (in MeV$^2$). Note the existence of a left cut up to 
	$(m_K-m_\pi)^2$ and a circular cut. We have shown the position
	of the poles associated to the $K^\ast(892)$ and the 
	$\kappa/K^\ast_0(800)$. \label{fig:kappaplane}}
 \end{figure}
%%%-----------------------------------

Note that in order to write Cauchy's Theorem as we described above we need to 
consider the amplitude as a one-variable function. For this, there are two 
approaches:
\begin{enumerate}
\item   Fix one variable and obtain a dispersion relation in terms of the other.
	This leads to the popular fixed-$t$ dispersion relations, although one 
	could also obtain  fixed-$u$ or even fixed-$s$ dispersion relations.
	But one can also fix one variable in terms of the other one by means of a
	constrain. For instance, this is the case of Hiperbolic Dispersion Relations,
	where one imposes a constraint $(s-a)(u-a)=b$. Complicated relations 
	like the latter are used to maximize the applicability region of the 
	Dispersion Relation but if that is not needed for a particular study a 
	simple choice could be more convenient. Of particular interest are 
	Forward Dispersion Relations (FDRs), $t=0$, since they yield rather 
	simple expressions and because the higher part of the integrals
	can be expressed in terms of total cross sections thanks to the Optical Theorem.
	These FDRs have also the advantage that they can be applied in principle up to
	arbitrarily high energies and that their analytic structure is very simple. This 
	makes it easier to use crossing symmetry in order to rewrite the 
	left-cut contributions in terms of physical amplitudes over the physical 
	cut. Their expressions are rather simple for $\pi K$ scattering and as 
	we will see below provide stringent constraints on the  description of 
	existing data. Unfortunately it is not possible to use the integral FDRs 
	to access the second Riemann sheet in search for poles.
\item   Partial-wave Dispersion Relations. Now the scattering angle dependence 
	is eliminated by projecting the amplitudes in partial waves, $t_J(s)$. 
	The main advantage is that the second Riemann sheet, where resonance 
	poles can appear, is easily accessible. The reason is that, due to 
	unitarity, for elastic partial waves the second Riemann sheet of the $S$ 
	matrix is just the inverse of the first, namely 
	$S_J^{II}(s)=1/S^I_J(s)$. Recalling that $S_J(s)=1-2 i \sigma(s) t_J(s)$ 
	with $\sigma(s)=k/2\sqrt{s}$, where $k$ is the CM-momentum, then the 
	Second Riemann Sheet partial wave amplitude can be obtained from the 
	first as follows:
\begin{equation}
	t_J^{II}(s)=\frac{t_J^I(s)}{1-2i \sigma(s)t_J^I(s)}
	\label{ec:seconsheet}
\end{equation}
	Unfortunately, when the scattering particles do not have equal masses, the 
	partial-wave projection leads to a more complicated analytic structure. 
	For $\pi K$ partial waves this structure was already shown in 
	Fig.~\ref{fig:kappaplane} and includes a new circular cut and a longer 
	left cut. The most difficult part of using Dispersion Relation is to 
	calculate the contributions along these unphysical cuts, and two main 
	approaches appear once again in the literature:
\begin{enumerate}
\item   Do not use crossing and just approximate the left cut. This leads to 
	simple dispersion relations (see for instance~\cite{Zheng:2003rwm}) and 
	has become very popular when combined with Chiral Perturbation Theory to 
	approximate the left cut and the subtraction  constants. Within this
	approach, the dispersion relations are written for the inverse partial-wave, 
	since unitarity fixes its imaginary part in the elastic region. This 
	approach is known as the Inverse  Amplitude Method~\cite{IAMm} that is 
	one instance of Unitarized Chiral Perturbation Theory. It  provides a 
	fairly good description of data and yields the poles of all resonances 
	that appear in the two-body scattering of pions and kaons below 1~GeV. 
	In particular, a pole for the kappa is found at~\cite{Pelaez:2004xpm}
	$(753\pm52)-i(235\pm33)\,$MeV. Despite not being good for precision, it 
	is very convenient to connect the meson scattering phenomenology with 
	QCD parameters like $N_c$ or the quark masses~\cite{IAMmqncm}
	and thus study the nature of resonances.
\item   Use crossing to rewrite the contributions from unphysical cuts in terms of
	Use crossing to rewrite the contributions from unphysical cuts in terms of
	partial waves on the physical cut. Unfortunately, when projecting into a partial 
	wave of the $s$-angle, one does not extract a single partial wave in the 
	$u$-channel angle, so that the whole tower of partial waves is still 
	present when rewriting the unphysical cuts into the physical region 
	using crossing. This leads to an infinite set of coupled dispersion 
	relations. These are called Roy-Steiner dispersion 
	relations~\cite{Roy-Steinerm} (different versions can be obtained 
	starting from fixed-$t$ or hyperbolic dispersion relations). They are more 
	cumbersome to use, and usually they are applied only to the lowest 
	partial waves, taking the rest as fixed input. However, they yield the 
	most rigorous determination of the $\kappa/K_0^\ast(800)$ pole so 
	far~\cite{Buettiker:2003ppm}: $(658\pm13)-i(278.5\pm12)\,$MeV.
	Note that in these works Roy-Steiner equations were {\it solved} in the elastic 
	region for the $S$ and $P$ waves, using data on higher energies and 
	higher waves as input, but { \it no data on the elastic region below 
	$\sim$1~GeV}. It was also shown that the pole appears within the
	Lehman-ellipse that ensures the convergence of the partial-wave 
	expansion (actually, even inside a more restrictive region that also 
	ensures the use of hyperbolic dispersion relations). Nevertheless the 
	RPP still considers that the existence of the $\kappa/K_0^\ast(800)$ `` 
	Needs Confirmation''.
\end{enumerate}

%%%-----------------------------------
\item \textbf{Amplitudes Constrained with Forward Dispersion Relations}

Hence, in order to provide the required confirmation of the 
$\kappa/K_0^\ast(800)$ existence and of the values of its parameters, we have 
recently performed a Forward Dispersion Relation study of $\pi K$ scattering 
data~\cite{Pelaez:2016tgim}. The explicit expressions of the FDRs for the 
symmetric amplitude $T^+$ and the antisymmetric one $T^-$ and all other details 
can be found in~\cite{Pelaez:2016tgim}. We took particular care to use simple 
parameterizations of data on partial waves that could be easily used later in 
further theoretical or experimental studies. The first relevant result, as shown 
in the left column of  Fig.~\ref{fig:FDR} is that the

Unconstrained Fits to Data (UFD) do not satisfy FDRs well, particularly at high 
energies and in the threshold region.
%%%-----------------------------------
  \begin{figure}[!htbp] %  figure placement: here, top, bottom, or page
    \centering
    \includegraphics[width=0.9\columnwidth]{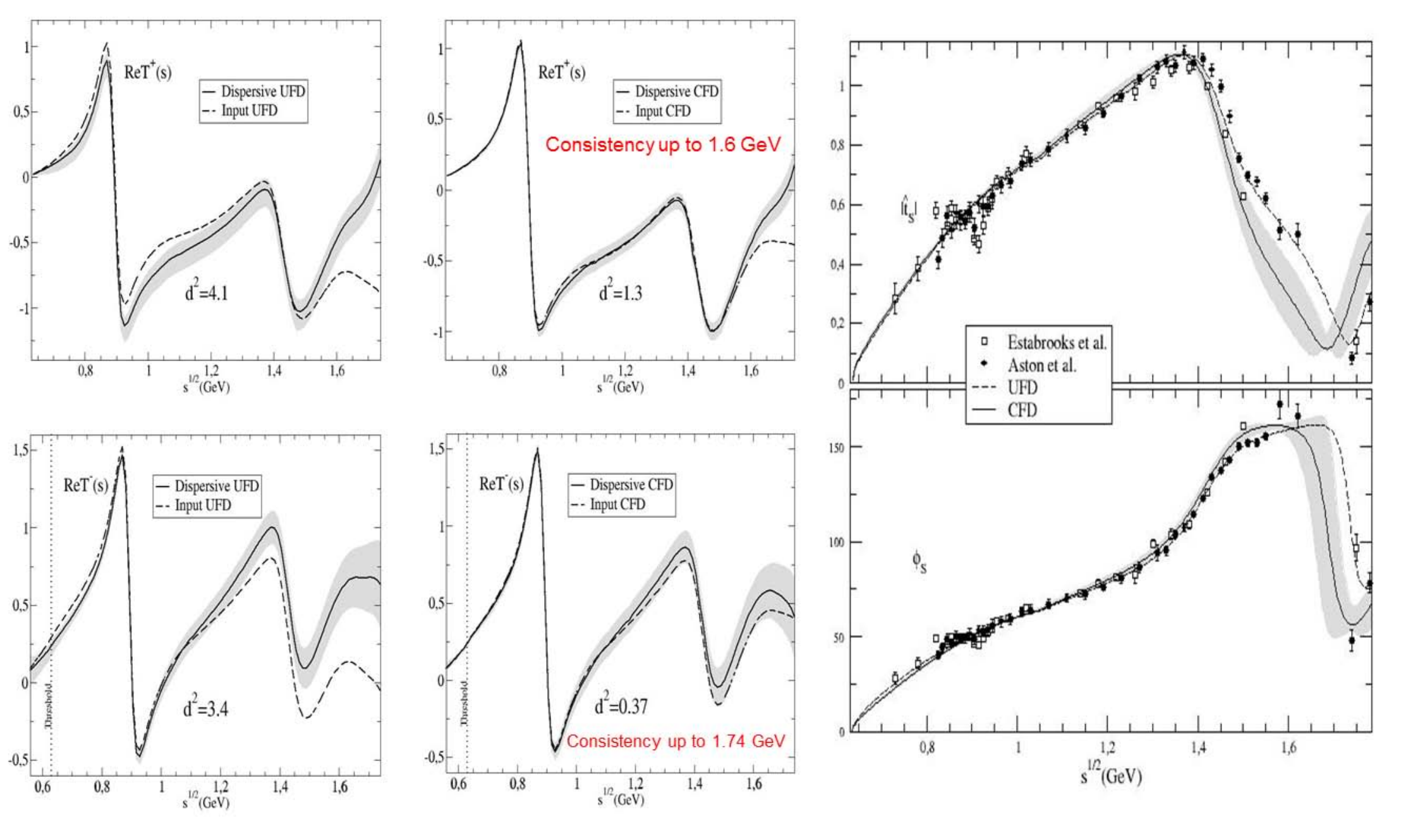}

	\caption{Left Column:  The input from Unconstrained Fits to Data (UFD)
	do not agree well with the Dispersive output of FDRs. Central Column:
	By imposing FDRs as constraints on data fits (CFD) they can be well 
	satisfied up to 1.6~GeV without spoiling the data description. Right 
	Column: The difference between UFD and CFD data descriptions. Data 
	comes from Ref.~\protect\cite{datapiKm}. \label{fig:FDR}}
 \end{figure}
%%%-----------------------------------
For this reason we {\it imposed} the FDRs  to obtain Constrained Fits to Data 
(CFD).  Once this is done the agreement is remarkable up to 1.6 GeV for the 
symmetric FDR and up to 1.74~GeV for the antisymmetric one, as seen in the 
central column of Fig.~\ref{fig:FDR}. In the right column of that figure, we 
show that the difference between the UFD and CFD is significant above 1.4~GeV 
and also in the threshold region for the phase, where there is very little 
experimental information. As we have already commented when discussing the 
$\kappa$ pole and Fig.\ref{fig:kappaplane}, this area is particularly relevant 
for the determination of the $\kappa/K_0^\ast(800)$ resonance from experimental 
data.  It would be very interesting to have new data to confirm our findings 
{\bf and this could certainly be achieved with the proposed kaon beam at JLab}.

Moreover, it can be seen in that last column of Fig.\ref{fig:kappaplane}, that 
the most reliable data source, which is coming from the LASS/SLAC 
spectrometer~\cite{datapiKm}, only provides a combination of isospins, but not 
the separated $I=1/2$ and $I=3/2$ partial waves. Actually, the information on 
$I=3/2$ is rather scarce and sometimes contradictory~\cite{datapiKm}.
This will be one of the main sources of uncertainty in our calculation of the 
$\kappa$ pole. {\bf As it has already been explained in other talks of this 
conference  that a clear isospin separation could be achieved with the proposed 
kaon beam facility}.

%%%-----------------------------------
\item \textbf{Analytic Methods to Extract Resonances}

Our parameterizations of the data are obtained from piece-wise analytic 
functions which are carefully matched to impose continuity. Therefore, none of 
the pieces by itself has all the relevant information to determine the resonance 
poles with precision. Nevertheless, the lowest energy piece was constructed by 
means of a conformal expansion, which exploits the analyticity of the partial 
wave on the full complex plane and has the correct analytic structure while 
satisfying the elastic unitarity constraint that allows for a  continuation
to the second Riemann sheet as explained in Eq.~\ref{ec:seconsheet}.
We have actually checked that this simple parameterization yields a pole
for the $\kappa/K_0^\ast(800)$ at: $(680\pm15)-i(334\pm7.5)$. However, this 
still makes use of a specific parameterization and, no matter how good it might 
be, is still model-dependent and the small uncertainties can only be understood 
within that model.

However, there is a procedure to extract poles~\cite{Padesm}, which is model 
independent in the sense that it does not rely on a specific parameterization 
choice. It is based on a powerful theorem of Complex Analysis, which ensures 
that in a given domain where a function is analytic except for a pole, one can 
construct a sequence of Pad\'e approximants that contains a pole that 
converges to the actual pole in the function. This sequence is built in terms of 
the values of the partial wave and the derivatives at a given point within that 
domain.  Thus,  using our amplitudes constrained with FDRs (CFD), we have 
searched~\cite{Pelaez:2016klvm} for a point in the real axis in which we can 
define a domain that includes the expected
$\kappa/K_0^\ast(800)$ pole and built a sequence of Pad\'es. In this way we have 
confirmed the existence of the $\kappa/K_0^\ast(800)$ 
pole~\cite{Pelaez:2016klvm} and its location at: $(670\pm18)-i(295\pm28)$. This 
time the errors are larger because the determination does not depend on a 
specific parameterization. Similar results have been obtained for all the 
resonances that appear in $\pi K$ scattering below 1.8~GeV (except for the 
$K^\ast(1680)$), thus providing further support for their existence and a 
determination of their parameters that do not depend on approximations or 
specific choices of parameterizations like Breit-Wigner parameterizations. 
Further details can be found in~\cite{Pelaez:2016klvm}.

Of course, the best would be a new analysis of the Roy-Steiner  type, 
independent from that of~\cite{Buettiker:2003ppm}. Actually our  Madrid group 
already has a {\it preliminary result} of this kind of analysis which finds a 
pole at $(662\pm13)-i(289\pm25)\,$MeV~\cite{inprepm}. For this we use 
partial-wave Roy-Steiner-like equations obtained from hyperbolic dispersion 
relation using as input the amplitudes constrained to satisfy 
FDRs~\cite{Pelaez:2016tgim}, and therefore using data below 1~GeV instead of 
solving the equations in the elastic region as in Ref~\cite{Buettiker:2003ppm}. 
We are showing this preliminary result in red in 
Fig.~\ref{fig:kappacompilation}, together with all the other poles mentioned in 
this mini-review.

Note however that there is considerable room for improvement by reducing the 
uncertainties. For this it would be of the uttermost importance not only to have 
more precise data, but also data in the regions where there is none nowadays, 
and particularly with a clear separation of partial waves with different 
isospin. {\bf  As we have seen in this workshop, all this could be achievable 
with the proposed kaon beam at JLab} that therefore could provide a sound 
experimental basis for the understanding of the spectroscopy of strange mesons, 
which nowadays relies on strong model-dependent assumptions and data sets with 
strong inconsistencies among themselves and with the fundamental dispersive 
constraints.
%%%-----------------------------------
  \begin{figure}[!htbp] %  
    \centering
    \includegraphics[width=0.7\columnwidth]{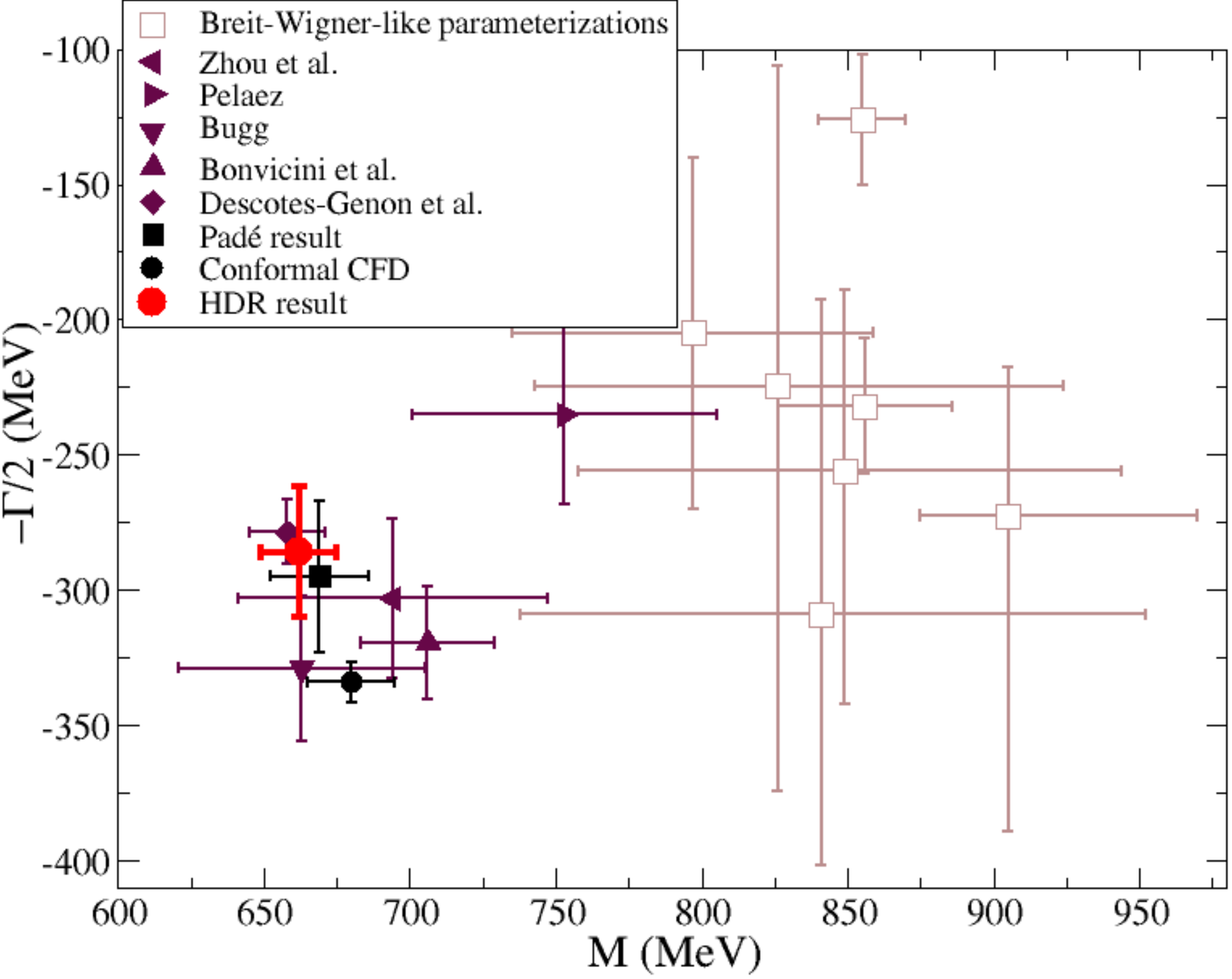}
\caption{Compilation of results for the $\kappa/K^\ast_0(800)$ resonance pole 
	positions. We list as empty squares all the determinations using some 
	for of Breit-Wigner shape listed in the RPP~\protect\cite{PDG}. 
	The rest of references correspond to: 
	Zhou \textit{et al.}~\protect\cite{Zheng:2003rwm}, 
	Pelaez~\protect\cite{Pelaez:2004xpm}, 
	Bugg~\protect\cite{Bugg:2003kjm}, 
	Bomvicini~\protect\cite{Bonvicini:2008jwm}, 
	Descotes-Genon \textit{et al.}~\protect\cite{Buettiker:2003ppm}, 
	Pad\'e result~\protect\cite{Pelaez:2016klvm}, 
	Conformal CFD~\protect\cite{Pelaez:2016tgim} and our preliminary 
	HDR result using Roy-Steiner equations~\protect\cite{inprepm}
	\label{fig:kappacompilation}}
\end{figure}
%%%-----------------------------------
\end{enumerate}

%%%-----------------------------------
\newpage
\item \textbf{Acknowledgments}

JRP and AR are supported by the Spanish Project FPA2016-75654-C2-2-P.
The work of JRE was supported by the Swiss National Science Foundation.
AR would also like to acknowledge financial support of the Universidad
Complutense de Madrid through a predoctoral scholarship.
\end{enumerate}

%%%-----------------------------------

%%%%%%%%%%%%%%%%%%%%%%%%%%%%%%%%%%%%%%%%%%%%%%%%%%%%%%%%%%%%%%%%%%%%%%%%%
\newpage
\subsection{Analyticity Constraints for Exotic Mesons}
\addtocontents{toc}{\hspace{2cm}{\sl Vincent Mathieu}\par}
\setcounter{figure}{0}
\setcounter{table}{0}
\setcounter{equation}{0}
\setcounter{footnote}{0}
\halign{#\hfil&\quad#\hfil\cr
\large{Vincent Mathieu}\cr
\textit{Thomas Jefferson National Accelerator Facility}\cr
\textit{Newport News, VA 23606, U.S.A.}\cr}

%%%%-----------------------------------
\begin{abstract}
Dispersive techniques have drastically improved the extraction of the
pole position of the lowest hadronic resonances, the $\sigma$ and $\kappa$ 
mesons. I explain how dispersion relations can be used in the search of the 
lowest exotic meson, the $\pi_1$  meson. It is shown that a combination of the 
forward and backward elastic $\pi\eta$ finite energy sum rules constrains the 
production of exotic mesons.
\end{abstract}

%%%-----------------------------------
\begin{enumerate}
\item \textbf{Introduction}

Despite the knowledge of the Quantum ChromoDynamics (QCD) Lagrangian for about 
50 years and the abundance of data, the pole position of the lowest QCD 
resonances, the $\sigma$ and $\kappa$ mesons also called $f_0(500)$ and 
$K^\ast_0(800)$ respectively, have remained inaccurate for decades. The 
situation of the $\sigma$ has, however, changed recently. The combined use of 
dispersion relations, crossing symmetry and experimental data have
led independent teams to a precise location of the complex 
pole~\cite{Caprini:2005zrn,GarciaMartin:2011jxn,Albaladejo:2012ten}. The 
interested readers can find a historical and technical review in 
Ref.~\cite{Pelaez:2015qban}.

The situation of the $\kappa$ is going in the same direction. A recent
analysis~\cite{DescotesGenon:2006ukn} using dispersion relation and crossing 
symmetry have determined its pole location accurately. Nevertheless, the lowest 
resonance with strangeness is currently not reported in the summary tables in 
the Review of Particle Properties~\cite{Patrignani:2016xqpn}. This situation 
might change when the results of Ref.~\cite{DescotesGenon:2006ukn} will be 
confirmed independently. Studies in this direction are in 
progress~\cite{Pelaez:2016klvn}.

With these recent confirmation of existence and precise location of the $\sigma$ 
and $\kappa$ poles, the next challenge in meson spectroscopy is the existence of 
exotic mesons. \footnote{By exotic, I mean in the quark model sense. A 
quark-antiquark pair cannot couple to the $\pi_1$ quantum number $I^GJ^{PC} = 
1^-1^{-+}$. A review of exotic quantum numbers and notation is presented in 
Ref.~\cite{Meyer:2010kun}.} The theoretical and experimental works tend to 
support the existence of an isovector with the quantum numbers $J^{PC} = 
1^{-+}$, denoted $\pi_1$, around 1400-1600~MeV. The experimental status is still 
nevertheless controversial~\cite{Meyer:2010kun}. As in the case 
of the $\sigma$ and the $\kappa$ meson, the (possible) large width of the 
$\pi_1$ prevents the use of standard partial-wave parametrizations, such as the 
Breit-Wigner formula, to extract the pole location. The parametrization of the 
exotic wave can, nevertheless, be constrained thanks to dispersive techniques.

Dispersion relations, in meson-meson scatterings, are typically used in their 
subtracted form to reduce the influence of the, mostly unknown, high energy 
part. One can alternatively write dispersion relations for moments of the 
amplitudes. Using a Regge form for the high energy part, they lead to the finite 
energy sum rules (FESR)~\cite{Dolen:1967jrn,Collins:1977jyn}. FESR were applied 
recently in hadro- and photo-production on a nucleon 
target~\cite{Mathieu:2015gxan, Nys:2016vjzn,Mathieu:2017butn} in which data in 
the low and high energy mass region are available. With the advent of high 
statistic generation of hadron spectroscopy meson, one now has access to data in 
a wide energy range in meson-meson interaction. The COMPASS collaboration has 
reported $\pi\eta$ and $\pi\eta'$ (acceptance corrected) partial waves from 
threshold to 3 GeV and has recorded data up to 5~GeV~\cite{Adolph:2014rppn}. 
CLAS12 and GlueX are currently taking data  on two mesons photo- and 
electro-production with an expected coverage up to at least 3~GeV. These 
data sets allow then to constrain the resonance region with the Regge region 
{\it via} FESR.

Schwimmer~\cite{Schwimmer:1970ykn} has demonstrated that duality and the 
non-existence of odd wave in elastic $\pi\eta$ scattering lead to a degeneracy 
relation between Regge exchange residues in the equal mass case. In order words, 
in the equal mass case, the equality between the forward and backward region at 
high energy is equivalent to the non-existence of odd waves. This results, 
showing the self-consistency of the quark model, was actually already included 
in the general results of Mandula, Weyers and Zweig~\cite{Mandula:1970wzn}
also in the $SU(3)$ symmetry limit.

Here I demonstrate how FESR in elastic $\pi\eta$ scattering can be used to 
constrain the exotic $P$-wave. I examine Schwimmer argument with exact 
kinematics and derive the forward and backward FESR for $\pi\eta$ scattering. It 
is shown how the deviation from forward/backward equality and the mass 
difference is related to the amount of exotic meson in $\pi\eta$ scattering.

The kinematics are reviewed in Section~2 and the FESR are written in Section~3. 
The main results are derived in Section~4.

%%%-----------------------------------
\item \textbf{Kinematics}

%%%-----------------------------------
\begin{figure}[htb]
\begin{center}
\includegraphics[width=.6\linewidth]{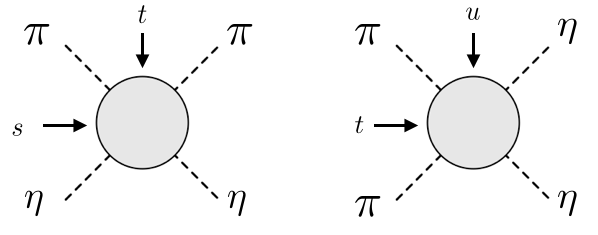}
\end{center}

\caption{\label{fig:1}The $s$-channel $\pi\eta \to \pi\eta$ and the $t$-channel 
	$\pi\pi\to\eta\eta$. }
\end{figure}
%%%-----------------------------------

The $s-$channel reaction $\pi \eta \to \pi\eta$. The $u-$channel is identical to 
the $s-$channel and the $t-$channel is $\pi\pi \to \eta\eta$, as depicted in 
Fig~\ref{fig:1}. All channels are described by a single function $A(s,t,u)$.
Let $m_\pi$ and $m_\eta $ be the masses of the $\pi$ and $\eta$ mesons, and 
$s+t+u=\Sigma = 2m_\pi^2+2m_\eta^2$. The center-of-mass momentum and scattering 
angle in the $s$-channel are
\begin{subequations}
\begin{align}
	q^2 &= \frac{1}{4s} \left( s-(m_\pi+m_\eta)^2\right) \left( 
	s-(m_\pi-m_\eta)^2\right),\\
	z_s &= 1+ \frac{t}{2 q^2} =-1  - \frac{u-u_0}{2 q^2},
	& u_0 & = \frac{(m_\eta^2-m_\pi^2)^2}{s}
\end{align}
\end{subequations}

The unitarity threshold of the $s$- and $u$-channels is $(m_\pi+m_\eta)^2$. In 
the $t$-channel the threshold is $4m_\pi^2$. I will do fixed-$t$ and fixed-$u$ 
dispersion relations and finite energy sum rules. Fixed-$s$ sum rules are 
identical to fixed-$u$ sum rules. The crossing variables are $\nu= (s-u)/2$ and
$\nu' = (s-t+4m_\pi^2-(m_\eta+m_\pi)^2)/2$. $\nu'$ is designed to symmetrize the 
two cuts at fixed $u$, {\it i.e.,} $\nu'(s= (m_\eta+m_\pi)^2) = - \nu'(t = 
4m_\pi^2)$. That will allow to place both cuts under the same integral in the 
fixed-$u$ sum rules.  I will need to trade the Mandelstam variables for the pair 
$(\nu,t)$ and $(\nu',u)$ {\it via}
\begin{subequations}
\begin{align}
	s(\nu,t) & = +\nu -(t - \Sigma)/2, & s(\nu',u) & = +\nu' -(u - 
	4m_\eta^2+(m_\eta-m_\pi)^2)/2, \\
	u(\nu,t) & = -\nu -(t - \Sigma)/2, & t(\nu',u) & = -\nu' -(u - 
	4m_\pi^2-(m_\eta-m_\pi)^2)/2,
\end{align}
\end{subequations}
so that $s(-\nu,t) = u(\nu,t)$.

%%%-----------------------------------
\item \textbf{Finite Energy Sum Rules}

%%%-----------------------------------
\begin{figure}[htb]
\begin{center}
\includegraphics[width=.33\linewidth]{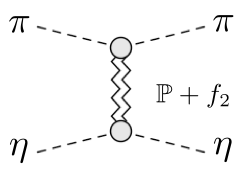}
\includegraphics[width=.33\linewidth]{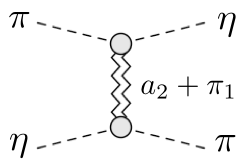}
\end{center}

\caption{\label{fig:2} Exchanges in forward (left) and backward (right) 
	$\pi\eta$ elastic scattering. }
\end{figure}
%%%-----------------------------------
I consider small $t=x$, {\it i.e.,} small angles with the pion going forward. 
The two exchanges at high energy are the Pomeron and the $f_2$ Regge pole. At high 
energy, the Regge form is $\im A(s,t,u) = \beta_{\mathbb P}(t) 
\nu^{\alpha_{\mathbb P}} + \beta_{f_2}(t)\nu^{\alpha_{f_2}}$ for $\nu>\Lambda$. The 
threshold is $\nu_0(x) = 2 m_\eta m_\pi + x/2$. The FESR at fixed $t$ reads
\begin{align} \nonumber
	\int_{\nu_0(t)}^\Lambda \im \left\{A\left[s(\nu,x),x,u(\nu,x)\right] + (-)^k 
	A\left[s(-\nu,x),
	x,u(-\nu,x)\right]  \right\} \nu^k d\nu \\  = \sum_\tau
	\left[1+ \tau(-)^k \right] \beta_\tau(t) 
	\frac{\Lambda^{\alpha_\tau(t)+k+1}}{\alpha_\tau(t)+k+1}
\label{eq:fesr1}
\end{align}
The $s$- and $u$-channel are identical, so $A(s,x,u) =  A(u,x,s) $ and there can 
only be even moments. This is consistent with the exchanges that can only have 
positive signature $\tau(\mathbb P) = \tau (f_2) = +1$. I obtain, with $n$ a 
positive integer
\begin{align}
	\int_{\nu_0(x)}^{\Lambda} \im A\left[s(\nu,x),x,u(\nu,x)\right] \nu^{2n} d\nu & 
	=
	\beta_{\mathbb P}(x) \frac{\Lambda^{\alpha_{\mathbb P}(x)+2n+1}}{\alpha_{\mathbb 
	P}(x)+2n+1} +
	\beta_f(x) \frac{\Lambda^{\alpha_{f_2}(x)+2n+1}}{\alpha_{f_2}(x)+2n+1}
\end{align}
The factorization of Regge pole allow write $\beta_{\mathbb P}(x) 
=\beta^{\pi\pi}_{\mathbb P}(x) \beta^{\eta\eta}_{\mathbb P}(x) $ and
$\beta_f(x) \equiv \beta^{\pi\pi}_{f_2}(x) \beta^{\eta\eta}_{f_2}(x)$, as 
represented in Fig.~\ref{fig:2}.

\medskip

I now consider small $u=x$, {\it i.e.,} small angles with the eta meson going 
forward. The threshold is $\nu'_{0}(x) = (4m_\pi^2-(m_\eta-m_\pi)^2 + x)/2$. 
Similarly to the fixed-$t$ case, the FESR at fixed $u$ read
\begin{align} \nonumber
	\int_{\nu'_{0}(x)}^{\bar \Lambda} \im \left\{ 
	A\left[s(\nu',x),t(\nu',x),x\right]+ (-)^k A\left[s(-\nu',x),t(-\nu',x),x\right]  \right\} \nu'^k d\nu' \\
	= \sum_\tau
	\left[1+ \tau(-)^k \right] \beta_\tau(x) \frac{\bar 
	\Lambda^{\alpha_\tau(x)+k+1}}{\alpha_\tau(x)+k+1}
\end{align}
There are two kinds of exchanges. The positive signature $a_2$ pole and the 
negative signature $\pi_1$ pole. So for $k$ even there is the $a_2$ pole and $k$ 
odd the $\pi_1$ (exotic) pole. I thus obtain two backward FESR:
\begin{align} \nonumber
	\frac{1}{2}\int_{\nu'_{0}(x) }^{\bar\Lambda} \im \left\{ A\left[s(\nu',x),t(\nu',x),x\right]
	+ A\left[s(-\nu',x),t(-\nu',x),x\right]  \right\} \nu'^{2n} d\nu'  &=
	\frac{\beta_{a_2}(x)\bar\Lambda^{\alpha_{a_2}(x)+2n+1}}{\alpha_{a_2}(x)+2n+1}\\
	\frac{1}{2}\int_{\nu'_{0}(x) }^{\bar\Lambda} \im \left\{ A\left[s(\nu',x),t(\nu',x),x\right]
	- A\left[s(-\nu',x),t(-\nu',x),x\right]  \right\}\nu'^{2n+1} d\nu' & =
	\frac{\beta_{\pi_1}(x) \bar\Lambda^{\alpha_{\pi_1}(x)+2n+2}}{\alpha_{\pi_1}(x)+2n+2}
	\label{eq:fesr2}
\end{align}
Again, the factorization of Regge pole leads to $\beta_{a_2}(x)= \left[  
\beta^{\pi\eta}_{a_2}(x) \right]^2$ and $ \beta_{\pi_1}(x) = \left[ 
\beta^{\pi\eta}_{\pi_1}(x) \right]^2 $ as displayed in Fig.~\ref{fig:2}.
For small $x$, the amplitude $A\left[s(\nu',x),t(\nu',x),x\right]$ represents 
the reaction $\pi\eta\to\pi\eta$ in the backward direction and 
$A\left[s(-\nu',x),t(-\nu',x),x\right]$ represents the $t$-channel reaction 
$\pi\pi \to \eta \eta$.

%%%-----------------------------------
\item \textbf{Constraint on Exotica Production}

Several models are available for the elastic $\pi\eta$ 
$S$-wave~\cite{Guo:2012ymn,Guo:2012ytn,Albaladejo:2015acan,Guo:2016zepn}. For 
higher waves, little information is known beside the resonance content. The 
elastic $\pi\eta$ $D$-wave is certainly dominated by the $a_2(1320)$ meson. At 
higher mass, the excitation $a_2'(1700)$ has been recently confirm by a joint
publication by the JPAC and COMPASS collaborations~\cite{Jackura:2017ambn}, but 
its branching fraction to $\pi\eta$ is not known. The branching ratio of the 
$a_4(2040)$ is not known either but quark model calculations yields the width 
$\Gamma(a_4\to \pi\eta) \sim (1.7-2.4)^2$~MeV~\cite{Godfrey:1985xjn,Kokoski:1985isn}.
\footnote{The $a_4$ was denoted $\delta$ in these publications.}

Experimental information on $\pi\pi\to \eta\eta$ are available above the 
$\eta\eta$ threshold~\cite{Binon:1983ny}. Below the physical threshold, one can 
use Watson's theorem and the $\pi\pi$ elastic phase shift from the unitarity 
threshold up to the $K \bar K$ threshold. Unfortunately between the $K \bar K$ 
and the $\eta\eta$ thresholds, little information is known about the phase shift 
and inelasticities of the $\pi\pi\to\eta\eta$ amplitude. Although some couple 
channels calculations exist, see for instance~\cite{Albaladejo:2008qan}.

The $a_2$, $f_2$ are lying on the trajectory degenerate with the $\rho$ and 
$\omega$ one with $\alpha_{a_2}(t) = \alpha_{f_2}(t) \equiv \alpha_N(t) \sim 0.9 
t  + 0.5$. The  exotic $\pi_1$ is lying on a trajectory below the leading 
natural exchange trajectory. A mass around 1.6~GeV yields the intercept 
$\alpha_{\pi_1}(0) \sim \alpha_N(t)-1$, which makes both right-hand-side
of Eq.~(6) for the same $n$ of the same order up to the residues $\beta_{a_2}$ 
and $\beta_{\pi_1}$. The existence of an exotic would make $\beta_{\pi_1}$ not 
zero. However since its pole would be far away from the physical region, there 
is a possibility of $\beta_{\pi_1}(x)$ (with $x<0$) being small. In this case, 
one could neglect the right-hand-side of the second sum rule in  
Eq.~\eqref{eq:fesr2} for any $n$, which would lead to
\begin{align}
%	\im  A\left[s(\nu',x),t(\nu',x),x\right] \approx \im 
%	A\left[t(\nu',x),s(\nu',x),x\right]
	\im  A\left[s(\nu',x),t(\nu',x),x\right] \approx \im A\left[s(-\nu',x),t(-\nu',x),x\right]
\end{align}
Using this approximation, I can subtract the forward $t=x$ and backward $u=x$ 
sum rules to obtain the constraint
\begin{align} \nonumber
	\int_{\nu_0(x)}^{\Lambda} \im \left\{
	A\left[s(\nu,x),x,u(\nu,x)\right] - A\left[s(\nu,x),t(\nu,x),x\right]
	\right\} \nu^{2n} \diff \nu  = \\
	\beta_{\mathbb P}(x) \frac{\Lambda^{\alpha_{\mathbb P}(x)+2n+1}}{\alpha_{\mathbb 
	P}(x)+2n+1} + \left[
	\beta_{f_2}(x)-  \beta_{a_2}(x) 
	\right]\frac{\Lambda^{\alpha_N(x)+2n+1}}{\alpha_N(x)+2n+1}
\label{eq:constraint1}
\end{align}
In term of partial waves, the forward and backward amplitudes read
\begin{subequations}
\begin{align}
	A\left(s,x,u\right) & =\sum_\ell (2\ell+1) t_\ell(s) P_\ell(z_1), &
	z_1 & = +1+ \frac{x}{ 2q^2}, \\
	A\left(s,t,x\right) & = \sum_\ell (2\ell+1) t_\ell(s) P_\ell(z_2), &
	z_2 & = -1- \frac{x-u_0}{ 2q^2},
\end{align}
\end{subequations}
with $q$ and $u_0$ evaluated at $s(\nu,x)$. If the forward and backward 
direction were exactly equal, {\it i.e.,} $u_0 =0$ the even waves would exactly 
cancel in Eq.~\eqref{eq:constraint1}. The reader can then see that the two 
components duality\footnote{In the two component duality hypothesis, one 
separate the background dual to the Pomeron and the resonance dual to the Regge 
exchange~\cite{Harari:1968jwn}.} and the absence of odd wave would then imply 
degeneracy between the $f_2$ and $a_2$ Regge exchanges as demonstrated by 
Schwimmer~\cite{Schwimmer:1970ykn}. In the presence of only one 
odd wave $\ell=1$, its absorptive part is constrained by
\begin{align} \nonumber
	3\int_{\nu_0(x)}^{\Lambda} \im t_1\left[ s(\nu,x) \right] \left[
	P_1(z_1) - P_1(z_2) \right] \nu^{2n} d\nu & =
	\beta_{\mathbb P}(x) \frac{\Lambda^{\alpha_{\mathbb P}(x)+2n+1}}{\alpha_{\mathbb 
	P}(x)+2n+1} \\ \nonumber
	&+ \left[
	\beta_{f_2}(x)-  \beta_{a_2}(x) 
	\right]\frac{\Lambda^{\alpha_N(x)+2n+1}}{\alpha_N(x)+2n+1} \\
	&- \sum_{\ell \in \text{even}} \int_{\nu_0(x)}^\Lambda \im t_\ell(s_\nu) 
	\Delta_\ell(\nu,x) \nu^{2n} d\nu
	\label{eq:last}
\end{align}
The even waves enter {\it via} the quantity $\Delta_\ell(\nu,x) =(2\ell+1) 
\left[ P_\ell(z_1) - P_\ell(z_2) \right]$. Note that $\Delta_\ell\propto u_0 $ 
(for $\ell$ even) vanishes in the limit $m_\eta \to m_\pi$. Of course the 
$S$-wave cancels $\Delta_0 =0$. We also have $\frac{1}{2} \left[P_1(z_1) - 
P_1(z_2)\right] = 1+ x/2 q^2 - u_0/4q^2$. The individual contribution of the 
right-hand-side of Eq.~\eqref{eq:last} can be evaluated.

%%%-----------------------------------
\item \textbf{Acknowledgments}

I thank M. Albaladejo, A. Jackura, and J. Ruiz de Elvira for providing their 
elastic $\pi\eta$ solutions. This work was supported by the U.S. Department of 
Energy under grants No. DE-AC05-06OR23177
\end{enumerate}

%%%-----------------------------------

%%%%%%%%%%%%%%%%%%%%%%%%%%%%%%%%%%%%%%%%%%%%%%%%%%%%%%%%%%%%%%%%%%%%%%%%%
\newpage
\subsection{Pion--Kaon Final-State Interactions in Heavy-Meson Decays}
\addtocontents{toc}{\hspace{2cm}{\sl Bastian Kubis and Franz Niecknig}\par}
\setcounter{figure}{0}
\setcounter{table}{0}
\setcounter{equation}{0}
\setcounter{footnote}{0}
\halign{#\hfil&\quad#\hfil\cr
\large{Bastian Kubis and Franz Niecknig}\cr
\textit{Helmholtz-Institut f\"ur Strahlen- und Kernphysik (Theorie) and}\cr
\textit{Bethe Center for Theoretical Physics}\cr
\textit{Universit\"at Bonn}\cr
\textit{53115 Bonn, Germany}\cr}

%%%-----------------------------------
\begin{abstract}
We discuss the description of final-state interactions in three-body hadronic decays 
based on Khuri--Treiman equations, in particular their application to the 
charmed-meson decays $D^+ \to \bar{K}\pi\pi^+$.  We point out that the knowledge of 
pion--pion and pion--kaon scattering phase shifts is of prime importance in this 
context, and that there is no straightforward application of Watson's theorem in the 
context of three-hadron final states.
\end{abstract}

%%%-----------------------------------
\begin{enumerate}
\item \textbf{What's not to Like about the Isobar Model?}

In experimental analyses, the Dalitz-plot distributions of hadronic three-body decays 
are still conventionally described in terms of the \textit{isobar model}: amplitudes 
are constructed in terms of subsequent two-body decays, and the resonant intermediate 
states mostly modeled in terms of Breit--Wigner line shapes.  This is problematic in 
several respects.  First of all, many resonances cannot be described by Breit--Wigner 
functions at all, be it due to the close proximity of thresholds (e.g.,\ the $f_0(980)$ 
or the $a_0(980)$ among the light mesons, and many of the newly found, potentially 
exotic states in the charmonium and bottomonium sectors), or because their poles lie 
too far in the complex plane: prime examples include the lightest scalar resonances,
the $f_0(500)$ or $\sigma$ (see the comprehensive review~\cite{Pelaez:2015qbai} and 
references therein) as well as, in the strange sector, the 
$K_0^\ast(800)$~\cite{DescotesGenon:2006uki,Pelaez:2016klvi}. These problems can be 
avoided by using known phase shifts as input, potentially including coupled channels;
see, e.g.,\ Refs.~\cite{Daub:2015xjai,Albaladejo:2016madi} for recent applications in 
the context of heavy-meson decays.  On the other hand, models of subsequent two-body 
decays ignore rescattering effects between all three strongly interacting particles in 
the final state, which will in general affect the phase motion of the amplitudes in 
question.  Close control over the various amplitude phases is, e.g.,\ important in the 
context of CP-violation studies: rapidly varying, resonant strong phases may 
significantly enhance small weak phases (originating in the Cabibbo--Kobayashi--Maskawa 
matrix) locally in the Dalitz plot, which may give the search for CP violation in 
three-body charmed-meson decays the edge over two-body decays, which occur at fixed 
energy.

A tool to consistently treat full iterated two-body final-state interactions
in three-body decays are the so-called Khuri--Treiman equations~\cite{Khuri:1960zzi}, 
originally derived for the description of $K\to3\pi$, and recently increasingly 
popular for the description of various low-energy decays such as
$\eta\to3\pi$~\cite{Anisovich:1996txi,Kambor:1995yci,Guo:2015zqai,Guo:2016wsii,
Colangelo:2016jmci,Albaladejo:2017hhji},
$\eta'\to\eta\pi\pi$~\cite{Isken:2017dkwi}, or 
$\omega/\phi\to3\pi$~\cite{Niecknig:2012sji,Danilkin:2014crai}.
We will illustrate the formalism with the last example in the following, before 
turning to the application in the more complex Dalitz plot studies of 
$D^+\to\bar{K}\pi\pi^+$~\cite{Niecknig:2015ijai,Niecknig:2017ylbi}.

%%%-----------------------------------
\item \textbf{Dispersion Relations for Three-Body Decays (1): \boldmath{$V\to 3\pi$}}

Particularly simple examples of three-body decays are those of isoscalar vector mesons 
into three pions, $V\to 3\pi$, $V=\omega,\,\phi$.  Pairs of pions can only be in odd 
relative partial waves; neglecting discontinuities in $F$- and higher partial waves, 
the decay amplitude can be decomposed into single-variable functions according to
\begin{equation}
	\mathcal{M}(s,t,u) = i\epsilon_{\mu\nu\alpha\beta} \,\epsilon_V^\mu p_{\pi^+}^\nu p_{\pi^-}^\alpha 
	p_{\pi^0}^\beta
	\big\{ \mathcal{F}(s) + \mathcal{F}(t) + \mathcal{F}(u) \big\},
\end{equation}
where $\epsilon_V^\mu$ denotes the vector meson's polarization vector.  The function 
$\mathcal{F}(s)$ obeys a discontinuity equation~\cite{Niecknig:2012sji,Hoferichter:2012pmi}
\begin{align}
	\mathrm{disc}\,\mathcal{F}(s) &= 2i\big[ \mathcal{F}(s) + \hat{\mathcal{F}}(s) \big] 
	\theta\big(s-4M_\pi^2\big)
	\sin\delta(s) e^{-\delta(s)}, \notag\\
	\hat{\mathcal{F}}(s) &= \frac{3}{2} \int_{-1}^1 \mathrm{d}z \big(1-z^2\big) \mathcal{F}\big(t(s,z)\big),
\label{eq:disc}
\end{align}
where $z = \cos\theta_s$ denotes the cosine of the $s$-channel scattering angle, 
and $\delta(s) \equiv \delta_1^1(s)$ is the isospin $I=1$ $P$-wave $\pi\pi$ 
scattering phase shift.  Without the inhomogeneity $\hat{\mathcal{F}}(s)$ in the 
discontinuity equation, the latter's solution could be given in terms of an Omn\`es 
function~\cite{Omnes:1958hvi} like a simple form factor, and the phase of 
$\mathcal{F}(s)$ would immediately coincide with $\delta(s)$ according to Watson's 
theorem~\cite{Watson:1954uci}.
%%%-----------------------------------
\begin{figure}[b!]
\includegraphics[width=\linewidth]{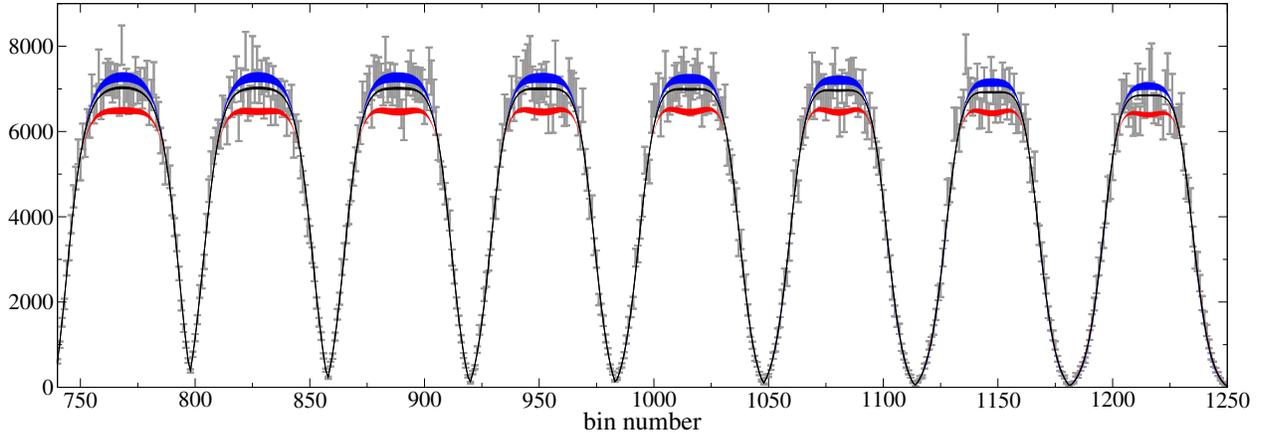}

\caption{Selected slices through the $\phi\to3\pi$ Dalitz plot measured by the KLOE 
	Collaboration~\protect\cite{Aloisio:2003uri}, compared to the theoretical 
	description of Ref.~\protect\cite{Niecknig:2012sji}. The fit with Omn\`es 
	functions only, neglecting crossed-channel rescattering, is shown in red, 
	once-subtracted  Khuri--Treiman amplitudes in blue, and twice-subtracted
	Khuri--Treiman solutions in black.  The widths of the uncertainty bands are 
	due to variation of the phase shift input as well as fit uncertainties. 
	\label{fig:phi3piDP}}
\end{figure}
%%%-----------------------------------

The presence of the partial-wave-projected crossed-channel contributions 
$\hat{\mathcal{F}}(s)$ renders the solution of Eq.~\eqref{eq:disc} slightly more 
complicated; it is given as
\begin{equation}
	\mathcal{F}(s) = \Omega(s)\biggl\{a + 
	\frac{s}{\pi}\int_{4M_\pi^2}^{\infty}\frac{\mathrm{d}x}{x}\frac{\sin\delta(x)
	\hat{\mathcal{F}}(x)}{|\Omega(x)|(x-s)}\biggr\} , \quad
	\Omega(s) = \exp\bigg\{ \frac{s}{\pi} \int_{4M_\pi^2}^{\infty}\frac{\mathrm{d}x}{x} \frac{\delta(x)}{x-s} \bigg\} ,
	\label{eq:inhomOmnes}
\end{equation}
where $\Omega(s)$ is the Omn\`es function, and $a$ a subtraction constant that 
has to be fixed phenomenologically.  Care has to be taken in performing the angular 
integral for $\hat{\mathcal{F}}(s)$ in Eq.~\eqref{eq:disc} as to avoid crossing the 
right-hand cut: in decay kinematics, $\hat{\mathcal{F}}(s)$ itself becomes complex,
signaling the appearance of three-particle cuts, and a simple phase relation between 
$\mathcal{F}(s)$ and the elastic scattering phase shift is lost.

The solution to Eq.~\eqref{eq:inhomOmnes} is compared to high-statistics Dalitz plot 
data for $\phi\to3\pi$ by the KLOE collaboration~\cite{Aloisio:2003uri} in 
Fig.~\ref{fig:phi3piDP}.  Different $\pi\pi$ phases have been employed, derived from 
Roy (and similar) equations~\cite{GarciaMartin:2011cni,Caprini:2011kyi}.  Non-trivial, 
crossed-channel rescattering effects improve the data description significantly; 
introducing a second subtraction constant in order to better suppress imperfectly 
known high-energy behavior of the dispersion integral leads to a perfect 
fit~\cite{Niecknig:2012sji}.  A first modern experimental investigation of the 
$\omega\to3\pi$ Dalitz plot by WASA-at-COSY~\cite{Adlarson:2016wkwi} is not yet 
accurate enough to discriminate these subtle rescattering effects.
%%%-----------------------------------
\begin{figure}
\includegraphics[width=0.49\linewidth]{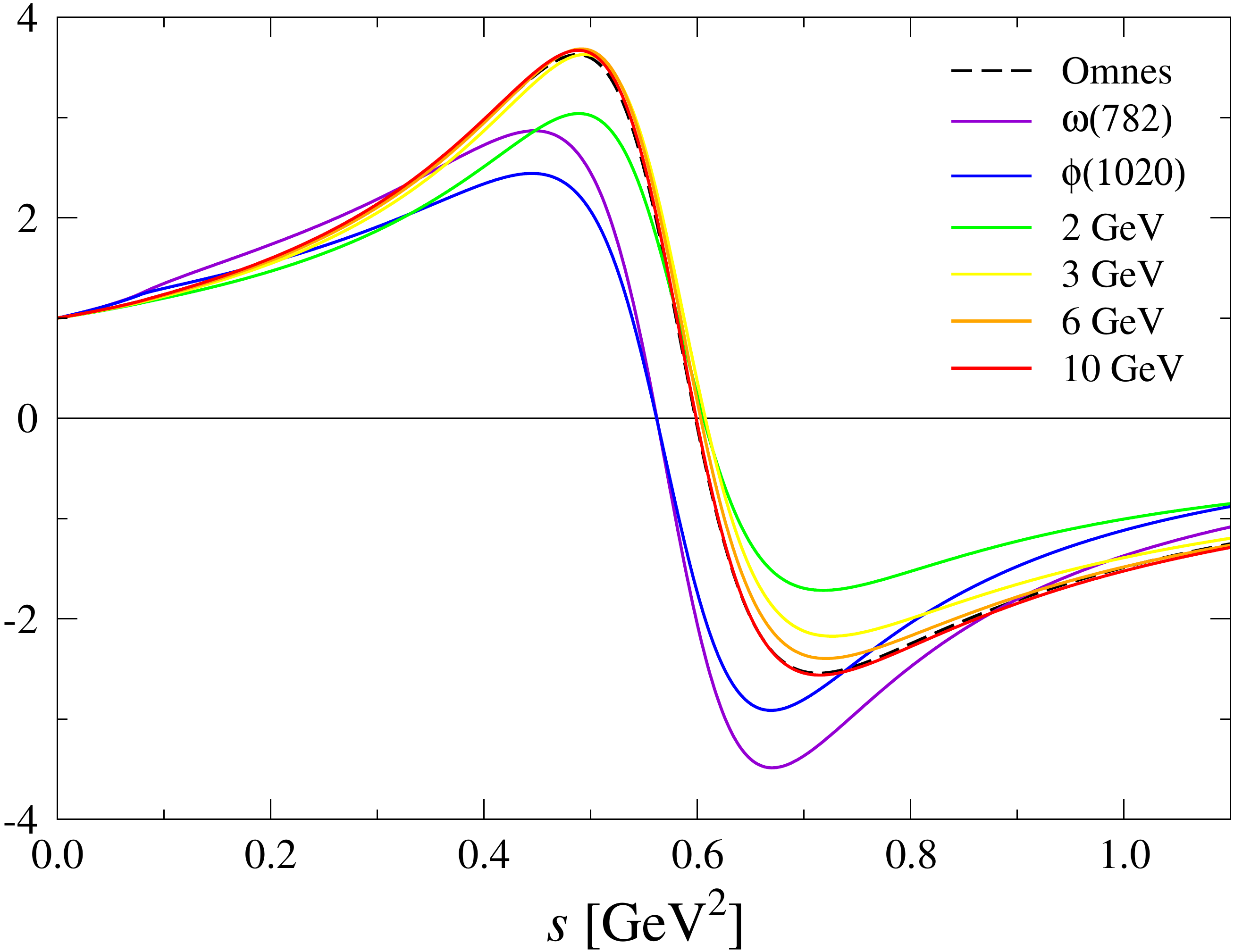} \hfill
\includegraphics[width=0.49\linewidth]{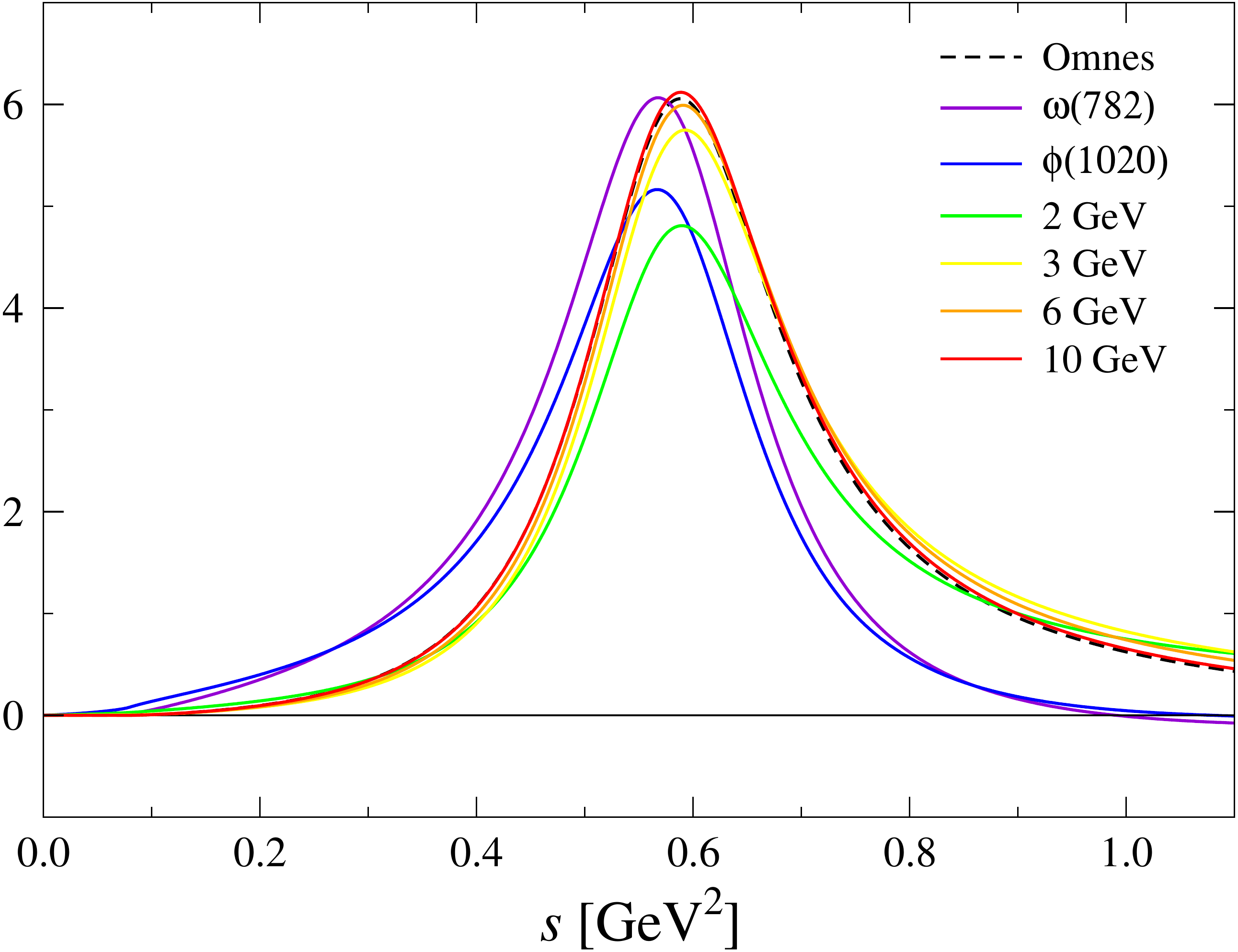}

\caption{Real (left) and imaginary (right) parts of single-variable 
	amplitudes $\mathcal{F}(s)$ for varying decay masses, compared 
	to the Omn\`es function. \label{fig:V3pi}}
\end{figure}
%%%-----------------------------------

It is also interesting to investigate the dependence of $\mathcal{F}(s)$ on the 
decay mass: it differs for $\omega$ and $\phi$ 
decays~\cite{Niecknig:2012sji,Danilkin:2014crai}, but even has a well-defined 
high-energy limit, as shown in Fig.~\ref{fig:V3pi}.  For high decay masses, 
$\mathcal{F}(s)$ approaches the input Omn\`es function, i.e.,\ crossed-channel
rescattering effects vanish~\cite{Niecknig:2016fvai}.  This conforms with physical 
intuition: if the third pion has a very large momentum relative to the other pair, 
its influence on the pairwise, resonant interaction should be small.  Note that 
this limit is only formal, as inelastic effects as well as higher partial waves
are not taken into account.  Such a partial wave for variable and higher decay 
masses has been employed both for the description of the reaction $e^+e^-\to 3\pi$, 
which serves as an input for a dispersive analysis of the $\pi^0$ transition form 
factor~\cite{Hoferichter:2014vrai}, and in a study of the $J/\psi\to\pi^0\gamma^\ast$ 
transition form factor~\cite{Kubis:2014gkai}, in analogy to preceding studies
of the conversion decays of the light vector mesons~\cite{Schneider:2012ezi}.

%%%-----------------------------------
\item \textbf{Dispersion Relations for Three-Body Decays (2): \boldmath{$D^+ \to 
	\bar{K}\pi\pi^+$}}

\begin{sloppypar}
In the first attempt to employ Khuri--Treiman equations to describe $D$-meson decays, 
we have chosen to consider the Cabibbo-favored processes $D^+\to K^-\pi^+\pi^+$ and 
$D^+\to \bar{K}^0\pi^0\pi^+$.  The data situation for these is rather good, with 
high-statistics Dalitz plot measurements available by the E791~\cite{Aitala:2005yhi}, 
CLEO~\cite{Bonvicini:2008jwi}, and FOCUS~\cite{Pennington:2007sei,Link:2009ngi}
Collaborations for the fully charged final state, and more recently by BESIII for 
the partially neutral one~\cite{Ablikim:2014ceai}.  Theoretical analyses of $D^+\to 
K^-\pi^+\pi^+$ have typically concentrated on improved descriptions of the $\pi K$ 
$S$-wave~\cite{Oller:2004xmi,Boito:2009qdi,Magalhaes:2011shi,Magalhaes:2015fvai,Guimaraes:2014ccai}, 
however neglecting rescattering with the third decay particle. Ref.~\cite{Nakamura:2015qgai} 
is the only previous combined study of both decay channels, using Faddeev equations 
to generate three-body rescattering effects.
\end{sloppypar}

An interesting aspect of these two decay channels is that they are coupled by a simple 
charge-exchange reaction, and can be related to each other by isospin; however, this 
relation is largely lost as long as only two-body rescattering is considered: e.g., the 
isospin $I=1$ $\pi\pi$ $P$-wave only features indirectly in the fully charged final 
state, where the $\rho(770)$ resonance can obviously not be observed.  Otherwise, we 
truncate the partial-wave expansion consistently beyond $D$-waves, but among the exotic, 
non-resonant ones only retain the $S$-waves (the $I=2$ $\pi\pi$ and the $I=3/2$ $\pi K$ 
$S$-waves).  This way, beyond the aforementioned $\rho(770)$, the following $\pi K$ 
resonances are included in the Dalitz plot description as the dominant structures:
$K_0^\ast(800)$, $K_0^\ast(1430)$, $K^\ast(892)$, [$K^\ast(1410)$,] and $K_2^\ast(1430)$.

The decomposition of the two decay amplitudes up-to-and-including $D$-waves has been 
derived~\cite{Niecknig:2015ijai} and proven in the sense of the \textit{reconstruction 
theorem}~\cite{Niecknig:2016fvai}.  With the exception of the $D$-wave, the number of 
subtractions has been determined by imposing the Froissart bound~\cite{Froissart:1961uxi};
note that some of them can be eliminated consistently due to the constraint 
$s+t+u=\text{const.}$.
The full system then reads
\begin{align}
	\mathcal{F}_0^2(u)&=\Omega^2_0(u)\frac{u^2}{\pi}\int_{u_\mathrm{th}}^\infty \frac{\mathrm{d} u'}{u'^2}
 	\frac{\hat{\mathcal{F}}_0^2(u')\sin\delta^2_0(u')}{\left|\Omega^2_0(u')\right|(u'-u)} ,\notag\\
	\mathcal{F}_1^1(u)&=\Omega^1_1(u)\Bigg\{c_0+c_1u+\frac{u^2}{\pi}\int_{u_\mathrm{th}}^\infty\frac{\mathrm{d} 
	u'}{u'^2}
 	\frac{\hat{\mathcal{F}}_1^1(u')\sin\delta^1_1(u')}{\big|\Omega^1_1(u')\big|(u'-u)}\Bigg\},\notag\\
	\mathcal{F}_0^{1/2}(s)&=\Omega^{1/2}_0(s)\Bigg\{c_2+c_3s+c_4s^2+c_5s^3
	+\frac{s^4}{\pi}\int_{s_\mathrm{th}}^\infty\frac{\mathrm{d} s'}{s'^4}
	\frac{\hat{\mathcal{F}}_0^{1/2}(s')\sin\delta^{1/2}_0(s')}{\big|\Omega^{1/2}_0(s')\big|(s'-s)}\Bigg\},\notag\\
	\mathcal{F}_0^{3/2}(s)&=\Omega^{3/2}_0(s)\Bigg\{\frac{s^2}{\pi}\int_{s_\mathrm{th}}^\infty\frac{\mathrm{d} s'}{s'^2}
	\frac{\hat{\mathcal{F}}_0^{3/2}(s')\sin\delta^{3/2}_0(s')}{\big|\Omega^{3/2}_0(s')\big|(s'-s)}\Bigg\},\notag\\
	\mathcal{F}_1^{1/2}(s)&=\Omega^{1/2}_1(s)\Bigg\{c_6+\frac{s}{\pi}\int_{s_\mathrm{th}}^\infty\frac{\mathrm{d} s'}{s'}
\frac{\hat{\mathcal{F}}_1^{1/2}(s')\sin\delta^{1/2}_1(s')}{\big|\Omega^{1/2}_1(s')\big|(s'-s)}\Bigg\},\notag\\
 	\mathcal{F}_2^{1/2}(s)&=\Omega^{1/2}_2(s) \Bigg\{c_7 + 
	\frac{s}{\pi}\int_{s_\mathrm{th}}^\infty\frac{\mathrm{d} s'}{s'}
 	\frac{\hat{\mathcal{F}}_2^{1/2}(s')\sin\delta^{1/2}_2(s')}{\big|\Omega^{1/2}_2(s')\big|(s'-s)}\Bigg\} ,\label{eq:fulleq}
\end{align}
where $s$ (and $t$) are the Mandelstam variables describing $\pi K$ invariant masses, 
while $u$ refers to the $\pi\pi$ system; the lower limits of the dispersion integrals 
are given by $s_\mathrm{th}=(M_K+M_\pi)^2$ and $u_\mathrm{th}=4M_\pi^2$. Explicit 
formulae for the various $\hat{\mathcal{F}}_L^I$ are given in 
Ref.~\cite{Niecknig:2015ijai}.  We have used the $\pi K$ phase shifts of 
Ref.~\cite{Buettiker:2003ppi} as input (note also other new~\cite{Pelaez:2016tgii} and 
ongoing~\cite{Colangelo:201xpiKi} analyses of $\pi K$ scattering).  Eq.~\eqref{eq:fulleq} 
therefore contains 8 complex subtractions constants, which (subtracting one overall 
normalization and one overall phase) need to be fitted to data. In particular, the 
subtraction in $\mathcal{F}_2^{1/2}(s)$ was introduced \textit{a posteriori}, as it 
turned out to be necessary to describe both Dalitz plots consistently~\cite{Niecknig:2017ylbi}.
As we work in the approximation of \textit{elastic} unitarity, we refrain from 
describing the parts of the Dalitz plot for which $\sqrt{s},\,\sqrt{t} \geq M_{\eta'}+M_K 
\approx 1.45\,\text{GeV}$, which is taken as the typical onset of significant 
inelasticities in particular in the $S$-wave.  Note that no isoscalar $\pi\pi$ $S$-wave 
can contribute in these decays, with its sharp onset of inelasticities around the 
$f_0(980)$.
%%%-----------------------------------
\begin{figure}
\centering
\includegraphics[width=0.9\linewidth]{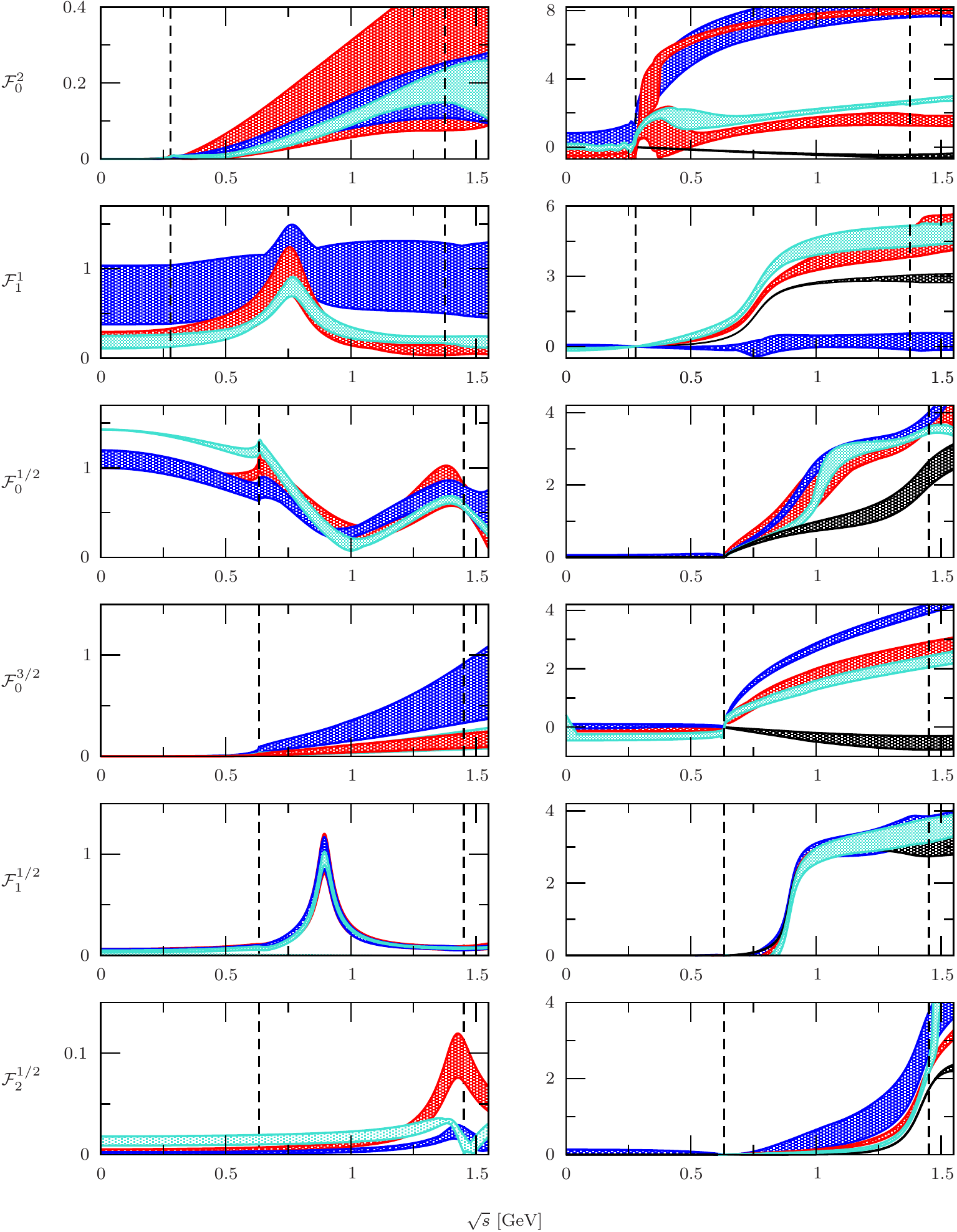}

\caption{\textit{Left:} Moduli of single-variable amplitudes, in arbitrary units: 
	CLEO/FOCUS fits (blue), BESIII (red), combined fit with improved $D$-wave 
	(turquoise).
	\textit{Right:} Phases of the single-variable amplitudes and input scattering 
	phases (black) in radiant. The dashed lines visualize the fitted area:  from 
	threshold to the $\eta' K$ threshold for the $\pi K$ amplitudes, the full 
	phase space for the $\pi\pi$ amplitudes. Figure taken from 
	Ref.~\protect\cite{Niecknig:2017ylbi}. \label{fig:DKpipi}}
\end{figure}
%%%-----------------------------------

Moduli and phases of the resulting single-variable amplitudes $\mathcal{F}_L^I$ are 
shown in Fig.~\ref{fig:DKpipi}. Obviously, the different data sets constrain the 
various amplitudes differently: the $\pi\pi$ $P$-wave is only well constrained when 
including the BESIII data on the partially neutral final state (in which it features 
directly), while in contrast, the $\pi\pi$ $I=2$ $S$-wave is determined with much 
higher precision in the $K^-\pi^+\pi^+$ final state.
%%%-----------------------------------
\begin{table}
 \centering
 \renewcommand{\arraystretch}{1.4}
 \begin{tabular}{c c c}
 \hline
  {}            &{BESIII$_\text{combined}$} &    {CLEO/FOCUS$_\text{combined}$} \\
 \hline
 $\text{FF}_0^2$                &$(1.7\pm 0.5)\%$       & $(5\pm 1)\%$          \\
 $\text{FF}_1^1$                &$(23\pm 3)\%$          &---                    \\
 $\text{FF}_0^{1/2}$            &$(36\pm 5)\%$          &$(46\pm 6)\%$          \\
 $\text{FF}_1^{1/2}$            &$(8.5\pm 0.4)\%$       &$(11.5\pm 0.5)\%$      \\
 $\text{FF}_0^{3/2}$            &$(6\pm 1)\%$           &$(0.6\pm 0.1)\%$       \\
 $\text{FF}_2^{1/2}$            &$(0.5\pm 0.1)\%$       &$(0.7\pm 0.1)\%$       \\
 \hline
 \end{tabular}
 \renewcommand{\arraystretch}{1.0}
 \caption{Fit fractions of the various partial waves for the best combined fit.
 The errors are evaluated by varying the basis functions within their uncertainty bands.
 The fit fractions of the $\pi K$ amplitudes in the $s$- and $t$-channel are summed 
 together.}
 \label{tab:tabFitfrac_Dwave}
\end{table}
%%%-----------------------------------

Given the relatively large number of subtraction constants compared to the $\phi\to3\pi$ 
analysis described in the previous section, the necessity to include Khuri--Treiman 
amplitudes as opposed to simple polynomial-times-Omn\`es functions is demonstrated less 
by the best $\chi^2/\text{d.o.f.}$, but rather by the resulting fit fractions, which 
become implausibly large in particular for the nonresonant waves in the case of Omn\`es 
fits~\cite{Niecknig:2015ijai}.  Similar observations are made when neglecting the $D$-wave.  
The fit fractions for the best combined fit to the CLEO, FOCUS, and BESIII data are
shown in Table~\ref{tab:tabFitfrac_Dwave}.  The total fit quality characterized by 
$\chi^2/\text{d.o.f.} \approx 1.2$ is at best satisfactory, but of similar quality as 
isobar fits performed by the experimental 
collaborations~\cite{Pennington:2007sei,Link:2009ngi,Ablikim:2014ceai}. We point out that 
three-body rescattering effects change the \textit{phases} of the amplitudes quite 
significantly in comparison to the input phase shifts, see Fig.~\ref{fig:DKpipi}:
in particular the $I=1/2$ $\pi K$ $S$-wave phase is seen to rise much
more quickly around $\sqrt{s} \approx 1\,\text{GeV}$.

%%%-----------------------------------
\item \textbf{Summary / Outlook}

We have demonstrated that dispersion relations constitute an ideal tool to analyze the 
final-state interactions of pions and kaons systematically.  While two-body form factors 
obey universal phase relations, in three-body decays non-trivial rescattering effects 
can affect phase motions and line shapes significantly. This was demonstrated succinctly 
for the ideal demonstration case $\phi\to3\pi$ that can be described in terms of one 
single partial wave only; an analysis of two coupled $D^+\to \bar{K}\pi\pi^+$ decay 
modes proved to be far more involved due to the proliferation of subtraction constants.

Future work could improve on the treatment of inelastic rescattering effects by using 
coupled channels, not the least in order to extend the amplitude description to the full 
$D^+\to \bar{K}\pi\pi^+$ Dalitz plots. The consistent inclusion of higher partial waves 
in Khuri--Treiman equations still requires further investigations, as does the role of 
three-body unitarity in constraining the phases of the subtraction constants.  Finally,
it should be attempted to match the subtraction constants to short-distance information 
on the weak transitions involved.

%%%-----------------------------------
\item \textbf{Acknowledgments}

B.K.\ thanks the organizers for the invitation to and support at this most 
enjoyable workshop.  Financial support by DFG and NSFC through funds provided to 
the Sino--German CRC~110 ``Symmetries and the Emergence of Structure in QCD'' is 
gratefully acknowledged.
\end{enumerate}

%%%-----------------------------------

%%%%%%%%%%%%%%%%%%%%%%%%%%%%%%%%%%%%%%%%%%%%%%%%%%%%%%%%%%%%%%%%%%%%
\newpage
\subsection{Using $\pi$ K to Understand Heavy Meson Decays}
\addtocontents{toc}{\hspace{2cm}{\sl Alessandro Pilloni and Adam Szczepaniak}\par}
\setcounter{figure}{0}
\setcounter{table}{0}
\setcounter{equation}{0}
\setcounter{footnote}{0}
\halign{#\hfil&\quad#\hfil\cr
\large{Alessandro Pilloni}\cr
\textit{Theory Center}\cr
\textit{Thomas Jefferson National Accelerator Facility}\cr
\textit{Newport News, VA 23606, U.S.A.}\cr\cr
\large{Adam Szczepaniak}\cr
\textit{Theory Center}\cr
\textit{Thomas Jefferson National Accelerator Facility}\cr
\textit{Newport News, VA 23606, U.S.A.~\&}\cr
\textit{Center for Exploration of Energy and Matter}\cr
\textit{Indiana University}\cr
\textit{Bloomington, IN 47403, U.S.A.~\&}\cr
\textit{Physics Department}\cr
\textit{Indiana University}\cr
\textit{Bloomington, IN 47405, U.S.A.}\cr}

%%%-----------------------------------
\begin{abstract}
Several of the mysterious $XYZ$ resonances have been observed in
3-body $B\to\pi K (c\bar c)$ decays. A better description of the
$\pi K$ dynamics is required to improve the understanding of these
decays, and eventually to confirm the existence of the exotic states.
As an example, we discuss  the  $B\to J/\psi\pi K$ decay, where the
$Z(4430)$ has been observed. We critically review the formalisms to
build amplitudes available in the literature.
\end{abstract}

%%%-----------------------------------
\begin{enumerate}
\item \textbf{Introduction}

The last decade witnessed the observation of many unexpected $XYZ$
resonances in the heavy quarkonium sector. Their spectrum and production
and decay rates are not compatible with a standard charmonium
interpretation~\cite{Esposito:2016nozp,Olsen:2017bmmp,Guo:2017jvcp}.
Some of these states have been observed in 3-body $B \to \pi K (c\bar
c)$ decays, with $(c \bar c) = J/\psi,\psi^\prime,\chi_{c1}$. 
Understanding the three-body dynamics, and specifically the effect of the 
$\pi K$ interaction, is mandatory to confirm the existence of these exotic
states, and to better establish their properties.

We focus here on the so-called $Z(4430)$. The state was claimed in 2007
by Belle, as a peak in the $B^0 \to K^+ (\psi^\prime\,\pi^-)$
channel~\cite{Choi:2007wgap}, and it was the first observation of a
charged charmoniumlike state. The rich structure of the $\pi K$ resonances 
has led to opposite claims about the need of an exotic state to describe 
data~\cite{Aubert:2008aap,Mizuk:2009dap,Chilikin:2013tchp}. The high-statistic 
analysis 4D analysis by LHCb provided further evidence for the existence 
of such state~\cite{Aaij:2014jqap}. In particular, the analysis of the 
Legendre moments in~\cite{Aaij:2015zxap} suggests that the $\pi K$ waves 
with $J \le 3$ are not able to describe the data, calling either for an
unexpected contribution of higher spin $K^\ast$, or for an exotic 
resonance in the crossed channel. This state is extremely interesting, 
because is far from any reasonable open-charm threshold with the correct 
quantum numbers~\cite{Maiani:2014ajap,Brodsky:2014xiap}. The averaged mass 
and width are $M=(4478 \pm 17)~MeV$ and $\Gamma =(180\pm31)~MeV$, 
whereas the favored signature is $J^{PC} = 1^{+-}$ (see 
Fig.~\ref{fig:z4430lhcb}).
%%%-----------------------------------
\begin{figure}[htb!]
  \begin{center}
  \includegraphics[width=.52\textwidth]{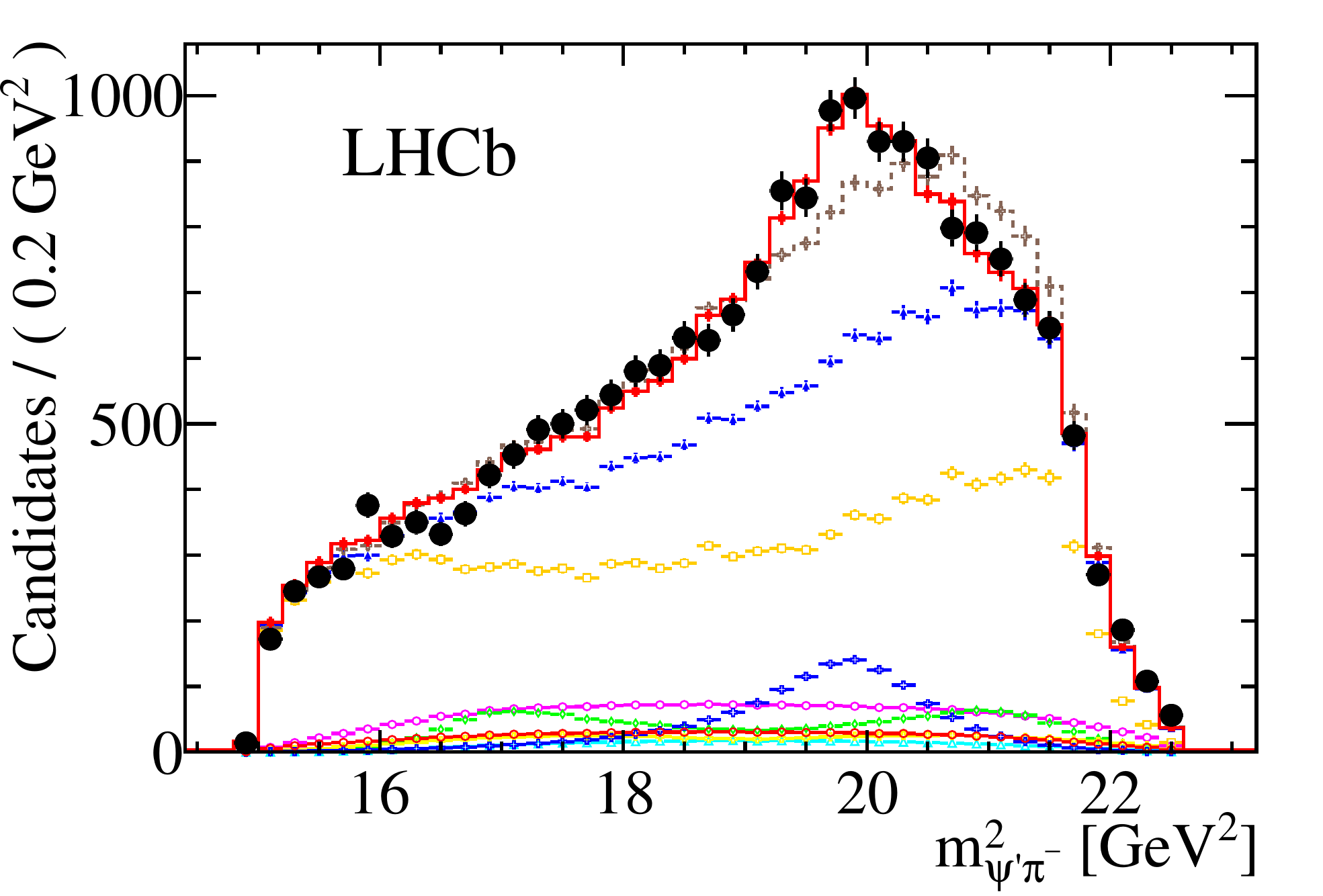}
  \includegraphics[width=.46\textwidth]{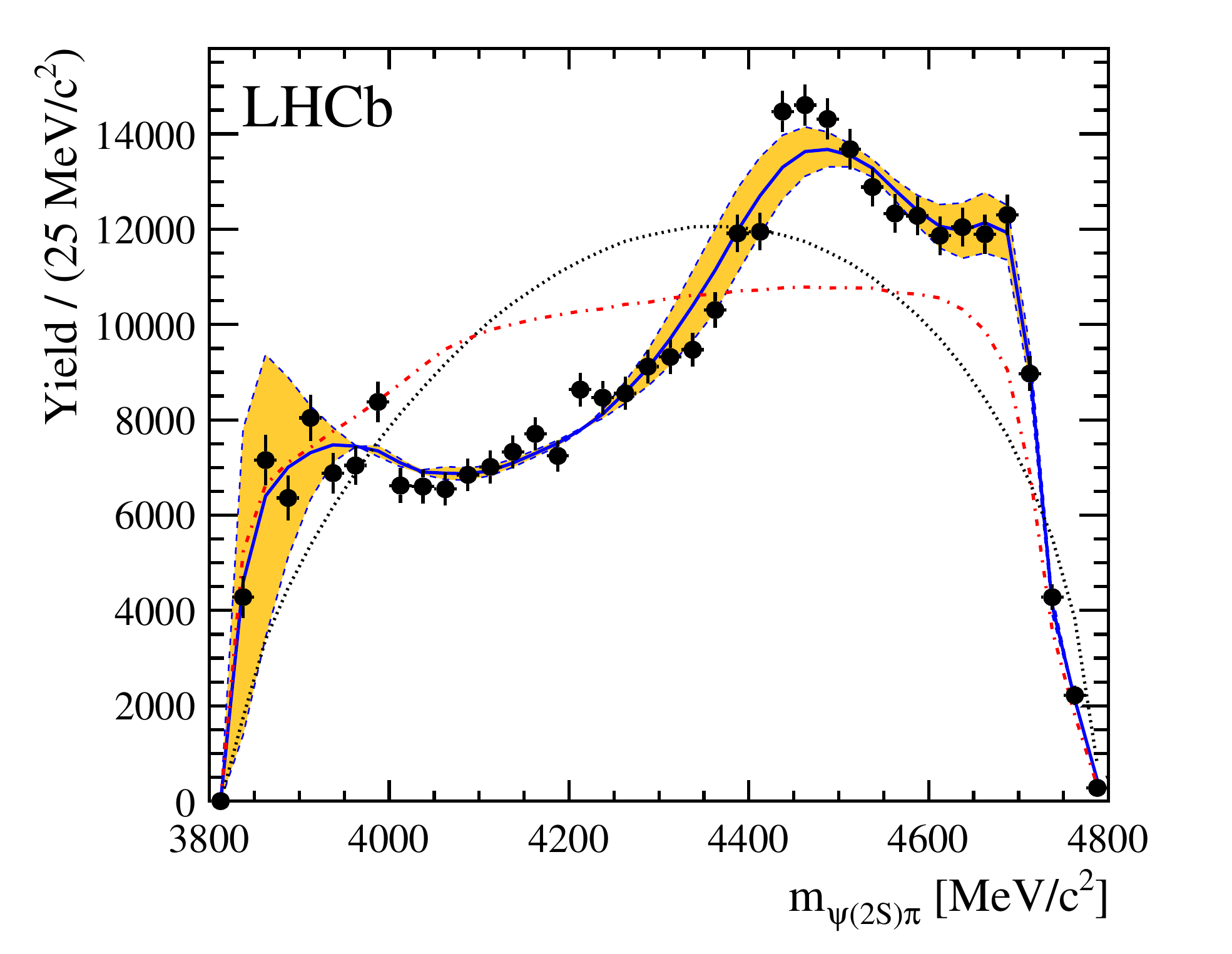}
  \end{center}

   \caption{Left panel: Invariant mass distributions in $\psi^\prime\,\pi^-$
	channel, from Ref.~\protect\cite{Aaij:2014jqap}. The red solid 
	(brown dashed) curve shows the fit with (without) the additional 
	$Z(4430)$ resonance. Right panel: the Legendre moments populated 
	by $\pi K$ waves with $J \le 3$ are not able to describe data 
	(from Ref.~\protect\cite{Aaij:2015zxap}). Although the agreement in 
	the figure may look good, the significance of higher momenta is 
	$8\sigma$.} \label{fig:z4430lhcb}
\end{figure}
%%%-----------------------------------

%%%-----------------------------------
\newpage
\item \textbf{Analyticity constraints for \texorpdfstring{$B\to \psi^\prime 
	\pi K$}{B\to\psi^'\pi K}}
\label{sec:s.channel}

In the modern literature, there seems to be a lot of confusion regarding
properties of the reaction amplitudes employed in analyses of such 
processes. This is often stated in the context of a potentially 
nonrelativistic character of certain approaches~\cite{Chung:1993dap,
Chung:2007nnp,Filippini:1995ycp,Anisovich:2006bcp,Anisovich:2011fcp,
Adolph:2015tqap}. As discussed in Ref.~\cite{Mikhasenko:2017rkhp}, rather 
than arising from relativistic kinematics, the differences between the 
various formalisms have a dynamical origin.

We consider the scattering process $\psi^\prime B \to \pi K$, related to 
the decay we are interested in via crossing symmetry. The spinless 
particles $B$, $\pi$, $K$ are stable against the strong interaction, and 
the $\psi^\prime$ is narrow enough to completely factorize its decay 
dynamics. We use $p_i$, $i=1\dots 4$ to label the momenta of 
$\psi^\prime$, $B$,
$\pi$, and $K$ respectively. We denote the helicity amplitude by 
$\mathcal{A}_\lambda(s,t)$, $\lambda$ being the helicity of $\psi^\prime$.
The amplitude depends on the standard Mandelstam variables $s =
(p_3+p_4)^2$, $t = (p_1 - p_3)^2$, and $u = (p_1 - p_4)^{2}$ with
$s + t + u = \sum_i m_i^2$.

We discuss the parity violating amplitude. We call $p$ ($q$) to the 
magnitude of the incoming (outgoing) three momentum in the $s$-channel 
center of mass, and $\theta_s$ the scattering angle. The quantities depend 
on the Mandelstam invariants through
\begin{align}
	\label{eq:zs}
	z_s &\equiv \cos \theta_s =\frac{s(t-u)+(m_1^2-m_2^2)(m_3^2-m_4^2)}{\lambda^{1/2}_{12} \lambda_{34}^{1/2}}
	, &
	p &= \frac{\lambda^{1/2}_{12}}{2\sqrt{s}},
	&  q &= \frac{\lambda^{1/2}_{34}}{2\sqrt{s}},
\end{align}
with $\lambda_{ik} = \left(s -  (m_i+m_k)^2\right)\left(s - (m_i-m_k)^2\right)$.
We assume the isobar model, and incorporate the $\pi K$ resonances
up to spin $J_\text{max}$ via
\begin{equation}
	\label{eq:isobar}
	A_\lambda(s,t,u) = \frac{1}{4\pi}\sum_{j=|\lambda|}^{J_\text{max}}
	(2j+1) A_{\lambda}^{j}(s) \,d_{\lambda0}^j(z_s),
\end{equation}
where $A^j_{\lambda}(s)$ are the helicity partial wave amplitudes in
the $s$-channel. In Eq.~\eqref{eq:isobar} the entire $t$ dependence enters 
though the $d$ functions. The $d$ functions have singularities in $z_s$ 
which lead to kinematical singularities in $t$ of the helicity amplitudes 
$A_\lambda$. An extensive discussion and the full characterization of the 
kinematical singularities can be found in 
Refs.~\cite{Hara:1964zzap,Wang:1966zzap,Jackson:1968rfnp,
Cohen-Tannoudji:1968kvrp,Martin:1970p,Collins:1977jyp}.
We recall that $d_{\lambda0}^j(z_s) = \hat d_{\lambda0}^j(z_s)
\xi_{\lambda 0}(z_s)$, where $\xi_{\lambda 0}(z_s) = \left(\sqrt{1
-z_s^2}\right)^{|\lambda|}$ is the so-called half angle factor that
contains  all the kinematical singularities in $t$. The reduced rotational
function $\hat d_{\lambda0}^j(z_s)$ is a polynomial in $s$ and $t$ of 
order $j - |\lambda|$ divided by the factor 
$\lambda_{12}^{(j-|\lambda|)/2}\lambda_{34}^{(j-|\lambda|)/2}$.
The helicity partial waves $A_\lambda^j(s)$ have singularities in $s$.
These have both  dynamical and kinematical origin. The former arise, for
example, from $s$-channel resonances. The kinematical singularities, just 
like the $t$-dependent kinematical singularities, arise because of 
external particle spin. We explicitly isolate the kinematic factors in 
$s$, and denote the kinematical singularity-free helicity partial wave
amplitudes by $\hat{A}_\lambda^j(s)$,
\begin{subequations}
\begin{align}
	A_0^j(s) &= K_{00}\, \left(p q\right)^j \,\hat{A}_0^j(s),\\
	A_\pm^j(s) &= K_{\pm0}\, \left(p q\right)^{j-1} \,\hat{A}_\pm^j(s),
\end{align}
\end{subequations}
with $K_{00} =  2m_1 /\lambda_{12}^{1/2}$, and $K_{\pm 0} =
\lambda_{34}^{1/2}/2\sqrt{s}$. The $j=0$ amplitude is exceptional,
$A_0^0(s) = \hat{A}_0^0(s) / K_{00}$. The $\hat A_\lambda^j(s)$ are left 
as the dynamical functions we are after, usually parameterized in terms of 
a sum of Breit-Wigner amplitudes with Blatt-Weisskopf barrier factors.
We now seek a representation of $A_\lambda(s,t)$ in terms of the scalar
functions,
\begin{equation}
	A_\lambda(s,t) = \epsilon_\mu(\lambda,p_1) \left[
	(p_3 - p_4)^\mu - \frac{m_3^2-m_4^2}{s} (p_3+p_4)^\mu
	\right] C(s,t)
	+ \epsilon_\mu(\lambda,p_1) (p_3 + p_4)^\mu B(s,t),\label{eq:cov}
\end{equation}
where the functions $B(s,t)$ and $C(s,t)$ are the kinematical singularity
free scalar amplitudes. We can match Eqs.~\eqref{eq:isobar} 
and~\eqref{eq:cov}, and express the scalar functions as a sum over 
kinematical singularity free helicity partial waves.
\begin{align}
	\label{eq:matching.C}
	\sqrt{2}\, C(s,t) &= \frac{1}{4\pi} \sum_{j>0} (2j+1) (pq)^{j-1}\hat{A}_{\pm}^{j}(s)\, \hat{d}_{10}^j(z_s),\\
    	\label{eq:matching.B}
	4\pi B(s,t) &= \hat{A}_{0}^{0}(s) + \frac{4m_1^2}{\lambda_{12}} \sum_{j>0} (2j+1) (pq)^{j} \nonumber \\
	&\qquad\times \left[\hat{A}_{0}^{j}(s) \hat{d}_{00}^j(z_s) +
	\frac{s+m_1^2-m_2^2}{\sqrt{2}m_1^2} \hat{A}_{+}^{j}(s)\, z_s\hat{d}_{10}^j(z_s)
	\right].
\end{align}

Neither $B(s,t)$ nor $C(s,t)$ can have kinematical singularities in $s$
or $t$. In  Eqs.~\eqref{eq:matching.C}-\eqref{eq:matching.B}, $\hat
d^j_{10}(z_s)$ is regular in $t$, and the $s$ singularities at
(pseudo)thresholds are canceled by the factor $(pq)^{j-1}$. For the same 
reason  the sum in Eq.~\eqref{eq:matching.B} has no kinematical 
singularities in $s$ and $t$, however the $1/\lambda_{12}$
factor in front of the sum generates two poles at $s_{\pm} =
(m_1\pm m_2)^2$, unless the expression in brackets vanishes at those
points. This means that the $\hat A^j_\lambda(s)$ with different
$\lambda$ cannot be  independent functions at  the (pseudo)threshold.
Explicitly, using the expansion of the Wigner $d$-function for $z_s
\to \infty$, we get
\begin{equation}
	\hat{A}_{0}^{j}(s) \frac{(z_s)^j}{\langle j-1,0; 1, 0| j, 0\rangle} -
	\frac{s+m_1^2-m_2^2}{\sqrt{2}\: m_1^2} \hat{A}_{+}^{j}(s)\,
	\frac{(z_s)^j}{\sqrt{2} \: \langle j-1,0; 1, 1| j, 1\rangle}.
\end{equation}

This combination has to vanish to cancel the $1/\lambda_{12}$, thus one 
finds (for $j>0$)
\begin{subequations}
\label{eq:general.form}
\begin{align}
	\hat{A}_{+}^j(s) &= \langle j-1,0; 1, 1| j, 1\rangle \: g_j(s)
	+ \lambda_{12}\: f_j(s),\\
	\hat{A}_{0}^j(s) &= \langle j-1,0; 1, 0| j, 0\rangle \frac{s+m_1^2
	-m_2^2}{2m_1^2} \: g'_j(s) + \lambda_{12} \: f_j'(s),
\end{align}
\end{subequations}
where $g_j(s)$, $f_j(s)$, $g'_j(s)$, and $f'_j(s)$ are regular functions
at $s=s_\pm$, and $g_j(s_\pm)=g'_j(s_\pm)$. Together with 
Eq.~\eqref{eq:general.form}, the expressions in Eqs.~\eqref{eq:cov}, 
\eqref{eq:matching.C} and \eqref{eq:matching.B} provide the most general 
parameterization of the amplitude that incorporates the minimal kinematic 
dependence that generates the correct kinematical singularities as 
required by analyticity.

Upon restoration of the kinematic factors, the original helicity partial
wave amplitudes read
\begin{subequations}
\label{eq:helicitypw}
\begin{align}
	A_{+}^j(s) &=  p^{j-1} q^j \bigg[\langle j-1,0; 1, 1| j, 1\rangle \: 
	g_j(s)+ \lambda_{12} \: f_j(s)\bigg], \\
	A_{0}^j(s) &=  p^{j-1} q^j \bigg[\langle j-1,0; 1, 0| j, 0\rangle \: 
	\frac{s+m_1^2-m_2^2}{2m_1\sqrt{s}}\: g'_j(s) + \frac{m_1}{\sqrt{s}} 
	\lambda_{12}\: f_j'(s)\bigg],
\end{align}
\end{subequations}
and $A_{0}^0(s) = \lambda_{12}^{1/2}/(2m_1)\,\hat A_{0}^0(s)$, where $\hat
A_{0}^0(s)$ is regular at (pseudo)threshold. A particular choice of the 
functions $g_j(s)$, $g_j'(s)$, $f_j(s)$ and $f_j'(s)$ constitutes a given 
hadronic model.

We now compare the general expression for the helicity partial waves
with the spin-orbit LS~partial waves, $\hat G^j_{L}(s)$. These match the 
general form in Eq.~\eqref{eq:general.form} when
\begin{subequations}
\label{eq:lsmatching}
\begin{align}
	g_j(s) &= \textstyle{\sqrt{\frac{2j-1}{2j+1}}}  \hat{G}^j_{j-1}(s), &
	f_j(s) &= \frac{1}{4s} \textstyle{\sqrt{\frac{2j+3}{2j+1}}} \langle j+1,0; 1, 1| j, 1\rangle \: \hat{G}^j_{j+1}(s), \label{eq:fj_s_pole}\\
	g^\prime_j(s) &= \frac{2m_1\sqrt{s}}{s+m_1^2-m_2^2}
	\textstyle{\sqrt{\frac{2j-1}{2j+1}}} \hat{G}^j_{j-1}(s), &
	f_j'(s) &= \frac{1}{4m_1\sqrt{s} } \textstyle{\sqrt{\frac{2j+3}{2j+1}}} \langle j+1,0; 1, 0| j, 
	0\rangle \: \hat{G}^j_{j+1}(s).
\end{align}
\end{subequations}

The common lore is that the LS formalism is intrinsically 
nonrelativistic. However, the matching in Eq.~\eqref{eq:lsmatching} proves 
that the formalism is fully relativistic, but care should be taken when 
choosing a parameterization of the LS amplitude so that the expressions 
in Eqs.~\eqref{eq:lsmatching} are free from kinematical singularities. 
For example, if one  takes the functions $\hat G^j_{j-1}(s)$ and $\hat 
G^j_{j+1}(s)$  to be proportional to Breit-Wigner functions with constant 
couplings, the amplitudes $g'_j(s)$ and $f'_j(s)$ would end up having a 
pole at $s = m_2^2 - m^2_1$. It is clear that using  Breit-Wigner 
parameterizations, or any other model for helicity amplitudes, i.e., the 
left-hand sides of Eq.~\eqref{eq:lsmatching}, instead of the LS 
amplitudes helps prevent unwanted singularities.

We also consider  the Covariant Projection Method (CPM) approach
of~\cite{Chung:1993dap,Chung:2007nnp,Filippini:1995ycp,Anisovich:2006bcp},
based on the construction of explicitly covariant expressions. We limit
ourselves to the special case of an intermediate $K^\ast$ with $j=1$. We 
start with the tensor amplitude for the scattering process $\psi B \to 
K^\ast\to\pi K$,
\begin{align}
	\nonumber A_\lambda(s,t) &= \epsilon_\mu(\lambda,p_1) \left(-g^{\mu\nu} + \frac{P^\mu P^\nu}{s}\right)X_\nu(q,P) g_S(s) \\
	&\qquad + \epsilon^\rho(\lambda,p_1) X_{\rho\mu}(p,P) \left(-g^{\mu\nu} + \frac{P^\mu P^\nu}{s}\right)X_\nu(q,P) g_D(s),
        \label{eq:cov.LS}
\end{align}
where $P$ is the $K^\ast$ momentum. The final P-wave orbital tensor
is $X_{\nu}(q,P) = q^\perp_\nu = q_\nu - P_\nu P\cdot q /s$. The D-wave 
orbital tensor $X^{\rho\mu}(p,P)=3p^\rho_\perp p^\mu_\perp/2
-g^{\rho\mu}_\perp p_\perp^2/2$, with 
$p_\perp^\mu = p^\mu - P^\mu \,P\cdot p/s$, and $g^{\rho\mu}_\perp = 
g^{\rho\mu} - P^\rho P^\mu/s$. Explicitly,
\begin{equation}
	\label{eq:tensor.A.scatt}
	A_+(s,\theta_s) = -q \frac{\sin\theta_s}{\sqrt{2}} \left[g_S(s)
	+ \frac{p^2}{2} g_D(s)\right], \quad
	A_0(s,\theta_s) = q \frac{E_1}{m_1} \cos\theta_s \left[g_S(s) -
	p^2 g_D(s) \right],
\end{equation}
and  matching with  Eq.~\eqref{eq:general.form} gives
\begin{align}
	g_1(s) &= g_1'(s) = \frac{4\pi}{3} g_S(s), &
	f_1(s) &= \frac{2\pi}{3s} g_D(s), &
	f_1'(s) &= -\frac{4\pi}{3s}\frac{s+m_1^2-m_2^2}{m_1^2} g_D(s).
\end{align}
%%%-----------------------------------
\begin{figure}[htb!]
\centering
\includegraphics[width=0.48\textwidth]{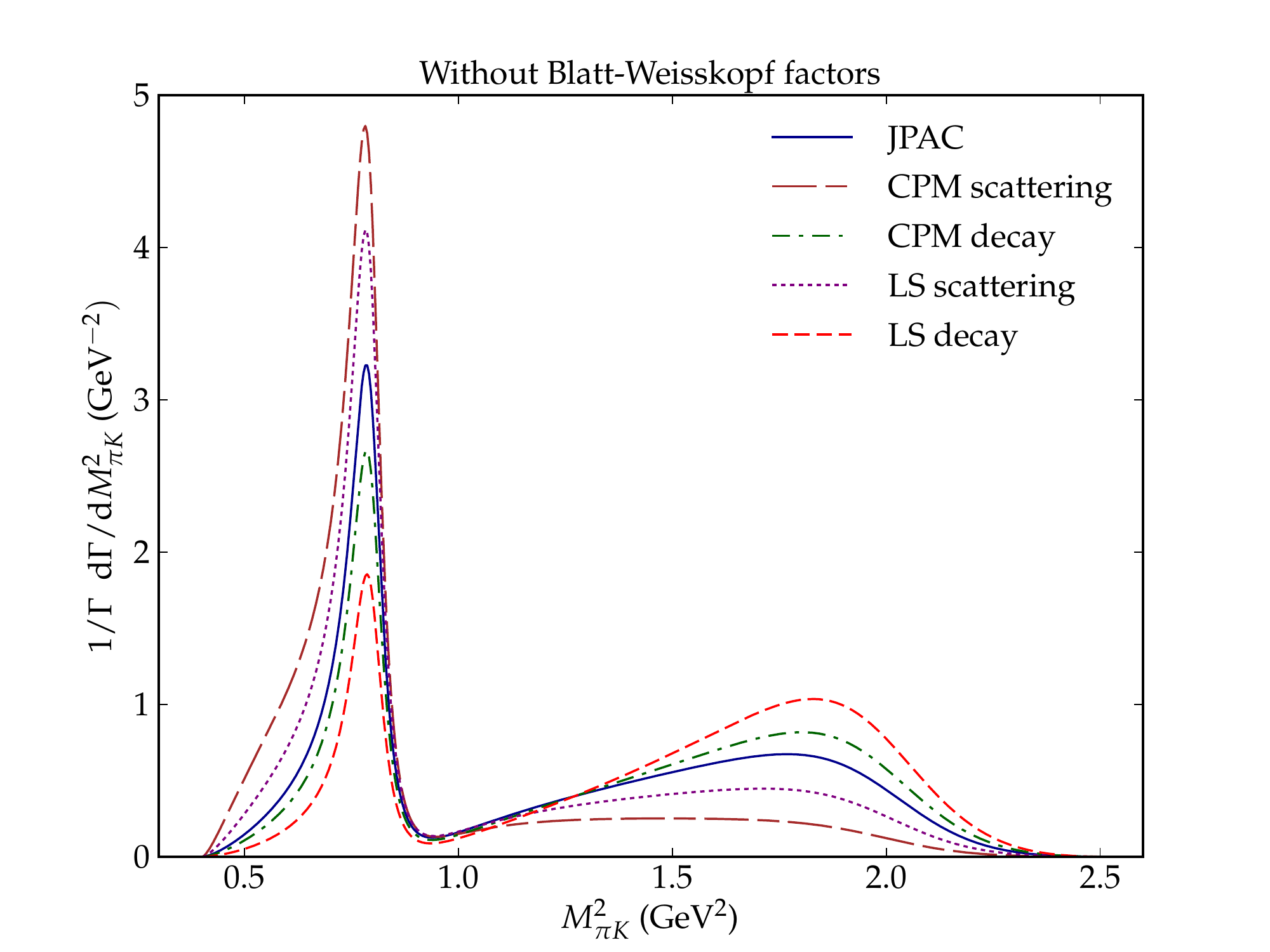}\quad
\includegraphics[width=0.48\textwidth]{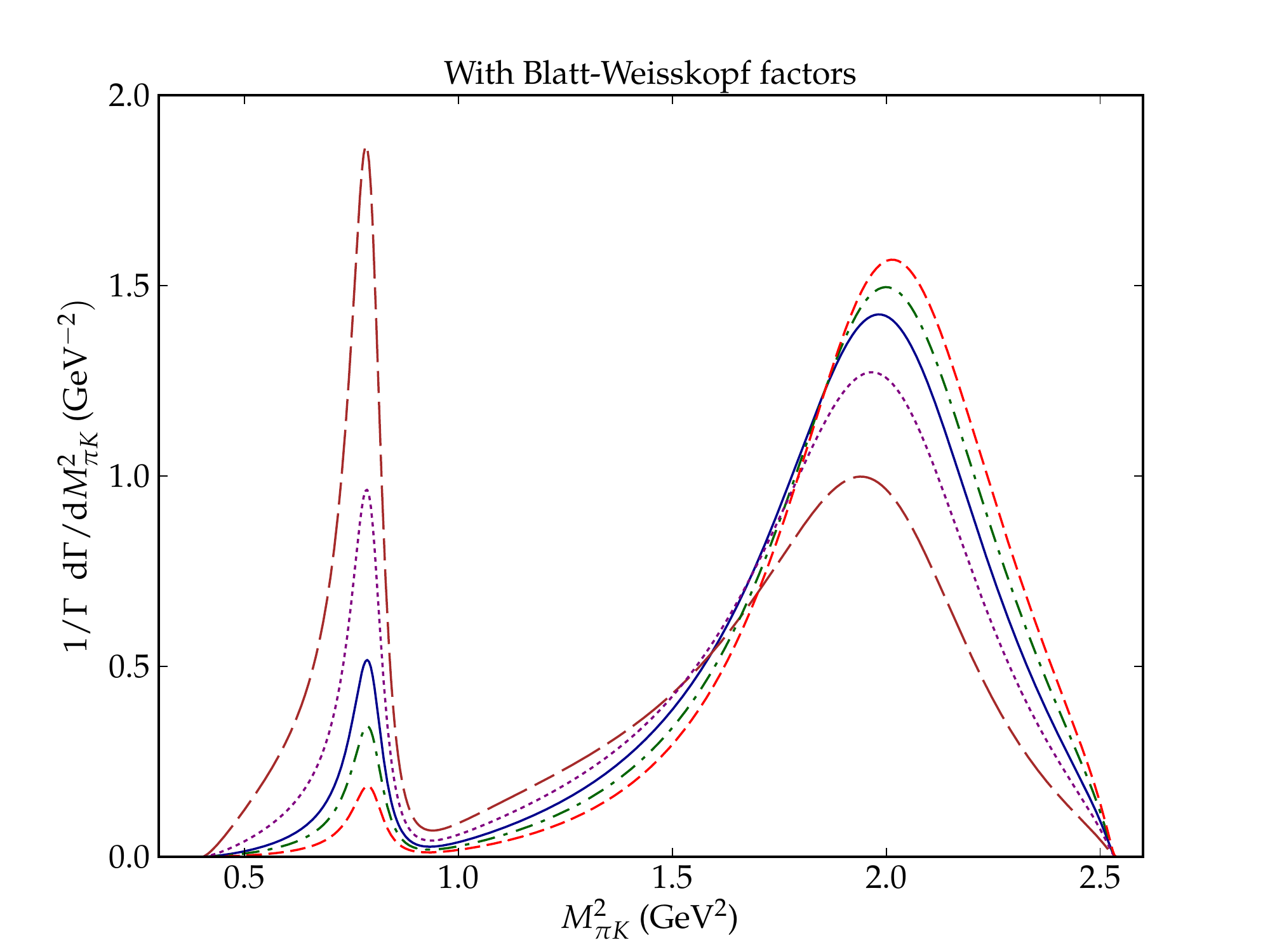}

\caption{Comparison of the lineshape of $K^\ast(892)$ and $K^\ast(1410)$
        in the $\pi K$-invariant mass distribution, constructed with the
        different formalisms. In the left panel we show the result with no
        barrier factors. In the right panel, we include the customary
        Blatt-Weisskopf factors. From~\protect\cite{Mikhasenko:2017rkhp}.}
        \label{fig:K1410.three.methods}
\end{figure}
%%%-----------------------------------

The threshold conditions $g_1(s_\pm) = g_1'(s_\pm)$ are satisfied, and
the functions $f_1(s)$ and $f_1'(s)$ are regular at the thresholds.
Finally, we show the relation between the CPM and the LS amplitudes:
\begin{subequations}
\label{eq:tensor.to.ls}
\begin{align}
	\frac{3}{4\pi}G_0^1(s) &=
	g_S(s)\: q \: \sqrt{\frac{1}{3}}\left(\frac{E_1}{m_1}+2\right)
	-g_D(s) \: q \: p^2 \: \sqrt{\frac{1}{3}}\left(\frac{E_1}{m_1}-1\right), 
	\\
	\frac{3}{4\pi} G_2^1(s) &=
	g_D(s) \: q \: p^2 \: \sqrt{\frac{1}{6}}\left(2\frac{E_1}{m_1}+1\right)
	-g_S(s) \: q \: \sqrt{\frac{2}{3}}\left(\frac{E_1}{m_1}-1\right).
\end{align}
\end{subequations}

Although the $g_S(s)$ and $g_D(s)$ of the CPM formalism, see
Eq.~\eqref{eq:cov.LS}, are typically interpreted as the $S$ and $D$  
partial wave amplitudes, we see that this is the case only at 
(pseudo)threshold $s=s_\pm$, where  the factor $E_1/m_1-1$
vanishes. An extensive discussion about the same calculation performed
in the decay kinematics, which turns out into an explicit violation of
crossing symmetry, can be found in~\cite{Mikhasenko:2017rkhp}.

To explore the differences between the various approaches,
we consider the example of two intermediate vectors in the $\pi K$
channel: the $K^\ast(892)$, with mass and width $M_{K^\ast}=892$~MeV,
$\Gamma_{K^\ast} = 50$~MeV, and the  $K^\ast(1410)$, with 
$M_{K^\ast}=1414$~MeV, $\Gamma_{K^\ast} = 232$~MeV.

In Fig.~\ref{fig:K1410.three.methods}, we show the results for the
differential decay width in five different scenarios. We consider the CPM 
formalism (for the scattering and decay kinematics, respectively), 
setting 
$g_S(s)=0$  and $g_D(s) = T_{K^\ast}(s)$, with $T_{K^\ast}(s)$ being the 
sum of Breit-Wigners for the two resonances. For the LS formalism, we 
choose the couplings to be $\hat{G}_0^1(s)=0$, 
$\hat{G}_2^1(s)=T_{K^\ast}(s)$. The LS amplitude in the decay kinematics
differs from the one in the scattering kinematics because of the breakup
momentum of $B \to \psi K^\ast$, calculated in the $B$ rest frame or in 
the $K^\ast$ rest frame, respectively. In Ref.~\cite{Mikhasenko:2017rkhp} 
we also propose an alternative model. We see in 
Fig.~\ref{fig:K1410.three.methods} that the $K\pi$ invariant mass squared 
distribution is significantly distorted in all models. This is important 
to extract the physical couplings, and may enhance the high mass 
contributions of the higher spin $K^\ast$, thus affecting the extraction 
of the properties of the $Z(4430)$.

%%%-----------------------------------
\item \textbf{Conclusions}
\label{sec:sc}

%%%-----------------------------------
\begin{figure}[htb!]
\centering
\includegraphics[width=0.6\textwidth]{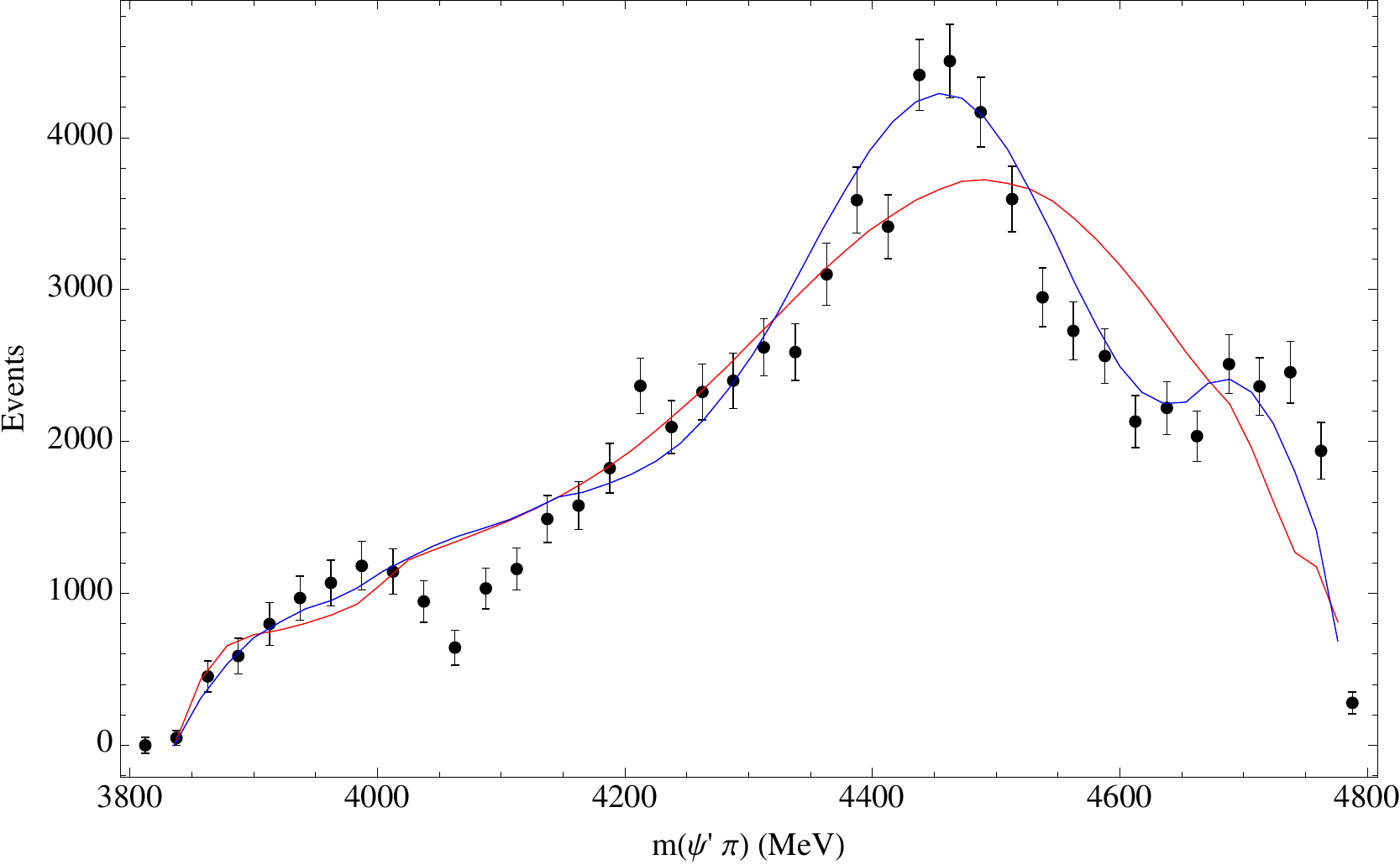}

\caption{Fit to the $\psi^\prime\pi$ invariant mass, in the $1~GeV \le
	m(\pi K) \le 1.39~GeV$ slice, where the signal of the $Z(4430)$ 
	is more prominent. The red curve includes the $K^\ast$ resonances 
	with $J \le 3$, whereas the blue curve includes the tails of the  
	$K_4^\ast(2045)$ and the $K_5^\ast(2380)$. Data 
	from~\protect\cite{Aaij:2015zxap}.}
\label{fig:fit}
\end{figure}
%%%-----------------------------------

The rich structure of the $\pi K$ system affects the extraction of 
several exotic candidates, as the $Z(4200)$ in $B\to J/\psi\pi
K$~\cite{Chilikin:2014bkkp}, the $Z(4050)$ and $Z(4250)$ in $B\to
\chi_{c1} \,\pi K$~\cite{Mizuk:2008mep,Lees:2011ikp}, and in particular
the $Z(4430)$ in $B\to\psi^\prime \,\pi K$ we have discussed
here~\cite{Choi:2007wgap,Aubert:2008aap,Mizuk:2009dap,Chilikin:2013tchp,
Aaij:2014jqap,Aaij:2015zxap}. As we show in a temptative fit in
Fig.~\ref{fig:fit}, the inclusion of higher spin $K^\ast$ resonances can
improve the description of the $\psi^\prime \pi$ invariant mass, even
without adding an exotic state. Although this may not be enough to
challenge the existence of the $Z(4430)$, this can dramatically affect
the estimate of the resonance parameters and of its quantum numbers.
This also calls for more refined analysis, which take into account
unitarity in the isobar model~\cite{Khuri:1960zzp,Niecknig:2012sjp,
Niecknig:2015ijap,Danilkin:2014crap,Guo:2015zqap,Pilloni:2016obdp,
Albaladejo:2017hhjp,Albaladejo:2018gifp}, or include the effect of
higher spin resonances~\cite{Szczepaniak:2014bsap}.

%%%-----------------------------------
\newpage
\item \textbf{Acknowledgments}

I wish to thank Igor Strakovsky for his kind invitation to this workshop,
and Tomasz Skwarnicki for the fruitful collaboration. My participation to
the conference was supported by the U.S. Department of Energy, Office of 
Science, Office of Nuclear Physics under contract DE--AC05--06OR23177.
\end{enumerate}

%%%-----------------------------------

%%%%%%%%%%%%%%%%%%%%%%%%%%%%%%%%%%%%%%%%%%%%%%%%%%%%%%%%%%%%%%%%%%%%%%%%%
\newpage
\subsection{Three Particle Dynamics on the Lattice}
\addtocontents{toc}{\hspace{2cm}{\sl Akaki Rusetsky}\par}
\setcounter{figure}{0}
\setcounter{table}{0}
\setcounter{equation}{0}
\setcounter{footnote}{0}
\halign{#\hfil&\quad#\hfil\cr
\large{Akaki Rusetsky}\cr
\textit{Helmoltz-Institut f\"ur Strahlen- und Kernphysik}\cr
\textit{Universit\"at Bonn~\&}\cr
\textit{Bethe Center for Theoretical Physics}\cr
\textit{Universit\"at Bonn}\cr
\textit{D-53115 Bonn, Germany}\cr}

%%%-----------------------------------
\begin{abstract}
In my talk, I review the recent progress in understanding the 
three-particle quantization condition, which can be used for 
the extraction of physical observables from the finite-volume 
spectrum of the three-particle system, measured on the lattice. 
It is demonstrated that the finite-volume energy levels allow 
for a transparent interpretation in terms of the three-body 
bound states, as well as the three-particle and particle-dimer 
scattering states. The material, covered by this talk, is mostly 
contained in the recent publications~\cite{pang1c,pang2c,Doring:2018xxxc}.
\end{abstract}

%%%-----------------------------------
\begin{enumerate}
\item \textbf{Introduction}

The L\"uscher approach~\cite{Luescher-torusc} has become a standard tool 
for the analysis of lattice data in the two-body scattering sector. It 
has been further generalized to study the scattering of particles with 
arbitrary spin, the scattering in the moving frames, as well as the
scattering in coupled two-body channels. Moreover, in their seminal
paper~\cite{Lellouch:2000pvc}, Lellouch and L\"uscher have discussed 
the application of the finite-volume formalism to the calculation of 
the two-body decay matrix elements. This approach has been also 
generalized to include multiple decay channels and applied to study of 
various formfactors in the timelike region and matrix elements with 
resonances (in the latter case, the analytic continuation in the complex 
plane to the resonance pole has to be 
considered)~\cite{Bernard:2012bic,Meyer:2011umc,Hansen:2012tfc,Kim:2005gfc}.

The formulation of the finite-volume approach for the three-body problem 
(in analogy with the L\"uscher approach) has however proven to be a 
challenging task. Despite the significant effort during the last few
years~\cite{pang1c,pang2c,Doring:2018xxxc,Polejaeva:2012utc,Briceno:2012rvc,
Hansen:2014ekac,Hansen:2016fzjc,Hansen-allc,Briceno:2018mlhc,Guo:2017crdc,
Mai:2017bgec,Meissner:2014deac,Kreuzer:2008bic,Bour:2012hnc}, the progress 
has been rather slow. Recently, the tree-body quantization condition,
which is a counterpart of the L\"uscher equation in the two-body case, has 
been derived in different settings~\cite{Hansen:2014ekac,Briceno:2012rvc,
pang1c,pang2c,Mai:2017bgec} (it has been shown~\cite{Doring:2018xxxc} that 
all these different settings are essentially equivalent to each other, so 
in practice all boils down to the choice of merely the most convenient one). 
On the basis of the quantization condition, the finite-volume energy levels 
have been calculated in some simple cases. Such studies are extremely
interesting because, at this stage, one does not yet have enough insight in 
the problem and lacks the intuition to predict the volume dependence of the 
three-body spectrum, which would emerge in lattice simulations. Moreover, 
such studies may facilitate the interpretation of this volume dependence in 
terms of the observable characteristics of the three-particle systems in a 
finite volume, in analogy with the two-particle case where, e.g., the avoided 
level crossing in the spectrum is often related to a nearby narrow resonance.

The study of the three-particle (and, in general, many-particle) systems on 
the lattice is interesting at least from two different perspectives. First, 
we would refer to the potential applications in study of the nuclear physics 
problems in lattice QCD. Second, it would be interesting to study the 
systems, whose decay into the final states with three and more particles 
cannot be neglected. In the meson sector, the simplest example could be the 
decay $K\to 3\pi$ (a counterpart of the process $K\to 2\pi$, which was 
considered in the original paper by Lellouch and 
L\"uscher~\cite{Lellouch:2000pvc}), but physically more interesting processes 
like the decays of $a_1(1260)$ or $a_1(1420)$ as well. In the baryon sector, 
the most obvious candidate is the Roper resonance. In order to obtain an 
analog of the Lellouch-L\"uscher formula for such systems, however, one has 
first to understand the final-state interactions in a finite volume, and it 
is where the study of the solutions of the three-body quantization condition 
might help.

In my talk, I shall briefly cover the formalism of 
Refs.~\cite{pang1c,pang2c}, which is based on the use of the effective field 
theory in a finite volume, and will further discuss the solution of these 
equations, which includes the projection of the quantization condition onto 
the different irreducible representations (irreps) of the octahedral
group~\cite{Doring:2018xxxc}. It will be shown that these solutions allow for 
a nice interpretation in terms of the three-particle bound states, as well as 
three-particle and particle-dimer scattering states.

%%%-----------------------------------
\item \textbf{The Formalism}
\label{sec:formalism}

In Refs.~\cite{pang1c,pang2c}, the three-particle quantization condition was 
obtained by using effective field theory approach in a finite volume. 
Moreover, it has been shown that it is very convenient to use the so-called 
particle-dimer picture. It should be stressed that this approach is not an 
approximation, but an equivalent description of the three-particle 
interactions. This is most easily seen in the path integral formalism, where
the introduction of a dimer amounts to using an additional dummy integration 
variable, without changing the value of the path integral. Moreover, using 
the particle-dimer picture does not necessarily imply the existence of a 
shallow bound state or a narrow resonance in the two-body sector.

Below we consider the case of three identical spinless bosons with the S-wave 
pair interactions and use, for simplicity, the non-relativistic kinematics. 
In order to derive the quantization condition, we consider the finite-volume 
Bethe-Salpeter equation for the particle-dimer amplitude ${\cal M}_L$
\eq\label{eq:ML}
	{\cal M}_L({\bf p},{\bf q};E)=Z({\bf p},{\bf q};E)+
	\frac{1}{L^3}\,\sum_{\bf k} ^\Lambda
	Z({\bf p},{\bf k};E)\,\tau_L({\bf k};E)\,
	{\cal M}_L({\bf k},{\bf q};E)\,.
\en
Here, ${\bf p}$ and {\bf q} are three-momenta of the incoming and outgoing
spectators and $E$ is the total energy of three particles. All momenta are 
discretized, e.g., ${\bf p}=2\pi{\bf n}/L,~{\bf n}\in \mathbb{Z}^3$ and, 
similarly, for other momenta. Further, $L$ is the spatial size of the cubic 
box and $\Lambda$ denotes the ultraviolet cutoff. The quantity $ \tau_L$ 
corresponds to the two-body scattering amplitude (the dimer propagator, in 
the particle-dimer language), and is given by
\eq\label{eq:tauL}
	8\pi\tau_L^{-1}({\bf k};E)&=&k^\ast\cot\delta(k^\ast)+S({\bf k},(k^\ast)^2)\, ,
	\nonumber\\[2mm]
	S({\bf k},(k^\ast)^2)&=&-\frac{4\pi}{L^3}\sum_{\bf l}\frac{1}{{\bf k}^2+{\bf 
	k}{\bf l}+{\bf l}^2-mE}\, ,
\en
where $k^\ast$ is the magnitude of the relative momentum of the pair in the rest frame,
\eq
	k^\ast=\sqrt{\frac{3}{4}\,{\bf k}^2-mE}\, .
\en
In Eq.~(\ref{eq:tauL}), unlike Eq.~(\ref{eq:ML}), the momentum sum is 
implicitly regularized by using dimensional regularization and $\delta(k^\ast)$ 
is the S-wave phase shift in the two-particle subsystem.
The effective range expansion for this quantity reads
\eq
	k^\ast\cot\delta(k^\ast)=-\frac{1}{a}+\frac{1}{2}\,r(k^\ast)^2+O((k^\ast)^4)\, ,
	\label{eq:ERE}
\en
where $a,r$ are the two-body scattering length and the effective range, 
respectively. In the numerical calculations, for illustrative purpose,
we shall use a simplified {\em model,} assuming that the effective range $r$ 
and higher-order shape parameters are all equal to zero, corresponding to the 
leading order of the effective field theory for short range-interactions.
The equation~(\ref{eq:ML}) is valid, however, beyond this approximation.

Finally, the quantity $Z$ denotes the kernel of the Bethe-Salpeter
equation. It contains the one-particle exchange diagram, as well as
the local term, corresponding to the particle-dimer interaction 
(three-particle force). In general, the latter consists of a string of 
monomials in the 3-momenta ${\bf p}$ and ${\bf q}$. In the numerical 
calculations, we shall again restrict ourselves to the {\em model,} where 
only the non-derivative coupling, which is described by a single constant 
$H_0(\Lambda)$, is non-vanishing. The kernel then takes the form
\eq\label{eq:Z}
	Z({\bf p},{\bf q};E)=\frac{1}{-mE+{\bf p}^2+{\bf q}^2+{\bf p}{\bf q}}
	+\frac{H_0(\Lambda)}{\Lambda^2}\, .
\en
The dependence of $H_0(\Lambda)$ on the cutoff is such that the 
infinite-volume scattering amplitude is cutoff-independent. In a finite 
volume, this ensures the cutoff-independence of the spectrum.

At the next step, the three-particle Green function can be expressed in 
terms of the particle-dimer scattering amplitude ${\cal M}_L$ by using the 
LSZ formalism, and hence the poles of the latter can be mapped onto the 
finite-volume energy spectrum of the three particle system. It can be 
shown~\cite{Doring:2018xxxc} that the poles arise at the energies, 
where the determinant of the linear equation (\ref{eq:ML}) vanishes.
This finally gives the quantization condition we are looking for
\eq\label{eq:master}
	\det(\tau_L^{-1}-Z)=0\, .
\en
The l.h.s. of the above equation defines a function of the total energy $E$,
which, for a fixed $\Lambda$ and $L$, depends both on the two-body input (the 
scattering phase $\delta$ both above and below the two-body threshold)
as well as the three-body input (the non-derivative coupling $H_0(\Lambda)$,
higher-order couplings). The former input can be independently determined
from the simulations in the two-particle sector and extrapolation below 
threshold. Hence, measuring the three-particle energy levels, one will be 
able to fit the parameters of the three-body force. Finally, using the same 
equations in the infinite volume with the parameters, determined on the 
lattice, one is able to predict the physical observables in the infinite 
volume.

%%%-----------------------------------
\item \textbf{The Projection of the Quantization Condition}

The three-body quantization condition, Eq.~(\ref{eq:master}), determines the 
entire finite-volume spectrum of the system. However, the eigenvalue problem 
in a cubic box has the octahedral symmetry, which is a remnant of the 
rotational symmetry in the infinite volume. This means that all energy levels 
can be assigned to one of the ten irreps 
$A_1^\pm,A_2^\pm,E^\pm,T_1^\pm,T_2^\pm$ of the octahedral group ${\cal G}$
and the spectrum in each irrep can be measured separately with a proper 
choice of the source/sink operators. It is possible to use this symmetry and 
to project the quantization condition onto the different irreps -- the 
obtained equations will determine the energy spectrum in each irrep 
separately~\cite{Doring:2018xxxc}.

In order to do this, we shall act in a close analogy with the partial-wave 
expansion in the infinite volume. A substitute for radial integration will be 
the sum over {\em shells}, defined as sets of momenta with equal magnitude, 
which can be obtained from any vector (referred hereafter as {\em the 
reference vector}), belonging to the same shell, by applying all 
transformations of the octahedral group. Note that the vectors with the same 
magnitude, which are {\em not} connected by a group transformation belong, by 
definition, to different shells. Furthermore, the integration over the solid 
angle in the infinite volume is replaced by the sum over all $G=48$ elements 
of the octahedral group.

On the cubic lattice, the analog of the partial-wave expansion is given by
\eq\label{eq:exp}
	f({\bf p})=f(g{\bf p}_0)=
	\sum_\Gamma\sum_{\rho\sigma}
	T^{\Gamma}_{\sigma\rho}(g)f^{\Gamma}_{\rho\sigma}(s)\, ,\quad\quad 
	\Gamma=A_1^\pm,A_2^\pm,E^\pm,T_1^\pm,T_2^\pm\, ,
\en
where $T^{\Gamma}_{\sigma\rho}(g)$ are the matrices of the irreducible 
representations, ${\bf p}_0$ denotes the reference momentum, and $s$ is the 
shell to which both ${\bf p}$ and ${\bf p}_0$ belong. Nothing depends on the 
choice of ${\bf p}_0$.

Using the orthogonality of the matrices of the irreducible representations, 
it is possible to project out the quantity $f^{\Gamma}_{\rho\sigma}(s)$:
\eq\label{eq:proj}
	\sum_{g\in {\cal G}}(T^{\Gamma}_{\lambda\delta}(g))^\ast f(g{\bf p}_0)
	=\frac{G}{s_\Gamma}\,f^{\Gamma}_{\delta\lambda}(s)\, ,
\en
where $s_\Gamma$ is the dimension of the irrep $\Gamma$, and the indices
$\lambda,\delta,\ldots$ run from 1 to $s_\Gamma$.

Next, we note that both the kernel of the Bethe-Salpeter equation and the 
dimer propagator are invariant with respect to the group ${\cal G}$:
\eq
	Z(g{\bf p},g{\bf q};E)=Z({\bf p},{\bf q};E)\, ,\quad\quad
	\tau_L(g{\bf k};E)=\tau_L({\bf k};E)\, ,\quad\quad\mbox{for all}~g\in {\cal 
	G}\, .
\en
Using this property, it can be shown that the projection of the quantization 
condition onto the irrep $\Gamma$ takes the form
\eq
	\det\biggl(\tau_L(s)^ {-1}\delta_{rs}\delta_{\sigma\rho}
	-\frac{\vartheta(s)}{GL^3}\,
	Z^{\Gamma}_{\sigma\rho}(r,s)\biggr)=0\, ,\label{eq:qc-gnl}
\en
where $\vartheta(s)$ is the multiplicity of the shell $s$ (the number of 
independent vectors in this shell), and
\eq
	Z^{\Gamma}_{\lambda\rho}(r,s)
	=\sum_{g\in {\cal G}}(T^{\Gamma}_{\rho\lambda}(g))^\ast Z(g{\bf p}_0,{\bf k}_0)\, .
\en
Here, the reference vectors ${\bf p}_0$ and ${\bf k}_0$ belong to the shells
$r$ and $s$, respectively. In the next section, we shall solve this equation
and obtain the spectrum of both the bound and scattering states.

%%%-----------------------------------
\item \textbf{The Energy Spectrum}

%%%-----------------------------------
\begin{enumerate}
\item \textbf{The Choice of the Model}

In order to gain insight into the volume-dependence of the three-particle 
spectrum, we shall solve the equation (\ref{eq:qc-gnl}) in the irrep 
$\Gamma=A_1^+$ in a simple model, described in section~\ref{sec:formalism}. 
The parameters of the model are fixed as follows. First, we take $m=a=1$. 
This means that there exists a bound dimer with the binding energy 
$B_2=\frac{1}{ma^2}=1$. Further, we fix the ultraviolet cutoff $\Lambda=225$ 
-- large enough so that all cutoff artifacts can be safely neglected. The 
last remaining parameter $H_0(\Lambda)$ is fixed from the requirement that 
there exists a three-body bound state with the binding energy $B_3=10$. This 
gives $H_0(\Lambda)=0.192$. Of course, for a different choice of $\Lambda$, 
one may adjust $H_0(\Lambda)$, so that all low-energy spectrum remains the 
same.

%%%-----------------------------------
\item \textbf{Bound States}

Except of the already mentioned deeply bound three-body state with $B_3=10$,
the model contains an extremely shallow bound state with $B_3=1.016$ --
just below the particle-dimer threshold. Since the characteristic size of 
such system is much larger than that of the dimer, it is conceivable that it 
should behave like a two-body bound state of a particle and a dimer. For the 
deep bound state, two-particle and three-particle bound-state scales are 
comparable in magnitude and hence one might expect a behavior that 
interpolates between the extreme cases of a three-particle shallow bound 
state and a particle-dimer bound state.

It is quite intriguing that the study of the finite-volume spectrum allows 
one to make a choice among the above alternatives. It is, in particular, 
well known that the L\"uscher equation leads to the finite-volume correction
$\propto\exp(-\Delta L)/L$ to the infinite-volume binding energy (here, 
$\Delta$ is some mass scale determined by the kinematics).
The functional dependence of the three-body shallow bound state
is different,\\ 
$\propto\exp(-\Delta L)/L^{3/2}$, see, e.g., Refs.~\cite{Meissner:2014deac}. 
So, the volume dependence of the binding energy contains information about 
the nature of the three-body bound states.

Among the solutions of the quantization condition (\ref{eq:qc-gnl}) one 
can readily identify the levels that tend to the infinite-volume binding 
energies. It is seen that the volume-dependence of the shallow bound state 
can be approximated by the function \\
$\propto\exp(-\Delta L)/L$ very well. For the deeper state, the situation 
is different and one needs a linear combination of the above two functions 
to get a decent fit. All this of course perfectly matches our expectations. 
More details can be found in Ref.~\cite{Doring:2018xxxc}.

%%%-----------------------------------
\item \textbf{Scattering States}

There are two types of the non-interacting energy levels, corresponding to 
the different asymptotic states in the three-body problem. In particular, we 
have the particle-dimer scattering states with different back-to-back 
momenta. The threshold for such states is given by the dimer binding energy 
and lies at $E_{th}=-1$. In addition to this, we have free 3-particle states 
with the threshold $E_{th}^0=0$. Of course, in the interacting theory, which 
we are considering, the energy levels are displaced from their ``free'' 
values, but the displacement is relatively small almost everywhere.
For this reason, one can identify the levels of the interacting theory with 
the different free levels.
%%%%%%%%%%%%%%%%%%%%%%%%%%%
\begin{figure}[t]
\begin{centering}
\includegraphics*{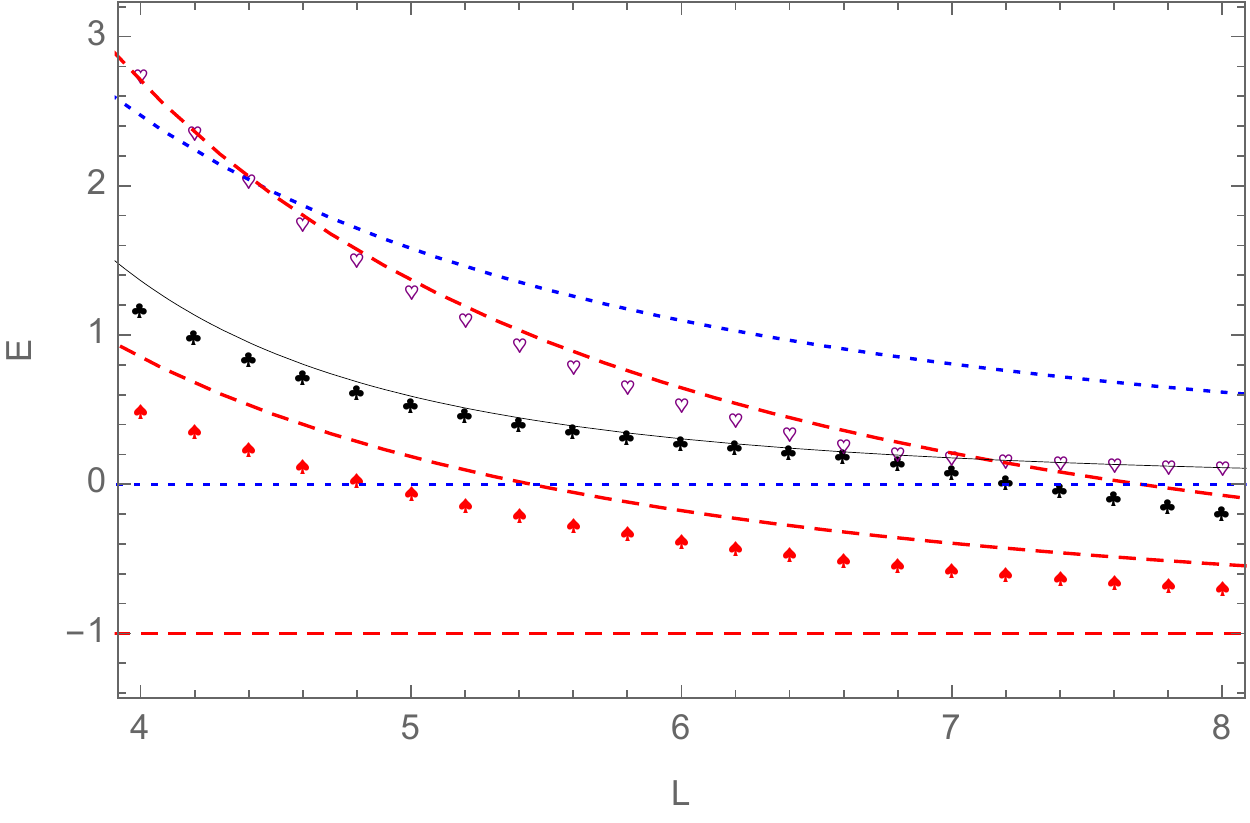}
\par\end{centering}

\caption{\label{fig:groundstate}
	The three lowest-lying states above threshold. The result obtained 
	using Eq.~(\protect\ref{eq:Beane-Sharpe}) is given by the black 
	solid curve. The red dashed curve shows the free particle-dimer 
	states with back-to-back momenta $(0,0,0)$, $(0,0,1)$ and $(1,1,0)$, 
	whereas the blue dotted lines denote the free three-particle states 
	(the lowest level at $E=0$ corresponds to the threshold, where all 
	three particles are at rest).}
\end{figure}
%%%%%%%%%%%%%%%%%%%%%%%%%%%

In Fig.~\ref{fig:groundstate}, the volume-dependence of the three lowest 
scattering states, obtained from the solution of Eq.~(\ref{eq:qc-gnl}), is 
shown. For comparison, in the same figure, we plot the free particle-dimer 
and three-particle energy levels.

Last but not least, we plot the displaced ground-state energy of three 
particles which, up to and including order $L^{-5}$, is given by the formula 
(see, e.g., Refs.~\cite{Beane:2007qrc,Hansen:2016fzjc})
\eq\label{eq:Beane-Sharpe}
	E=\frac{12\pi a}{L^3}-\frac{12 a^2}{L^4}\,{\cal I}
	+\frac{12 a^3}{\pi L^5}({\cal I}^2+{\cal J})+O(L^{-6})\, ,
\en
where ${\cal I}\simeq -8.914$ and ${\cal J}\simeq 16.532$ are numerical 
constants. It is seen that, at this accuracy, the energy shift from the
unperturbed value $E=0$ contains only the single parameter $a$ which is 
fixed from the beginning. Thus, Eq.~(\ref{eq:Beane-Sharpe}) constitutes a 
prediction that allows to identify the three-particle ground state.

The interpretation of the levels, shown in Fig.~\ref{fig:groundstate} is 
crystal clear. The lowest level is a displaced particle-dimer ground state. 
Its relatively large displacement is caused by the existence of a very 
shallow three-body bound state, which pushes all other particle-dimer levels 
up. At small values of $L$, the next level is the displaced three-body ground 
state. Around $L\simeq 7$, free particle-dimer and three-particle levels
cross each other. This fact leads to the avoided level crossing in the 
interacting spectrum, the second level goes down to the particle-dimer 
threshold and the third level becomes the displaced ground-state 
three-particle level. This pattern repeats itself each time, when we have a 
crossing of two levels corresponding to the different asymptotic states.
For more details, see Ref.~\cite{Doring:2018xxxc}.
\end{enumerate}

%%%-----------------------------------
\item \textbf{Conclusions}

\begin{itemize}

\item[i)] We use the effective field theory approach in a finite volume
	to obtain the three-particle quantization condition. The fit of 
	the solutions of this equation to the three-particle finite-volume 
	spectrum determines the particle-dimer coupling constant(s), which 
	can be further used in the Bethe-Salpeter equation to reconstruct 
	the infinite-volume scattering amplitudes in the three-particle 
	sector. This is the essence of the approach proposed in
	Refs.~\cite{pang1c,pang2c}.
\item[ii)] Using the octahedral symmetry of the problem in a cubic box, the 
	quantization condition was projected onto the different irreps of 
	this symmetry group.
\item[iii)] The equation was numerically solved in the $A_1^+$ irrep, both 
	for the bound states and scattering states. It was shown that a nice 
	interpretation of the energy levels is possible in terms of the 
	different asymptotic states of the three-particle system. In 
	particular, it was shown that the avoided level crossing occurs at 
	those values of $L$, when the free levels, corresponding to these 
	states, have the same energy.
\end{itemize}

%%%-----------------------------------
\item \textbf{Acknowledgments}

The author acknowledges the support from the CRC 110 ``Symmetries 
and the Emergence of Structure in QCD'' (DFG grant no. TRR~110 and 
NSFC grant No. 11621131001).  This research was also supported in 
part by Volkswagenstiftung under contract no. 93562 and by Shota 
Rustaveli National Science Foundation (SRNSF), grant No. 
DI--2016--26.
\end{enumerate}

%%%-----------------------------------

%%%%%%%%%%%%%%%%%%%%%%%%%%%%%%%%%%%%%%%%%%%%%%%%%%%%%%%%%%%%%%%%%%%%%%%%%
\newpage
\subsection{S-matrix Approach to Hadron Gas: a Brief Review}
\addtocontents{toc}{\hspace{2cm}{\sl Pok Man Lo}\par}
\setcounter{figure}{0}
\setcounter{table}{0}
\setcounter{equation}{0}
\setcounter{footnote}{0}
\halign{#\hfil&\quad#\hfil\cr 
\large{Pok Man Lo}\cr
\textit{Institute of Theoretical Physics}\cr
\textit{University of Wroclaw}\cr
\textit{PL-50204 Wroc\l aw, Poland~\&}\cr
\textit{Extreme Matter Institute (EMMI), GSI}\cr
\textit{D-64291 Darmstadt, Germany}\cr}

%%%-----------------------------------
\begin{abstract}
I briefly review how the S-matrix formalism can be applied to analyze 
a gas of interacting hadrons.
\end{abstract}

%%%-----------------------------------
\begin{enumerate}
\item \textbf{Introduction}

The S-matrix formulation of statistical mechanics by Dashen, Ma, and 
Bernstein~\cite{dmbd} expresses the grand canonical potential in terms of 
the scattering matrix elements. When applied to describe the system of 
interacting hadrons~\cite{Prakashd}, the logarithm of the partition 
function can be written as a sum of two pieces~\footnote{For simplicity, 
we present the formulae for the case of vanishing chemical potentials 
and Boltzmann statistics. In practical calculations, suitable fugacity 
factors and the correct quantum statistics have to be implemented.}:
\begin{align}
        \ln Z &= \ln Z_0 + \Delta \ln Z,
\end{align}
\noindent where
\begin{align}
	\ln Z_0 &=  V \times \sum_{i \in {\rm gs}} d_i \int \frac{d^3 k}{(2 
	\pi)^3} \, e^{-\beta \sqrt{k^2 + m_i^2}}
\end{align}
\noindent is the grand potential for an uncorrelated gas of particles that 
do not decay under the strong interaction (i.e., ground-state particles), 
such as pions, kaons, and nucleons. The interacting part of the grand 
potential, $\Delta \ln Z$, can be written in the form
\begin{align}
	\label{eq:smat1}
	 \Delta \ln Z &=  V \times \int d \sqrt{s} \, \, \frac{d^3 P}{(2 \pi)^3} 
	\, e^{-\beta \sqrt{P^2 + s}} \,
	\rho_{\rm eff}(\sqrt{s}),
\end{align}
\noindent where $\sqrt{s}$ is the invariant mass of the relevant 
scattering system. The quantity $\rho_{\rm eff}(\sqrt{s})$ can be 
understood as an effective level density due to the interaction.
A key step of the S-matrix approach to study thermodynamics is to 
identify such effective level density, in the low density limit where only 
binary collisions are important, with 
\begin{align}
	\label{eq:smat2}
	\rho_{\rm eff}(\sqrt{s}) \rightarrow \rho_{\rm smat}(\sqrt{s}) = 
	\sum_{\rm int} d_{IJ} \times \frac{d}{d \sqrt{s}}
	\, \left( \frac{1}{\pi} \, \delta_{IJ}(\sqrt{s}) \right).
\end{align}
\noindent Here the sum is over all interaction channels, $d_g$ is the 
relevant degeneracy factor, and $\delta_{IJ}(\sqrt{s})$ is the scattering 
phase shift. Note that the standard Hadron Resonance Gas (HRG)
model~\cite{BraunMunzinger:2003zdd} can also be expressed in this form, 
with the replacement
\begin{align}
	\rho_{\rm eff}(\sqrt{s}) \rightarrow \rho_{\rm HRG}(\sqrt{s}) = 
	\sum_{\rm res} d_{IJ} \times \frac{d}{d \sqrt{s}}
	\, \left( \theta(\sqrt{s}-m_{\rm res}) \right),
\end{align}
\noindent where the sum is now over all resonances, treated as point-like 
particles.
%-----------------------------------------
\begin{figure}[ht!]
\begin{centering}
  \includegraphics[width=0.9\textwidth]{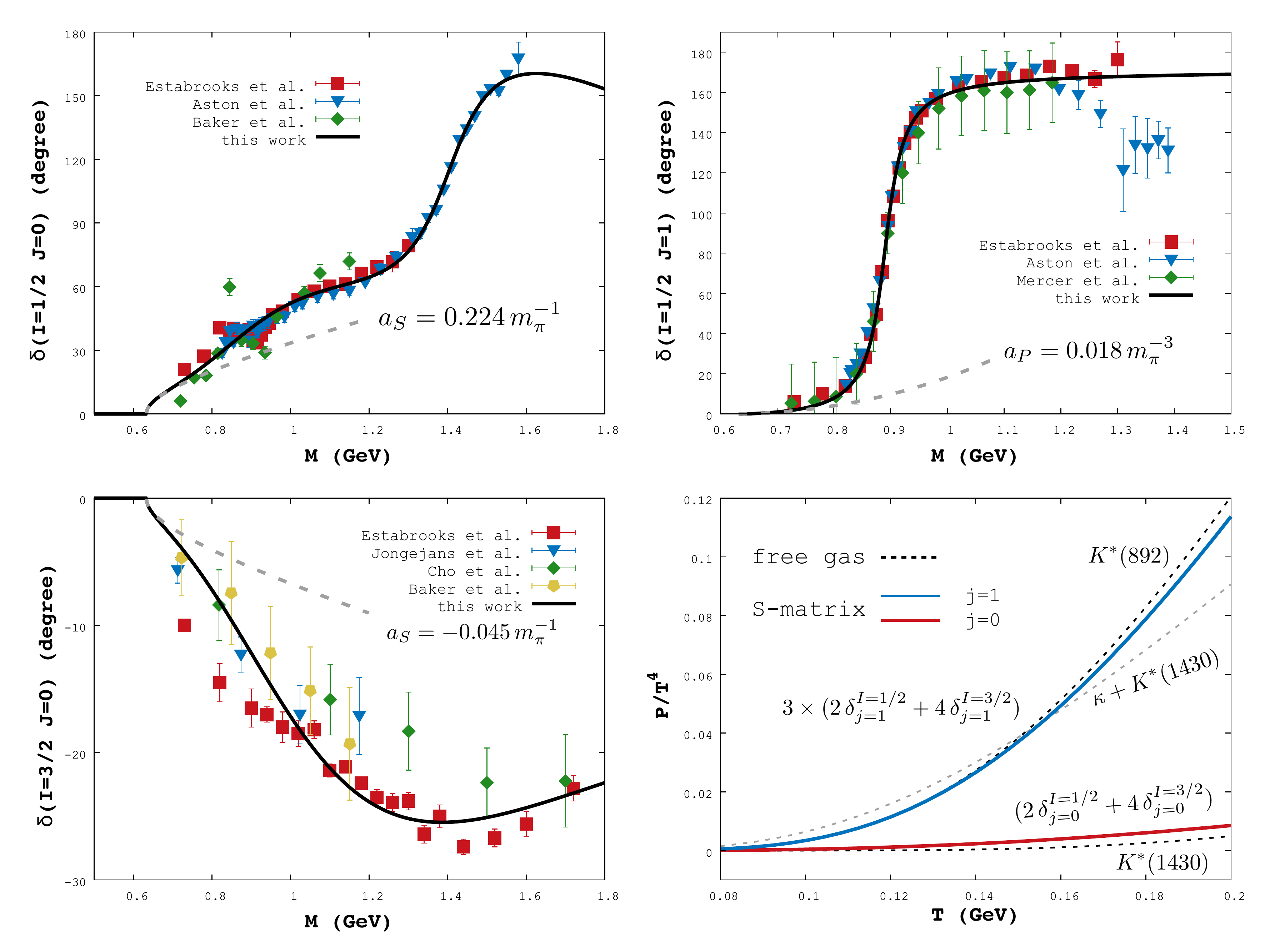}
\end{centering}

  \caption{$\pi K$ scattering phase shifts for the S-wave ($I = 1/2,
        3/2$) and the P-wave ($I=1/2$), with their contributions to
        thermodynamic pressure.}
        %\end{center} \label{fig:one}
\end{figure}
%----------------------------------------
%----------------------------------
\begin{figure}[ht!]
\begin{centering}
  \includegraphics[width=0.9\textwidth]{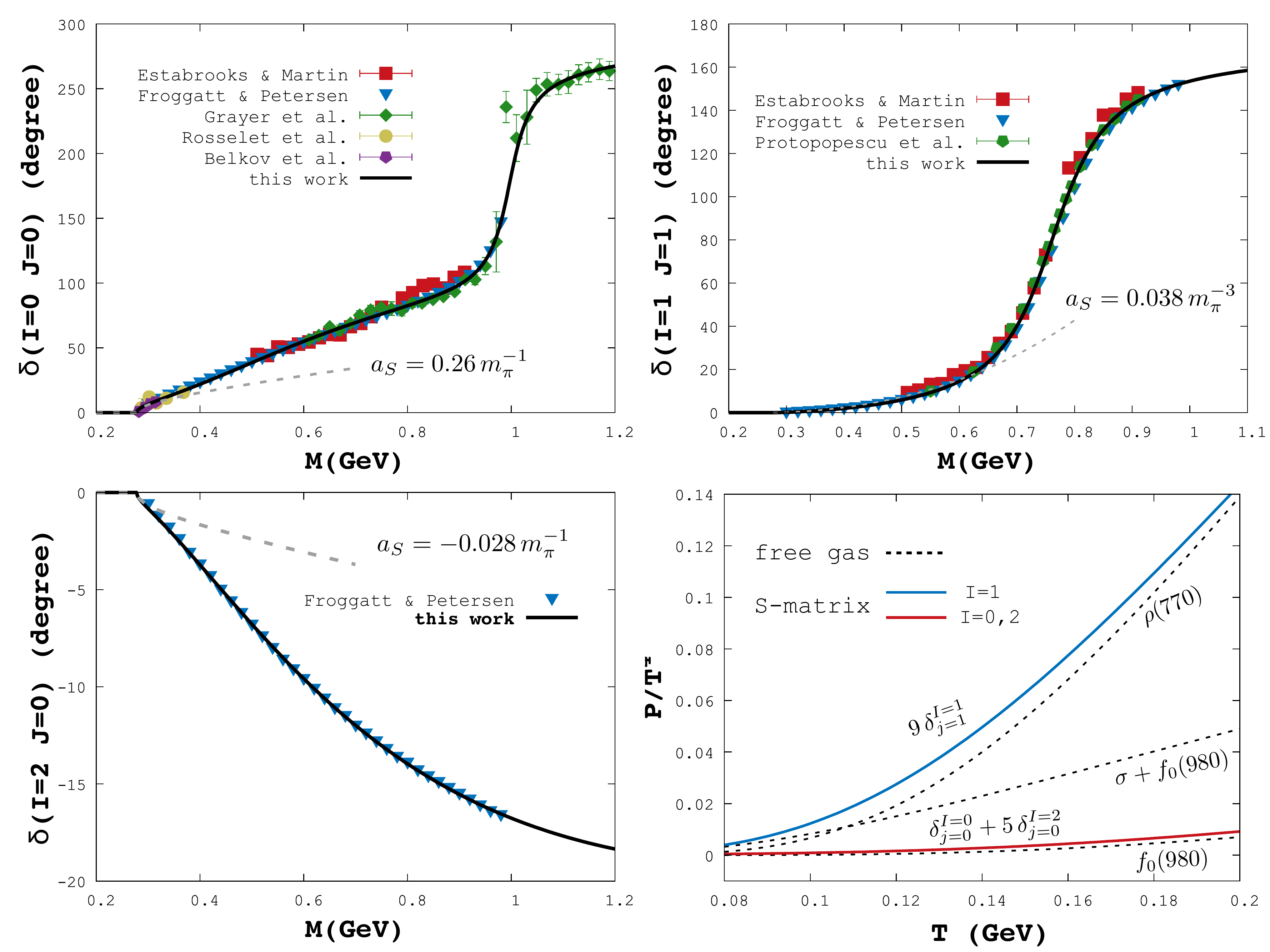}
\end{centering}

  \caption{$\pi \pi$ scattering phase shifts for $I = 0, 1, 2$ and their
        contributions to thermodynamic pressure.}
        %\end{center} \label{fig:one}
\end{figure}
%----------------------------------

When used in conjunction with the empirical phase shifts from scattering 
experiments, the S-matrix approach offers a model independent way to 
consistently incorporate both the attractive and repulsive
forces between hadrons for the study of thermodynamics. Indeed, a very 
detailed picture of hadronic interactions has emerged from the impressive 
volume of experimental data~\cite{Olive:2016xmwd}, carefully analyzed by 
theory such as chiral perturbation 
theory~\cite{Gasser:1983ygd,Oller:1997ngd},lattice 
QCD~\cite{Shepherd:2016dnid}, effective hadron models~\cite{Rapp:1999ejd}
and the classic potential models~\cite{Godfrey:1985xjd,Barnes:1991emd}.
Consequently we acquire very precise information on particle spectra, 
production mechanisms and decay properties of typical hadrons and even the 
exotics~\cite{eric:exoticad}. The method presented here is ideal for 
tapping into these resources in the field of hadron physics for the study 
of heavy ion collisions~\cite{Friman:2011zzd} and thermal properties of 
hadronic medium.

In the following, we employ the approach to study some simple hadronic 
systems.

%%%-----------------------------------
\item \textbf{Thermal System of Pions and Kaons}

In Figs.~1 and 2, we present results for the interaction contribution 
to the thermodynamic pressure, computed within the S-matrix approach for 
$\pi K$ and $\pi \pi$ scatterings respectively. Also shown are the 
corresponding results from the HRG model.

The key input here is the scattering phase shifts. Extensive 
experimental~\cite{Estabrooks:1974vud,Protopopescu:1973shd,Froggatt:1977hud, 
compilationd, Baker:1974krd,Estabrooks:1977xed,Aston:1987ird}
and theoretical~\cite{Ishida:1997nzd,Oller:1998hwd,Wilson:2015dqad,Briceno:2017maxd} 
efforts are devoted to study these quantities. An efficient way to compute 
the effective level density in Eq.~\eqref{eq:smat2} is to perform a 
phenomenological fit to the phase shift data~\cite{Ishida:1997nzd,kappad}.
This offers some insights into the relative importance of resonant and 
non-resonant contributions in an interaction channel. Nevertheless, more 
fundamental approaches such as chiral perturbation 
theory~\cite{Oller:1998hwd} and
LQCD~\cite{Wilson:2015dqad,Briceno:2017maxd} are in better position to 
implement known theoretical constraints and predict phase shifts for 
channels that are not yet measured.

Let us now discuss several key features of the results in  Figs.~1 and 2.

%%%-----------------------------------
\begin{enumerate}
\item \textbf{S-wave Channel}

In the $I = 1/2$ ($I = 0$) channel of $\pi K$ ($\pi \pi$) scattering
we observe a slowly rising phase shift in the low-energy region. This 
portion of the phase shift can reflect the physical properties of the 
unconfirmed $\kappa$- (confirmed $\sigma$-) meson. In particular, the 
phase shift does not reach $180^\circ$ before $K^\ast(1430)$ ($f_0(980)$) 
emerges. In the same energy range we also observe a major cancellation 
effect from the $I=3/2$ ($I=2$) channel after the multiplication of an 
appropriate degeneracy factor. The two effects combined severely suppress 
the overall pressure from these channels, down to the value of a free gas 
of $K^\ast(1430)$ ($f_0(980)$). This underpins the proposed prescription 
that $\kappa$- ($\sigma$-) meson should not be included in the HRG 
particle sum~\cite{kappad,Broniowskid}.

%%%-----------------------------------
\item \textbf{P-wave Channel}

For the $\pi K$ system, the S-matrix approach yields a similar result
on the thermodynamic pressure as the HRG model. The latter is based on a 
point-like treatment $K^\ast(892)$. On the other hand, the approach gives 
an enhanced effect from the two-body scattering beyond the free gas result 
for the $\pi \pi$ system.
%-------------------------------------------
\begin{figure*}[ht!]
\begin{centering}
 \includegraphics[width=0.48\textwidth]{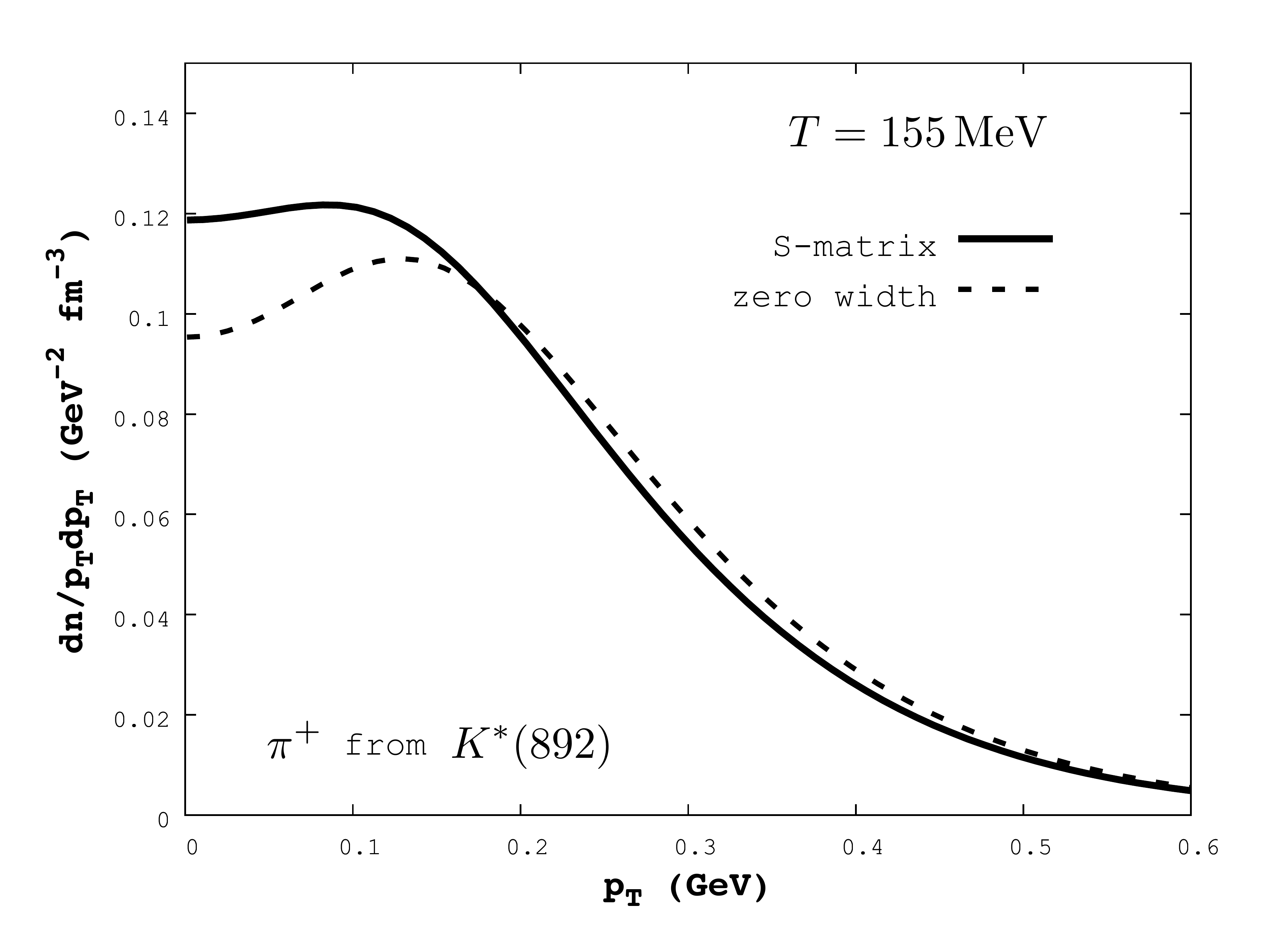}
 \includegraphics[width=0.48\textwidth]{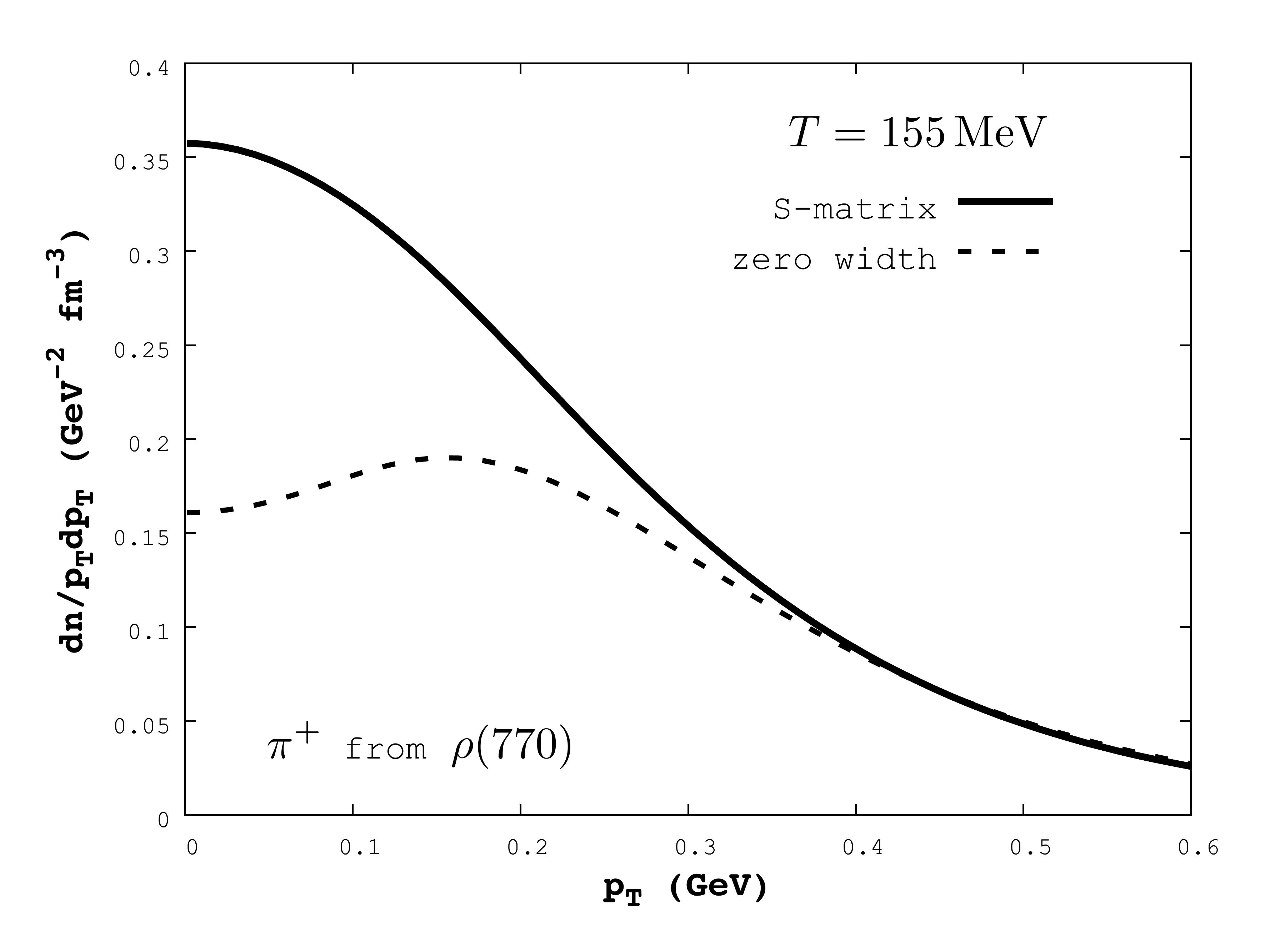}
\end{centering}

 \caption{The rapidity-integrated $p_T$ spectra of pions originated
        from the decay of $K^\ast(892)$ and $\rho(770)$ resonances (
        static source, $T= 155\,$ MeV ).
        The low-$p_T$ enhancement in the S-matrix treatment is clearly
        visible in both cases. For details see
        Refs.~\protect\cite{rhod,Lo:2017suxd,smatd}.}
        %\end{center} \label{fig:three}
\end{figure*}
%-------------------------------------------

It may appear that the point-particle treatment of the $K^\ast(892)$ 
resonance is a good approximation. However, this conclusion depends 
strongly on the observable of interest. For example, when the transverse 
momentum ($p_T$) distribution of the pions originating from a resonance
decay is studied~\cite{rhod,Lo:2017suxd}, differences between the two 
approaches become appreciable. This is shown in Fig.~3. The enhancement in 
the S-matrix approach at the low-momentum region of the spectra 
originates from a non-resonant threshold effect described in 
Ref.~\cite{Weinhold:1997igd,rhod,Lo:2017suxd,smatd}, which is 
commonly neglected in the standard HRG-based thermal models. This effect 
may help to account for the excess of soft pions ($p_T \leq 0.3$~GeV) 
observed in recent measurements of the LHC~\cite{Abelev:2013vead}
over the predictions by conventional fluid-dynamical 
calculations~\cite{rhod}.
\end{enumerate}

%%%-----------------------------------
\item \textbf{Baryon Sector}

The S-matrix approach has also been applied to study the pion-nucleon 
system~\cite{chiBQd}.  Using the empirical phase shifts from the SAID PWA 
database~\cite{Workman:2012hxd}, it is demonstrated that the natural 
implementation of the repulsive forces and the consistent treatment of 
broad resonances can improve our understanding of fluctuation observables 
computed in lattice QCD, such as the baryon electric charge correlation.

Lattice study of thermal QCD also indicates a larger interaction strength 
in the strange-baryon channel than that predicted by the HRG model using 
only the list of confirmed resonances~\cite{missingSd}. A simple extension 
of the HRG model~\cite{Lo:2015ccad}, which uses the lattice results to 
guide the incorporation of some extra hyperon states, suggests that the 
``missing'' states required is roughly consistent with the trend of the 
observed but unconfirmed resonances (1 and 2 stars) in the PDG. The 
corresponding S-matrix based analysis is currently pursued. It is clear 
that a detailed experimental knowledge of the hyperon spectrum is 
critical for this task.

%%%-----------------------------------
\item \textbf{Summary}

The S-matrix approach offers a consistent way to incorporate attractive 
and repulsive forces between hadrons. Using the input of empirical phase 
shifts from hadron scattering experiments, the important physics of 
resonance widths and the contribution from purely repulsive channels are 
naturally included. Further research in extending the scheme~\cite{smatd} 
to include inelastic effects and three-body 
scatterings~\cite{Kaminski:1996dad,Ronchen:2012egd,Mai:2017bged} has 
begun.

The proposed new $K_L$-beam facility can potentially strengthen our 
understanding of the hadron spectroscopy, such as the investigation of 
the ``missing" hyperon states. This is crucial for a reliable thermal 
description of the hadronic medium within the S-matrix approach.
We expect interesting results when applying the approach to explore 
various phenomena of heavy ion collisions.

%%%-----------------------------------
\item \textbf{Acknowledgments}

I would like to thank Igor I. Strakovsky for the invitation to the 
Workshop. I am also grateful to Ted Barnes, Michael D{\"o}ring, Jose 
R. Pelaez and James Ritman for the encouraging remarks and the fruitful 
discussion. This research benefits greatly from the discussion with Hans 
Feldmeier in GSI. I would also like to thank Bengt Friman, Krzysztof 
Redlich, and Chihiro Sasaki for the productive collaboration. Support 
from the Extreme Matter Institute EMMI at the GSI is gratefully 
acknowledged. This work was partially supported by the Polish National 
Science Center (NCN), under Maestro grant DEC--2013/10/A/ST2/00106.
\end{enumerate}

%%%-----------------------------------

%%%%%%%%%%%%%%%%%%%%%%%%%%%%%%%%%%%%%%%%%%%%%%%%%%%%%%%%%%%%%%%%%%%%%%%%%
\newpage
\subsection{Measurement of Hadronic Cross Sections with the 
	BaBar Detector}
\addtocontents{toc}{\hspace{2cm}{\sl Alessandra Filippi (for the BaBar Collaboration)}\par}
\setcounter{figure}{0}
\setcounter{table}{0}
\setcounter{equation}{0}
\setcounter{footnote}{0}
\halign{#\hfil&\quad#\hfil\cr 
\large{Alessandra Filippi}\cr
\textit{I.N.F.N. Sezione di Torino}\cr
\textit{10125 Torino, Italy}\cr}

%%%-----------------------------------
\begin{abstract}
An overview of the measurements performed by BaBar of $e^+e^-$ 
annihilation cross sections in exclusive channels with kaons and 
pions in the final state is given. The Initial State Radiation 
technique, which allows to perform cross section measurements in 
a continuous range of energies, was employed.
\end{abstract}

%%%-----------------------------------
\begin{enumerate}
\item \textbf{Introduction}

The BaBar experiment~\cite{re:babarh} at the SLAC National 
Accelerator Laboratory has performed, over the last decade, a 
complete set of measurements of $e^+e^-$ annihilation cross 
sections of exclusive hadronic channels at low energies,
exploiting the Initial State Radiation (ISR) technique. The 
main purpose of this study is to provide the most precise and 
complete input for the calculation of the muon anomalous 
magnetic moment $a_\mu = (g_\mu -2)/2$ and the running 
electromagnetic constant $\alpha_{EW}$. However, also several 
important and useful indications on the composition of the light
and charmed meson spectrum could be obtained as by-products, by 
the inspection of the trends of the cross sections and the 
composition of the intermediate states of the studied reactions, 
as well as their decay branching fractions.

The knowledge of the total hadronic cross section is fundamental 
for the determination of the hadronic vacuum polarization 
contribution, at leading order, to $a_\mu$~\cite{re:gminus2h}. 
It is well known that a discrepancy at the level of 3.5$\sigma$ 
exists between the value of the muon anomolous magnetic moment 
expected from the Standard Model and what is experimentally 
observed, and the hadronic term of the sum adding up to $a_\mu$ 
is the one that bears the largest uncertainty. The dominant 
contribution to this term is played by the hadronic cross sections 
below 2~GeV, that until BaBar were measured mainly inclusively. 
Only recently measurements of exclusive channels could be performed
accurately enough. The goal of BaBar, almost completely 
accomplished, was to  provide measurements for all the
exclusive channels below 2~GeV, comparing them with previous 
inclusive measurements and pQCD predictions.

In the following a description of the measurements performed for
all channels containing kaons in final states composed by a total of 
three or four particles will be described.

%%%-----------------------------------
\item \textbf{The Initial State Radiation (ISR) Technique and Kaon 
	Detection with BaBar}

BaBar operated at the PEP-II machine at SLAC and had been taking 
data for about 10 years, up to 2008, integrating a luminosity of 
about 500 fb$^{-1}$ of $e^+e^-$ annihilations taken at a few fixed 
energies values corresponding mainly to the excitation of 
$\Upsilon(4S)$, with smallest samples at the peaks of 
$\Upsilon(2S)$ and $\Upsilon(3S)$.

In spite of the fixed energy in the center of mass, resorting to 
the ISR technique it was however possible to perform cross section
measurements in a continuous range of energies, potentially
up to at least 8~GeV. Given a hadronic system $f$, the differential
cross section of the process $e^+e^-\rightarrow \gamma f$ in which a 
photon is radiated from the electron or positron  before their interaction 
can be related to that of the non-radiative
process by means of the equation:
\begin{equation}
	\frac{d\sigma_{e^+e^-\rightarrow\gamma f}(s,m_f)}{dm_f 
	d\cos\theta^\ast_\gamma} =
	\frac{2m_f}{s}W(s,x,\theta^\ast_\gamma)\sigma_{e^+e^-\rightarrow f}(m_f)
\end{equation}
where $\sqrt{s}$ is the $e^+e^-$ center-of-mass energy, $E_\gamma$ and 
$\theta^\ast_\gamma$ are the energy and the center-of-mass polar angle of 
the emitted ISR photon, and $x = 2E_\gamma/\sqrt{s}$. 
$W(s,x,\theta^\ast_\gamma)$ is a QED radiation function known with an 
accuracy better than 0.5\%. Therefore, the measurement of reactions in 
which an additional photon, of variable energy, is detected together
the hadronic system may allow to infer the non radiative cross-section in 
a continuous range of energies. The ISR photon is generally emitted along 
the $e^+e^-$ collision axis, and the hadronic system is mainly produced 
back-to-back with respect to it. Due to the limited detector acceptance, 
the mass region below 2~GeV can only be studied if the ISR photon is 
detected: to perform exclusive cross section measurements the events are 
required to feature a photon with a center-of-mass energy larger than 
3~GeV, and a fully reconstructed and identified recoiling hadronic system. 
In order for the latter to be fully contained in the detector fiducial 
volume, the ISR photon must be emitted at large angles.

Among the main advantages of the large-angle ISR method over conventional 
$e^+e^-$ measurements one can count on a weak dependence of the detection 
efficiency on the dynamics of the hadronic system and on its invariant
mass; therefore, measurements in wide energy ranges can be performed
applying the same selection criteria. The exclusivity of the final states 
moreover allows the application of stringent kinematic fits, which can 
largely improve the mass resolution and provide effective background
suppression.

The identification of charged kaons in BaBar was performed through 
standard techniques based on specific energy loss, time of flight and 
Cerenkov radiation, combined to provide a likelihood value for each
particle identification hypothesis. The identification efficiency of 
charged kaons was as large as 89\%, with a $K\pi$ misidentification rate 
not larger than 2\%. Neutral $K^0_S$'s were identified through their
$\pi^+\pi^-$ decay reconstucting a displaced vertex formed by two 
oppositely charged tracks, while neutral $K^0_L$'s were identified 
following their nuclear interaction with the material of the 
electromagnetic calorimeter, which produced an energy cluster with a shape 
not consistent with that typical of a photon. The minimum required energy 
deposition per cluster was typically 200~MeV. The $K^0_L$ detection 
efficiency was measured through the $e^+e^-\rightarrow \phi\gamma$ events, 
selected without the detection of the $K^0_L$~\cite{re:KsKpih}.

By means of charged and neutral kaon identification the measurement of 
cross sections of basically all exclusive channels with kaons and pions 
(except those with two $K^0_L$'s) could be performed:; in this way the use 
of isospin relationships to assess the value of the total cross section 
could be avoided.

The measurement of cross sections in a given channel proceeds from 
counting of the number of events in intervals of total center-of-mass 
energy, after proper background subtraction (usually based on Monte Carlo
simulations, normalized to the existing data in rate and shape); the 
number of events is then normalized to the detection efficiency, and to 
the integrated luminosity in the selected energy range.

%%%-----------------------------------
\newpage
\item \textbf{Three Particles Final States: $K\overline K\pi$}
%%%-----------------------------------
\begin{enumerate}
\item \textbf{Measurement of $e^+e^-\rightarrow K^0_SK^\pm K^\mp$ Cross 
	Section}

The purpose of this analysis was to look for possible signatures of the 
$Y(4260)$ state~\cite{re:KsKpih}. As can be seen by the trend of the cross 
section reported in Fig.~\ref{fig:xsec_KsKpi}(a) a clean signal due
to the $J/\psi$ can be observed but, as shown in the inset, just an excess 
of events is present at about 4.2~GeV, with a significance of 3.5$\sigma$ 
only. This prevents to assess the existence of any new state, but just 
allows to quote an upper limit, at 90\% C.L., for the electronic width of 
the possible resonance times its branching ratio in the studied channel:
$U.L.(\Gamma^{Y(4260)}_{ee}\mathcal{B}^{Y(4260)}_{K^0_SK^\pm\pi^\mp})$ = 
0.5~eV. The cross section reaches a maximum of about 4~nb at 
$\sim$1.7~GeV, indicating a dominant contribution from the $\phi(1680)$ 
intermediate state decaying in $K^\ast K$.
%%%-----------------------------------
\begin{figure}[h]
\centering
\begin{tabular}{lr}
\includegraphics[height=5.5truecm,clip=true]{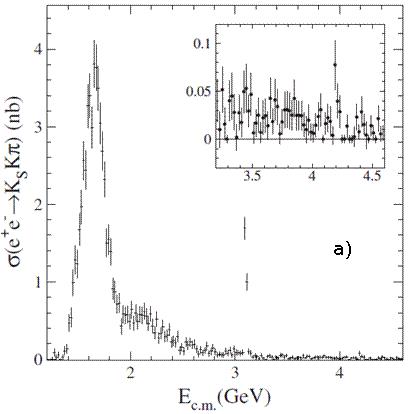} &
\includegraphics[width=6.truecm, height=5.8truecm, clip=true, trim= 14 14 0 0]{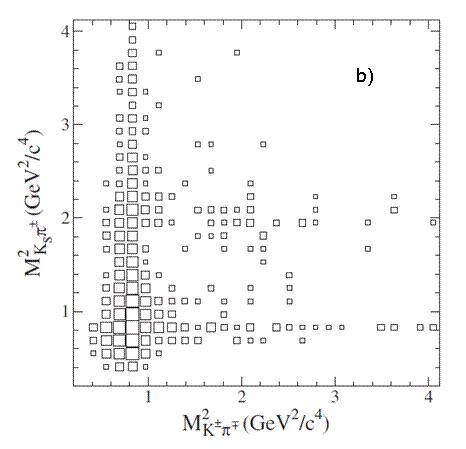}
\end{tabular}

\caption{a) The $e^+e^-\rightarrow K^0_SK^\pm \pi^\mp$ cross section 
	measured by BaBar. b) Dalitz plot of the $e^+e^-\rightarrow 
	K^0_SK^\pm K^\mp$ reaction: squared invariant masses of 
	the $(K^0_S\pi^\pm)$ versus the $(K^\pm\pi^\mp)$ systems.}
	\label{fig:xsec_KsKpi}
\end{figure}
%%%-----------------------------------

From the inspection of the Dalitz plot shown in 
Fig.~\ref{fig:xsec_KsKpi}(b), which is fairly asymmetric, it is evident
that the main contribution to the intermediate state is given by 
$K^\ast(892)^\pm K^\mp$ and $K^\ast(892)^0 K^0_S$ and, at masses above 
2~GeV, by $K^\ast_2(1430)^\pm K^\mp$ and $K^\ast_2(1430)^0 K^0_S$.
A Dalitz plot analysis was performed~\cite{re:KsKpih}  to extract the 
isovector and isoscalar contributions to the cross section: interference 
effects model differently the production of charged and neutral $K^\ast$,
which is mirrored in the asymmetry of the Dalitz plot. The dominant 
component was found to be the isoscalar one, with a clear resonant 
behavior to be identified with the mentioned $\phi(1680)$, while a broad 
contribution of marginal importance, that could be identified as the 
$\rho(1450)$, emerged from the fit.

%%%-----------------------------------
\item \textbf{Measurement of $e^+e^-\rightarrow K^+K^- \pi^0$ Cross 
	Section}

As the previous analysis, also in this case the main target was the 
possible identification of the Y(4260) state~\cite{re:KsKpih}. The trend 
of the measured cross section, reported in Fig.~\ref{fig:xsec_KpKmpi0}(a), 
shows even less evidence than in the previous case for the existence of an 
event excess, and again only un upper limit could be quoted for the decay 
branching ratio to this channel times the electronic width of the 
tentative signal, 0.6~eV at 90\% C.L. A clear indication for $J/\psi$ 
production emerges from the cross section, whose maximum, $\sim 1$~nb, is 
reached at about 1.7~GeV.
%%%-----------------------------------
\begin{figure}[h]
\centering
\begin{tabular}{cc}
\includegraphics[height=5.5truecm,clip=]{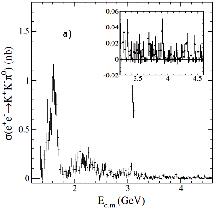} &
\includegraphics[width=6.truecm, height=5.6truecm, clip=true, trim= 25 25 0 0]{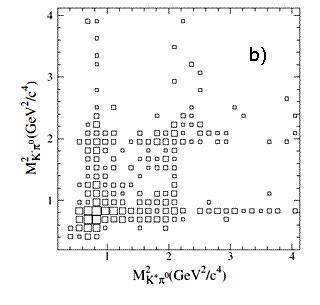}
\end{tabular}

\caption{a) The $e^+e^-\rightarrow K^+K^- \pi^0$ cross section measured by 
	BaBar. b) Dalitz plot of the $e^+e^-\rightarrow K^+K^- \pi^0$ 
	reaction: squared invariant masses of the $(K^-\pi^0)$ versus
	the $(K^+\pi^0)$ systems.}
	\label{fig:xsec_KpKmpi0}
\end{figure}
%%%-----------------------------------

The reaction Daliz plot, shown in Fig.~\ref{fig:xsec_KpKmpi0},
differently from the previous analysis is symmetric and features a 
dominant production of the final state through $K^\ast(892)^\pm K^\mp$ and 
$K^\ast_2(1430)^\pm K^\mp$. Selecting the $K^+K^-$ invariant mass system 
in the $\phi(1020)$ mass window, the $\phi\pi^0$ system could be studied: 
even if the available statistics was limited, as can be seen by the green 
points in Fig.~\ref{fig:phipi0}, some indications for the existence of 
stuctures like $\rho(1700)/\rho(1900)$, observed as a dip, and a possible 
enhancement at about 1480~MeV (a possible hint for the long-sought isospin 
1 $C(1480)$?~\cite{re:landsbergh}) are present.
%%%-----------------------------------
\begin{figure}[h]
\centering
\includegraphics[width=6truecm, height=6.0truecm,clip=true]{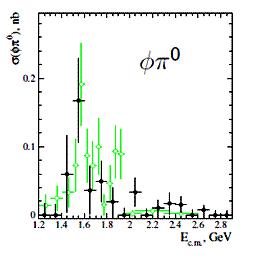}

\caption{$(\phi\pi^0)$ cross section for events selected from the
	$e^+e^-\rightarrow K^+K^- \pi^0$ reaction (green points) and 
	from the $e^+e^-\rightarrow K^0_SK^0_L \pi^0$ reaction (black 
	points).}\label{fig:phipi0}
\end{figure}
%%%-----------------------------------

%%%-----------------------------------
\item \textbf{Measurement of $e^+e^-\rightarrow K^0_SK^0_L \pi^0$ Cross 
	Section}

Fig.~\ref{fig:xsec_KSKLpi0} shows the trend of the $e^+e^-\rightarrow 
K^0_SK^0_L \pi^0$ cross section for this channel~\cite{re:KSKLpi0h}, whose 
maximum is about 3~nb corresponding to the excitation energy of the 
$\phi(1680)$. The systematic uncertainty of the measurement is 10\% at the 
peak, and increases up to 30\% at 3~GeV. For the first time a clear 
signal of $J/\psi$ is observed in the $K^0_SK^0_L\pi^0$  decay channel.  
The dominant intermediate state is $K^\ast(892)^0 K^0$, which almost 
saturates the channel intensity: a clear signal is obtained for 
$K^\ast(892)^0$ for the combinations with both $K^0_S$ and $K^0_L$. Some 
hints for the production of the $K^\ast_2(1430)^0 K^0$ are also observed.
%%%-----------------------------------
\begin{figure}[h]
\centering
\begin{tabular}{c}
\includegraphics[width=5.5truecm,clip=]{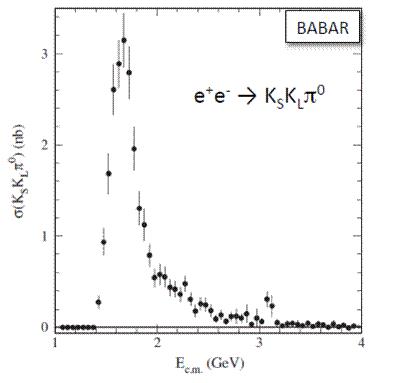}
\end{tabular}

\caption{The $e^+e^-\rightarrow K^0_SK^0_L \pi^0$ cross section measured 
	by BaBar.} \label{fig:xsec_KSKLpi0}
\end{figure}
%%%-----------------------------------

The invariant mass of the $\phi(1020)\pi^0$ system, with the $\phi(1020)$ 
selected through its $K^0_SK^0_L$ decay, is shown by the black points in 
Fig.~\ref{fig:phipi0}: even in this case some hints for a resonant
activity at about 1.6~GeV can be observed, which could be a signature for 
an isospin 1 exotic structure.
\end{enumerate}

%%%-----------------------------------
\item \textbf{Four Particles Final States}
%%%-----------------------------------
\begin{enumerate}
\item \textbf{Measurement of $e^+e^-\rightarrow K^0_SK^0_L \pi^+ \pi^-$ 
	sross section}

In the $e^+e^-\rightarrow K^0_SK^0_L \pi^+ \pi^-$ channel the contribution 
of the background, coming both from ISR and non-ISR multihadronic events, 
is rather sizeable. Also the systematic uncertainty of the cross section, 
shown in Fig.~\ref{fig:xsec_KSKL2pi}(a), is larger, rising from about 10\% 
at the peak, about 1~nb at $\sim$2~GeV, to 30\% at 2.5--3~GeV, and up to 
100\% for energies in the center of mass larger than 
3.4~GeV~\cite{re:KSKL2pih}. A clean signal of $J/\psi$  can be observed.
%%%-----------------------------------
\begin{figure}[h]
\centering
\begin{tabular}{cc}
\includegraphics[height=5.5truecm,clip=]{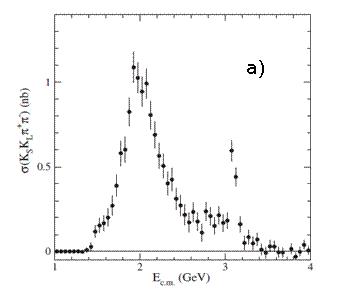} &
\includegraphics[width=5.9truecm, height=5.5truecm,clip=true, trim= 8 8   0 0]{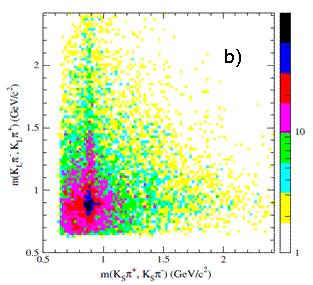}
\end{tabular}

\caption{a) The $e^+e^-\rightarrow K^0_SK^0_L \pi^+ \pi^-$ cross section 
	measured by BaBar. b) Scatter plot of the $e^+e^-\rightarrow 
	K^0_SK^0_L \pi^+ \pi^-$ reaction: invariant masses of the
	$(K^0_L\pi^\mp)$ versus the $(K^0_S\pi^+)$ systems.}
	\label{fig:xsec_KSKL2pi}
\end{figure}
%%%-----------------------------------

From the reaction scatter plot, shown in Fig.~\ref{fig:xsec_KSKL2pi}(b), 
clear bands of $K^\ast(892)^\pm$ can be observed with a weak indication for 
$K^\ast_2(1430)^\pm$. A strong correlated production of $K^\ast(892)^+
K^\ast(892)^-$ and $K^\ast(892)^\pm K^\ast_2(1430)^\mp$ appears.

%%%-----------------------------------
\item \textbf{Measurement of $e^+e^-\rightarrow K^0_SK^0_S \pi^+ \pi^-$ 
	Cross Section}

The cross section for the $e^+e^-\rightarrow K^0_SK^0_S\pi^+\pi^-$ 
reaction features a maximum of about 0.5~nb, affected by a 5\% error, at 
about 2~GeV~\cite{re:KSKL2pih} (see Fig.~\ref{fig:xsec_2KS2pi}(a)). From 
the inspection of the scatter plot shown in Fig. ~\ref{fig:xsec_2KS2pi}(b) 
one can note a clear correlated production of $K^\ast(892)$, dominant 
below 2.5~GeV, and a mild indication for $K^\ast_2(1430)^\pm$. There is 
basically no correlated production of $K^\ast(892)^\pm K^\ast_2(1430)^\mp$, 
and just a small strenght for the $K^\ast_2(1430)^\pm K_S\pi^\mp$ intermediate 
state can be observed.
%%%-----------------------------------
\begin{figure}[h]
\centering
\begin{tabular}{cc}
\includegraphics[height=5.5truecm,clip=]{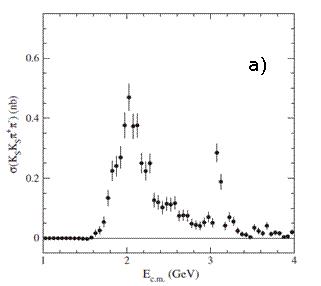} &
\includegraphics[width=6.0truecm, height=5.4truecm,clip=true, trim= 6 6 0 0]{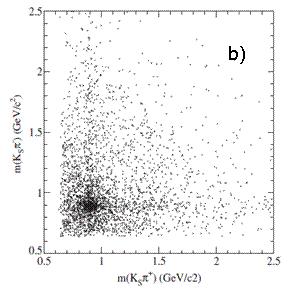}
\end{tabular}

\caption{a) The  $e^+e^-\rightarrow K^0_SK^0_S \pi^+ \pi^-$ cross section 
	measured by BaBar. b) Scatter plot of the $e^+e^-\rightarrow K^0_SK^0_S 
	\pi^+ \pi^-$ reaction: invariant masses of the $(K^0_S\pi^-)$ versus
	the $(K^0_S\pi^\mp)$ systems.} \label{fig:xsec_2KS2pi}
\end{figure}
%%%-----------------------------------

%%%-----------------------------------
\newpage
\item \textbf{Measurement of $e^+e^-\rightarrow K^+K^- \pi^+ \pi^-$ 
	Cross Section}

The cross section for this channel was measured several years ago by 
the DM1 Experiment~\cite{re:dm1h}, up to an energy of about 2.5~GeV: 
the comparison with the most recent measurements by BaBar~\cite{re:KpKm2pih}, 
reported in Fig.~\ref{fig:xsec_KPKM2pi}(a), show that the early results, 
represented by the red open points, were most probably overestimated and
affected by a sizeable systematic error. In the present case, the systematic
uncertainty was evaluated as about 20\% below 1.6~GeV, lowering to as little 
as 2\% in the region around 2~GeV to rise again up to about 10\% above 3~GeV. 
Narrow peaks from the formation of charmonium ($J/\psi$ and $\psi(2S)$) and 
possibly other structures which may be produced upon the opening of reaction 
thresholds are visible in the cross section, whose maximum is $\sim$4~nb at 
about 2~GeV.
%%%-----------------------------------
\begin{figure}[h]
\centering
\begin{tabular}{cc}
\includegraphics[height=5.5truecm,clip=]{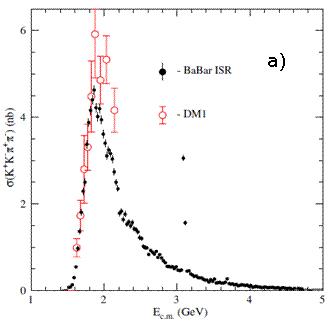} &
\includegraphics[width=11truecm, height=5truecm,clip=true, trim= 10 10 0 0]{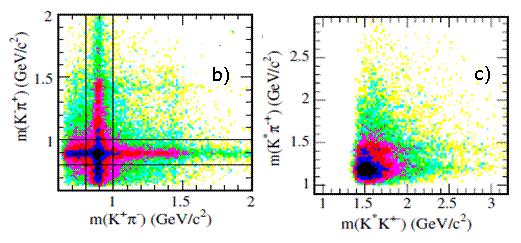}
\end{tabular}

\caption{a) The  $e^+e^-\rightarrow K^+K^- \pi^+ \pi^-$ cross section measured by 
	BaBar and DM1 (red open points~\protect{\cite{re:dm1h}}). 
	b)-c) Scatter 
	plots of the $e^+e^-\rightarrow K^+K^- \pi^+ \pi^-$ reaction:
	b) invariant masses of the $(K^-\pi^+)$ versus the $(K^+\pi^-)$ systems, 
	c) invariant masses of the $(K^\ast\pi^\mp)$ versus the $(K^\ast K^\pm)$ 
	systems.} \label{fig:xsec_KPKM2pi}
\end{figure}
%%%-----------------------------------

The scatter plots of the invariant masses of particles' pairs are displayed in
Fig.~\ref{fig:xsec_KPKM2pi}(b) and c): they show the presence of plenty of 
intermediate 
states, with clear signatures from $K^\ast(892)^0$ and $K^\ast_2(1430)^0$, and the 
evidence, seen as horizontal bands in Fig.~\ref{fig:xsec_KPKM2pi}(c), of 
contributions 
from the $K_1(1270)$ and $K_1(1400)$ axial excitations decaying in $K^\ast(892)^0\pi^\pm$.  
The $(\pi^+\pi^-)$ invariant mass spectrum (not shown) displays a clear enhancement
corresponding to the $\rho(770)$ signal, which can be an evidence for the possible 
decay of the two axial mesons in $K\rho$. No other signal is observed in the 
$(\pi^+\pi^-)$ system.

%%%-----------------------------------
\item \textbf{Measurement of $e^+e^-\rightarrow K^+K^- \pi^0 \pi^0$ 
	Cross Section}

The cross section for the $e^+e^-\rightarrow K^+K^-\pi^0\pi^0$ reaction is shown 
in Fig.~\ref{fig:xsec_KpKm2pi0}: it does not exceed 1 nb and the maximum is located 
around 2~GeV~\cite{re:KpKm2pih}. A clear charmonium signal is present.
The systematic uncertainty of the cross section is about 7\% at low energies,
rising to 18\% above 3~GeV.  The dominant contribution to the intermediate states 
is played by $K^\ast(892)$.
%%%-----------------------------------
\begin{figure}
\centering
\begin{tabular}{cc}
\includegraphics[height=5.5truecm,clip=]{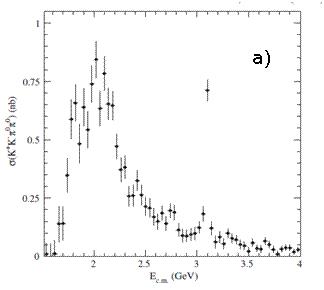} &
\includegraphics[width=6.2truecm, scale=1.5,clip=true, trim= 0 17 0 0]{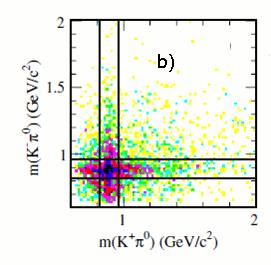}
\end{tabular}

\caption{a) The  $e^+e^-\rightarrow K^+K^- \pi^0 \pi^0$ cross section measured by 
	BaBar. b) Scatter plot of the $e^+e^-\rightarrow K^+K^- \pi^0 \pi^0$ reaction:
	invariant masses of the $(K^-\pi^0)$ versus the $(K^+\pi^0)$ systems.}
	\label{fig:xsec_KpKm2pi0}
\end{figure}
%%%-----------------------------------

The corresponding scatter plot of the invariant masses of the $(K^-\pi^0)$
versus the $(K^+\pi^0)$ systems is shown in Fig. 
\ref{fig:xsec_KpKm2pi0}(b), 
and displays a clear correlated production of $K^\ast(892)$ pairs as well as
$K^\ast(892)^+K^\ast_2(1430)^-$; however, no evidence of resonant production 
in the three particles systems $(K^+K^-\pi^0)$ or $(K^\pm\pi^0\pi^0)$
can be observed.

%%%-----------------------------------
\item \textbf{Measurement of $e^+e^-\rightarrow K^0_SK^0_L \pi^0 \pi^0$ 
	Cross Section}

The $e^+e^-\rightarrow K^0_SK^0_L \pi^0 \pi^0$ cross section was found to be 
relatively small, reaching at most 0.6~nb at its maximum at $\sim$ 1.7~GeV, 
with a systematic uncertainty of about 25\% at the peak, increasing to 60\% 
above 2~GeV. For the first time the $J/\phi$ was observed in this decay 
channel~\cite{re:KSKL2pi0h}. The cross section is shown in 
Fig.~\ref{fig:xsec_KSKL2pi0}.
%%%-----------------------------------
\begin{figure}
\centering
\begin{tabular}{c}
\includegraphics[width=6truecm,clip=]{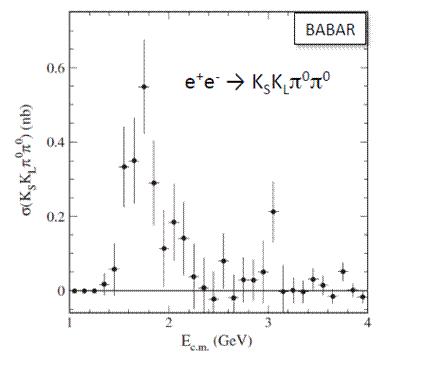}
\end{tabular}

\caption{The  $e^+e^-\rightarrow K^0_SK^0_L \pi^0 \pi^0$ cross section 
	measured by BaBar.} \label{fig:xsec_KSKL2pi0}
\end{figure}
%%%-----------------------------------

While the dominant contribution to the intermediate states is played, as 
usual, by $K^\ast(892)$, there is no significant contribution to this 
channel of the correlated production of two $K^\ast$'s.

%%%-----------------------------------
\item \textbf{Measurement of $e^+e^-\rightarrow K^0_SK^\pm \pi^\mp \pi^0$ 
	Cross Section}

The $K^0_SK^\pm \pi^\mp \pi^0$ channel features a rich intermediate state
composition~\cite{re:KSK2pih}. Apart from the clean $J/\psi$ signal, 
whose strength exceeds the maximum of the cross section (about 2~nb, measured 
with a systematic uncertainty of $\sim$ 7\% below 3~GeV), as shown in 
Fig.~\ref{fig:xsec_KSK2pi}(a), several intermediate states can be observed 
from the inspection of the scatter plots reported in 
Fig.~\ref{fig:xsec_KSK2pi}(b).
%%%-----------------------------------
\begin{figure}
\centering
\begin{tabular}{cc}
\includegraphics[height=5.5truecm,clip=]{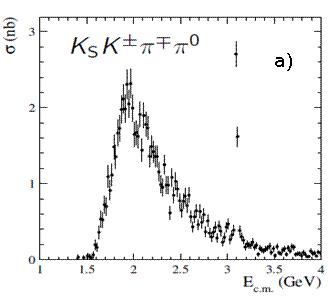} &
\includegraphics[width=11truecm, scale=1.7,clip=true, trim= 10 10 0 0]{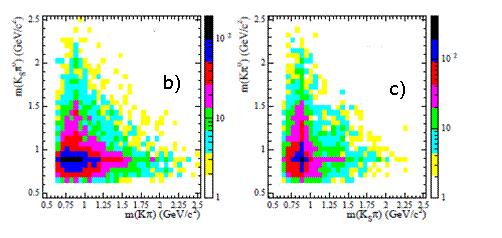}
\end{tabular}

\caption{a) The  $e^+e^-\rightarrow K^0_SK^\pm \pi^\mp \pi^0$
	cross section measured by BaBar. b)-c) Scatter plots of the
	$e^+e^-\rightarrow K^0_SK^\pm \pi^\mp \pi^0$ reaction:
	b) invariant masses of the $(K^0_S\pi^0)$ versus
	the $(K\pi)$ systems, c) invariant masses of the
	$(K\pi^0)$ versus the $(K^0_S\pi)$ systems.}
	\label{fig:xsec_KSK2pi}
\end{figure}
%%%-----------------------------------

The dominant ones are $K^\ast(892)\overline K\pi$ and $K^0_SK^\pm\rho^\mp$.
This last channel is partly fed by the decay of the axial excitations
$K_1(1270),\; K_1(1400)$ and $K_1(1650)$.  The correlated $K^\ast\overline 
K$ production is relatively small (less than 15\%), and almost saturated 
by the $K^{\ast +}K^{\ast -}$ charged mode.
\end{enumerate}

%%%-----------------------------------
\item \textbf{Summary and Conclusions}

The total cross sections for $e^+e^-$ annihilation in $K\overline K\pi$ and
 $K\overline K 2\pi$, thanks to the new BaBar measurements described above,
can now be evaluated basically without any model assumptions, nor resorting to
isospin symmetry relationships~\cite{re:druzininh}.  In the comparison of all 
the exclusive hadronic cross section measured by BaBar, the dominant 
contribution is played by full pionic channels: in particular between 1 and 
2~GeV the four pion production dominates, while above 2~GeV the six pion 
production gives the largest contribution.

Adding up all channels with kaons and a total of three particles in the final 
state, the cross section amounts to about 12\% of the total hadronic cross
section, at the maximum located at about 1.6--1.7~GeV.

Concerning the four particle final state, the largest contribution is played 
by the $K^+K^-\pi^+\pi^-$ and $K^0_SK^+\pi^-\pi^0$ channels;
the total cross section at 2~GeV is about 1/4 of the total hadronic 
cross section.

The precise knowledge of the hadronic cross sections, for which BaBar
contributed providing an unprecedented bulk of new results in more than 20
channels (including those with pions only and $\eta$'s not described in this 
paper), could improve the accuracy of new evaluations of the $a^{had,LO}_\mu$ 
term of the muon anomalous magnetic moment by 21\% as compared to previous 
assessments~\cite{re:davier2h}.

Moreover, the program of ISR measurements with BaBar could provide new 
interesting information about hadronization at low energies and the properties 
of the light meson spectrum.

New more precise results are awaited soon from the application of this 
technique at the new generation of charm and B-Factories.  
\end{enumerate}

%%%-----------------------------------

%------------------------------------
\newpage
\section{List of Participants of PKI2018 Workshop}

\begin{itemize}
\item Miguel Albaladejo Serrano, U. Murcia        <albaladejo{\_}at{\_}um.es>
\item Moskov Amaryan, ODU       		  <mamaryan{\_}at{\_}odu.edu>
\item Alexander Austregesilo, JLab		  <aaustreg{\_}at{\_}jlab.org>
\item Marouen Baalouch, ODU/JLab		  <baalouh{\_}at{\_}jlab.com
\item Ted Barnes, DOE-NP			  <ted.barnes{\_}at{\_}science.doe.gov>
\item Silas R. Beane, University of Washington	  <silas{\_}at{\_}uw.edu>
\item Raul Briceno, ODU			  	  <rbriceno{\_}at{\_}jlab.org>
\item Eugene Chudakov, JLab			  <gen{\_}at{\_}jlab.org>
\item Michael D\"oring, GW/JLab			  <doring{\_}at{\_}gwu.edu>
\item Josef Dudek, W\&M				  <dudek{\_}at{\_}jlab.org>
\item Robert Edwards, JLab			  <edwards{\_}at{\_}jlab.org>
\item Rolf Ent, JLab				  <ent{\_}at{\_}jlab.org>
\item Denis Epifanov, BINP, NSU                   <d.a.epifanov{\_}at{\_}inp.nsk.su>
\item Stuart Fegan, GW/JLab			  <sfegan{\_}at{\_}jlab.org>
\item Alessandra Filippi, INFN Torino		  <filippi{\_}at{\_}to.infn.it>
\item Jos\'e Goity, JLab			  <goity{\_}at{\_}jlab.org>
\item Boris Grube, Technical University Munich	  <bgrube{\_}at{\_}tum.de>
\item Zhihui Guo, Hebei Normal University	  <zhguo{\_}at{\_}hebtu.edu.cn>
\item Avetik Hayrapetyan, JLU Giessen		  <Avetik.Hayrapetyan{\_}at{\_}uni-giessen.de>
\item Charles E. Hyde, ODU			  <chyde{\_}at{\_}odu.edu>
\item Vyacheslav Ivanov, Budker Institute of Nuclear Physics <vyacheslav-lvovich-ivanov{\_}at{\_}mail.ru>
\item Tanjib Khan, W\&M				  <mkhan01{\_}at{\_}email.wm.edu>
\item Chan Kim, GW				  <kimchanwook{\_}at{\_}gwu.edu>
\item Bastian Kubis, Bonn U.                      <kubis{\_}at{\_}hiskp.uni-bonn.de>
\item Luka Leskovec, University of Arizona	  <leskovec{\_}at{\_}email.arizona.edu>
\item Wenliang Li, W\&M        			  <wli08{\_}at{\_}wm.edu>
\item Pok Man Lo, University of Wroclaw		  <pmlo{\_}at{\_}ift.uni.wroc.pl>
\item Maxim Mai, GW				  <maxim.mai{\_}at{\_}posteo.net>
\item Vincent Mathieu, JLab			  <vmathieu{\_}at{\_}jlab.org>
\item Curtis Meyer, CMU				  <cmeyer{\_}at{\_}cmu.edu>
\item Daniel G. Mohler, Helmholtz-Institut Mainz  <damohler{\_}at{\_}uni-mainz.de>
\item Colin Morningstar, CMU			  <colin-morningstar{\_}at{\_}cmu.edu>
\item Bachir Moussallam, Universite Paris-Sud	  <moussall{\_}at{\_}ipno.in2p3.fr>
\item Leonid Nemenov, JINR			  <leonid.nemenov{\_}at{\_}cern.ch>
\item Jos\'e Antonio Oller, Universidad de Murcia <oller{\_}at{\_}um.es>
\item Bilas K. Pal, University of Cincinnati	  <palbs{\_}at{\_}ucmail.uc.edu>
\item Antimo Palano, INFN and University of Bari  <antimo.palano{\_}at{\_}ba.infn.it>
\item Emilie Passemar, IU/JLab			  <epassema{\_}at{\_}indiana.edu>
\item Jos\'e R. Pelaez, Universidad Complutense de Madrid <jrpelaez{\_}at{\_}fis.ucm.es>
\item William Phelps, GW			  <wphelps{\_}at{\_}jlab.org>
\item Alessandro Pilloni, JLab			  <pillaus{\_}at{\_}jlab.org>
\item Jianwei Qiu, JLab				  <jqiu{\_}at{\_}jlab.org>
\item David Richards, JLab			  <dgr{\_}at{\_}jlab.org>
\item James L. Ritman, Forschungszentrum J\"ulich GmbH <j.ritman{\_}at{\_}fz-juelich.de>
\item Torry Roak, ODU 				  <roark{\_}at{\_}jlab.org>
\item Patrizia Rossi, JLab			  <rossi{\_}at{\_}jlab.org>
\item Jacobo Ruiz de Elvira, Bern University      <elvira{\_}at{\_}itp.unibe.ch>
\item Akaki Rusetsky, University of Bonn	  <rusetsky{\_}at{\_}hiskp.uni-bonn.de>
\item Amy M. Schertz, W\&M 			  <amschertz{\_}at{\_}email.wm.edu>
\item Rafael Silva Coutinho, Universit\"at Z\"urich <rsilvaco{\_}at{\_}cern.ch>
\item Justin R. Stevens, W\&M			  <jrstevens01{\_}at{\_}wm.edu>
\item Igor Strakovsky, GWU                        <igor{\_}at{\_}gwu.edu>
\item Adam Szczepaniak, IU/JLab		  	  <aszczepa{\_}at{\_}indiana.edu>
\item Tyler Viducic, W\&M			  <tyler.viducic{\_}at{\_}gmail.com>
\item Nilanga Wickramaarachchi, ODU 		  <nwick001{\_}at{\_}odu.edu>
\end{itemize}
%%%%%%%%%%%%%%%%%%%%%%%%%%%%%%%%%%%%%%%%%%%%%%%%%%%
\end{document}